\documentclass{usbthesis}

\usepackage{epsf}
\usepackage{amsmath,amssymb,array,calc,rotating,epsfig,psfrag}
\usepackage[nosort]{cite}

\title{QCD Resummation Techniques}

\author{Tibor K\'ucs}

\month{May} \year{2004 \\ $\mbox{}$
\\ Dissertation Director: Prof. George Sterman} 

\director{George Sterman}{Professor, C.N. Yang Institute for 
Theoretical Physics, SBU}
\chairman{Jack Smith}{Chair, C.N. Yang Institute for Theoretical 
Physics, SBU }
\fstmember{Michael Rijssenbeek}{Professor, Department of Physics,
SBU }
\outmember{Werner Vogelsang}{Department of Physics, 
Brookhaven National Laboratory}

\begin{document}
\ifx\href\undefined\else\hypersetup{linktocpage=true}\fi

\maketitle
\makeapproval

\begin{abstract}
The primary aim of high-energy QCD phenomenology is the determination
of cross sections for particle collisions. One of the fundamental 
properties of QCD, the asymptotic freedom, suggests that 
the coupling constant in this high-energy regime is small.
Provided that the  coefficients of the perturbative expansion are small 
enough, the perturbation theory should give reliable results.
However, in many quantities of interest the smallness of the 
expansion coefficients is violated due to large logarithmic enhancements.
In this case the perturbation series cannot be truncated at fixed order 
and, instead, it must be resummed. The development of such resummation 
algorithms is the main subject of the research presented in this thesis.

In the first part, we propose a resummation technique applicable 
to the Regge limit, which is defined for elastic scattering as the region
of large energies and small momentum transfer. We develop a new 
systematic procedure for this limit in perturbative QCD to arbitrary 
logarithmic order.
The formalism relies on the IR structure and the gauge symmetry of the theory.
We identify leading regions in loop momentum space responsible for the
singular structure of the amplitudes and perform power counting to determine 
the strength of these divergences. 
Using a factorization procedure introduced by Sen, we derive a sum of 
convolutions in transverse momentum space over soft and jet functions, 
which approximate the amplitude up to power-suppressed corrections. 
A set of evolution equations generalizing the BFKL equation and 
controlling the high energy behavior of the amplitudes to arbitrary 
logarithmic accuracy is derived. The general method is illustrated in the 
case of leading and next-to-leading logarithmic gluon reggeization
and BFKL equation.
We confirm the standard results at LL accuracy. 
At NLL order, we find an agreement with the reggeization conjecture 
up to two loops. 
However, starting at three loop order, we identify contributions violating the 
Regge ansatz. In addition, we calculate the evolution kernel determining the 
high-energy behavior of the non-reggeized term in the scattering amplitude.

In the second part, we focus our attention to another intriguing problem 
of high-energy QCD, the resummation associated with soft radiation in 
dijet events which is complicated by the presence of non-global logarithms.
We introduce a set of correlations between energy flow
and event shapes that are sensitive to the flow
of color at short distances in jet events.  These correlations are
formulated for a general set of
event shapes, which includes jet broadening and thrust as special cases.
We illustrate the method for $e^+e^-$ dijet events, and calculate the
correlation at leading logarithm in the energy flow and
at next-to-leading-logarithm in the event shape.

\end{abstract}

\tableofcontents

\begin{acknowledgements}
It is my pleasure to thank all the people who directly or indirectly influenced
the course of my research and without whom this work would have never been 
completed.

I wish to thank my adviser George Sterman for his patient guidance and  
invaluable help during my graduate career. 
I would also like to thank my collaborators Carola F. Berger and 
Maarten Boonekamp for working with me on many challenging problems. 
I am grateful to Zurab Kakushadze, Peter van Nieuwenhuizen, 
Martin Rocek, Robert Shrock, Jack Smith and George Sterman 
for wonderful lectures on quantum field theory, particle physics and string 
theory that I was lucky to attend.
I am very thankful to Michael Rijssenbeek and Werner Vogelsang for 
making it possible to spend very fruitful research periods at CERN 
and BNL, respectively. 
My first steps in particle physics started in Prague, where
Jiri Formanek, Jiri Horejsi and Jiri Chyla had very positive influence 
on me. 
I have also benefited from many physics and non-physics related conversations
with my friends Lilia Anguelova, Nathan Clisby, Olindo Corradini, 
Alberto Iglesias, Peter Langfelder, Valer Zetocha and Kostas Zoubos 
during the years spent at Stony Brook. 

Last but definitely not least, I am indebted to my parents and to my 
wife Simona for their love and overall support during my educational period.

\end{acknowledgements}

\pagenumbering{arabic}



\newcommand \be {\begin{equation}}
\newcommand \ee {\end{equation}}
\newcommand \bay {\begin{array}}
\newcommand \eay {\end{array}}
\newcommand \bea{\begin{eqnarray}}
\newcommand \eea{\end{eqnarray}}

\def \kp {|k_{\perp}|}
\def \lp {|l_{\perp}|}
\def \qkp {|q_{\perp}-k_{\perp}|}
\def \qlp {|q_{\perp}-l_{\perp}|}
\def \qklp {|q_{\perp}-k_{\perp}-l_{\perp}|}


\def \ba  {\begin{eqnarray}}
\def \ea  {\end{eqnarray}}
\def \baa {\begin{eqnarray*}}
\def \eaa {\end{eqnarray*}}
\def \bb  {\begin{thebibliography}}
\def \eb  {\end{thebibliography}}
\def \cusp {{\rm cusp}}
\newcommand \ci [1] {\cite{#1}}
\newcommand \bi [1] {\bibitem{#1}}
\def \lab #1 {\label{#1}}
\newcommand\re[1]{(\ref{#1})}
\def \qqquad {\qquad\quad}
\def \qqqquad {\qquad\qquad}
\newcommand\lr[1]{{\left({#1}\right)}}
\def \Tr {\mbox{Tr\,}}
\def \tr {\mbox{tr\,}}
\newcommand \vev [1] {\langle{#1}\rangle}
\newcommand \VEV [1] {\left\langle{#1}\right\rangle}
\newcommand \ket [1] {|{#1}\rangle}
\newcommand \bra [1] {\langle {#1}|}
\def \e {\mbox{e}}
\def \CO {{\cal O}}
\def \CP {{\cal P}}
\def \CT {{\cal T}}
\def \CF {{\cal F}}
\def \W {\Sigma}
\def \PT {{\rm PT}}
\def \pint {\int\hspace{-1.19em}\not\hspace{0.6em}}

\def \fracs #1#2 {\mbox{\small $\frac{#1}{#2}$}}
\newcommand \partder [1] {{\partial \over\partial #1}}
\def \bin #1#2 {{\left({#1}\atop{#2}\right)}}
\def \as {\relax\ifmmode\alpha_s\else{$\alpha_s${ }}\fi}
\def \alpi {\frac \as \pi}
\def \al #1 {\frac {\as({#1})}{\pi} }
\def \ds #1 {\ooalign{$\hfil/\hfil$\crcr$#1$}}
\def \MS {\overline{\rm MS}}
\def \QCD {\mbox{{\tiny QCD}}}
\def \GeV {\mbox{GeV}}

\newcommand \eh {\hat{\eta}}
\newcommand \eo {{\eta}_0}
\newcommand \ej {{\eta}_{JJ}}
\newcommand \eu {{\eta}_1}
\newcommand \bit {\begin{itemize}}
\newcommand \eit {\end{itemize}}
\newcommand \emi {{\eta}_{min}}
\newcommand \ema {{\eta}_{max}}
\newcommand \fmi {{\phi}_{min}}
\newcommand \fma {{\phi}_{max}}

\newcommand \ok {{\omega}_k}
\newcommand \ol {{\omega}_l}
\newcommand \sfi{\sin \phi}
\newcommand \cfi{\cos \phi}
\newcommand \qb {{\bar Q}}
\newcommand \cd {\cos \delta}
\newcommand \sd {\sin \delta}
\newcommand \ep {\epsilon}
\newcommand \de {\Delta \eta}
\newcommand \xic {{\hat \xi}_c}
\newcommand \Ncol {{\mathcal{N}}_C}

\def \tomega {W}
\def \d {{\rm d}}
\def \bi {\bibitem}
\def \CO {{\cal O}}
\def \O {\Omega}
\def \o {\omega}

\def\hepph  #1 {{\tt hep-ph/#1}}

\def \mc {\mathcal}

\chapter{Introduction}

The birth of modern theory of strong interactions dates back to the early
sixties when Gell-Mann and Zweig, \cite{gellmann} proposed the quark model 
to describe the properties of hadrons. This model suggests that all strongly 
interacting objects are composed of more elementary constituents named quarks.
The first experimental support of this idea came in the late sixties at 
SLAC in a Deeply Inelastic Scattering. After this, the top priority became 
the hunt for a theory describing the dynamics of quarks. 
The distinctive feature of quarks is that they carry internal 
quantum number called color. Each quark can be in three color states.
In the early seventies, Fritzsch, Gell-Mann and Leutwyler, 
Ref. \cite{fritzsch}, proposed a gauge theory describing the interaction 
between these quarks, called Quantum Chromodynamcis (QCD). 
The mediator of this interaction, analogous to a 
photon in QED, is a gluon. At the classical level, QCD is invariant 
under the color $SU(3)$ local gauge transformations. The quarks transform 
in the fundamental representation and the gluons in the adjoint representation 
of this group. The process of quantization introduces a gauge-fixing and 
the ghost terms into the Lagrangian. It satisfies a global residual 
gauge symmetry, called BRS symmetry, Ref. \cite{brs1}.
This symmetry proves to be very powerful in deriving identities 
between Green functions which are exact, i.e. they hold to all orders in 
perturbation theory. 

The use of QCD as a perturbation theory 
is justified due to the property of asymptotic freedom discovered by 
't Hooft, Gross, Wilczek and Politzer \cite{asymfreed}.
This property says that with increasing energies the coupling characterizing 
the interaction between the quarks and gluons decreases.

At the Ultra-Violet (UV) spectrum of internal momenta, the 4-D field 
theory is plagued by infinities. 
They appear since the current theory is only an effective theory
valid in a certain energy regime. At sufficiently high energies it must, 
presumably, be embedded into some more fundamental theory. 
Nevertheless the theory is internally consistent, since we can remove 
these infinities by the process of renormalization, \cite{uvrenorm}. 

On the opposite end of momentum spectra, we encounter another type of 
divergences due to the Infra-Red (IR) region of soft momenta and 
collinear momenta. These divergences occur only for Green functions
with external particles on-shell, when at least one of the 
particles is massless. Their origin is due to the degeneracy of states 
occurring in the soft and collinear limit, since we 
cannot possibly distinguish soft emissions and collinear splittings
from situations when these emissions and splittings are absent.
This observation suggests, that although IR divergences appear on a 
diagram by diagram basis, they cancel in properly averaged quantities. 
Namely, one needs to sum over all indistinguishable states.
Already in 1937, Bloch and Nordsieck, Ref. \cite{bloch}, showed that this is, 
indeed, the case in QED when the summation over final states is performed.

In QCD the situation is more complicated due to the self-coupling of gluons.
In this case the KLN theorem, Ref. \cite{kln}, 
which extends the summation over final state degeneracies to initial states 
as well, comes to our rescue. 
It is these quantities, which are free of IR divergences in QCD. 
From the experimental point of view this solution to the IR problem has a 
caveat, however.
We can hardly expect to prepare our initial states in collision experiments 
to satisfy the conditions of the KLN theorem. 
Instead, we take one step further and invoke factorization theorems, 
Ref. \cite{pQCD}.
It is exactly these theorems that enable us to turn perturbative QCD to 
a predictive calculational tool. 

Factorization theorems claim that it is possible 
to separate short and long distance physics in physical quantities. 
A cross section for hadronic collision can 
be written as a convolution over functions describing long distance dynamics, 
which can be either Parton Distribution Functions (PDFs) or 
Fragmentation Functions (FFs), and functions describing short distance 
physics, which are partonic cross sections. The former have a physical 
interpretation of probabilities to find partons inside hadrons in case of 
PDFs, or to find hadrons inside partons in the case of FFs. These quantities 
cannot be completely determined from perturbative QCD and need to be 
fitted to experimental data. However, perturbative QCD enables us to find 
the evolution of the density functions with energy scale, Ref. \cite{dglap}. 
These evolution equations resum large logarithmic corrections in this energy 
scale. They are the consequence of the factorized form for the cross section 
and the renormalization group equation stating that the physical quantities 
cannot depend on the scale at which we make the separation of short and long 
distance dynamics. Actually, under very general assumptions, we can claim 
that whenever there is a factorization, there is a corresponding resummation.
We will encounter concrete examples of this statement in later chapters. 
The main feature of the distribution functions is 
that they are process independent. This suggests that after fitting long 
distance functions in one process, we can use them in certain other processes 
involving the same type of hadron.
 
The second ingredient of the factorized cross section is the partonic cross 
section, which quantifies the interaction of the underlying partons in the 
hard process. This quantity is IR safe 
and calculable within perturbation theory provided the coefficients of the 
perturbative expansion are small. This is, however, not always the 
case. The coefficients are usually enhanced at kinematic edges of phase 
space. The reason for this is easily understood. The cancellation of IR 
divergences happens between virtual and real 
corrections. The integration in the virtual corrections spans all the energy 
scales: soft, hard and UV. The last one is removed by renormalization, so 
only the region between the soft and the hard scale remains. The phase space 
for real corrections depends on the kinematics considered. If we are 
completely inclusive then the integration region is the same as in the 
case of virtual corrections and there is a perfect cancellation between 
the two. However, once we impose kinematic constraints on the final state, 
then the cancellation between the two is incomplete and we can be left with 
large logarithms spoiling the perturbative expansion. 
In this case we need to resum the perturbation series. 

The development of resummation procedures for various processes is the 
main topic of this work.
It consists of two parts. In the first part we propose a systematic 
resummation technique applicable in the Regge limit, Ref. \cite{tibor1} - 
\cite{tibor2}. 
This limit concerns almost forward elastic scattering of particles. 
In the second part of this thesis we pursue resummation of soft 
radiation accompanying final state jets in lepton collisions, Ref. \cite{bks}.

\part{Regge Resummation}

\chapter{The Method} 
\label{ch1}

\section{Introduction}

The study of semi-hard processes within the framework of gauge quantum field
theories has a long history. For reviews see Refs.
\cite{lipatov}-\cite{ross}.
The defining feature of such processes is that they involve two or more hard
scales, compared to $\Lambda_{\rm QCD}$, which are strongly ordered relative
to each other.
The perturbative expansions of scattering amplitudes for these processes must
be resummed since they contain logarithmic enhancements
due to large ratios of the scales involved. One of the most important
examples is elastic $2 \rightarrow 2$
particle scattering in the Regge limit, $s \gg |t|$ (with $s$ and $t$ the
usual Mandelstam variables).
It is this process that we investigate in the next two chapters.
We extend the techniques developed in Refs. \cite{sen83} and
\cite{jaroszewicz} and devise a new systematic method for evaluation of
QCD scattering amplitudes in the Regge limit to arbitrary logarithmic
accuracy, Ref. \cite{tibor1}.

The problem of the Regge limit in quantum field theory
was first tackled in the case of fermion exchange amplitude
within QED in Ref. \cite{gellMann}.
Here it was found that the positive signature amplitude takes a reggeized
form up to the two loop level in Leading Logarithmic
(LL) approximation. In Ref. \cite{wuMcCoy} the calculations were extended to
higher loops, and the imaginary
part of the Next-to-Leading Logarithms (NLL) was also obtained.
The analysis in Refs. \cite{gellMann} and \cite{wuMcCoy} was
performed in Feynman gauge.
It was realized in Ref. \cite{mason} that a suitable choice of gauge can
simplify the class of diagrams contributing at LL.
The common feature of all this work was the use of fixed order calculations.
To verify that the pattern of low order calculations survives at higher orders,
a method to demonstrate
the Regge behavior of amplitudes to all orders is necessary. This analysis was
provided  by A. Sen in Ref. \cite{sen83},
in massive QED. Sen developed a systematic way to control the high
energy behavior of fermion and photon exchange amplitudes to arbitrary
logarithmic accuracy. The formalism relies heavily
on the IR structure and gauge invariance of QED and provides a proof of the
reggeization of a fermion at NLL to all orders in
perturbation theory.

The resummation of color singlet exchange amplitudes
in non-abelian gauge theories in LL was achieved in the pioneering work
of Ref. \cite{bfkl}, where the reggeization of a gluon in LL was also
demonstrated.
The evolution equations resumming LL in the case of three gluon exchange was
derived in Ref. \cite{kwiecinski}.
In Ref. \cite{jaroszewicz}, $n$-gluon exchange amplitudes in QCD at LL level
were studied and a set of evolution
equations governing the high energy behavior of these amplitudes was
obtained at LL. A different approach was undertaken
in Ref. \cite{bartels}.
Here $n \rightarrow m$ amplitudes were studied in
SU(2) Higgs model with spontaneous symmetry breaking.
Starting with the tree level amplitudes,  an iterative
procedure was developed, which generates a minimal set of terms
in the perturbative expansion that have to be taken into account in order to
satisfy the unitarity requirement of the theory. See also Ref. \cite{cheng}.
The extension of the BFKL formalism to NLL spanned over a decade. For a
review see Ref. \cite{salam}. The building blocks
of NLL BFKL are the emissions of two gluons or two quarks along the
ladder, Ref. \cite{emission},
one loop corrections to the emission of a gluon along the ladder, Ref.
\cite{oneLoop}, and the two loop gluon trajectory,
Refs. \cite{gluonTraj}, \cite{korchemsky}, \cite{ducaGlover} and \cite{bff}.
The particular results were put together in Ref. \cite{nllBFKL}. In Ref.
\cite{ducaFadin},
the trajectory for the fermion at NLL was evaluated by taking the Regge
limit of the explicit
two loop partonic amplitudes, Ref. \cite{tejeda}.

Besides the NLO perturbative corrections to the BFKL kernel a variety of
approaches have been developed for unitarization corrections,
Refs. \cite{smallx, balitski,largeN}, which extend
the BFKL formalism by incorporating selected higher-order corrections.
The procedure proposed in this work, Refs. \cite{tibor1} and \cite{tibor2}, 
in a way, places these approaches in an even more general context. 
In principle, it makes it possible to find the scattering
amplitudes to arbitrary logarithmic accuracy and to determine
the evolution kernels to arbitrary fixed order in the coupling constant.
The formalism contains all color structures and, of course, the
construction of the amplitude to any given level requires 
the computation of the kernels and the solution of 
the relevant equations.

The first part of the thesis is organized as follows. 
In Chapter \ref{ch1}, we develop the general algorithm. 
In Sec. \ref{kinem} we discuss the kinematics of the partonic process 
under study and the gauge used. 
In Sec. \ref{leadingRegions} we identify the leading regions in internal 
momentum space, which produce logarithmic enhancements in the perturbation 
series. 
After identifying these regions, we perform power counting to verify that 
the singularity structure
of individual diagrams is at worst logarithmic.
The leading regions lead to a factorized form for the amplitude (First
Factorized Form).
It consists of soft and jet functions, convoluted over soft loop momenta,
which can still produce logarithms of $s/|t|$.
In Sec. \ref{jet functions} we study the properties of the jet functions
appearing in the factorization formula for the amplitude.
We show how the soft gluons can be factored from the jet functions. In Sec.
\ref{evolution equations} we demonstrate how to
express systematically  the amplitude as a convolution in transverse momenta.
In this form
all the large logarithms are organized in jet functions and the soft
transverse momenta integrals do not introduce
any logarithms of $s/|t|$ (Second Factorized Form).
We derive evolution equations that enable us to control the high energy
behavior of
the scattering amplitudes. 
 
In Chapter \ref{ch2}, we illustrate the general method
valid to all logarithmic accuracy in the case of LL and NLL in the amplitude
and we examine the evolution equations at the same level. 
In Sec. \ref{llAmplitude}, we resum the amplitude at LL and we find 
the LL gluon Regge trajectory. 
In Sec. \ref{nllAmplitude}, we analyze the amplitude at NLL order. 
We confirm the LL BFKL evolution equation. 
In Sec. \ref{gluonReg}, we address the problem of NLL evolution equations 
and the gluon reggeization at this accuracy. We confirm the Regge hypothesis 
at two loop level. However, we identify contributions violating the Regge 
ansatz starting at three loop order.

Some technical details are discussed in
appendices \ref{contracted vertices} - \ref{appb}.
The first appendix treats power counting for regions of integration space where
internal loop momenta become much larger than the momentum transfer.
In Appendix \ref{variation} we illustrate the origin of special vertices
encountered due to the resummation.
In Appendix \ref{tulipGarden} we show a systematic expansion for the
amplitude leading to the first factorized form.
In Appendix \ref{feynmanRules} we list the Feynman rules used throughout
the text. In Appendix \ref{glauberRegion}
we demonstrate the origin of extra soft momenta configurations (Glauber
region) which need to be considered in the analysis of amplitudes in the 
Regge limit. 
In Appendix \ref{appa}, we study some symmetry properties of jet functions 
and finally in Appendix \ref{appb}, we give details on the derivation 
of the color octet NLL evolution equations.

\section{Kinematics and Gauge} \label{kinem}

We analyze the amplitude for the elastic scattering of massless quarks
\be
q (p_A, r_A, \lambda_A) + q'(p_B, r_B, \lambda_B) 
\rightarrow q (p_A -q, r_1, \lambda_1) + q' (p_B + q, r_2, \lambda_2),
\label{qqqq}
\ee
within the framework of perturbative QCD in the kinematic region $s \gg -t$
({\it Regge limit}),
where $s=(p_A + p_B)^2$ and $t=q^2$ are the usual Mandelstam variables.
We stress, however, that the results obtained below apply to arbitrary
elastic two-to-two partonic process.
We pick process (\ref{qqqq}) for concreteness only.
The arguments in Eq. (\ref{qqqq}) label the momenta, $p_i$, 
the colors, $r_i$, and polarizations, $\lambda_i$, for $i = A,B,1,2$, 
respectively, of the initial and final state quarks. 
We choose to work in the center-of-mass (c.m.) where the momenta of the
incoming quarks and the momentum transfer have the following components
\footnote{We use light-cone coordinates,
$v=(v^+,v^-,v_{\perp})$, $v^{\pm}=(v^0 \pm v^3)/\sqrt{2}$.}
\bea \label{momentaDefinition}
p_A & = & \left(\sqrt{\frac{s}{2}}, 0^-, 0_{\perp}\right), \nonumber \\
p_B & = & \left(0^+,\sqrt{\frac{s}{2}}, 0_{\perp}\right), \nonumber \\
q   &=& (0^+, 0^-, q_{\perp}).
\eea
Strictly speaking $q^{\pm}= \pm |t| / \sqrt{2s}$, so the $q^{\pm}$
components vanish in the Regge limit only.

In the color basis
\bea \label{qqBasis}
b_{\bf 1} & = & \delta_{r_A,\,r_1} \delta_{r_B,\,r_2}, \nonumber  \\
b_{\bf 8} & = & - \frac{1}{2 N_c} \delta_{r_A,\,r_1} \delta_{r_B,\,r_2} +
\frac{1}{2} \delta_{r_A,\,r_2} \delta_{r_B,\,r_1},
\eea
with $N_c$ the number of colors, we can view the amplitude for process
(\ref{qqqq}) as a two dimensional vector in color space
\be
A = \left( \bay{c} A_{\bf 1} \\ A_{\bf 8} \eay \right),
\ee
where $A_{\bf 1}$ and $A_{\bf 8}$ are defined by the expansion
\be
A_{\, r_A \, r_B, \; r_1 \, r_2} = A_{\bf 1} \, (b_{\bf 1})_{r_A \, r_B, \;
r_1 \, r_2} + A_{\bf 8} \,
(b_{\bf 8})_{r_A \, r_B, \; r_1 \, r_2}.
\ee
Since the amplitude is dimensionless and all
particles are massless, its components can depend, in general,
on the following variables
\be
A_i \equiv A_i \left (\frac{s}{{\mu}^2},
\frac{t}{{\mu}^2},{\alpha}_s({\mu}^2),\epsilon \right) \hspace{2cm}
\mbox{for} \; i={\bf 1}, {\bf 8} ,
\ee
where $\mu$ is a scale introduced by regularization.
We use dimensional regularization in order to regulate both
infrared (IR) and ultraviolet (UV) divergences with $D = 4 - 2\varepsilon$ the
number of dimensions. Choosing the scale ${\mu}^2 = s$, the
strong coupling,
${\alpha}_s(\mu)$, is small. However, in general, an individual Feynman
diagram contributing to the process (\ref{qqqq}) at $r$-loop order can give
a contribution as singular as
$(s/t)\,{\alpha}_s^{r+1}{\ln}^{2r}(-s/t)$. In Sec.
\ref{doubleLogCancellation} we will confirm that there is a cancellation of
all terms proportional to the $i$-th logarithmic power for $i=r+1,\ldots ,
2r$ at order ${\alpha}_s^{r+1}$ in the perturbative expansion of the
amplitude. Hence at $r$ loops the amplitude is enhanced by a factor
$(s/t)\,{\alpha}_s^{r+1}\ln^r(-s/t)$, at most.
In order to get reliable results in perturbation theory we must,
nevertheless, resum these large contributions. In the $k$-th non-leading
logarithmic approximation one needs to resum all the terms proportional to $
(s/t)\,{\alpha}_s^{r+1}\ln^{r-j}(-s/t), \;\; j=0,\ldots , k $ at $r$-loop
level.

We perform our analysis in the Coulomb gauge, where the propagator of a
gluon with momentum $k$ has the form
\be \label{propagator}
-i \, {\delta}_{ab} \; \frac{N_{\alpha \beta}(k,{\bar k})}{k^2 + i\epsilon}
\equiv -i \, {\delta}_{ab} \; \frac{1}{k^2 + i\epsilon}
\left(g_{\alpha \beta} - \frac{k_{\alpha} \, {\bar k}_{\beta} + {\bar 
k}_{\alpha} \, k_{\beta}
  -k_{\alpha} \, k_{\beta}}{k \cdot {\bar k}}\right),
\ee
in terms of the vector
\be \label{barVector}
\bar k = k - (k \cdot {\eta}) \, \eta,
\ee
with
\be \label{eta}
\eta = \left(\frac{1}{\sqrt{2}},\frac{1}{\sqrt{2}},0_{\perp}\right),
\ee
an auxiliary four-vector defined in the partonic c.m. frame.
The numerator of the gluon propagator satisfies the following identities
\bea \label{gluonProperties}
k^{\alpha} \, N_{\alpha \beta}(k,{\bar k}) & = & k^2 \, \frac{k_{\beta} -
{\bar k}_{\beta}}{k \cdot {\bar k}}, \nonumber \\
{\bar k}^{\alpha} \, N_{\alpha \beta}(k,{\bar k}) & = & 0.
\eea
The first equality in Eq. (\ref{gluonProperties}) is the statement that the
nonphysical degrees of freedom do not propagate in
this gauge. For use below, we list the components of the gluon propagator:
\bea \label{propagComp}
N^{+ -} (k)  =  N^{- +}(k) & = & \frac{k^+ k^- - k_{\perp}^2}{k \cdot \bar
k}, \nonumber \\
N^{+ +} (k)  =  N^{- -}(k) & = & \frac{k^+ k^-}{k \cdot \bar k}, \nonumber
\\
N^{\pm \; i} (k) = N^{i \; \pm}(k) & = & \pm \frac{(k^- - k^+) k^i}{2 k
\cdot \bar k}, \nonumber \\
N^{i \; j} (k) = N^{j \; i}(k) & = & g^{i j} - \frac{k^i k^j}{k \cdot \bar
k}.
\eea
We note that these are symmetric functions under the transformation $k^{\pm}
\rightarrow - k^{\pm}$, except for the components
$N^{\pm \; i} = N^{i \; \pm}$, which are antisymmetric under this
transformation.
It was demonstrated in Ref. \cite{zwanziger} that QCD is renormalizable in
Coulomb gauge, by considering a class
of gauges which interpolates between the covariant (Landau) and the physical
(Coulomb) gauge.

\section{Leading Regions, Power Counting} \label{leadingRegions}

In order to resum the Regge logarithms, we need to identify the regions of
integration in the loop momentum space that give rise to singularities
in the limit $t/s\rightarrow 0$. We follow the
method developed in Refs. \cite{sterman78, powerCounting},
which begins with the identification of the relevant
regions in momentum space.

\subsection{Singular contributions and reduced diagrams}

The singular contributions of a Feynman integral come from the points in
loop momentum space where the integrand becomes singular due to the
vanishing of propagator denominators. However, in order to give a true
singularity the integration variables must be trapped at such a singular
point. Otherwise we can deform the integration contour away from the
dangerous region. These singular points are called pinch singular points.
They can be identified with the following regions of integration in momentum
space,
\begin{enumerate}
\item soft momenta, with scaling behavior $k^{\mu} \sim \sigma \, \sqrt{s}$
for all components \\ ($\sigma \ll 1$),
\item momenta collinear to the momenta of the external particles, with
scaling behavior \\ $k^+ \sim \sqrt{s}, \; k^- \sim \lambda \, \sqrt{s}, \;
|k_{\perp}| \sim {\lambda}^{1/2} \, \sqrt{s}$ for the particles moving in
the $+$ direction and \\ $k^+ \sim \lambda \, \sqrt{s}, \; k^- \sim
\sqrt{s}, \; |k_{\perp}| \sim {\lambda}^{1/2} \, \sqrt{s}$ for the particles
moving in the $-$ direction,
\item so-called Glauber or Coulomb momenta, Ref. \cite{glauber}, with scaling
behavior $k^{\pm} \sim {\sigma}^{\pm} \, \sqrt{s}$, \, $|k_{\perp}| \sim
\sigma \, \sqrt{s}$, where $\lambda \lesssim {\sigma}^{\pm} \lesssim
\sigma$, and where the scaling factors $\lambda, \sigma$ satisfy the strong
ordering $ \lambda \ll \sigma \ll 1$ (The origin of this region is
illustrated in Appendix \ref{glauberRegion}.),
\item hard momenta, having the scaling behavior $k^{\mu} \sim \sqrt{s}$ for
all components.
\end{enumerate}
The extra gauge denominators $1/(k \cdot {\bar k})$ originating from the
numerators of the gluon propagator, Eq. (\ref{propagator}),
do not alter the classification of the pinch singular points mentioned
above. Actually, only the subsets 1 and 3 in the above classification
can be produced due to the extra gauge denominators.

\begin{figure} \center
\scalebox{0.85}[0.85]{\includegraphics*{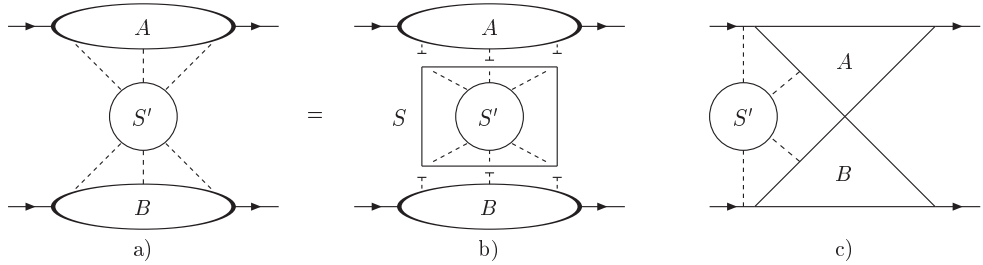}}
\caption{\label{reduced} The reduced diagrams a) and c) contributing to the
amplitude. Diagram b) represents a decomposition
of diagram a) for the purpose of power counting.}
\end{figure}

With every pinch singular point, we may associate a reduced diagram, which
is obtained from the original diagram by contracting all hard lines (subset
4) at the particular singular point. As shown in Refs. \cite{sterman78, powerCounting, cono} 
the reduced diagram corresponding to a given pinch singular
point must describe a real physical process, with each vertex of the reduced
diagram representing a real space-time point. This physical interpretation
suggests two types of reduced diagrams contributing to the process
(\ref{qqqq}), shown in Fig. \ref{reduced}.

The jet $A$($B$) contains lines whose momenta represent motion in the $ + \;
(-) $ direction. The lines included in the blob $S'$
and the lines coming out of it are all soft (configurations 1 and 3 in the
classification of loop momenta described above). These two oppositely moving
(virtual) jets may interact through the exchange of soft lines, Fig.
\ref{reduced}a, and/or they can meet at one or more
space-time points, Fig. \ref{reduced}c.

Having found the most general reduced diagrams giving the leading behavior
of the amplitude for process (\ref{qqqq}) in the Regge limit, we can
estimate the strength of the IR divergence of the integral near a given
pinch singular point. First we restrict ourselves to
cases involving  subsets 1 and 2 from the classification of loop momenta above.
To do so, we count powers in the scaling variables $\lambda$
and $\sigma$.

The scaling behavior of these loop momenta implies that every soft loop
momentum contributes a factor ${\sigma}^4$, every jet loop momentum gives
rise to the power ${\lambda}^2$, every internal soft boson (fermion) line
provides a contribution ${\sigma}^{-2}$ (${\sigma}^{-1}$) and every internal
jet line (fermionic or bosonic) scales as ${\lambda}^{-1}$. In addition,
there can be suppression factors arising from the numerators of the
propagators associated with internal lines and from internal vertices.
As pointed out in Ref. \cite{sterman78}, in physical gauges each three-point
vertex connecting three jet lines is associated with a numerator factor that
vanishes  at least linearly in the components of the transverse jet momenta,
and therefore provides a suppression ${\lambda}^{1/2}$.

We are now ready to estimate the power of divergence corresponding to the
reduced diagrams describing our process. First we restrict ourselves to the
case shown in Fig. \ref{reduced}a.
As indicated schematically in Fig. \ref{reduced}b, we can perform the power
counting for the jets and for the soft part separately.
All soft propagators and all soft loop momenta are included in the soft
subdiagram $S$.
The superficial degree of IR divergence of the reduced diagram $R$ from Fig.
\ref{reduced}a and Fig. \ref{reduced}b
can then be written as
\be \label{omegaR}
\omega(R) = \omega(A) + \omega(B) + \omega(S),
\ee
where the external lines and loops of $S'$ are included in $S$.
For $\omega(R) > 0$ the overall integral is finite, while $\omega(R) \leq 0$
corresponds to an IR divergent integral.  
When $\omega(R) = 0$, the integral diverges logarithmically.
Here we set $\lambda\sim\sigma$ for power counting purposes.  
We come back to the effect of relaxing this condition 
in connection with a discussion of item 3, Glauber regions, 
in our list of singular momentum configurations.

\subsection {Power counting} \label{elementary vertices}

In this subsection, we consider the case when
all vertices in a diagram are elementary only, that is, without contracted
sub-diagrams carrying large loop momenta.
In Appendix \ref{contracted vertices} we show that our conclusions are
unchanged by contracted vertices.

We perform the power counting for the soft part $S$ first. Let $f, b$ be the
number of fermion, boson lines external to $S^{'}$
and let $E=f+b$. The superficial degree of divergence for $S$, found by
summing powers of $\sigma$, can be written
\be \label{os1}
\omega(S) = 4(E-2) - 2b - f + 2 + \omega(S^{'}),
\ee
where the first term is due to loop integrations linking $S'$ to the jets,
while the second and the third terms originate from
propagators associated with the bosonic and fermionic lines, respectively,
connecting the jets $A$, $B$ and the soft part $S'$. The term $+2$ is
introduced because we are resumming only leading power corrections
proportional to $s/t$ and therefore we exclude the overall factor $s/t$ from
the power counting.
Since the lines entering $S'$ are soft, we obtain the superficial degree of
divergence for $S'$ simply from dimensional analysis.
It is given by
\be \label{osp}
\omega(S^{'}) = 4 - b - 3f/2.
\ee
Combining Eqs. (\ref{os1}) and (\ref{osp}), the superficial degree of infrared divergence 
for the soft part $S$ is then
\be \label{os}
\omega(S) = b + 3f/2 - 2.
\ee

Before carrying out the jet power counting, we introduce some notation.
Let $E_A$ be the number of soft lines attached to jet $A$; $I$ is the total
number of jet internal lines; $v_{\alpha}$ is the number of $\alpha$-point
vertices connecting jet lines only; $w_{\alpha}$ has a meaning similar to
$v_{\alpha}$, with the difference that every vertex counted by $w_{\alpha}$
has at least one soft line attached to it.
These are the vertices that connect the jet $A$ to the soft part $S$.
Finally, $L$ denotes the number of loops internal to jet $A$.
As noted above, we will perform the power counting for the case when 
the scaling factor for
the soft momenta, $\sigma$, is of the same order as the scaling factor for
jet $A$ momenta. When the scaling factors are different we encounter
subdivergencies, which can be analyzed the same way
as described below.
We also assume that there are no internal and external ghost lines included
in the jet function.
Later we will discuss the effect of adding ghost lines.

The superficial degree of divergence for jet $A$ can now be expressed as
\be \label{oa1}
\omega(A) = 2L - I + v_3/2.
\ee
The last term represents the suppression factor associated with  the three
point vertices.
We denote the total number of vertices internal to jet $A$ by
\be \label{v}
v = \sum_{\alpha}(v_{\alpha}+w_{\alpha}).
\ee
Next we use the Euler identity relating the number of loops, internal lines
and vertices of jet $A$
\be \label{euler}
L = I - v + 1,
\ee
and the relation between the number of lines and the number of vertices
\be \label{live}
2I + E_A + 2 = \sum_{\alpha} \alpha (v_{\alpha} + w_{\alpha}).
\ee

Using Eqs. (\ref{oa1})-(\ref{live}) we arrive at the following form for the
superficial degree of divergence for jet $A$
\be \label{oa2}
\omega(A) = 1 - (E_A  + w_3)/2  + \sum_{\alpha \ge 5}(\alpha - 4)(v_{\alpha}
+ w_{\alpha})/2.
\ee
Since every vertex counted by $w_{\alpha}$ connects at least one external
soft line, we have the condition
\be \label{ea}
E_A \ge w_3 + \sum_{{\alpha}\ge 4} w_{\alpha}.
\ee
The equality holds when there is no vertex with two or more soft lines
attached to it. Combining Eqs. (\ref{oa2})-(\ref{ea}) we arrive at the
following lower bound on the superficial degree of divergence for jet $A$:
\be \label{oa3}
\omega(A) \ge 1 - E_A +\sum_{{\alpha} \ge 4} w_{\alpha}/2 + \sum_{\alpha \ge
5}(\alpha - 4)(v_{\alpha} + w_{\alpha})/2.
\ee
The third and the last term in Eq. (\ref{oa3}) are always positive or zero
and hence
\be \label{oa4}
\omega(A) \ge 1 - E_A.
\ee
A similar result holds for jet $B$, and therefore the superficial degree of
collinear divergence for jets $A$ and $B$ is
\be \label{oa}
\omega(A) + \omega (B) \ge 2 - E,
\ee
with $E = E_A + E_B$ as in Eq. (\ref{os1}).
Combining the results for soft and jet power counting, Eqs. (\ref{os}) and
(\ref{oa}), respectively in Eq. (\ref{omegaR}),
we finally obtain the superficial degree of IR divergence for the
reduced diagram in Fig. \ref{reduced}a,
\be \label{or}
\omega(R) \ge f/2.
\ee
This condition says that we can have at worst logarithmic divergences,
provided no soft fermion lines are exchanged between the jets $A$ and $B$.
We can therefore conclude that a reduced diagram from Fig. \ref{reduced}a
containing elementary vertices can give at worst logarithmic enhancements in
perturbation theory. In order for the divergence to occur, the following set
of conditions must be satisfied:
\begin{enumerate}
\item There is an exchange of soft gluons between the jets $A$ and $B$ only,
with no soft fermion lines attached to the jets.
\item The jets $A$ and $B$ contain $3$ and $4$ point vertices only, see Eq.
(\ref{oa3}).
\item Soft gluons are connected to jets only through $3$ point vertices, Eq.
(\ref{oa3}), and at most one soft line is attached to each vertex inside the
jets, Eq. (\ref{ea}).
\item In the reasoning above we have assumed that there is no suppression
factor associated with the vertices where soft and jet lines meet. In order
for this to be true, the soft gluons must be connected to the jet $A$($B$)
lines via the $+(-)$ components of
the vertices.
\end{enumerate}

Next we consider adding ghost lines to the jet functions. As we review in
Appendix \ref{feynmanRules}, the propagator for a
ghost line with momentum $k$ is proportional to $1/(k \cdot {\bar k})$.
Hence every internal ghost line belonging to the jet gives a contribution
which is power suppressed as $1/s$. Since the numerator factors do not
compensate for this suppression, we can immediately conclude that the jet
functions cannot contain internal or external ghost lines at leading power.

So far we have not taken into account the possibility when the soft loop
momenta are pinched by the singularities of the jet lines. This situation
allows different components of soft momenta to scale differently. For
example, a minus component of soft momentum can scale as the minus component
of jet $A$ momentum $\lambda$, while the rest of the soft momentum
components may scale as $\sigma$, where $ \lambda \ll \sigma \ll 1$. The
origin of these extra pinches is illustrated in Appendix
\ref{glauberRegion}.

Let us see what happens when we attach the ends of a gluon line with this
extra pinch to jet $A$ at one end and the soft
subdiagram $S$ at the other end. The integration volume for this soft loop
momentum scales as $\lambda{\sigma}^3$. The soft gluon denominator gives a
factor ${\sigma}^{-2}$. If this soft gluon is connected to the soft part at
a $4$-point vertex, there is no new denominator in the soft part. On
the other hand, if the soft gluon is attached to the soft part via a
$3$-point vertex then the extra denominator including the numerator
suppression factors scales as ${\sigma}^{-1}$. The new jet line scales as
${\lambda}^{-1}$ as long as the condition ${\lambda}^{1/2} \gtrsim \sigma$
is obeyed; otherwise, we have the scaling ${\sigma}^{-2}$ for the extra jet
line. For ${\lambda}^{1/2} \gtrsim \sigma$ the Glauber region produces
logarithmic infrared divergence. When ${\lambda}^{1/2} \lesssim \sigma$, the
overall scaling factor $\lambda / {\sigma}^2$
indicates power suppressed contribution.

Let us now investigate another possibility, when the soft gluon connects jet
$A$ and jet $B$ directly and its momentum is pinched by the singularities of
the jet $A$ and the jet $B$ lines. Denoting the scaling factors of jet $A$
and jet $B$ as ${\lambda}_A$ and ${\lambda}_B$, respectively, the
integration volume provides the factor ${\lambda}_A
{\lambda}_B {\sigma}^2$ and the soft gluon denominator contributes the power
${\sigma}^{-2}$. The extra jet $A$ and jet $B$ denominators scale as
${\lambda}_A^{-1}$ and
${\lambda}_B^{-1}$, provided ${\lambda}_A^{1/2} \gtrsim \sigma$ and
${\lambda}_B^{1/2} \gtrsim \sigma$. For ${\lambda}_{A,B}^{1/2} \lesssim
\sigma$ both extra jet denominators provide the scaling factor
${\sigma}^{-2}$. When ${\lambda}_{A, B}^{1/2} \gtrsim \sigma$, the
power counting suggests logarithmically divergent integrals.

We have therefore verified that when the softest component of a soft line
satisfies the ordering
${\sigma}^2 \lesssim \lambda \lesssim \sigma$, the Glauber (Coulomb) momenta
produce logarithmically IR divergent integrals and need to
be taken into an account when identifying enhancements in perturbation
series.  The analysis demonstrated above for the case of one Glauber 
gluon can be
extended to the situation with arbitrary
number of Glauber gluons. This follows from dimensional
analysis, in a similar fashion as the treatment of purely soft loop momenta above.

We conclude that the reduced diagram in Fig. \ref{reduced}a is at most
logarithmically IR divergent, modulo the factor $s/|t|$. The reduced diagram
in Fig. \ref{reduced}b looses one small denominator compared to the reduced
diagram in Fig. \ref{reduced}a and since we are working in physical gauge,
this loss cannot be compensated by a large kinematical factor coming from
the numerator. Hence the reduced diagram in Fig. \ref{reduced}b is power
suppressed compared to the reduced diagram in Fig. \ref{reduced}a, and we do
not need to consider it at leading power.

Finally, let us discuss the scale of the soft momenta. In the case of soft
exchange lines, each gluon propagator supplies a factor $1/({\sigma}^2 \,
{s})$, which we want to keep at or below the order $t$ in the leading power
approximation.   Thus the size of the scale is fixed to be $\sigma \sim
\sqrt{|t|}/\sqrt{s}$. In the case of soft lines which are attached to jet
$A$ or to jet $B$ only, the scaling factor lies in the interval
$(\sqrt{|t|}/\sqrt{s}, 1)$. In the case of Glauber momenta, we again
need $\sigma \sim  \sqrt{|t|}/\sqrt{s}$. Then the condition
${\lambda}^{1/2} \gtrsim \sigma$, which is necessary for the logarithmic
enhancement, implies that the scaling factors for $+$ and $-$
components of the Glauber (Coulomb) momenta can go down to $|t|/s$, the scale
of the small components of jet momenta.
Additionally, we should note that soft and jet sub-diagrams that do 
not carry the momentum
transfer may approach the mass shell ($\lambda,\ \sigma\rightarrow 0$).
Such lines produce true infrared divergences, which we assume
are made finite by dimensional regularization to preserve the gauge
properties that we will use below.  The same power counting as above shows that
these divergences are also at worst logarithmic.

\begin{figure} \center
\includegraphics*{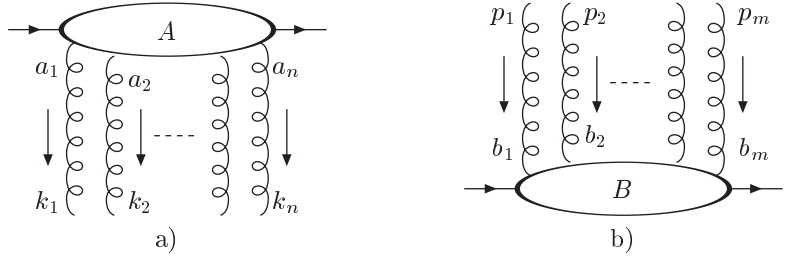}
\caption{\label{jetA} Jet $A$ moving in the $+$ direction (a) and jet $B$ moving in the $-$ direction (b).}
\end{figure}

\subsection{First factorized form} \label{first factorized form}

The analysis of the previous subsection suggests the following decomposition
of the leading reduced diagram from Fig. \ref{reduced}a. Let us denote the
$(n+2)$-point and $(m+2)$-point  Green
functions, 1PI in external soft gluon lines, corresponding to jet 
$A$, $J_{(A) \, {\mu}_1 \ldots \,
{\mu}_n}^{(n) \, a_1 \ldots \, a_n} (p_A, q, \eta; k_1, \ldots, k_n)$, Fig.
\ref{jetA}a, and to jet $B$, $J_{(B) \, {\nu}_1 \ldots \, {\nu}_m}^{(m) \,
b_1 \ldots \, b_m}(p_B, q, \eta; p_1, \ldots, p_m)$, Fig. \ref{jetA}b,
respectively. The jet function $J_{(A)}^{(n)}$ ($J_{(B)}^{(m)}$) also
depends on the color of the incoming and outgoing partons $r_A$, $r_1$
($r_B$, $r_2$), as well as on their
polarizations ${\lambda}_A$, ${\lambda}_1$ (${\lambda}_B$, ${\lambda}_2$),
respectively. In order to avoid making the notation even more cumbersome we do
not exhibit this dependence explicitly. In addition the dependence of
$J_{(A)}^{(n)}$ and $J_{(B)}^{(m)}$ on the renormalization scale $\mu$ and
the running coupling ${\alpha}_s(\mu)$ is understood. The jet functions also
depend on the following parameters: the gauge fixing vector $\eta$, Eq. (\ref{eta}), of the
Coulomb gauge, the four momenta of the external soft gluons attached to jet
$A$ ($B$), $k_1, \ldots, k_n$ ($p_1, \ldots, p_m$), and the Lorentz and
color indices of the soft gluons attached to the jet $A$ ($B$), ${\mu}_1,
\ldots, {\mu}_n$; $a_1, \ldots, a_n$ (${\nu}_1, \ldots, {\nu}_m$; $b_1,
\ldots, b_m$). The momenta of the soft gluons attached to the jets $A$ and
$B$ satisfy the constraints $\sum_{i=1}^{n} k_i = q$ and
$\sum_{j=1}^{m} p_j = q$.

According to the results of the power counting, the soft gluons couple to
jet $A$ via the minus components of their polarizations, and
to jet $B$ via the plus components of their polarizations.
Therefore, only the following components survive in the leading power
approximation
\bea \label{jetab}
J_{A}^{(n) \, a_1 \ldots \, a_n} (p_A, q, \eta, v_B; k_1, \ldots, k_n) &
\equiv & \left(\prod^n_{i=1} v_B^{{\mu}_i} \right) J^{(n) \, a_1 \ldots \,
a_n}_{(A) \, {\mu}_1 \ldots \, {\mu}_n}(p_A, q, \eta; k_1, \ldots, k_n),
\nonumber \\
J_{B}^{(m) \, b_1 \ldots \, b_m} (p_B, q, \eta, v_A; p_1, \ldots, p_m) &
\equiv & \left(\prod^m_{i=1} v_A^{{\nu}_i} \right) J^{(m) \, b_1 \ldots \,
b_m}_{(B) \, {\nu}_1 \ldots \, {\nu}_m}(p_B, q, \eta; p_1, \ldots, p_m),
\nonumber \\
\eea
where we have defined light-like momenta in the plus direction $v_A = (1, 0,
0_{\perp})$ and in the minus direction $v_B = (0, 1, 0_{\perp})$. We can now
write the contribution to the reduced diagram in Fig. \ref{reduced}a, and
hence to the amplitude for process (\ref{qqqq}), in the form
\bea \label{fact1}
A & = & \sum_{n, m} \int \left(\prod^{n-1}_{i=1} \mathrm{d}^D k_i \right) \,
\int \left(\prod^{m-1}_{j=1} \mathrm{d}^D p_j \right) \, J_A^{(n) \, a_1
\ldots \, a_n} (p_A, q, \eta, v_B; k_1, \ldots, k_n) \nonumber \\
& \times & S^{(n, m)}_{a_1 \ldots \, a_n, b_1 \ldots \, b_m} (q, \eta, v_A,
v_B; k_1, \ldots, k_n; p_1, \ldots, p_m) \nonumber \\
& \times & J_B^{(m) \, b_1 \ldots \, b_m} (p_B, q, \eta, v_A; p_1, \ldots,
p_m),
\eea
where the sum over repeated color indices is understood.
Corrections to Eq. (\ref{fact1}) are suppressed by positive powers of
$t/s$.
The jet functions $J_{A,B}$ are defined in Eq. (\ref{jetab}) in the leading
power accuracy.
The internal loop momenta of the jets $A$, $B$ and of the soft
function $S$ are integrated over. The soft function will, in general,
include delta functions setting some of the momenta $k_1, \ldots \, , k_n$
and color indices $a_1, \ldots \, , a_n$ of jet function $J_A$ to the
momenta $p_1, \ldots \, , p_m$ and to the color indices $b_1, \ldots \, ,
b_m$ of jet function $J_B$. The construction of the soft function $S$ is
described in Appendix \ref{tulipGarden}. For a given Feynman diagram there
exist many reduced diagrams of the type shown in Fig. \ref{reduced}a, and one
has to be careful in systematically expanding this diagram into the terms
that have the form of Eq. (\ref{fact1}). This systematic method can be
achieved using the ``tulip-garden'' formalism first introduced in Ref. \cite{coso} and
used in a similar context in Ref. \cite{sen83}. For convenience of the
reader we summarize this procedure in
Appendix \ref{tulipGarden}.

Let us now identify the potential sources of the enhancements
in $\ln(s/|t|)$ of the amplitude given by Eq. (\ref{fact1}). If we 
integrate over the
internal momenta of the jet functions then we can get $\ln((p_A \cdot
\eta)^2 / |t|)$ from $J_A$ and $\ln((p_B \cdot \eta)^2/|t|)$ from $J_B$. In
addition, according to the results of the power-counting, Eq. (\ref{oa4}),
we know that the jet function with $n$ external soft gluons diverges as
$1/{\lambda}^{n-1}$. After performing the integrals over the minus
components of the external soft gluon lines attached to jet $A$ and over the
plus components of the external soft gluons connected to jet $B$, these
divergent factors are
potentially converted into logarithms of $\ln((p_A \cdot \eta)^2/|t|)$
and $\ln((p_B \cdot \eta)^2/|t|)$, respectively. Our goal will be to separate
the full amplitude into a convolution over parameters that do not introduce
any further logarithms of the form $\ln(s/|t|)$. This task will be achieved
in Sec. \ref{secondFactorizedForm}. In the following section,
we analyze the characteristics of the jet functions.

\section{The Jet Functions} \label{jet functions}

In this section we study the properties of the jet functions $A$, $B$ given
by Eq. (\ref{jetab}) since, as Eq. (\ref{fact1}) suggests, they will play an
essential role in later analysis. Since the methods for both jet functions
are similar we restrict our analysis to jet $A$ only; jet $B$ can be worked
out in the same way.
In Sec. \ref{jet1} we examine the properties of jet $A$ when the minus
component of one of its external soft gluon momenta is of order
$\sqrt{|t|}$. In Sec. \ref{jet2} we find the variation of jet $A$ with
respect to the gauge fixing vector $\eta$, and finally in Sec. \ref{jet3} we
examine the dependence of jet $A$ on the plus component of a soft gluon
momentum attached to this jet.

\subsection{Decoupling of a soft gluon from a jet} \label{jet1}

According to the results of power counting above, soft gluons attach 
to lines in
jet $A$ via the minus components of their polarization. Following the
technique of Grammer and Yennie \cite{graye} we decompose the vertex at
which the $j$th gluon is connected to jet A.
We start with a trivial rewriting of $J_A$ in Eq. (\ref{jetab})
\begin{figure} \center
\scalebox{0.85}[0.85]{\includegraphics*{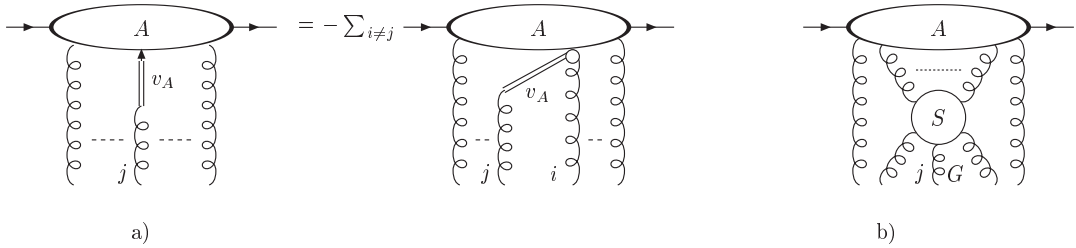}}
\caption{\label{decoupling} a) Decoupling of a $K$ gluon from jet $A$.
b) Leading contributions resulting from the attachment of a $G$ gluon to
jet $A$.}
\end{figure}
\be \label{gy1}
J_A^{(n) \, a_1 \ldots \, a_n} = \left( \prod^{n}_{i \ne j} v_B^{{\mu}_i}
\right) \,
v_B^{{\mu}_j} \, g_{{\mu}_j}^{\;\;\; {\nu}_j} \, J_{(A) \; {\mu}_1 \ldots \,
{{\nu}_j} \ldots \, {{\mu}_n}}^{(n) \, a_1 \ldots \, a_n}.
\ee
We now decompose the metric tensor into the form $g^{\mu\nu} = K^{\mu\nu}(k_j)
+ G^{\mu\nu}(k_j)$ where for a gluon with
momentum $k_j$ attached to jet $A$, $K^{\mu\nu}$ and $G^{\mu\nu}$ are 
defined by
\bea \label{kgDef}
K^{\mu\nu}(k_j) & \equiv & \frac{v_A^{\mu} \, k_j^{\nu}}{v_A \cdot k_j -
i\epsilon} \nonumber \\
G^{\mu\nu}(k_j) & \equiv & g^{\mu\nu} - K^{\mu\nu} (k_j) \label{gy2}.
\eea
The $K$ gluon carries scalar polarization. Since the jet $A$ function has no
internal tulip-garden subtractions (they are contained in the soft function
$S$), we can use the Ward identities of the theory \cite{thooft},
which are readily derived  from its underlying BRS symmetry
\cite{brs}, to decouple this gluon from the rest of the jet $A$ after we sum
over all possible insertions of the gluon. The result is
\bea \label{gy2}
J_A^{(n) \, a_1 \ldots \, a_j \ldots \, a_n} (p_A, q, v_B, \eta; k_1,
\ldots, k_i, \ldots, k_j, \ldots, k_n)
& = & - \frac{1}{v_A \cdot k_j - i\epsilon} \nonumber \\ 
& & \hspace*{-10cm} \times \sum^n_{i \neq j} \left(- ig_s
f^{c_i a_i a_j}\right) \, J_A^{(n-1) \, a_1 \ldots \, c_i \ldots \,
{\underline a}_j \, \ldots \, a_n}
(p_A, q, v_B, \eta; k_1, \ldots, k_i + k_j, \ldots, {\underline k}_j, \ldots
k_n). \nonumber \\
\eea
The notation ${\underline a}_j$ and ${\underline k}_j$ indicates that the
jet function $J_A^{(n-1)}$ does not depend on the color index $a_j$ and the
momentum $k_j$, because they have been factored out. In Eq. (\ref{gy2}),
$g_s$ is the QCD coupling constant and $f^{c_i a_i a_j}$ are the structure
constants of the $SU(3)$ algebra. The pictorial representation of this
equation is shown in Fig. \ref{decoupling}a. The arrow represents a scalar
polarization and the double line stands for the eikonal line. The Feynman
rules for the special vertices
and the eikonal lines in Fig. \ref{decoupling}a are listed in Appendix
\ref{feynmanRules}. Strictly speaking the right-hand side
of Eq. (\ref{gy2}) and Fig. \ref{decoupling}a contain contributions
involving external ghost lines.
However, from the power counting arguments of Sec. \ref{elementary vertices}
we know that when all lines inside of the jet are jet-like, the jet function
can contain neither external nor
internal ghost lines. Therefore Eq. (\ref{gy2}) is valid up to power
suppressed corrections for this momentum configuration.

The idea behind the $K$-$G$ decomposition is that the contribution of the
soft $G$ gluon attached to the jet line in the leading power is proportional
to $v_B^{\mu} G_{\mu\nu} v_A^{\nu} = 0$.
In order to avoid this suppression, the $G$ gluon must be attached to a soft
line. The general reduced diagram corresponding to the $G$ gluon attached to
jet $A$ is depicted in Fig. \ref{decoupling}b.
The lines coming out of $S$ as well as the lines included in it are soft.
The letter $G$ next to the $j$th gluon in Fig. \ref{decoupling}b reminds us
that this gluon is a $G$-gluon attaching to jet
$J_{(A) \, \mu}$ via the $G^{+\mu}(k_j)$ vertex.

The reasoning described above applies to the case when all components of
soft momenta are of the same order. In the situation of Coulomb (Glauber)
momenta, this picture is not valid anymore, since the large ratio
$k_{\perp}/k^-$ coming from the $G^{+\perp}$ component can compensate for
the suppression due to the attachment of the $G$ part to a jet $A$ line via
the transverse components of the vertex.

\subsection{Variation of a jet function with respect to a gauge fixing vector $\eta$} \label{jet2}

In this subsection we find the variation of the jet function $J_A^{(n)}$ 
with respect to a gauge fixing vector $\eta$. 
The motivation to do this can be easily understood. 
We consider the jet function with one soft gluon attached to it only, $J_A^{(1)}(p_A, q, v_B, \eta)$. 
Let us define
\be \label{xidef}
{\xi}_A \equiv p_A \cdot \eta \;\; \mbox{and} \;\; {\zeta}_B \equiv \eta \cdot v_B.
\ee
In these terms, jet function $J_A^{(1)}$ can depend on the following 
kinematical combinations: 
$J_A^{(1)}(p_A, q, v_B, \eta) = J_A^{(1)}({\xi}_A, \, p_A \cdot v_B, \, {\zeta}_B, \, t)$. Using the identity 
$p_A \cdot v_B = 2 \, {\xi}_A {\zeta}_B$ and the fact, 
that the dependence of $J_A$ on the vector $v_B$ is introduced trivially via Eq. (\ref{jetab}), we conclude that 
\be \label{jetDependence}
J_A^{(1)}(p_A, q, v_B, \eta) = {\zeta}_B \, {\bar J}_A^{(1)}({\xi}_A, \, t).
\ee
Our aim is to resum the large logarithms of $\ln (p_A^+)$ that appear in the perturbative expansion of the jet $A$ function. 
In order to do so, we shall derive an evolution equation for $p_A^+ \, \partial J^{(1)}_A / \partial p_A^+$. 
Since $p_A$ appears in combination with $\eta$ only, we can trace out the $p_A^+$ dependence of $J_A^{(1)}$ 
by tracing out its dependence on $\eta$. This can be achieved by varying the gauge fixing vector $\eta$. 
The idea goes back to Collins and Soper \cite{coso} and Sen \cite{sen81}. We will generalize the result to $J_A^{(n)}$ 
in Sec. \ref{evoleq}. 

We consider a variation that corresponds to an infinitesimal Lorentz boost in a positive $+$ direction with velocity $\delta \beta$. 
Thus, for the gauge fixing vector $\eta =(1,0,0,0)$ \footnote{For the moment we use Cartesian coordinates.}, Eq. (\ref{eta}), 
the variation is: $\delta \eta \equiv {\tilde \eta} \, \delta \beta \equiv 
(0,0,0,1) \, \delta \beta$. It leaves invariant the norm ${\eta}^2 = 1$ to order ${\cal O}(\delta \beta)$.
The precise relation between the variation of the jet $A$ function with respect to $p_A^+$ and $\delta {\eta}^{\alpha}$ is
\be \label{evolut}
p_A^+ \frac{\partial J_A^{(1)}}{\partial \, p_A^+} = - {\tilde \eta}^{\alpha}\frac{\partial J_A^{(1)}}{\partial 
\, \eta^{\alpha}} + {\zeta}_B\frac{\partial J_A^{(1)}}{\partial {\zeta}_B} = - 
{\tilde \eta}^{\alpha}\frac{\partial J_A^{(1)}}{\partial \, \eta^{\alpha}} + J_A^{(1)}. 
\ee
We have used the chain rule in the first equality and the simple relation 
${\zeta}_B \, \partial J_A^{(1)}/ \partial {\zeta}_B = J_A^{(1)}$, following from Eq. (\ref{jetDependence}), in the second one.
 
\begin{figure} \center 
\includegraphics*{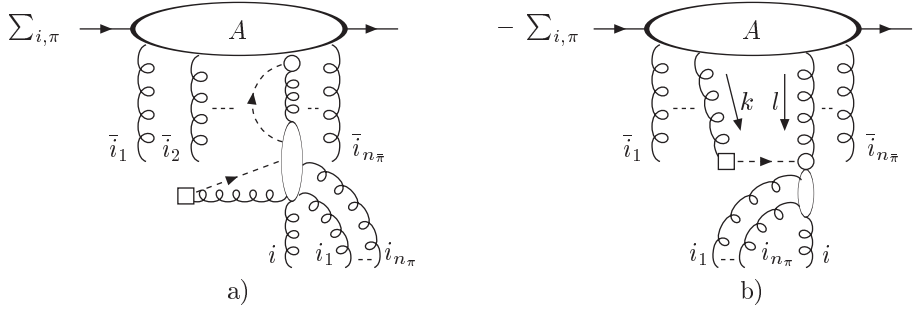}
\caption{\label{variationJ} The result of a variation of jet function $J_A^{(n)}$ with respect to a gauge fixing vector.} 
\end{figure}
 
In order for Eq. (\ref{evolut}) to be useful, we need to know what the variation of jet $A$ with respect to the gauge fixing 
vector $\eta$ is. The result of this variation for $J_A^{(n)}$ is shown in Fig. \ref{variationJ}. 
It can be derived using either the formalism of the 
effective action, Ref. \cite{nielsen}, or a diagrammatic approach first suggested in Ref. \cite{coso} and performed in axial gauge. 
We give an argument how Fig. \ref{variationJ} arises in Appendix \ref{variation}. 
Here we only note that the form of the diagrams in Fig. \ref{variationJ}
is a direct consequence of a 1PI nature of the jet functions. The explicit form of the boxed vertex 
\be \label{boxedVertex}
-i \, S^{\alpha}(k) \equiv -i \left( \eta \cdot k \, {\tilde \eta}^{\alpha} + {\tilde \eta} \cdot k \, \eta^{\alpha} \right),
\ee
as well as of the circled vertex is given in Fig. \ref{feynmanRulesF} of Appendix \ref{feynmanRules}, 
while their origin is demonstrated in Appendix \ref{variation}.
The dashed lines in Fig. \ref{variationJ} represent ghosts, and these are also given in Fig. \ref{feynmanRulesF} 
of Appendix \ref{feynmanRules}.
The four vectors $\eta$, given in Eq. (\ref{eta}), and 
\bea \label{etaTilde}
{\tilde \eta} & = & \left(\frac{1}{\sqrt{2}},- \frac{1}{\sqrt{2}},0_{\perp}\right),
\eea
appearing in Eq. (\ref{boxedVertex}) are defined in the partonic c.m. frame, Eq. (\ref{momentaDefinition}). 
We list the components of $S_{\mu} \, N^{\mu \, \alpha}(k)$
\bea \label{sCompon}
S_{\mu}(k) \, N^{\mu \, \pm}(k) & = & k^{\mp} \left( \frac{k_+^2 - k_-^2}{2 \, k \cdot \bar k} \pm 1 \right), \nonumber \\
S_{\mu}(k) \, N^{\mu \, i}(k) & = & \frac{k_-^2 - k_+^2}{2 k \cdot \bar k} \, k^i, 
\eea
for later reference.

In Fig. \ref{variationJ}, we sum over all external gluons. This is 
indicated by the sum over $i$.
In addition, we sum over all possible 
insertions of external soft gluons 
$\{i_1, \ldots, i_{n_{\pi}}\} \in 
\{1, \ldots, n\} \backslash \{i\}$. 
This summation is denoted by the 
symbol $\pi$.
We note that at lowest order, with only a gluon $i$ 
attached to the vertical blob in 
Fig. \ref{variationJ}b, this 
vertical blob denotes
the transverse tensor structure depending on the momentum $k_i$
of this gluon
\be \label{invPropag}
i\left(k_i^2 g^{\alpha\beta} - k_i^{\alpha} 
k_i^{\beta}\right).
\ee
It is labeled by a gluon line which is 
crossed by two vertical lines, Fig. \ref{feynmanRulesF}. 
The ghost 
line connecting the boxed and the circled vertices in Fig. 
\ref{variationJ}b can interact with jet $A$ 
via the exchange of an 
arbitrary number of soft gluons.
We do not show this possibility in 
Fig. \ref{variationJ}b for brevity.  

Let us now examine what the 
important integration regions for a loop with momentum $k$ in Fig. 
\ref{variationJ}b are. 
The presence of the ghost line and of the 
nonlocal boxed vertex requires that in the leading power
the loop momentum $k$ must be soft. It can be neither collinear nor 
hard. This will enable us to
factor the gluon with momentum $k$ from the rest of the jet according 
to the procedure described in
Sec. \ref{jet1}.

\subsection{Dependence of a jet function on the plus component of a 
soft gluon's momentum attached to it} \label{jet3}

\begin{figure} \center
\includegraphics*{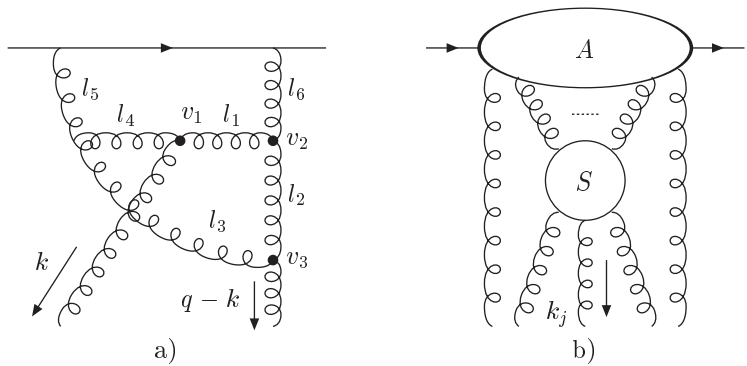}
\caption{\label{plusVar} a) Momentum flow of the external soft gluon 
inside of jet $A$.
b) Typical contribution to $k_j^+ {\partial J_A^{(n)}} / {\partial} k_j^+$.}
\end{figure}

In this subsection we want to find the leading regions of the object 
$k_j^+ {\partial J_A^{(n)}} / {\partial} k_j^+ $.
This information will be essential for the analysis pursued in the 
next sections. For a given diagram contributing to $J_A^{(n)}$ we can
always label the internal loop momenta in such a way that the 
momentum $k_j$ flows along a continuous path connecting the vertices 
where the
momentum $k_j$ enters and leaves the jet function $J_A^{(n)}$. When 
we apply the operation $k_j^+ \partial / \partial k_j^+$ on a 
particular
graph corresponding to $J_A^{(n)}$, it only acts on the lines and 
vertices which form this path.  The idea is illustrated in Fig.
\ref{plusVar}a. The gluon with momentum $k$ attaches to jet $A$ via 
the three-point vertex $v_1$. Then the momentum $k$ flows through the
path containing the vertices $v_1, v_2, v_3$ and the lines $l_1, 
l_2$. The action of the operator $k^+ \partial / \partial k^+$ on a 
line or
vertex which carries jet-like momentum  gives a negligible 
contribution, since the $+$ component of this lines momentum will be 
insensitive
to $k^+$. In order to get a non-negligible contribution, the 
corresponding line must be soft. In Fig.\ \ref{plusVar}a, lines $l_1$ 
and $l_2$
must be soft in order to get a non-suppressed contribution from the 
diagram after we apply the $k^+ \partial / \partial k^+$ operation on 
it.
This, with the fact that the external soft gluons carry soft momenta, 
also implies that the lines $l_3,
\ldots, \, l_6$ must be soft. This reasoning suggests that in general 
a typical contribution to $k_j^+ {\partial
J_A^{(n)}} / {\partial} k_j^+$ comes from the configurations shown in 
Fig. \ref{plusVar}b. It can be represented as
\bea \label{varPlus1}
J_A^{(n) \, a_1 \, \ldots \, a_n} & = & \int \left( 
\prod_{i=1}^{n'-1} {\mathrm d}^D k'_i \right) \, \,  j^{(n, n') \, 
a_1 \, \ldots
\, a_n, \, a'_1 \, \ldots \, a'_{n'}} (v_A, q, \eta; k_1, \ldots, 
k_n; k'_1, \ldots, k'_{n'}) \nonumber \\
  & \times & J_A^{(n') \, a'_1 \, \ldots \, a'_{n'}} (p_A, q, \eta, 
v_B; k'_1, \ldots, k'_{n'}).
\eea
The function $j^{(n, n')}$ contains the contributions from the soft 
part $S$ and from the gluons
connecting the jet $J_A^{(n')}$ and $S$ in Fig. \ref{plusVar}b. The 
jet function $J_A^{(n')}$ has fewer  loops
than the original jet function $J_A^{(n)}$. Now applying the 
operation $k_j^+ {\partial} / {\partial} k_j^+ $ to Eq.
(\ref{varPlus1}), the operator $k_j^+ {\partial} / {\partial} k_j^+ $ 
acts only to the function $j^{(n, n')}$. Hence we
can write
\bea \label{varPlus2}
k_j^+ \frac{\partial}{k_j^+} J_A^{(n) \, a_1 \, \ldots \, a_n} & = &
  \int \left( \prod_{i=1}^{n'-1} {\mathrm d}^D k'_i \right) 
k_j^+\frac{\partial}{\partial k_j^+} \, j^{(n, n') \, a_1 \, \ldots \,
a_n, \, a'_1 \, \ldots \, a'_{n'}} (v_A, q, \eta; \nonumber \\ 
& & k_1, \ldots, k_n; k'_1, \ldots, k'_{n'}) \, 
J_A^{(n') \, a'_1 \, \ldots \, a'_{n'}} (p_A, q, \eta, 
v_B; k'_1, \ldots, k'_{n'}). \nonumber \\
\eea
We conclude that the contribution to $k_j^+ {\partial J_A^{(n)}} / 
{\partial} k_j^+$ can be expressed in
terms of jet functions
$J_A^{(n')}$ which have fewer loops than the original jet function.

\section{Factorization and Evolution Equations} \label{evolution equations}

We are now ready to obtain evolution equations which will enable us 
to resum the large logarithms.
First, in Sec. \ref{secondFactorizedForm}, we will put Eq. 
(\ref{fact1}) into what we call the second factorized form.
Then,  in Sec. \ref{evoleq}, we derive the desired evolution 
equations. In Sec. \ref{doubleLogCancellation}, we will show the cancellation of 
the double logarithms and finally in Sec. \ref{solEvolEq}, we demonstrate that the evolution equations
derived in Sec. \ref{evoleq} are sufficient to determine the high-energy behavior of the scattering 
amplitude.

\subsection{Second factorized form} \label{secondFactorizedForm}

The goal of this subsection is to rewrite Eq. (\ref{fact1}) into the 
following form \cite{sen83}
\bea \label{fact2}
A & = & \sum_{n,m} \int \left( \prod_{i=1}^{n-1} \mathrm{d}^{D-2} 
k_{i \perp} \right)
\left( \prod_{j=1}^{m-1} \mathrm{d}^{D-2} p_{j \perp} \right) \nonumber \\ 
& \times & {\Gamma}_A^{(n) \, a_1 \ldots \, a_n}
(p_A, q, \eta, v_B; k_{1 \perp}, \ldots, k_{n \perp}; M) \nonumber \\
& \times & S^{' \, (n,m)}_{a_1 \ldots \, a_n, \, b_1 \ldots \, 
b_m}(q, \eta, v_A, v_B; k_{1 \perp}, \ldots, k_{n \perp}; p_{1 
\perp}, \ldots, p_{m \perp}; M) \nonumber \\
& \times & {\Gamma}_B^{(m) \, b_1 \ldots \, b_m}(p_B, q, \eta, v_A; 
p_{1 \perp}, \ldots, p_{m \perp}; M),
\eea
where ${\Gamma}_A^{(n)}$ and ${\Gamma}_B^{(m)}$ are defined as the 
integrals of the jet
functions $J_A^{(n)}$ and $J_B^{(m)}$, over the minus
and plus components, respectively, of their external soft momenta, 
with the remaining light-cone
components of soft momenta set to zero,
\bea \label{gammaDef}
{\Gamma}_A^{(n) \, a_1 \ldots \, a_n} (p_A, q, \eta, v_B; k_{1 
\perp}, \ldots, k_{n \perp}; M) & \equiv & \prod_{i=1}^{n-1} \left( 
\int_{-M}^{M} \mathrm{d} k^-_i \right ) \nonumber \\ 
& & \hspace{-8cm} \times \, J_A^{(n) \, a_1 \ldots \, a_n}
(p_A, q, \eta, v_B; k_{1\perp}, 
\ldots, k_{n\perp}, k_1^+ = 0, \ldots, k_n^+ = 0, 
k_1^-, \ldots, k_n^-), \nonumber \\
{\Gamma}_B^{(m) \, b_1 \ldots \, b_m} (p_B, q, \eta, v_A; p_{1 
\perp}, \ldots, p_{m \perp}; M) & \equiv & \prod_{i=1}^{m-1} \left( 
\int_{-M}^{M} \mathrm{d} p^+_i \right ) \nonumber \\ 
& & \hspace{-8cm} \times \, J_B^{(m) \, b_1 \ldots \, b_m}
(p_B, q, \eta, v_A; p_{1\perp}, 
\ldots, p_{m\perp}, p_1^- = 0, \ldots, p_m^- = 0, p_1^+, \ldots, p_m^+).
\nonumber \\
\eea
In Eq. (\ref{fact2}), $S'$ is a calculable function of its arguments 
and $M$ is an arbitrary scale of the order $\sqrt{|t|}$. The 
functions ${\Gamma}_{A,B}$ and $S'$ depend individually on this 
scale, but the final result, of course, does not. Based on the 
discussion at the end of Sec. \ref{first factorized form}, one can 
immediately recognize that all the large logarithms are now contained 
in the functions ${\Gamma}_A$ and ${\Gamma}_B$. The convolution of 
${\Gamma}_A, {\Gamma}_B$ and $S'$ is over the transverse momenta of 
the exchanged soft gluons. Since these momenta are restricted to be 
of the order $\sqrt{|t|}$, the integration over transverse momenta 
cannot introduce $\ln(s/|t|)$.
This indicates that at leading logarithm approximation the 
factorized diagram with the exchange of one gluon only contributes.
In general, when we consider a contribution to the amplitude at $ L 
= L_A + L_B+L_{S'}$ loop level, where $L_A, L_B$ and $L_{S'}$ is the 
number of loops in ${\Gamma}_A, {\Gamma}_B$ and $S'$,
respectively, we can get $L - L_{S'}$ logarithms of $s/|t|$ at most.
Hence, the investigation of the $s/t$ dependence of the full 
amplitude reduces to the study of the $p_A^+$ and $p_B^-$ dependence 
of ${\Gamma}_A$ and ${\Gamma}_B$, respectively. We formalize this 
statement at the end of Sec. \ref{doubleLogCancellation} after we 
have proved that $\Gamma_A$ ($\Gamma_B$) contains one logarithm of 
$p_A^+$ ($p_B^-$) per loop.

Let us now show how we can systematically go from Eq. (\ref{fact1}) 
to Eq. (\ref{fact2}). We follow the method
developed in
Ref. \cite{sen83}.
We start from Eq. (\ref{fact1}) and consider the $k_i^-$ integrals 
over the jet function $J_A$ for fixed $k^+_i, \, k_{i \perp}$:
\be \label{minusint}
A = \sum_n \int \prod_{i=1}^{n-1} {\mathrm d}k_i^- \; R_A^{\, a_1 
\ldots \, a_n}(k_1^-, \ldots, k_n^-; \ldots) \, J_A^{(n) \, a_1 
\ldots \, a_n}(p_A, q, \eta, v_B; k_1, \ldots, k_n),
\ee
where $R_A$ is given by the soft function $S$ and the jet function $J_B$,
\bea \label{ra}
R_A^{\, a_1 \ldots \, a_n}(k_1^-, \ldots, k_n^-; \ldots) & = & 
\sum_{m} \int \left(\prod^{m-1}_{j=1} \mathrm{d}^D p_j \right) 
\nonumber \\
& & \hspace{-3cm} \times \, 
S^{(n, m)}_{a_1 \ldots \, a_n, b_1 \ldots \, b_m} (q, \eta, v_A, v_B; 
k_1, \ldots, k_n; p_1, \ldots, p_m) \nonumber \\
& & \hspace{-3cm} \times \, J_B^{(m) \, b_1 \ldots \, b_m} 
(p_B, q, \eta, v_A; p_1, \ldots, p_m).
\eea
We next use the following identity for $R_A$: \footnote{Recall that 
$k_n = q - (k_1 + \ldots \, +k_{n-1})$, so $k_n$ is
not an independent momentum.}
\bea \label{iden}
R_A(k_1^-, \ldots, k_{n-1}^-) & = & R_A(k_1^-=0, \ldots, k_{n-1}^-=0) 
\, \prod_{i=1}^{n-1} {\theta}(M-|k_i^-|) \nonumber \\
& + &
\sum^{n-1}_{i=1} \; \left[ R_A(k_1^-, \ldots, k_i^-, k_{i+1}^-=0, 
\ldots, k_{n-1}^- =0) \right.
\nonumber \\
& - & \left.
R_A (k_1^-, \ldots, k_{i-1}^-, k_{i}^-=0, \ldots, k_{n-1}^- =0) \, 
{\theta}(M-|k_i^-|) \right] \nonumber \\
& \times & \prod_{j=i+1}^{n-1} {\theta} (M-|k_j^-|).
\eea
We have suppressed the dependence on the color indices and other 
possible arguments in $R_A$ for brevity. The scale $M$ can be 
arbitrary, but, as above, we take it to be of the order of 
$\sqrt{|t|}$. The first term on the right hand side of Eq. 
(\ref{iden}) has all $k_i^- = 0$. The rest of the terms can be 
analyzed using the $K$-$G$ decomposition discussed in Sec. 
\ref{jet1}. Consider the ($i=1$) term, say, in the square bracket of 
Eq. (\ref{iden}) inserted in Eq. (\ref{minusint}). Let us denote it 
$A_1$. In the region $|k_1^-| \ll M$ the integrand vanishes. On the 
other hand, for $|k_1^-| \sim M$ we can use the $K$-$G$ decomposition 
for the gluon with momentum
$k_1$. The contribution from the $K$ part factorizes and the integral 
over the component $k_1^-$ has the form
\bea \label{factorize}
A_1 & =  & \int \frac{{\mathrm d} k_1^-}{v_A \cdot k_1} \, 
\left[R_A^{ \, a_1 \ldots \, a_n}(k_1^-,k_2^-=0,
\ldots,k_{n-1}^-=0) \right. \nonumber \\
& - & \left. \theta (M - |k_1^-|) \, R_A^{ \, a_1 
\ldots \, a_n}(k_1^-=0, \ldots,k_{n-1}^-=0)\right] \nonumber \\
& \times & \sum_{i=2}^{n-1} \left( i g_s f^{a_1 c_i a_i} \, 
\int_{-M}^{M} \prod_{j=2}^{n-1}
{\mathrm d}k^-_j \, J_A^{(n-1) \, a_2 \ldots \, c_i \ldots \, 
a_n}(p_A, q, \eta, v_B; \right. \nonumber \\ 
& & \left. k_2, \ldots, k_1 + k_i, \ldots, k_n) \right).
\eea
Eq. (\ref{factorize}) is valid when all the lines inside the jet are 
jet-like. In that case the contributions from the ghosts are power 
suppressed.
The contribution corresponding to a $G$ gluon comes from the region 
of integration shown in
Fig. \ref{decoupling}b. It can be expressed in the form of Eq. 
(\ref{minusint}) involving some $J_A^{(n')}$
with fewer loops than in the original $J_A^{(n)}$, and an $R_A'$ with 
more loops than in the original $R_A$.
Then we can repeat the steps described above with this new integral.

Every subsequent term in the square bracket of Eq. (\ref{iden}) can 
be treated the same way as the first term. This allows us to express 
the integral in Eq. (\ref{minusint}) in terms of $k_i^-$ integrals 
over some $J_A^{(n')}$s, which have the same or fewer number of loops 
than the original $J_A^{(n)}$,
\bea \label{minusint2}
& & {\Gamma}_A^{(n') \, a'_1 \ldots \, a'_{n'}} \left( p_A, q, \eta, 
v_B; \, k_1^{' \; +}, \ldots, k_{n'}^{' \; +}; \, k'_{1 \perp},
\ldots, k'_{n' \perp}; M \right) \equiv  \nonumber \\
& & \int_{-M}^{M} \prod_{i=1}^{n'-1} {\mathrm d}k^{' \; -}_i \, 
J_A^{(n') \, a'_1 \ldots \, a'_{n'}}
\left( p_A, q, \eta, v_B; k'_1, \ldots, k'_{n'} \right).
\eea
We now want to set $k_i^{' \; +} =0$ in order to put Eq. 
(\ref{minusint}) into the
form of Eq. (\ref{fact2}). To that end,
we employ an identity for $J_A^{(n')}$ (we again suppress the dependence on the
color indices for brevity)
\bea \label{iden2}
J_A^{(n')}(p_A, q, \eta, v_B; k'_1, \ldots, k'_{n'}) & = & \nonumber \\
& & \hspace{-5cm} 
J_A^{(n')} \left(p_A, q, \eta, v_B; k_1^{' \; +}=0, \ldots, k_{n'}^{' 
\; +} = 0, k_1^{' \; -}, \ldots, k_{n'}^{' \; -},
k'_{1 \perp}, \ldots, k'_{n' \perp} \right) \nonumber \\
& & \hspace{-5cm} + \sum_{i=1}^{n'-1} \int_0^{k_i^{' \; +}} {\mathrm d}
l^+_i \frac{\partial}{\partial l^+_i} \, J_A^{(n')}\left( p_A, q, 
\eta, v_B; k'_{1 \perp}, \ldots, k'_{n' \perp}, k_1^{' \; -},
\ldots, k_{n'}^{' \; -}, \right. \nonumber \\
& & \hspace{-5cm} \left. k_1^{' \; +}, \ldots, k^{' \; +}_{i-1}, \, l_i^+, 
k_{i+1}^{' \; +}=0, \ldots, k_{n'}^{' \; +}=0 \right).
\eea
Substituting the first term of Eq. (\ref{iden2}) into Eq. 
(\ref{minusint2}), we recognize the definition for ${\Gamma}_A$, Eq. 
(\ref{gammaDef}). We have shown in Sec. \ref{jet3} that the 
contributions from the terms proportional to
$\partial J_A^{(n')} / \partial l^+_i$ in Eq. (\ref{iden2}) can be 
expressed as soft-loop integrals of some $J_A^{(n'')}$, again with 
fewer loops than in $J_A^{(n')}$. When we substitute this into Eq. 
(\ref{minusint2}) we may express the resulting contribution in terms 
of integrals which have the form of Eq. (\ref{minusint}). We can now 
repeat all the steps mentioned so far, with this new integral. By 
this iterative procedure we can transfer the $k_i^-$ integrals
in Eq. (\ref{minusint}) to $J_A^{(n)}$ and also set $k_i^+ = 0$ 
inside $J_A^{(n)}$. In a similar manner, we can analyze
the $p_j^+$ integrals in Eq. (\ref{fact1}), and express them in terms 
of ${\Gamma}_B$ defined in Eq. (\ref{gammaDef}).
This algorithm, indeed, leads from the first factorized form of the 
considered amplitude, Eq. (\ref{fact1}), to the second factorized 
form, Eq. (\ref{fact2}).
   
\subsection{Evolution equation} \label{evoleq}

We have now collected all the ingredients necessary to derive the 
evolution equations for quantities defined in Eq. (\ref{gammaDef}). 
Consider ${\Gamma}_A^{(n)}$. We aim to find an expression for 
$p_A^+ \partial {\Gamma}^{(n)}_A / \partial p_A^+$. As discussed in 
Sec. \ref{jet2} this will enable us to resum the large logarithms of 
$\ln(p_A^+)$ and eventually the logarithms of $\ln(s/|t|)$. According 
to Eq. (\ref{gammaDef}), in order to find $p_A^+ \partial 
{\Gamma}^{(n)}_A / \partial p_A^+$, we need to study $p_A^+ \partial 
J^{(n)}_A / \partial p_A^+$. Using the identities
$p_A \cdot v_B = 2 \, {\xi}_A {\zeta}_B$, $p_A \cdot k_i = 2 \, 
{\xi}_A {\xi}_i$, where ${\xi}_i \equiv k_i^- {\eta}^+$ and
${\xi}_A, {\zeta}_B$ are defined in Eq. (\ref{xidef}), we conclude that
\be
J_A^{(n)} = {\zeta}_B^n \, {\bar J}_A^{(n)}\left({\xi}_A, \, 
\{{\xi}_i\}_{i=1}^{n-1}, \, t, \,
\{q_{\perp} \cdot k_{i\perp}\}_{i=1}^{n-1}, \, \{k_{i\perp} \cdot 
k_{j\perp}\}_{i,j=1}^{n-1}\right).
\ee
 From this structure, using the chain rule, we derive the following 
relation satisfied by $J_A^{(n)}$, which generalizes Eq. 
(\ref{evolut}) to $J^{(n)}_A$ with arbitrary number of external 
gluons,
\be \label{evolJn}
p_A^+ \frac{\partial J_A^{(n)}}{\partial \, p_A^+} = - {\tilde 
\eta}^{\alpha}\frac{\partial J_A^{(n)}}{\partial \, {\eta}^{\alpha}} +
\sum_{i=1}^{n-1}k_i^- \frac{\partial J_A^{(n)}}{\partial k_i^-}
+ {\zeta}_B \frac{\partial J_A^{(n)}}{\partial {\zeta}_B}.
\ee
Now, we integrate both sides of Eq. (\ref{evolJn}) over
$\prod_{j=1}^{n-1} \left( \int_{-M}^{M} \mathrm{d} k^-_j \right )$
and set all $k_j^+=0$. Then, using the definition for
${\Gamma}_A^{(n)}$, Eq. (\ref{gammaDef}), the left hand side is 
nothing else but $p_A^+ \partial
{\Gamma}^{(n)}_A / \partial p_A^+$. The first term on the right hand side of 
Eq. (\ref{evolJn}) is simply $ - {\tilde
\eta}^{\alpha} \partial \, {\Gamma}_A^{(n)} / \partial \, 
{\eta}^{\alpha}$. Noting that ${\zeta}_B
\partial J_A^{(n)}/\partial {\zeta}_B = n \, J_A^{(n)}$, the last 
term gives simply $n \,
{\Gamma}_A^{(n)}$. For the middle term, we use integration by parts
\bea \label{byParts}
& &\prod_{j=1}^{n-1} \left( \int_{-M}^{M} \mathrm{d} k^-_j \right )
\sum_{i=1}^{n-1} \, k_i^- \frac{\partial J_A^{(n)}}{\partial k_i^-} \, = \,
\prod_{j=1}^{n-1} \left( \int_{-M}^{M} \mathrm{d} k^-_j \right )
\sum_{i=1}^{n-1} \left[ \frac{\partial}{\partial k_i^-} ( k_i^- \, 
J_A^{(n)} ) -
J_A^{(n)} \right] \, = \nonumber \\
& & \sum_{i=1}^{n-1} \int_{-M}^{M} \left(\prod_{j \neq i}^{n-1} { \mathrm d}
k_j^- \right) \, M \, \left[ \, J_A^{(n)}(k_i^- = +M, \ldots) + 
J_A^{(n)}(k_i^- = -M, \ldots) \, \right] \nonumber \\
& & - \; (n-1) \, {\Gamma}_A^{(n)}.
\eea
Combining the partial results, Eqs. (\ref{evolJn}) and 
(\ref{byParts}), we obtain the
following evolution equation
\bea \label{evolG}
p_A^+ \, \frac{\partial \, {\Gamma}_A^{(n)}}{\partial \, p_A^+} & = &
\sum_{i=1}^{n-1} \int_{-M}^{M} \left(\prod_{j \neq i}^{n-1} { \mathrm d}
k_j^- \right) \, M \, \left[ J_A^{(n)}(k_i^- = +M, \ldots) + 
J_A^{(n)}(k_i^- = -M, \ldots) \right]
  \nonumber \\
& & + \, {\Gamma}_A^{(n)} \, - \, {\tilde \eta}^{\alpha} \frac{\partial \,
{\Gamma}_A^{(n)}}{\partial \, {\eta}^{\alpha}}.
\eea
The jet function $J_A^{(n)}$ in the first term of Eq. (\ref{evolG})
is evaluated at $\{k_i^+=0\}_{i=1}^{n}$ and the $k_j^-$s are 
integrated over for $j=1,\ldots,n-1$
and $j \neq i$.
The first term in Eq. (\ref{evolG}) can be analyzed using the $K$-$G$ 
decomposition for gluon $i$
since the $k_i^-$ is evaluated at the scale $M \sim \sqrt{|t|}$. The 
outcome of the last term in Eq.
(\ref{evolG}) has been determined in Sec. \ref{jet2}, Fig. 
\ref{variationJ} \footnote{ Strictly
speaking we have analyzed
${\tilde \eta}^{\alpha} \partial \, J_A^{(n)} / \partial \, 
{\eta}^{\alpha}$, but because of the relationship between $J_A^{(n)}$ 
and ${\Gamma}_A^{(n)}$ given by Eq. (\ref{gammaDef}), once we know 
${\tilde \eta}^{\alpha} \partial \, J_A^{(n)} / \partial \, 
{\eta}^{\alpha}$ we also know
${\tilde \eta}^{\alpha} \partial \, {\Gamma}_A^{(n)} / \partial \, 
{\eta}^{\alpha}$.}.
As a result we have all the tools necessary to determine the 
asymptotic behavior of the high energy amplitude for process 
(\ref{qqqq}).
To demonstrate this, we will rewrite Eq. (\ref{evolG}) into the form 
where on the right hand side there will be a sum of terms involving
${\Gamma}_A^{(n')}$s convoluted with functions which do not depend on 
$p_A^+$. Let us proceed term by term.

Again, the $K$-$G$ decomposition applies to the first term in Eq. (\ref{evolG})
because the external momenta are fixed with
$k_i^- = \pm M$. Using the factorization of a $K$ gluon given in
Eq. (\ref{gy2}) it is clear that the contributions from the $K$ gluons cancel
for $J_A^{(n)}$s evaluated at $k_i^- = + M$ and $k_i^- = - M$. Hence 
only the $G$ gluon contribution
survives in this term. Its most general form is shown in
Fig. \ref{decoupling}b. Before writing it down let us
introduce the following notation. For a set of indices $\{1, 2, 
\ldots , n\} \backslash \{i\}$
consider all the possible subsets of this set, with $1, 2, \ldots, (n 
- 1)$ number of elements. Let
us denote a given subset by $\pi$, its complementary
subset $\bar{\pi}$, the number of elements in this
subset as $n_{\pi}$ and in its complementary as $n_{\bar{\pi}} \equiv 
(n - 1) - n_{\pi}$.  With this
notation, we can write the $i$th contribution to the first term in 
Eq. (\ref{evolG}) in the form
\bea \label{Gcontrib}
  & & J_A^{(n) \, a_1 \, \ldots \, a_n} \left( k_i^- = +M, \ldots 
\right) + J_A^{(n) \, a_1 \, \ldots \, a_n}
\left( k_i^- = -M, \ldots \right) =
\sum_{\pi} \int \prod_{j = 1}^{N - 1} \frac{{\mathrm d}^D l_j}{(2 
\pi)^D} \nonumber \\
  & & S^{\mu_1 \ldots \, \mu_N}_{a_i \, a_{i_1} \ldots \, 
a_{i_{n_\pi}} \, b_1 \ldots \, b_N}
\left( k_i^- = + M, k_{i_1}^-, \ldots, k_{i_{n_{\pi}}}^-; k_i^+ = 0, 
k_{i_1}^+ = 0, \ldots,
k_{i_{n_{\pi}}}^+ = 0; \right. \nonumber \\ 
& & \left. k_{i \perp}, k_{i_1 \, \perp}, \ldots, k_{i_{n_{\pi}} \, \perp}; 
l_1, \ldots, l_N; q, \eta \right) \nonumber \\ 
& \times & J_{A \; \mu_1 \ldots \, \mu_N}^{(n_{\bar{\pi}} + N) \,
a_{\bar{i}_1} \, \ldots \, a_{\bar{i}_{n_{\bar{\pi}}}} \, b_1 \ldots 
\, b_N} \left (k_{\bar{i}_1}^-,
\ldots, k_{\bar{i}_{n_{\bar{\pi}}}}^-; k_{\bar{i}_1}^+ = 0, \ldots, 
k_{\bar{i}_{n_{\bar{\pi}}}}^+ = 0; \right. \nonumber \\
& & \left. k_{\bar{i}_1 \, \perp}, \ldots, 
k_{\bar{i}_{n_{\bar{\pi}}} \, \perp}; 
l_1, \ldots, l_N; p_A, q, \eta \right) + (k_i^- \rightarrow - M).
\eea
In Eq. (\ref{Gcontrib}), the summation over repeated indices is understood.
We sum over all possible subsets $\pi$. In other words, we sum over 
all possible attachments of
external gluons to jet function $J_A$ and to the soft function $S$. 
The elements of  a given set
$\pi$ are denoted $i_1, i_2, \ldots, i_{n_{\pi}}$. The elements of a 
complementary set $\bar{\pi}$
are labeled
$\bar{i}_1, \bar{i}_2, \ldots, \bar{i}_{n_{\bar{\pi}}}$. The number 
of gluons connecting $S$ and
$J_A^{(n_{\bar{\pi}} + N)}$ is $N$.

Following the procedure described in Sec. \ref{secondFactorizedForm} 
with $R_A$ in Eq. (\ref{minusint}) replaced by $S$ in Eq. 
(\ref{Gcontrib}), we can express the contribution from a $G$ gluon in 
the first term of Eq. (\ref{evolG}) in a form
\bea \label{evol1}
\lefteqn{ \sum_{i=1}^{n-1} \int_{-M}^{M} \left(\prod_{j \neq i}^{n-1} 
{ \mathrm d}
k_j^- \right) \, M \, \left[ J_A^{(n) \, a_1 \, \ldots \, a_n}(k_i^- 
= +M, \ldots) + J_A^{(n) \, a_1 \, \ldots \, a_n}
(k_i^- = -M, \ldots)\right] } \nonumber \\
& = & \sum_{m} \int \prod_{j = 1}^{m} \, {\mathrm d}^{D-2} l_{j \perp} 
\, {\cal K}^{(n, m)}_{a_1 \ldots \, a_n; \, b_1 \ldots \, b_m}(k_{1 
\perp}, \ldots, k_{n \perp}, l_{1 \perp}, \ldots, l_{m \perp}; q, 
\eta; M) \nonumber \\
& \times & \Gamma^{(m) \, b_1 \ldots \, b_m}_A (p_A, q, \eta; l_{1 
\perp}, \ldots, l_{m \perp}; M).
\eea
The function ${\cal K}^{(n, m)}$ does not contain any dependence on $p_A$.
It can contain delta functions setting some of the color indices $b_i$,
as well as transverse momenta $l_{i \perp}$ of $\Gamma_A^{(m)}$ equal 
to color indices $a_i$ and transverse momenta $k_{i \perp}$
of $\Gamma_A^{(n)}$.

Next we turn our attention to the last term appearing in Eq. 
(\ref{evolG}). The contribution to this term has been depicted 
graphically
in Fig. \ref{variationJ}. Consider the term in Fig. 
\ref{variationJ}a. It can be written in a form
\bea \label{virtualVar}
& & \int_{-M}^M \left(\prod_{j = 1}^{n - 1} {\mathrm d} \, k_j^- 
\right) \, \left( \mathrm{Fig}. \,
\ref{variationJ} \mathrm{a} \right) =
\int_{-M}^M \left(\prod_{j = 1}^{n - 1} { \mathrm d} \, k_j^- \right)
\nonumber \\ 
& & \times \sum_{{\pi}} S'_{a_i \, a_{i_1} \ldots \, a_{i_{n_\pi}} \, b} 
\, (k_i^-, k_{i_1}^-, \ldots, k_{i_{n_{\pi}}}^-; k_i^+ = 0, k_{i_1}^+ = 0, 
\ldots, k_{i_{n_{\pi}}}^+ = 0; \nonumber \\
& & \hspace*{3.5cm} k_{i \, \perp}, k_{i_1 \, \perp}, \ldots, 
k_{i_{n_{\pi}} \, \perp}; l = k_i + k_{i_1} + \ldots + 
k_{i_{n_{\pi}}}; q, \eta) \nonumber \\
& & \times \; J_{A}^{(n_{\bar{\pi}} + 1) \, a_{\bar{i}_1} \, \ldots 
a_{\bar{i}_{n_{\bar{\pi}}}} \, b} (k_{\bar{i}_1}^-, \ldots, 
k_{\bar{i}_{n_{\bar{\pi}}}}^-; k_{\bar{i}_1}^+ = 0, \ldots, 
k_{\bar{i}_{n_{\bar{\pi}}}}^+ = 0; k_{\bar{i}_1 \, \perp}, \ldots, 
k_{\bar{i}_{n_{\bar{\pi}}} \, \perp}; \, \nonumber \\
& & \hspace*{3.5cm} l = k_i + k_{i_1} + \ldots + k_{i_{n_{\pi}}}; 
p_A, q, \eta).
\eea
In Eq. (\ref{virtualVar}), we have used the same notation as in Eq. 
(\ref{Gcontrib}). Momentum $l$ connects $S'$ with
$J_A^{(n_{\bar{\pi}} + 1)}$. Following the same procedure as in Sec. 
\ref{secondFactorizedForm} with $R_A$ appearing in Eq. 
(\ref{minusint}) replaced by $S'$ introduced in Eq. 
(\ref{virtualVar}), we can express this contribution in a form given 
by Eq. (\ref{evol1}) with a different kernel ${\cal K}^{(n, m)}$.

The contribution from Fig. \ref{variationJ}b can be written
\bea \label{realVar}
& & \int_{-M}^{M} \left(\prod_{j = 1}^{n - 1} {\mathrm d} \, k_j^- 
\right) \, \left( \mathrm{Fig}. \,
\ref{variationJ}\mathrm{b} \right)
\, = \, \int_{-M}^{M} \left(\prod_{j = 1}^{n - 1} { \mathrm d} \, 
k_j^- \right) \sum_{{\pi}}
\int \frac{\mathrm{d}^D k}{(2\pi)^D} \nonumber \\ 
& & \times \, S''_{a_i \, a_{i_1} \ldots \, a_{i_{n_\pi}} \, b \, c} \,
(k_i^-, k_{i_1}^-, \ldots, k_{i_{n_{\pi}}}^-; k_i^+ = 0, k_{i_1}^+ = 0, 
\ldots, k_{i_{n_{\pi}}}^+ = 0; \nonumber \\ 
& & \hspace*{3.0cm} k_{i \, \perp}, k_{i_1 \, \perp}, 
\ldots, k_{i_{n_{\pi}} \, \perp}; k, l; q, \eta) \nonumber \\
& & \times \, J_{A}^{(n_{\bar{\pi}} + 2) \, a_{\bar{i}_1} \, \ldots \, 
a_{\bar{i}_{n_{\bar{\pi}}}} \, b \, c}
\left(k_{\bar{i}_1}^-, \ldots, k_{\bar{i}_{n_{\bar{\pi}}}}^-; 
k_{\bar{i}_1}^+ = 0, \ldots,
k_{\bar{i}_{n_{\bar{\pi}}}}^+ = 0; \right. \nonumber \\ 
& & \hspace*{3.8cm} \left. k_{\bar{i}_1 \, \perp}, \ldots, 
k_{\bar{i}_{n_{\bar{\pi}}} \, \perp}; k, l; p_A, q, \eta \right). 
\nonumber \\
\eea
The flow of momenta $k$ and $l$ is exhibited in Fig. 
\ref{variationJ}b. The momentum $k$ flows through the boxed vertex and
the ghost line shown in Fig. \ref{variationJ}b which forces this 
momentum to be soft,
so that lines $k$ and $l$
are part of the function $S''$. Since the line with momentum $k$ is 
soft, then all
gluons attaching to $J_{A}^{(n_{\bar{\pi}} + 2)}$ in Eq. 
(\ref{realVar}) are soft and we can again
apply the procedure described in Sec. \ref{secondFactorizedForm} to 
bring the contribution in Fig.
\ref{variationJ}b into the form given by Eq. (\ref{evol1}) with a 
different kernel,  of course.

In summary, we have demonstrated that all the terms on the right hand 
side of Eq. (\ref{evolG}) can
be put into the form given by Eq. (\ref{evol1}). This indicates that 
Eq. (\ref{evolG}), indeed,
describes the evolution of $\Gamma_A^{(n)}$ in $\ln p_A^+$ since it 
can be written as
\bea \label{evol}
\lefteqn{ \left(p_A^+ \, \frac{\partial}{\partial \, p_A^+} - 1 
\right) \, {\Gamma}_A^{(n) \; a_1 \ldots \, a_n}(p_A, q, \eta;
k_{1 \perp}, \ldots, k_{n \perp}) = } \nonumber \\
& & \sum_{m} \int \prod_{j = 1}^{m} \, {\mathrm d}^{D-2} l_{j \perp} 
\, {\cal K}^{(n, m)}_{a_1 \ldots \, a_n; \, b_1 \ldots \, b_m}(k_{1 
\perp}, \ldots, k_{n \perp}, l_{1 \perp}, \ldots, l_{m \perp}; q, 
\eta) \nonumber \\
& & \times \, \Gamma^{(m) \; b_1 \ldots \, b_m}_A(p_A, q, \eta; l_{1 
\perp}, \ldots, l_{m \perp}).
\eea
The kernels ${\cal K}^{(n, m)}$ do not depend on $p_A^+$.  As 
indicated above, they can contain delta
functions setting  some of the color indices $b_i$, as well as 
transverse momenta $l_{i \perp}$ of
$\Gamma_A^{(m)}$ equal to color indices
$a_i$ and transverse momenta $k_{i \perp}$ of $\Gamma_A^{(n)}$. The 
systematic use of this evolution equation enables us to resum
large logarithms $\ln (p_A^+)$ at arbitrary level of logarithmic 
accuracy. Analogous equation is satisfied by $\Gamma_B$. It resums 
logarithms of $\ln(p_B^-)$.

\subsection{Counting the number of logarithms} \label{doubleLogCancellation}

Having derived the evolution equations for ${\Gamma}_A^{(n)}$, Eqs. 
(\ref{evolG}) and (\ref{evol}), it does not take too much effort to 
show that at $r$-loop order the amplitude contains at most $r$ powers 
of $\ln(s/|t|)$. We follow the method of Ref. \cite{sen83}. We have 
argued in Sec. \ref{secondFactorizedForm} that the power of 
$\ln(s/|t|)$ in the overall amplitude corresponds to the power of 
$\ln(p_A^+)$ in ${\Gamma}_A^{(n)}$. So we have to demonstrate that at 
$r$-loop order ${\Gamma}_A^{(n,r)}$, where ${\Gamma}_A^{(n,r)}$ 
represents a contribution to ${\Gamma}_A^{(n)}$ at $r$-loop level, 
does not contain more than $r$ logarithms of $\ln(p_A^+)$. We prove 
this statement by induction. First of all, the tree level 
contribution to ${\Gamma}_A^{(n,0)}$ is proportional to the expression
\be \label{treeG}
\int_{-M}^M \left( \prod_{i=1}^{n-1} {\mathrm d} k_i^- \right)
\sum_{\{i_1, \ldots, i_n\}} \, \prod_{j=1}^{n-1} \, \frac{1}{(p_A - 
\sum_{l=1}^{j} k_{i_l})^2 + i\epsilon} \,
\left( \prod_{j=n}^{1} \, t^{a_{i_j}} \right)_{r_1, r_A},
\ee
where $t^{a_{i_j}}$s are the generators of the $SU(3)$ algebra in the 
fundamental representation. The sum over $\{i_1,\ldots,i_n\}$ 
indicates that we sum over all possible insertions of the external 
soft gluons. Eq. (\ref{treeG}) is evaluated at $\{k_i^+=0\}_{i=1}^n$. 
Expanding the denominators in Eq. (\ref{treeG}) we obtain the 
expression $-2 p_A^+ (k_{i_1}^- + \ldots + k_{i_j}^-) - (k_{i_1} + 
\ldots + k_{i_j})^2_{\perp} + i\epsilon$. We see that the poles in 
$k_i^-$ planes are not pinched and therefore the $k_i^-$ integrals 
cannot produce $\ln(p_A^+)$ enhancements.

Next we assume that the statement is true at $r$-loop order, and show that
it then also holds at $(r+1)$-loop level. To this end we consider the 
evolution equation, Eq.
(\ref{evolG}), and  examine $(p_A^+ \, \partial/\partial \, p_A^+ -1) 
\, {\Gamma}_A^{(n,r+1)}$. Its
contribution is given by the first and the third term on the right 
hand side of Eq. (\ref{evolG}).
As already mentioned, the first term in  Eq. (\ref{evolG}) can be 
analyzed using $K$-$G$
decomposition. The contributions from the $K$ terms cancel each other 
while the contribution from the
$G$ gluons are given by the kind of diagram shown in Fig. 
\ref{decoupling}b. The latter, however,
can be written as a sum of soft loop integrals over
$J_A^{(n',r')}$ with $r' \le r$, since we loose at least one loop in 
the original $J_A^{(n, r+1)}$
due to the soft momentum integration. This is demonstrated in  Eq. 
(\ref{Gcontrib}). Following the
procedure described in Sec. \ref{secondFactorizedForm}, we may 
express these contributions as
transverse momentum integrals of some ${\Gamma}_A^{(n',r')}$, see Eq. 
(\ref{evol1}). These contain
at most $r'\le r$ logarithms of $\ln(p_A^+)$. The contribution from 
the third term in the evolution
equation, Eq. (\ref{evolG}), is given by the diagrams depicted in 
Fig. \ref{variationJ}. These are
again soft loop integrals of some $J_A^{(n',r')}$ with $r' \le r$, 
and they can be expressed as
transverse momentum integrals of ${\Gamma}_A^{(n',r')}$, see Eqs. 
(\ref{virtualVar}) and
(\ref{realVar}), which have, therefore, at most $r$ logarithms of 
$\ln(p_A^+)$. Since both terms on
the right hand side of Eq. (\ref{evolG}) have at most $r$ logarithms 
of $\ln(p_A^+)$, then also
$p_A^+ \partial \, {\Gamma}_A^{(n,r+1)} / \partial \, p_A^+$ has at 
most $r$ logarithms of
$\ln(p_A^+)$ at $(r+1)$-loop level. This immediately shows that 
${\Gamma}_A^{(n,r+1)}$ itself cannot
have more than $(r+1)$ logarithms of $\ln(p_A^+)$ at $(r+1)$-loop 
level.  

This result enables us to formally classify the types of diagrams 
which contribute to the amplitude at the $k$-th nonleading 
logarithm level. As has been shown in Sec. 
\ref{secondFactorizedForm}, we can write an arbitrary
contribution to the amplitude for process (\ref{qqqq}) in the Regge 
limit in the second factorized
form given by Eq. (\ref{fact2}). Consider an $r$-loop contribution 
to the amplitude and let $L_A$,
$L_B$ and $L_S$ be the number of loops contained in $\Gamma_A$, 
$\Gamma_B$ and $S$. Since $\Gamma_A$
($\Gamma_B$) can contain $L_A$ ($L_B$) number of logarithms of 
$p_A^+$ ($p_B^-$) at most, the
maximum number of logarithms, $N_{\mathrm{maxLog}}$, we can get is
\be \label{logCount}
N_{\mathrm{maxLog}} = r - L_S.
\ee
This indicates that when evaluating the amplitude at the $k$-th 
nonleading approximation, we need to consider diagrams where $1, 2, 
\ldots, (k+1)$ soft gluons are exchanged between the jet functions 
$J_A$ and $J_B$.

\subsection{Solution of the evolution equations} \label{solEvolEq}

Having obtained the evolution equations, Eqs. (\ref{evolG}) and
(\ref{evol}), we discuss how to construct their solution. 
Our starting point is Eq. (\ref{evol}). In shorthand notation it reads
\bea \label{evolr}
p_A^+ \, \frac{\partial}{\partial \, p_A^+} \, {\Gamma}_A^{(n, r)} =
\sum_{r'=0}^{r - 1} \sum_{n'} \, 
{\cal K}^{(n, n'; r - r')} \otimes \, \Gamma_A^{(n', r')},
\eea
at $r$-loop level. Indices $n$ and $n'$, besides denoting the number of
external gluons of the jet function, 
also label the transverse momenta and the color
indices of these gluons. The symbol $\otimes$ in Eq. (\ref{evolr}) denotes
convolution over the transverse momenta and the color indices.
Note that Eq. (\ref{evolr}) holds for $\Gamma_A$ with the overall factor
$p_A^+$ divided out $(\Gamma_A \equiv \Gamma_A / p_A^+)$. 
We have proved, in Sec. \ref{doubleLogCancellation}, that
$\Gamma_A^{(n,r)}$ can contain at most $r$ logarithms of $\ln(p_A^+)$ at 
$r$-loop level. Therefore the most general expansion for $\Gamma_A$ is
\be \label{gammaExpand}
\Gamma_A^{(n,r)} \equiv \sum_{j = 0}^{r} c_j^{(n, r)} \, \ln^j(p_A^+).
\ee
If we want to know $\Gamma_A^{(n,r)}$ at ${\rm N}^k$LL accuracy ($k = 0$ is
LL, $k = 1$ is NLL, etc.), we need to find all
$c_j^{(n,r)}$ such that $r - j \le k$.     
The coefficients $c_j^{(n,r)}$ in Eq. (\ref{gammaExpand}) depend on the
transverse momenta and the color indices of the 
external gluons. Using the expansion for $\Gamma_A^{(n,r)}$ and
$\Gamma_A^{(n',r')}$, Eq. (\ref{gammaExpand}), in Eq. (\ref{evolr}) 
and comparing the coefficients with the same power of $\ln(p_A^+)$, we
obtain the recursive relation satisfied by the 
coefficients $c_j^{(n,r)}$
\be \label{recurC}
j \, c_j^{(n,r)} = \sum_{r' = j-1}^{r-1} \sum_{n' = 1}^{n+r-r'} {\cal
K}^{(n,n';r-r')} \otimes c_{j-1}^{(n',r')}.
\ee
In Eq. (\ref{recurC}), we have used that, in general, $1 \le n' \le n + r
- r'$.
 
We now show that Eq. (\ref{recurC}) enables us to determine all the
relevant coefficients $c_j^{(r,n)}$ of $\Gamma_A^{(n)}$
order by order in perturbation theory at arbitrary logarithmic accuracy.
We start at LL, $k = 0$, and consider $n = 1$. At $r$-loop level we need
to find the coefficient $c_r^{(1,r)}$. It can 
be expressed in terms of lower loop coefficients using Eq. (\ref{recurC})
and setting $j = r$ and $n = 1$
\be \label{coeff1}
r \, c_r^{(1,r)} = \sum_{n' = 1}^{2} {\cal K}^{(1,n';1)} \otimes
c_{r-1}^{(n',r-1)}.
\ee
In Sec. \ref{llAmplitude} we will prove that the one loop kernel satisfies
${\cal K}^{(1,2;1)} = 0$, Eq. (\ref{k1}). 
This implies that in Eq. (\ref{coeff1}) the coefficient $c_r^{(1,r)}$ is
expressed in terms of lower loop 
coefficient $c_{r-1}^{(1,r-1)}$ and hence, we can construct the
coefficients at arbitrary loop level once 
we compute $c_0^{(1,0)}$, the coefficient corresponding to the tree level
jet function $\Gamma_A^{(1,0)}$.   

Next we construct all $\Gamma_A^{(n)}$ for $n > 1$ at LL accuracy. Let us
assume that we know all $c_{r}^{(n',r)}$ for all $r$ 
and for $n' < n$. We apply Eq. (\ref{recurC}) for $j = r$
\be \label{coeffr}
r \, c_r^{(n,r)} = \sum_{n' = 1}^{n+1} {\cal K}^{(n,n';1)} \otimes
c_{r-1}^{(n',r-1)}.
\ee
In Sec. \ref{nllAmplitude} we will show that the evolution kernel in Eq.
(\ref{coeffr}) obeys 
${\cal K}^{(n,n';1)} = \theta(n-n') \, {\tilde {\cal K}}^{(n,n';1)}$,
Eq. (\ref{ktheta}), where $\theta(n-n')$ is the step function.
This implies that the sum over $n'$ in Eq. (\ref{coeffr}) terminates at
$n' = n$. 
Isolating this term in Eq. (\ref{coeffr}), we can write
\be \label{coeff2}
r \, c_r^{(n,r)} = {\cal K}^{(n,n;1)} \otimes c_{r-1}^{(n,r-1)} + \sum_{n'
= 1}^{n-1} {\cal K}^{(n,n';1)} \otimes c_{r-1}^{(n',r-1)}.
\ee
So after we calculate the tree level coefficient $c_0^{(n,0)}$, we can
construct all the coefficients 
$c_r^{(n,r)}$ using Eq. (\ref{coeff2}) order by order in perturbation
theory, since according to the
assumption we know
$c_{r}^{(n',r)}$ for all $r$ and for $n' < n$. This proves that we can
construct the jet functions at LL, $k = 0$, 
for all $n$ to all loops.  

We now assume that we have constructed all the jet functions at the ${\rm
N}^k$LL accuracy for a
given
$k \ge 0$ and we will show that  we can determine all the jet functions at
the ${\rm N}^{k+1}$LL
level. We start with $n = 1$. Using Eq. (\ref{recurC})  with $n = 1$, $j =
r - (k+1)$, isolating the
term with $r' = r-1$ in the sum over $r'$ and using
${\cal K}^{(1,n';1)} = \delta_{1 \, n'} \, {\cal K}^{(1,1;1)}$, we arrive
at
\be \label{coeff3}
(r-k-1) \, c_{r-k-1}^{(1,r)} = {\cal K}^{(1,1;1)} \otimes
c_{r-k-2}^{(1,r-1)} + 
\sum_{r' = r-k-2}^{r-2} \sum_{n' = 1}^{1+r-r'} 
{\cal K}^{(1,n';r-r')} \otimes c_{r-k-2}^{(n',r')}.
\ee
After we evaluate the coefficient $c_{0}^{(1,k+1)}$ (impact factor), Eq.
(\ref{coeff3}) implies that
we can calculate  the coefficients $c_{r-k-1}^{(1,r)}$ order by order in
perturbation theory,
because, according to the induction assumption,  we know all the
coefficients $c_{r-k-2}^{(n',r')}$
since they are at most ${\rm N}^k$LL.  Once the coefficients of
$\Gamma_A^{(1)}$ are determined at 
${\rm N}^{k+1}$LL level, we assume that we know all the coefficients of
$\Gamma_A^{(n')}$s 
for $n' < n$. We want to show
that we can now construct all the coefficients for $\Gamma_A^{(n)}$ at
${\rm N}^{k+1}$LL accuracy.
First we need to calculate
$c_0^{(n,k+1)}$. Then we use Eq. (\ref{recurC}) to express the coefficient
$c_{r-k-1}^{(n,r)}$, isolating the terms with
$r' = r - 1$ and $n' = n$, as
\bea \label{coeff4}
(r-k-1) \, c_{r-k-1}^{(n,r)} & = & {\cal K}^{(n,n;1)} \otimes
c_{r-k-2}^{(n,r-1)} + 
\sum_{n' = 1}^{n-1} {\cal K}^{(n,n';1)} \otimes c_{r-k-2}^{(n',r-1)}
\nonumber \\
& + & \sum_{r' = r-k-2}^{r-2} \sum_{n' = 1}^{n+r-r'} {\cal
K}^{(n,n';r-r')} \otimes c_{r-k-2}^{(n',r')}.
\eea
The terms appearing in the sum over $r'$ in Eq. (\ref{coeff4}) are known
according to the assumptions 
since for them $r' - (r - k - 2) \le k$. We also know, according to the
induction assumptions, 
the contributions to the second term of Eq. (\ref{coeff4}), since they
have $n' < n$. 
Therefore, we can construct $c_{r-k-1}^{(n,r)}$
order by order in perturbation theory. This finishes our proof that we can
determine the high energy behavior of 
$\Gamma_A^{(n)}$ at arbitrary logarithmic accuracy.  Note that to any
fixed accuracy only a finite number
of fixed-order calculations of kernels and  coefficients $c_0^{(n,r)}$
must be carried out.
In a similar way we can construct a solution for $\Gamma_B^{(m)}$.

Once we know the high energy behavior for $\Gamma_A^{(n)}$ and
$\Gamma_B^{(m)}$, 
then the second factorized form, Eq. (\ref{fact2}), 
implies that we also know the high energy behavior for the overall
amplitude. 
Because a jet function $\Gamma^{(n)}$ is always
associated with at least $n-1$ soft loop momentum integrals in the amplitude, we infer from
Eq. (\ref{logCount}) 
that if we want to know
this amplitude at ${\rm N}^{K}$LL accuracy, it is sufficient to know
$\Gamma_A^{(n)}$ ($\Gamma_B^{(m)}$) at 
${\rm N}^{K+1 - n}$LL (${\rm N}^{K+1 - m}$LL) level for $n \le K + 1$ ($m
\le K + 1$). 
We note, however, that to construct these functions according to the
algorithm above,
it may be necessary to go to slightly larger, although always finite,
values of $n$ and $m$.
Let us describe how this comes about, starting with the basic
recursion relations for coefficients,  Eq. (\ref{recurC}).  

We assume that for fixed $n$ on
the left-hand side of Eq. (\ref{recurC}), the logarithmic accuracy $k$ is
bounded by
the value necessary to determine the 
overall amplitude to $K$th nonleading logarithm:
$k=r-j\le K+1-n$, which we may rewrite as $n + r - (K+1) \le j \le r$.
On the right-hand side of Eq.\ (\ref{recurC}) we encounter the coefficients of the jet functions
with $n'$ external lines,
satisfying the inequality 
$n' \le n + r - r' \le n + r - (j - 1)$. 
Combining these two inequalities, we immediately obtain 
that $n' \le K + 2$.  Then, for any given number of external gluons $n'$
on the right-hand side, 
we encounter a level of logarithmic accuracy $k'=r' - (j-1) \le n + r - n'
-(j-1) \le K + 2 - n'$. 
This reasoning indicates that, in general, we will
need all $\Gamma_A^{(n')}$ ($\Gamma_B^{(m')}$) at ${\rm N}^{K + 2 - n'}$LL
(${\rm N}^{K + 2 - m'}$LL) level 
for $n' \le K+2$ ($m' \le K+2$), when evaluating the amplitude at ${\rm
N}^{K}$LL accuracy.  
We note that for fermion exchange in QED it was shown in Ref. \cite{sen83}
that only contributions with $n'\le K+1$ are nonzero. 
As we will witness in Chapter \ref{ch2}, two-loop calculations appear 
to indicate, Ref. \cite{tibor2}, that QCD requires the full range of $n'$ 
identified above, starting at NLL.

\section{Conclusions}

We have established a systematic method that shows that
it is possible to resum the large logarithms appearing in the 
perturbation series of scattering amplitudes for $ 2 \rightarrow 2 $ 
partonic processes to arbitrary logarithmic accuracy in the Regge limit. 
Up to corrections suppressed by powers of $|t|/s$, the amplitude can
be expressed as a sum of convolutions in transverse momentum space
over soft and jet functions, Eq. (\ref{fact2}). All the large
logarithms are organized in the jet functions, Eq. (\ref{gammaDef}).
They are resummed using Eqs. (\ref{evolG}) and/or (\ref{evol}).
The evolution kernel $\cal K$ in Eq. (\ref{evol}) is a calculable
function of its arguments order by order in perturbation theory.
This is the central result of our analysis.

The derivation of the evolution equations and the procedure for
finding the kernels were given above in Coulomb gauge.  
Clearly, it will be useful and interesting to reformulate our
arguments in covariant gauges.  
In addition, the connection of our formalism to the 
effective action approach to small-$x$ and the Regge limit, 
Refs. \cite{smallx,balitski} should provide further insight.


\chapter{The Applications} \label{ch2}

In the previous chapter, we have developed the general formalism for
obtaining the high-energy behavior of the scattering amplitude for 
process (\ref{qqqq}) at arbitrary
logarithmic accuracy. In this chapter, we apply these techniques to study 
this amplitude at LL and NLL level.

\section{Amplitude at LL} \label{llAmplitude}

\begin{figure} \center
\includegraphics*{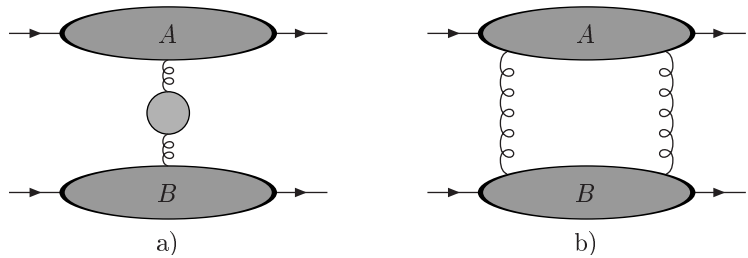}
\caption{\label{nllContribution} Diagrams contributing to the 
amplitude at NLL approximation:
factorized one gluon exchange diagram (a) and non-factorized two gluon 
exchange diagram (b).}
\end{figure}

\begin{figure} \center
\scalebox{0.85}[0.85]{\includegraphics*{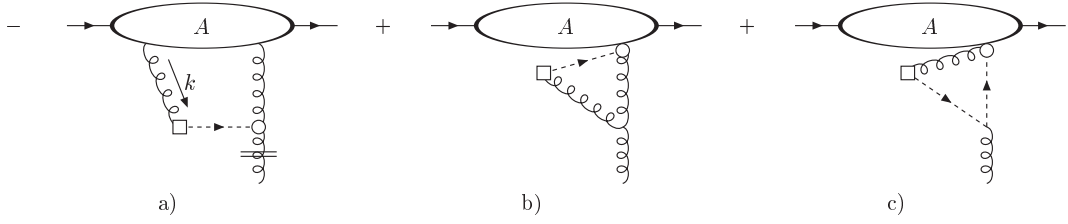}}
\caption{\label{lleq} Diagrammatic representation of the evolution 
equation for jet $J_A^{(1)}$ at LL.}
\end{figure}

According to Eq. (\ref{logCount}), the amplitude at LL comes solely 
from the factorized diagram
shown in Fig. \ref{nllContribution}a, but without any gluon 
self-energy corrections. The jet $A$,
containing lines moving in the plus direction, and jet $B$, 
consisting of lines moving in the minus
direction, interact via the exchange of a single soft gluon. This 
gluon couples to jet $A$ via the $-$
component of its polarization and to jet $B$ via the $+$ component of 
its polarization. Since
$v_A^{\alpha} \, N_{\alpha \, \beta}(q, \eta) \, v_B^{\beta} = 1$, 
we can write at LL
\be
A_{\bf 8} \, b_{\bf 8} = - \frac{1}{t} J_{A}^{(1) \, a} (p_A, q, \eta) \,
J_{B}^{(1) \, a} (p_B, q, \eta),
\ee
where $b_{\bf 8}$ is the color basis vector corresponding to the 
octet exchange, defined in Eq. (\ref{qqBasis}).
Using $s = 2p_A^+ \, p_B^-$, the logarithmic derivative of the 
amplitude can be expressed as
\be \label{ev}
\frac{\partial A_{\bf 8}}{\partial \ln s} \, b_{\bf 8} = -\frac{1}{t}
\frac{\partial J_A^{(1) \, a}}{\partial \ln p_A^+} \, J_B^{(1) \, a}
  = -\frac{1}{t} \, J_A^{(1) \, a} \, \frac{\partial J_B^{(1) \, 
a}}{\partial \ln p_B^-} \; .
\ee
In Sec. \ref{jet2}, Eq. (\ref{evolut}), we have derived an evolution 
equation resumming
$\ln(p_A^+)$ in $J_A^{(1)}$.  We note that $J_A^{(1)}=\Gamma_A^{(1)}$, and that
(\ref{evolut}) is a special case of the evolution equation (\ref{evolG}).
The diagrammatic representation of the first term on the far right 
hand side of Eq. (\ref{evolut}), which follows from
Fig. \ref{variationJ} in the case when we have one external soft 
gluon attached to a jet function,
is given by the diagrams in Fig. \ref{lleq}. Diagram in Fig. 
\ref{lleq}a corresponds to Fig. \ref{variationJ}b and the
diagrams in Figs. \ref{lleq}b and \ref{lleq}c correspond to Fig. 
\ref{variationJ}a for $n=1$.
The diagrams in Figs. \ref{lleq}b and \ref{lleq}c are in the 
factorized form, while the one in Fig. \ref{lleq}a is not.

As discussed in Sec. \ref{jet2}, power counting shows that the loop 
momentum $k$
in Fig. \ref{lleq}a must be soft. This implies that we can make the 
following approximations.
First, since at LL all internal lines of the
jet $A$ are collinear to the $+$ direction, we can neglect the $k^+$ 
dependence of $J_A^{(2)}$,
i.e. we may set $k^+ = 0$
inside $J_A^{(2)}$. Also, we can pick the plus components of the 
vertices where the soft
gluons attach to the jet $J_A^{(2)}$.
A short calculation, which uses the Feynman rules for special lines 
and vertices listed in
Appendix \ref{feynmanRules},
gives the contribution to Fig. \ref{lleq}a in a form
\bea \label{eikonalContrib1}
\mathrm{Fig.} \; \ref{lleq} \mathrm{a} & = & - {\bar g}_s \, t \, 
f_{acb} \, \int \frac{{\mathrm d}^D k}{(2 \pi)^D} \,
\frac{1}{k^2 (k-q)^2
\, k \cdot {\bar k}} \, v_A^{\rho} N_{\rho \mu}(k) S^{\mu}(k) \nonumber \\ 
& \times & v_B^{\alpha} N_{\alpha \nu} (q-k) v_A ^{\nu} \, 
v_B^{\beta} \, v_B^{\gamma} \; J_{(A) \; \beta \, 
\gamma}^{(2) \, bc}
\left(p_A, q, \eta; k^+=0, k^-, k_{\perp}\right), \nonumber \\
\eea
\begin{figure} \center
\includegraphics*{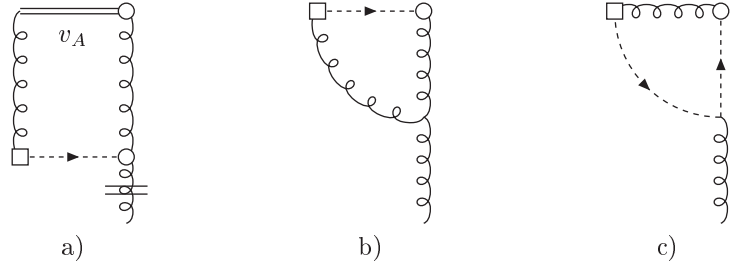}
\caption{\label{trajll} Diagrams determining the contributions to the 
gluon trajectory at the order ${\alpha}_s$.}
\end{figure}
where we have defined ${\bar g}_s \equiv g_s {\mu}^{\epsilon}$.
Using Eqs. (\ref{propagComp}) and (\ref{sCompon}) for the components 
of the gluon
propagator and the boxed vertex,
respectively, it is easy to see that in the Coulomb (Glauber) region, 
$k^- \ll k^+ \sim k_{\perp}$,
the integrand in Eq. (\ref{eikonalContrib1}) becomes an antisymmetric 
function of $k^+$ and
that therefore the integration over $k^+$ vanishes in this region.

In the soft region, where all the components of soft momenta are of 
the same size $\sqrt{-t}$, we can use the $K$-$G$ decomposition for 
the soft gluon with momentum $k$ attached to $J_A^{(2)}$. At LL, 
however, there cannot be any soft internal lines in $J_A^{(2)}$ in 
Eq. (\ref{eikonalContrib1}),
since, as discussed in Sec. \ref{doubleLogCancellation}, only 
integrals over collinear momenta can produce powers of $\ln p_A^+$.
Therefore, at LL, only the $K$ gluon contributes, because the $G$ gluon
must be attached to a soft line.
The $K$ gluon can be decoupled from the rest of the jet $J^{(2)}_A$ 
using the Ward identities, Eq. (\ref{gy2}).
Their application in Eq. (\ref{eikonalContrib1}) gives
\bea \label{eikonalContrib2}
\mathrm{Fig.} \; \ref{lleq} \mathrm{a} & = & - i {\bar g}_s^2 C_A t 
\, \int \frac{{\mathrm d}^D k}{(2 \pi)^D} \,
\frac{1}{k^2 (k-q)^2 \, k \cdot {\bar k} \, v_A \cdot k} \, 
v_A^{\rho} N_{\rho \mu}(k) S^{\mu}(k) \nonumber \\
& \times & v_B^{\alpha} N_{\alpha \nu} (q-k) v_A ^{\nu} \, 
J_{A}^{(1) \, a}(p_A, q, \eta).
\eea
We have used the identity $f_{acb} \, f_{dcb} = N_c \, {\delta}_{ad} 
\equiv C_A \, {\delta}_{ad}$ in Eq. (\ref{eikonalContrib2}).
Eq. (\ref{eikonalContrib2}) now gives a factorized form for Fig. \ref{lleq}a.
Since the contributions in Figs. \ref{lleq}b and \ref{lleq}c are 
already in the factorized form,
we can immediately infer that the gluon reggeizes
at LL. Combining the terms from Fig. \ref{lleq} in Eq. 
(\ref{evolut}), we obtain the evolution equation at leading logarithm
\be \label{ev1}
p_A^+ \frac{\partial}{\partial p_A^+} J_A^{(1) \, a}(p_A, q, \eta) = 
\alpha(t) \, J_A^{(1) \, a}(p_A, q, \eta).
\ee
Using the notation for evolution kernels introduced in Sec. \ref{solEvolEq}, 
Eq. (\ref{ev1}) implies that
\be \label{k1}
{\cal K}^{(1,2;1)} = 0.
\ee
In Eq. (\ref{ev1})
\be \label{trajectory}
\alpha (t) \equiv 1 + {\alpha}^{(1)}_{a}(t) + {\alpha}^{(1)}_{b}(t) + 
{\alpha}^{(1)}_{c}(t),
\ee
is the gluon trajectory up to the order ${\alpha}_s$, and 
${\alpha}^{(1)}_a (t)$, ${\alpha}^{(1)}_b (t)$ and ${\alpha}^{(1)}_c 
(t)$ are
its contributions given in Figs. \ref{trajll}a - \ref{trajll}c, respectively,
\bea \label{al1}
{\alpha}^{(1)}_{a}(t) & \equiv & - i {\bar g}_s^2 C_A t \, \int 
\frac{{\mathrm d}^D k}{(2 \pi)^D} \,
\frac{1}{k^2 (k-q)^2 \, k \cdot {\bar k} \, v_A \cdot k} \, 
v_A^{\rho} N_{\rho \mu}(k) S^{\mu}(k) \nonumber \\
& \times & v_B^{\beta} N_{\beta \nu} (q-k) v_A ^{\nu}, \nonumber \\
{\alpha}^{(1)}_{b}(t) & \equiv & i {\bar g}_s^2 C_A \, \int 
\frac{{\mathrm d}^D k}{(2 \pi)^D} \, \frac{1}{k^2 (k-q)^2 \,
k \cdot {\bar k}} \, S_{\alpha}(k) N^{\alpha \, \mu}(k) \; v_A^{\rho} 
N_{\rho}^{\;\; \nu} (q-k) \nonumber \\
& \times & V_{\mu \beta \nu}(k, -q, q-k) v_B^{\beta},\nonumber \\
{\alpha}^{(1)}_{c}(t) & \equiv & - i {\bar g}_s^2 C_A \, \int 
\frac{{\mathrm d}^D k}{(2 \pi)^D} \,
\frac{1}{k^2 \, k \cdot {\bar k} \, (q-k) \cdot ({\bar q}-{\bar k})} 
\, v_A^{\rho} N_{\rho \mu}(k) S^{\mu}(k) \;
(v_B \cdot {\bar k}) \, . \nonumber \\
\eea

In Eq. (\ref{al1}), $V_{\mu \beta \nu}(k, -q, q-k)$ stands for the 
momentum part
of the three-point gluon vertex.
After contracting the tensor structures in Eq. (\ref{al1}), using the explicit
form for $V_{\mu \beta \nu}$, $v_A$, $v_B$, $S^{\mu}$
(Eq. (\ref{boxedVertex})) and for the components of the gluon propagator,
Eq. (\ref{propagComp}), we obtain for ${\alpha}^{(1)}_{a,b,c}(t)$,
\bea \label{al2}
{\alpha}^{(1)}_{a} (t) & = & - i {\bar g}_s^2 C_A \frac{t}{2} \, \int 
\frac{{\mathrm d}^D k}{(2 \pi)^D}
\frac{[k_{\perp}^2 k_0 + k^2 k_3][(k-q)^2 - (k-q)^2_{\perp}]}{(k_0 + 
k_3) \, k^2 \, (k-q)^2 \,
(k \cdot {\bar k})^2 \, (k-q) \cdot ({\bar k} - {\bar q})} \; , \nonumber \\
{\alpha}^{(1)}_{b} (t) & = & i {\bar g}_s^2 C_A \frac{1}{2} \, \int 
\frac{{\mathrm d}^D k}{(2 \pi)^D}
\frac{1}{k^2 \, (k-q)^2 \, (k \cdot {\bar k})^2 \, (k-q) \cdot ({\bar 
k} - {\bar q})} \nonumber \\
& & \hspace{-2cm} \times \; [k_{\perp}^2 \, {\bar k}^2 \, 
(k-q)^2 + 2 k^2 \, k_3^2 \, (k-q)_{\perp} \cdot q_{\perp}
+ 2 k_0^2 \, k_3^2 \, k_{\perp} \cdot (k-q)_{\perp} + 2 k_0^2 \, 
k_{\perp}^2 \, (k-q)_{\perp}^2] \; , \nonumber \\
{\alpha}^{(1)}_{c} (t) & = & i {\bar g}_s^2 C_A \frac{1}{2} \, \int 
\frac{{\mathrm d}^D k}{(2 \pi)^D}
\frac{k_3^2}{(k \cdot {\bar k})^2 \, (k-q) \cdot ({\bar k} - {\bar q})} \; .
\eea
Next, we perform the $k^0$ and $k^3$ integrals in Eq. (\ref{al2}). 
For ${\alpha}^{(1)}_a(t)$, these integrals are UV/IR finite.
However in the case of ${\alpha}^{(1)}_{b,c}(t)$, the $k^0$ integral 
is linearly UV divergent.
In order to regularize this energy integral, we invoke split 
dimensional regularization introduced in Ref. \cite{leibbrandt96}.
The idea is to regularize separately the energy and the spatial 
momentum integrals, i.e. to write ${\mathrm d}^4 k_E \rightarrow
{\mathrm d}^{D_1} k_4 \, {\mathrm d}^{D_2} {\vec k}$ for Euclidean 
loop momenta $k_E$. The dimensions $D_1$ and $D_2$ are given by
$D_1 = 1 - 2 {\varepsilon}_1$ and $D_2 = 3 - 2 {\varepsilon}_2$, with 
${\varepsilon}_j \rightarrow 0+$ for $j = 1, 2$.
Since the energy integral for ${\alpha}^{(1)}_c(t)$ is scaleless, it 
vanishes in this split dimensional regularization.
The energy integrals in ${\alpha}^{(1)}_{a,b}(t)$ are straightforward.

All the $k^3$ integrals can be expressed as derivatives
with respect to $k_{\perp}^2$ and/or $(k-q)_{\perp}^2$ of a single integral
\be
I(a,b) \equiv \int_{0}^{\infty} {\mathrm d} k^3 \frac{1}{\sqrt{k^2_3 
+ a^2} \; (k^2_3 + b^2)} = \frac{1}{b\sqrt{b^2 - a^2}} \,
\ln \left(\frac{b + \sqrt{b^2 - a^2}}{a}\right).
\ee
The result of these integrations over $k^3$ is
\bea \label{al3}
{\alpha}^{(1)}_{a} (t) & = & {\alpha}_s {\mu}^{2 \epsilon} \, C_A \, 
t  \, \int \frac{{\mathrm d}^{D-2} k_{\perp}}{(2 \pi)^{D-2}} \nonumber \\
& \times & \left(I(|k_{\perp}|, |k_{\perp}-q_{\perp}|) \, 
\frac{k_{\perp}^2}{[(k-q)_{\perp}^2 - k_{\perp}^2]^2} +
\frac{2 (k - q)_{\perp}^2 - 3 k_{\perp}^2}{k_{\perp}^2 \, 
[(k-q)_{\perp}^2 - k_{\perp}^2]^2} \right), \nonumber \\
{\alpha}^{(1)}_{b} (t) & = & - {\alpha}_s {\mu}^{2 \epsilon} \, C_A 
\, t  \, \int \frac{{\mathrm d}^{D-2} k_{\perp}}{(2 \pi)^{D-2}} \nonumber \\
& \times & \left(I(|k_{\perp}|, |k_{\perp}-q_{\perp}|) \, 
\frac{k_{\perp}^2}{[(k-q)_{\perp}^2 - k_{\perp}^2]^2} -
\frac{1}{[(k-q)_{\perp}^2 - k_{\perp}^2]^2} \right), \nonumber \\
{\alpha}^{(1)}_{c} (t) & = & 0.
\eea
Combining the results of Eq. (\ref{al3}) and Eq. (\ref{trajectory}), 
we obtain the standard expression for the gluon trajectory at LL
\be \label{trajec1}
\alpha (t) = 1 + C_A {\alpha}_s {\mu}^{2 \epsilon} \, \int
\frac{{\mathrm d}^{D-2} k_{\perp}}{(2 \pi)^{D-2}} \frac{t}{k_{\perp}^2 \,
(k-q)_{\perp}^2} \; .
\ee
We can now simply solve the evolution equation (\ref{ev}), to derive 
the factorized (reggeized)
form for the amplitude in the color octet
\be
A_{\bf 8}(s,t,{\alpha}_s) = s^{\alpha (t)} \, {\tilde A}_{\bf 8} (t, 
{\alpha}_s).
\ee
The amplitude factorizes into the universal factor $s^{\alpha (t)}$, 
which is common for all processes involving two partons
in the initial and final state and dominated by the gluon exchange, 
and the part ${\tilde A}_{\bf 8}$, the so-called impact factor, which 
is specific to the process under consideration.

\section{Amplitude at NLL} \label{nllAmplitude}

At NLL level the contribution to the amplitude comes from both the 
one gluon exchange diagram,
Fig. \ref{nllContribution}a, and from the two gluon exchange diagram, 
Fig. \ref{nllContribution}b.
At this level, both singlet and octet color exchange are possible in 
the latter.
Including the self-energy corrections to the propagator of the 
exchanged gluon (taking into account
the corresponding counter-terms), we can write the contribution from 
the diagram in Fig. \ref{nllContribution}a as follows,
\be \label{a1}
A^{(1)} \equiv - \frac{1}{t} \, J_{(A) \, \alpha}^{(1) \, a} \, (p_A, 
q, \eta) \,
\left( N^{\alpha \beta}(q,\eta) + \frac{1}{t} \, v_{B}^{\alpha} \, 
v_A^{\mu} \, {\Pi}_{\mu \, \nu}(q,\eta)
\, v_B^{\nu} \, v_A^{\beta} \right) \, J_{(B) \, \beta}^{(1) \, a} \, 
(p_B, q, \eta),
\ee
where ${\Pi}_{\mu \, \nu}(q, \eta)$ stands for the one loop gluon self-energy.
We now put this contribution into the first factorized form, Eq. (\ref{fact1}), 
isolating the plus polarization for jet A, and the minus polarization for jet B.  
At NLL in the amplitude, we need 
the soft function $S^{(1,1)}$, Eq. (\ref{fact1}) with $n = m = 1$, to accuracy ${\cal O}(\alpha_s)$. 
Using the tulip-garden formalism described in Appendix 
\ref{tulipGarden}, the contribution to the first term on 
the right hand side of Eq. 
(\ref{a1}) is given by the subtractions shown in Fig. 
\ref{amplSubtract}.
In accordance with the notation adopted in Appendix 
\ref{tulipGarden}, the dashed lines indicate that we have made soft 
approximations
on gluons that are cut by them. A dashed line cutting a gluon 
attached to jet $A$($B$) means that the gluon is attached to the
corresponding jet through minus(plus) component of its polarization. 
Since $q^{\pm} = 0$ in the Regge limit,
Eq. (\ref{momentaDefinition}), we have $N^{\mu \pm}(q) = g^{\mu \pm}$.
This implies that the contributions between the diagrams in Fig. 
\ref{amplSubtract}c and in
Fig. \ref{amplSubtract}d as well as between the diagrams in Fig. 
\ref{amplSubtract}e and in Fig. \ref{amplSubtract}f cancel each other.
Therefore only the zeroth-order soft function diagram in Fig. \ref{amplSubtract}b 
survives in the factorized form, Eq.\ (\ref{fact1}).
\begin{figure} \center
\scalebox{0.85}[0.85]{\includegraphics*{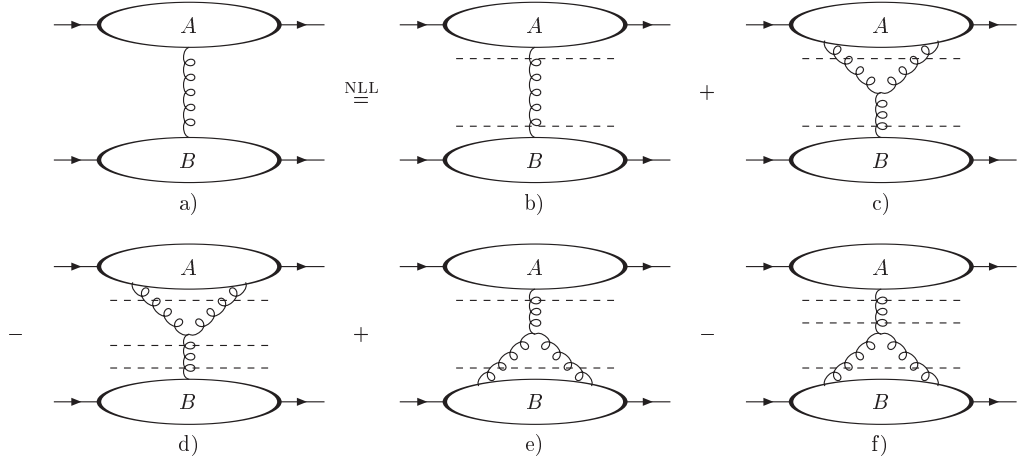}}
\caption{\label{amplSubtract} Expansion of the one gluon exchange 
amplitude at NLL using the tulip garden formalism.}
\end{figure}

For the two gluon exchange, Fig.\ \ref{nllContribution}b, 
we only need the lowest order soft function at
NLL in the amplitude (and LL in singlet exchange).
The expression for the two gluon exchange diagram in Fig. 
\ref{nllContribution}b takes the form, Eq. (\ref{fact1}),
\be \label{a2}
A^{(2)} = \int \frac{{\rm d}^{D} k}{(2 \pi)^D} \, J_A^{(2) \; a \, b} 
(k^+ = 0, k^-, k_{\perp}) \, S(k^+, k^-, k_{\perp}) \,
J_B^{(2) \; a \, b} (k^- = 0, k^+, k_{\perp}),
\ee
where $S(k)$ is given by
\be \label{sk}
S(k) \equiv \frac{i}{2!} \, \frac{N^{- \, +}(k)}{k^2+i\epsilon} \, 
\frac{N^{- \, +}(q - k)}{(q-k)^2+i\epsilon}.
\ee
We have suppressed the dependence of the functions appearing in Eq. 
(\ref{a2}) on other arguments for brevity.
At NLL accuracy we are entitled to pick the plus Lorentz indices for 
jet function $J_A$ and the minus indices for jet function
$J_B$ only. We can also set $k^+ = 0$ in $J_A$ and $k^- = 0$ in $J_B$ 
since all the loop momenta inside the jets are collinear.
Eq. (\ref{a2}) represents the first factorized form, Eq. 
(\ref{fact1}), for the amplitude $A^{(2)}$.

Next, we follow the procedure described in Sec. 
\ref{secondFactorizedForm} to bring the amplitude into the second 
factorized form,
Eq. (\ref{fact2}).
We employ an identity based on Eq. (\ref{iden}), for the function 
$S(k)$ defined in Eq. (\ref{sk})
\bea \label{sIdentity}
S(k^+,k^-) & = & S(k^+ = 0, k^- = 0) \, \theta(M-|k^+|) \, 
\theta(M-|k^-|) \nonumber \\
& + & [S(k^+, k^- = 0) - S(k^+ = 0, k^- = 0) \, \theta(M-|k^+|) ] \, 
\theta(M-|k^-|) \nonumber \\
& + & [S(k^+=0, k^-) - S(k^+ = 0, k^- = 0) \, \theta(M-|k^-|)]  \, 
\theta(M-|k^+|) \nonumber \\
& + & [ \{S(k^+, k^-) - S(k^+, k^- = 0) \, \theta(M-|k^-|)\} - \nonumber \\
& & \{S(k^+ = 0, k^-) - S(k^+ = 0, k^- = 0) \, \theta(M-|k^-|)\} \, 
\theta(M-|k^+|)]. \nonumber \\
\eea
The contribution from the first term in Eq. (\ref{sIdentity}) gives 
immediately the second factorized form with $\Gamma^{(2)}_A$ and
$\Gamma^{(2)}_B$ defined in Eq. (\ref{gammaDef}) for $n=m=2$.

We now discuss the rest of the terms in Eq. (\ref{sIdentity}), which 
can be analyzed using the $K$-$G$ decomposition, since,
by construction, there is no contribution from the Glauber region. At 
the current accuracy only the $K$-gluon contributes.
After substituting the second term of Eq. (\ref{sIdentity}) into Eq. 
(\ref{a2}), we can factor the gluon with momentum $k$ from
jet $J_B^{(2)}$. However,
it is easy to verify, using the definitions for $K$ and $G$ gluons,
Eq. (\ref{kgDef}), the Ward identities, Eq.\ (\ref{gy2}), and the explicit components of 
the gluon propagator, Eq.\ (\ref{propagComp}), that
the $k^+$ integral is over an antisymmetric function.  As a result, 
this contribution vanishes.
In a similar fashion, the contribution from the third term in Eq. 
(\ref{sIdentity}), after used in Eq. (\ref{a2}), vanishes, since now
we can factor the soft gluon with momentum $k$ from jet $J_A^{(2)}$ 
and the $k^-$ integral is over an antisymmetric function.

In the case of the last term in Eq. (\ref{sIdentity}), after used in 
Eq. (\ref{a2}), we can factor the soft gluon with
momentum $k$ from both jets $J_A^{(2)}$ and $J_B^{(2)}$. The 
integrals of the soft function $S(k)$ over $k^+$ and $k^-$
are then
\be \label{sInteg}
\tilde{S}(k_{\perp}, q; M) \equiv  C_A \, \frac{g_s^2}{(2\pi)^2} \, 
\int_{-M}^{M} \frac{{\rm d} k^+}{k^+} \, \frac{{\rm d} k^-}{k^-}
\, S(k^+, k^-, k_{\perp}, q).
\ee
As usually, we leave the transverse momentum integral undone.
The $1/k^+$ and $1/k^-$ in the integral above are given by the 
Principal Value prescription because there is no contribution from
the Glauber region. Since the amplitude is independent on the choice 
of scale $M$, we can evaluate it
at arbitrary scale. We choose to work in the limit $M \rightarrow 0$. 
In this limit the contribution to the integral comes from the imaginary parts 
of the gluon propagators in Eq. (\ref{sk}), 
$-i \pi \delta(k^2)$ and $-i \pi \delta((k-q)^2)$. The integration 
is then trivial and Eq. (\ref{sInteg}) becomes
\be \label{sm0}
\tilde{S}(k_{\perp}, q) \equiv \lim_{M \rightarrow 0} 
\tilde{S}(k_{\perp}, q; M) \equiv  - C_A \, \frac{i g_s^2}{8}
\, \frac{1}{k_{\perp}^2 \, (k-q)^2_{\perp}}.
\ee
Combining the partial results of the analysis described above in Eq. 
(\ref{a2}),
we arrive at the second factorized form for the double gluon exchange 
amplitude,
Fig. \ref{nllContribution}b,
\bea \label{a2factored}
A^{(2)} & = & \int \frac{{\rm d}^{D-2} k_{\perp}}{(2 \pi)^{D-2}} \, 
{\Gamma}_A^{(2) \; a \, b} (k_{\perp}) \, \frac{1}{(2 \pi)^2} \,
S(k^+ = 0, k^- = 0, k_{\perp}) \,
{\Gamma}_B^{(2) \; a \, b} (k_{\perp}) \nonumber \\
& + & \int \frac{{\rm d}^{D-2} k_{\perp}}{(2 \pi)^{D-2}} \, 
{\Gamma}_A^{(1) \; a} (p_A, q) \; \tilde{S}(k_{\perp}, q) \;
{\Gamma}_B^{(1) \; a} (p_A, q).
\eea
Using Eq. (\ref{a1}) for $A^{(1)}$ and Eq. (\ref{a2factored}) for $A^{(2)}$,
we obtain the amplitude for the process (\ref{qqqq}) at NLL accuracy
\bea \label{ampNll}
A^{({\rm NLL})} & = & - \frac{1}{t} \, \Gamma_A^{(1) \, a} \, (p_A, q, \eta) \,
\left(1 + \frac{1}{t} \, {\Pi}_{+ \, -}(q,\eta) + \frac{i \pi}{2} \, 
\alpha^{(1)}(t) \right) \,
\Gamma_B^{(1) \, a} \, (p_B, q, \eta) \nonumber \\
& + & \int \frac{{\rm d}^{D-2} k_{\perp}}{(2 \pi)^{D-2}} \, 
{\Gamma}_A^{(2) \; a \, b} (k_{\perp}) \,
\frac{i}{8 \pi^2} \, \frac{1}{k_{\perp}^2 \, (k-q)^2_{\perp}} \, 
{\Gamma}_B^{(2) \; a \, b} (k_{\perp}).
\eea
In Eq. (\ref{ampNll}), we have used the explicit form for $S(k^+ = 0, 
k^- = 0, k_{\perp})$, which can be easily
identified from Eq. (\ref{sk}). We have also used the integral 
representation of the gluon trajectory given in Eq. (\ref{trajec1}).

In order to determine the high energy behavior of the amplitude in 
Eq. (\ref{ampNll}), we need to examine the high energy behavior
of $\Gamma_A^{(1)}$ or $\Gamma_B^{(1)}$ at NLL and the evolution of 
$\Gamma_A^{(2)}$ or $\Gamma_B^{(2)}$ at LL.
In this section, we restrict the discussion of evolution equations to 
LL level, and hence we analyze the behavior of $\Gamma_A^{(2)}$ only.
We will address the study of NLL jet evolution, and gluon 
reggeization at this level in the next section.

We use the evolution equation given by Eq. (\ref{evolG}) in order to 
determine the LL dependence
of $\Gamma_A^{(2)}$ on $\ln(p_A^+)$.
In our special case of the two gluon exchange amplitude, it reads
\bea \label{evolGamma2}
\left(p_A^+ \, \frac{\partial}{\partial \, p_A^+} - 1 \right) 
{\Gamma}_A^{(2) \; a\, b} & = &
M \, \left[ J_A^{(2) \; a\, b}(k^- = +M, k^+ = 0, k_{\perp}) \right. 
\nonumber \\ 
& + & \left. J_A^{(2) \; a\, b}(k^- = -M, k^+ = 0, k_{\perp})\right] 
- {\tilde \eta}^{\alpha} \frac{\partial}{\partial \, 
{\eta}^{\alpha}} \, \Gamma_A^{(2) \; a\, b}. \nonumber \\ 
\eea
\begin{figure} \center
\scalebox{0.9}[0.9]{\includegraphics*{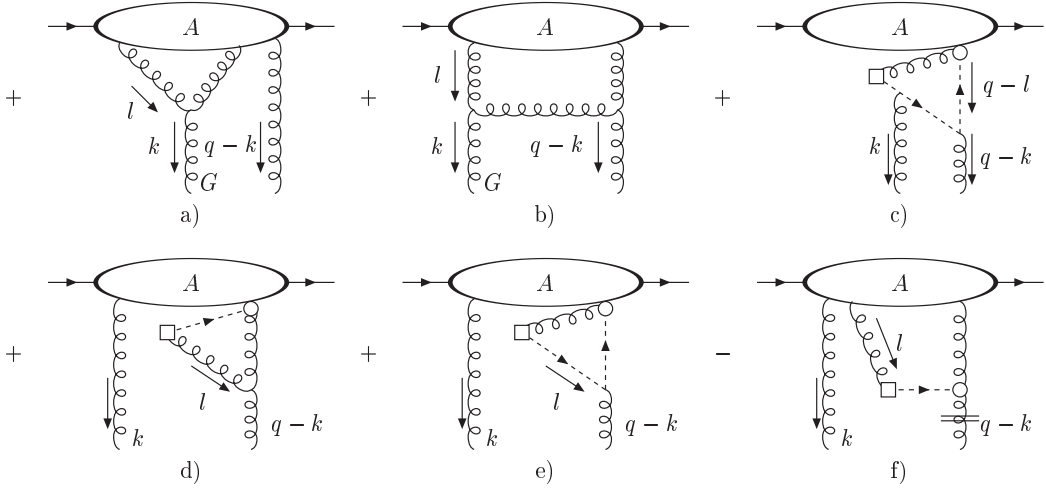}}
\caption{\label{bfklDiagrams} Diagrams determining the evolution of 
$\Gamma_A^{(2)}$.}
\end{figure}
The first term in Eq. (\ref{evolGamma2}) can be analyzed using the 
$K$-$G$ decomposition. The contributions from the $K$-gluon cancel
between the $J_A^{(2)}(k^- = +M)$ and $J_A^{(2)}(k^- = -M)$. The 
contributions from the $G$ gluon, which we now discuss,
are shown in Figs. \ref{bfklDiagrams}a and \ref{bfklDiagrams}b.

Since the gluon with momentum $q-k$ in Fig. \ref{bfklDiagrams}a 
cannot be in the Glauber region, we can use $K$-$G$
decomposition on it. The $K$ part factors from $J_A^{(3)}$, while the 
$G$ part does not contribute at LL.
After factoring out the gluon
with momentum $q-k$ and performing the approximations on the jet 
function $J_A^{(2)}$, the contribution to Fig. \ref{bfklDiagrams}a
for $k^- = + M$ is
\bea \label{fig10a}
{\rm Fig. \; \ref{bfklDiagrams}a} & = & -i g_s^2 f_{aec} f_{deb} \, 
\frac{1}{M} \int \frac{{\rm d}^D l}{(2 \pi)^D} \, S_1(k^+ = 0,
k^- = + M, k_{\perp}, l) \nonumber \\ 
& \times & J_A^{(2) \, c \, d}(l^+ = 0, l^-, l_{\perp}),
\eea
where we have defined
\be
S_1(k, l) \equiv \frac{N^{- \, \mu}(l)}{l^2} \, \frac{N^{- \, 
\nu}(k-l)}{(k-l)^2} \, V_{\mu \, \rho \, \nu}(l, -k, k-l) \,
\left( g^{\rho \, +} - \frac{k^{\rho}}{M} \right).
\ee
Next we follow the established procedure. First, we write
\be \label{s1Ident}
S_1(k,l) = S_1(k, l^- = 0) \, \theta(M-|l^-|) + \left[S_1(k,l) - 
S_1(k, l^- = 0) \, \theta(M-|l^-|)\right].
\ee
When we use the second term of Eq. (\ref{s1Ident}) in Eq. 
(\ref{fig10a}), we can factor the gluon with momentum $l$ from
$J_A^{(2)}$. Since the resulting integrand is an antisymmetric 
function under the simultaneous transformation $M \rightarrow -M$,
$l^{\pm} \rightarrow - l^{\pm}$, the contributions on the right hand 
side of Eq. (\ref{evolGamma2})
evaluated for $k^- = + M$ and $k^- = - M$ cancel each other.
Therefore we can write, using Eq. (\ref{s1Ident}) in Eq. (\ref{fig10a}),
\bea \label{fig10afact}
{\rm Fig. \; \ref{bfklDiagrams}a} & = & -i g_s^2 \, f_{aec} f_{deb} 
\, \frac{1}{M} \int \frac{{\rm d}^{D-2} l_{\perp}}{(2 \pi)^D}
\, \int {\rm d} l^+ \nonumber \\ 
& \times & S_1(k^+ = 0, k^- = + M, k_{\perp}, l^- = 0, l^+, l_{\perp}) \, 
{\Gamma}_A^{(2) \, c \, d}(l_{\perp}) + \ldots, \nonumber \\
\eea
where by dots we mean the term which is canceled after we take into 
account the contributions to both $J_A^{(2)}(k^- = + M)$
and $J_A^{(2)}(k^- = - M)$ on the right hand side of Eq. (\ref{evolGamma2}).

Next, we perform the $l^+$ integral in Eq. (\ref{fig10afact}). As we 
have already mentioned above,
since the final result does not depend on the scale $M$, we can choose
arbitrary value of $M$. We have chosen to perform the calculation in 
the limit $M \rightarrow 0$.
Then the only non-vanishing contribution comes from the imaginary part 
of the propagator $1/[(l-k)^2 + i \epsilon]$,
$-i \pi \, \delta(2 M l^+ + (l-k)^2_{\perp})$. For this term the 
$l^+$ integration is trivial and we obtain
\be \label{fig10afactInteg}
M \, ( {\rm Fig. \; \ref{bfklDiagrams}a} ) = - \alpha_s \, f_{aec} 
f_{deb} \, \int \frac{{\rm d}^{D-2} l_{\perp}}{(2 \pi)^{D-2}}
\, \frac{2 k_{\perp} \cdot l_{\perp}}{l_{\perp}^2 \, (k-l)_{\perp}^2} 
\; {\Gamma}_A^{(2) \, c \, d}(l_{\perp}) + \ldots \, ,
\ee
which gives an $M$-independent contribution to the right hand side of 
Eq. (\ref{fig10a}).

We follow the same steps when dealing with the diagram in Fig. 
\ref{bfklDiagrams}b, whose soft sub-diagram is given by
\bea \label{s2}
S_2 (k, l) & \equiv & \frac{N^{- \, \mu}(l)}{l^2} \, \frac{N^{- \, 
\nu}(q-l)}{(q-l)^2} \, V_{\mu \, \rho \, \gamma}(l, -k, k-l) \,
\left( g^{\rho \, +} - \frac{k^{\rho}}{M} \right) \frac{N^{\gamma 
\delta}(l - k)}{(l-k)^2} \nonumber \\
& \times & V_{\nu \, \delta \, -}(q - l, l - k, k - q).
\eea
First we use the identity (\ref{s1Ident}) for $S_2$.
The contribution due to the second term in Eq. (\ref{s1Ident}) 
vanishes, after the gluon with momentum $l$ has been factored
from $J_A^{(2)}$, due to the antisymmetry of the integrand.
Hence again, as in the case discussed above, only the term given by 
$S_2(l^- = 0, l^+, l_{\perp}, k)$ contributes.
In the limit $M \rightarrow 0$, the contribution comes from the 
imaginary part of the same denominator as in the case of Fig.
\ref{bfklDiagrams}a. The result is
\bea \label{fig10bfactInteg}
M \, ( {\rm Fig. \; \ref{bfklDiagrams}b} ) = & - & \alpha_s \, 
f_{aec} f_{deb} \, \int \frac{{\rm d}^{D-2} l_{\perp}}{(2 \pi)^{D-2}}
\, \frac{2}{l_{\perp}^2 \, (l-q)^2_{\perp} \, (k-l)_{\perp}^2} \nonumber \\
& \times & \left( k_{\perp}^2 l_{\perp}^2 - k_{\perp} \cdot l_{\perp} 
l_{\perp}^2 - k_{\perp} \cdot q_{\perp} l_{\perp}^2
- k_{\perp}^2 l_{\perp} \cdot q_{\perp} + 2 k_{\perp} \cdot l_{\perp} 
l_{\perp} \cdot q_{\perp} \right) \nonumber \\ 
& \times & {\Gamma}_A^{(2) \, c \, d}(l_{\perp}) + \ldots \, .
\eea
Combining the results of Eqs. (\ref{fig10afactInteg}) and 
(\ref{fig10bfactInteg}), we obtain the expression for the surface 
term in
Eq. (\ref{evolGamma2})
\bea \label{surfaceTerm}
M \, [\, J_A^{(2) \; a\, b}(k^- = +M, k^+ = 0, k_{\perp})
& + & J_A^{(2) \; a\, b}(k^- = -M, k^+ = 0, k_{\perp}) \, ] \nonumber \\ 
& & \hspace{-7cm} = \, 2 \alpha_s \, f_{aec} f_{bed} \, 
\int \frac{{\rm d}^{D-2} l_{\perp}}{(2 \pi)^{D-2}} \left(
\frac{k_{\perp}^2}{l_{\perp}^2 \, (k-l)_{\perp}^2} + 
\frac{(k-q)_{\perp}^2}{(l-q)_{\perp}^2 \, (k-l)_{\perp}^2}
- \frac{q_{\perp}^2}{l_{\perp}^2 \, (q-l)_{\perp}^2} \right) \nonumber \\ 
& & \hspace{-7cm} \times \, {\Gamma}_A^{(2) \, c \, d}(l_{\perp}).
\eea

Next, we analyze the contributions to the term ${\tilde 
\eta}^{\alpha} \partial / \partial \, {\eta}^{\alpha} \, 
\Gamma_A^{(2)}$
in the evolution equation (\ref{evolGamma2}).
The contributing diagrams are shown in Figs. \ref{bfklDiagrams}c - 
\ref{bfklDiagrams}f. Note that for every diagram
in Figs. \ref{bfklDiagrams}c - \ref{bfklDiagrams}f, we have also 
diagrams when a loop containing the boxed vertex
is attached to the external gluon with momentum $k$, instead of to 
the external gluon with momentum
$q-k$.

In Fig. \ref{bfklDiagrams}c, we have to consider all the possible 
insertions of external gluons
with momenta $k$ and $q-k$. We have six possibilities. The 
contribution shown in Fig. \ref{bfklDiagrams}c is proportional to
(omitting the color factor)
\be \label{fig10c}
\int_{-M}^{M} {\rm d}k^- \, ({\rm Fig. \; \ref{bfklDiagrams}c}) 
\propto \int_{-M}^{M} {\rm d}k^- \,
\int \frac{{\rm d}^D l}{(2 \pi)^D} \frac{N^{- \, \mu}(l)}{l^2} 
\frac{S_{\mu}(l)}{l \cdot {\bar l}} \,
\frac{(\bar l - \bar k)^+}{(\bar l - \bar k)^2} \, \frac{(\bar q - 
\bar l)^+}{(\bar q - \bar l)^2} \, (k^+ = 0).
\ee
Since the integrand is an antisymmetric function under $k^- 
\rightarrow - k^-$ and $l^{\pm} \rightarrow - l^{\pm}$,
the integral in Eq. (\ref{fig10c}) vanishes. The same antisymmetry 
property holds for the remaining five diagrams and therefore,
there is no contribution from them.
\begin{figure} \center
\scalebox{0.9}[0.9]{\includegraphics*{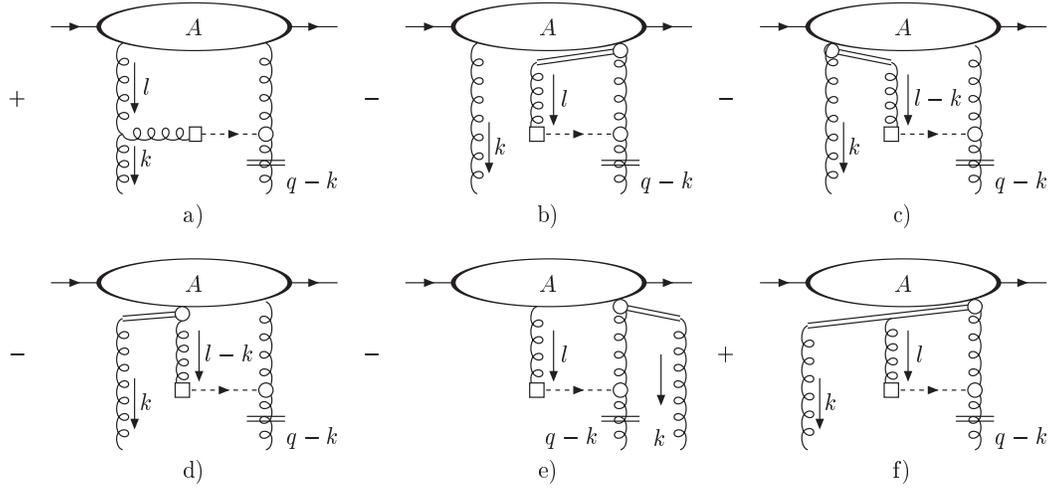}}
\caption{\label{bfklFact} Contributions to the diagram in Fig. 
\ref{bfklDiagrams}f when the gluon coming out of the boxed vertex
is attached to the soft line (a) and when either or both gluons with 
momenta $k$ and $l$ are $K$ gluons and they are factored
from the jet (b - f).}
\end{figure}

Let us next focus on the diagram in Fig. \ref{bfklDiagrams}f.
When the gluon with momentum $l$ attaches to a soft line inside of 
the jet $J_A^{(3)}$, the contribution takes the form shown in
Fig. \ref{bfklFact}a.
If it attaches to a jet line, its contribution can be written as
\bea \label{fig10f}
{\rm Fig. \; \ref{bfklDiagrams}f} & = & - g_s f_{b c d} \, \int 
\frac{{\rm d}^D l}{(2 \pi)^D} \, S_3(k^+ = 0,
k^-, k_{\perp}, l) \nonumber \\ 
& \times & J_A^{(3) \, a \, c \, d}(k^+ = 0, k^-, k_{\perp}, 
l^+ = 0, l^-, l_{\perp}),
\eea
with the soft function
\be \label{s3}
S_3(k,l) \equiv (q-k)^2 \, \frac{N^{- \mu}(l)}{l^2} \, 
\frac{S_{\mu}(l)}{l \cdot \bar l} \, \frac{N^{- +}(q-k-l)}{(q-k-l)^2}.
\ee
We use the identity for this soft function $S_3$, obtained from Eq. 
(\ref{sIdentity}) by the replacement $k^+ \rightarrow l^-$,
\bea \label{s3Ident}
S_3(l^-,k^-) & = & S_3(l^- = 0, k^- = 0) \, \theta(M-|l^-|) \, 
\theta(M-|k^-|) \nonumber \\
& + & [S_3(l^-, k^- = 0) - S_3(l^- = 0, k^- = 0) \, \theta(M-|l^-|) ] 
\, \theta(M-|k^-|) \nonumber \\
& + & [S_3(l^-=0, k^-) - S_3(l^- = 0, k^- = 0) \, \theta(M-|k^-|)] 
\, \theta(M-|l^-|) \nonumber \\
& + & [ \{S_3(l^-, k^-) - S_3(l^-, k^- = 0) \, \theta(M-|k^-|)\} \nonumber \\
& - & \{S_3(l^- = 0, k^-) - S_3(l^- = 0, k^- = 0) \, 
\theta(M-|k^-|)\} \, \theta(M-|l^-|)], \nonumber \\
\eea
to treat the soft gluons with momenta $k$ and $l$ attached to jet $J_A^{(3)}$.
The contribution from the first term in Eq. (\ref{s3Ident}), when 
used in Eq. (\ref{fig10f}), vanishes since the integrand
$S_3(k^+=k^- = 0, k_{\perp}, l^- = 0, l^+, l_{\perp})$ is an 
antisymmetric function of $l^+$, as can be easily checked
using Eqs. (\ref{propagComp}), (\ref{sCompon}) and (\ref{s3}).
We can apply the $K$-$G$ decomposition on the gluon with momentum $l$ 
when treating the
second term in Eq. (\ref{s3Ident}) used in Eq. (\ref{fig10f}). At LL 
only the $K$ gluon contributes. It can be factored from
the jet function $J^{(3)}_A$ with the result shown in Figs. 
\ref{bfklFact}b and \ref{bfklFact}c.
In a similar way we can treat the gluon with momentum $k$ in the 
third term of Eq. (\ref{s3Ident}).
After we factor this gluon from the jet $J_A^{(3)}$, we obtain the 
contributions shown in Figs. \ref{bfklFact}d and
\ref{bfklFact}e. In the case of the last term in Eq. (\ref{s3Ident}), 
we can factor out both soft gluons with momenta
$k$ and $l$ from jet $J_A^{(3)}$. The result of this factorization is 
shown in Fig. \ref{bfklFact}f.

Next, we note that the combination of the diagrams in Figs. 
\ref{bfklDiagrams}d, \ref{bfklDiagrams}e and \ref{bfklFact}b is the 
same
as the result encountered in the analysis of the LL amplitude, Fig. 
\ref{trajll}. We write
\be \label{bfklTraj2}
\int_{-M}^{M} {\rm d} k^- \left({\rm Fig. \; \ref{bfklDiagrams}d} + 
{\rm Fig. \; \ref{bfklDiagrams}e} + {\rm Fig. \; \ref{bfklFact}b}
\right) = \alpha^{(1)}(q_{\perp}-k_{\perp}) \, \Gamma_A^{(2) \; a \, 
b} (p_A, q, k_{\perp}).
\ee
where $\alpha^{(1)}(q-k)$ in Eq. (\ref{bfklTraj2}) is given by the 
diagrams in Fig. \ref{trajll} with an external momentum
$q-k = (0^+, 0^-, q_{\perp} - k_{\perp})$.
In the case when the gluon coming out of the boxed vertex attaches to 
an external gluon with momentum $k$, we evaluate
the one loop trajectory $\alpha^{(1)}(k_{\perp})$ in Eq. (\ref{bfklTraj2}).

To complete the analysis, we have to discuss the diagrams in Figs. 
\ref{bfklFact}a and \ref{bfklFact}c - \ref{bfklFact}f.
In the region $l^{\pm} \sim l_{\perp}$, we can factor the gluon with 
momentum $l$ from the jet function
$J_A^{(2)}(l^+ = 0, l^-, l_{\perp})$ in the case of the diagram in 
Fig. \ref{bfklFact}a.
The resulting $k^-$ and $l^{\pm}$ integral is over an antisymmetric 
function of $k^-$ and $l^{\pm}$, and therefore
it vanishes. So the only contribution comes from the Glauber region, 
where we can set $l^- = 0$ outside
$J_A^{(2)}(l^+ = 0, l^-, l_{\perp})$. As above, we perform the $l^+$ 
and $k^-$ integrals in the limit $M \rightarrow 0$.
The integrand does not develop a singularity in $k^-$ and/or $l^+$ 
strong enough to compensate
for the shrinkage of the integration region $\int_{-M}^{M} {\rm 
d}k^-$ when $M \rightarrow 0$. Hence the diagram
in Fig. \ref{bfklFact}a does not contribute in the limit $M 
\rightarrow 0$. In a similar way as for the diagram in
Fig. \ref{bfklFact}a, none of the diagrams in Figs. \ref{bfklFact}c - 
\ref{bfklFact}f contribute. The diagrams
in Figs. \ref{bfklFact}c - \ref{bfklFact}e vanish in the $M 
\rightarrow 0$ limit, while in the case of the diagram
in Fig. \ref{bfklFact}f the $k^-$ and $l^{\pm}$ integral is over an 
antisymmetric function of $k^-$ and $l^{\pm}$.

At this point we have discussed all the contributions appearing on 
the right hand side of the evolution equation (\ref{evolGamma2}).
Combining the partial results given by Eqs. (\ref{bfklTraj2}) and 
(\ref{surfaceTerm}) in Eq. (\ref{evolGamma2}),
we arrive at the evolution equation governing the high energy 
behavior of $\Gamma_A^{(2)}$
\bea \label{evolGamma2Result}
\left(p_A^+ \, \frac{\partial}{\partial \, p_A^+} - 
1\right){\Gamma}_A^{(2) \; a\, b} (p_A^+, q, k_{\perp}) & = &
2 \alpha_s \, f_{aec} f_{bed} \, \int \frac{{\rm d}^{D-2} 
l_{\perp}}{(2 \pi)^{D-2}} \;
\Gamma_A^{(2) \; c\, d} (p_A^+, q, l_{\perp}) \nonumber \\
& & \hspace{-2cm} \times \, \left( \frac{k_{\perp}^2}{l_{\perp}^2 
\, (k-l)_{\perp}^2} 
+ \frac{(k-q)_{\perp}^2}{(l-q)_{\perp}^2 \, (k-l)_{\perp}^2}
- \frac{q_{\perp}^2}{l_{\perp}^2 \, (q-l)_{\perp}^2} \right) \nonumber \\
& & \hspace{-2cm} + \, \left( \alpha^{(1)}(k_{\perp}) + 
\alpha^{(1)}(q_{\perp}-k_{\perp}) \right) \times \Gamma_A^{(2) \; a\, 
b} (p_A^+, q, k_{\perp}). \nonumber \\
\eea
Projecting out onto the color singlet in Eq. 
(\ref{evolGamma2Result}), we immediately recover the celebrated BFKL 
equation
\cite{bfkl}.

\subsection{Evolution of $\Gamma^{(n)}$ at LL} \label{gamman}

We can now generalize Eq. (\ref{evolGamma2Result}) to the case of 
$\Gamma_A^{(n)}$.
The evolution kernel in this case contains, besides a piece diagonal 
in the number of external gluons,
also contributions which relate jet functions with different number 
of external gluons
\bea \label{evolGammaN}
\lefteqn{\left(p_A^+ \, \frac{\partial}{\partial \, p_A^+} - 1\right) 
{\Gamma}_A^{(n) \; a_1 \ldots \, a_n} (p_A^+, q, k_{1 \, \perp},
  \dots, k_{n \, \perp}) = } \nonumber \\
& & 2 \alpha_s \, \sum_{i<j}^{n} f_{a_i \, e \, b_i} f_{a_j \, e \, 
b_j} \, \int \frac{{\rm d}^{D-2} l_{i \, \perp}}{(2 \pi)^{D-2}} \,
\frac{{\rm d}^{D-2} l_{j \, \perp}}{(2 \pi)^{D-2}} \, 
\delta^{(2)}(l_{i \, \perp} + l_{j \, \perp} - k_{i \, \perp} - k_{j 
\,
\perp}) \nonumber \\
& & \times \; \left( \frac{k_{i \, \perp}^2}{l_{i \, \perp}^2 \, 
(k_i-l_i)_{\perp}^2} + \frac{k_{j \, \perp}^2}{l_{j \, \perp}^2
\, (k_j - l_j)_{\perp}^2} - \frac{(k_i + k_j)_{\perp}^2}{l_{i \, 
\perp}^2 \, l_{j \, \perp}^2} \right) \nonumber \\
& & \times \; \Gamma_A^{(n) \; a_1 \ldots \, b_i \ldots \, b_j \ldots 
\, a_n} (p_A^+, q, k_{1 \, \perp}, \ldots,
l_{i \, \perp}, \ldots, l_{j \, \perp}, \ldots, k_{n \, \perp}) \nonumber \\
& & + \; \sum_{i = 1}^{n} \left( \alpha^{(1)}(k_{i \, \perp}) \right) \times
{\Gamma}_A^{(n) \; a_1 \ldots \, a_n} (p_A^+, q, k_{1 \, \perp}, 
\dots, k_{n \, \perp}) \nonumber \\
& & + \; \sum_{n' = 1}^{n-1} {\cal K}^{(n,n')}_{a_1 \ldots \, a_n; \; 
b_1 \ldots \, b_{n'}} \, {\otimes}_{\perp}
\, {\Gamma}_A^{(n')\; b_1 \ldots \, b_{n'}},
\eea
where ${\otimes}_{\perp}$ denotes a convolution in transverse 
momentum space. The last term in Eq. (\ref{evolGammaN}) corresponds 
to the
configurations when one or more external gluons attach to a gluon or 
a ghost lines forming the one loop kernel derived
for $\Gamma_A^{(2)}$.
Using the notation of Sec. \ref{solEvolEq}, we can write Eq. 
(\ref{evolGammaN}) at $r$-loop order in a form
\be \label{evolGamman}
\left(p_A^+ \frac{\partial}{\partial p_A^+} - 1 \right) \Gamma_A^{(n,r)}
= \sum_{n'=1}^{n} {\cal K}^{(n,n';1)} \otimes \Gamma_A^{(n',r-1)}.
\ee
It corresponds to Eq. (\ref{coeff2}) of Sec. \ref{solEvolEq} when written
in terms of the coefficients $c_r^{(n,r)}$
introduced in Eq. (\ref{gammaExpand}).
From Eq. (\ref{evolGamman}) we immediately see that the following 
property of the one loop kernel
\be \label{ktheta}
{\cal K}^{(n,n';1)} = \theta(n-n') \, {\tilde {\cal K}}^{(n,n';1)},
\ee
is satisfied. We recall that this step was essential in demonstrating 
that the set of evolution equations, Eq. (\ref{evolG}),
forms a consistent system, refer to the paragraph above Eq. (\ref{coeff2}).

The term diagonal in the number of external gluons in Eq. 
(\ref{evolGammaN}) coincides
with the evolution equation derived in Ref. \cite{jaroszewicz}.
Our formalism, besides enabling us to go systematically beyond LL 
accuracy, indicates that even at LL, in addition to the kernels found 
in Ref. \cite{jaroszewicz}, the kernel has contributions which relate 
jet functions with different number of external gluons.

\section{Gluon reggeization at NLL} \label{gluonReg}
In this section, we study the NLL evolution equation and the gluon 
reggeization. In Sec. \ref{secSignat}, we identify the contributions to the 
amplitude at NLL and negative signature channel. Sec. \ref{secNll} 
gives a detailed derivation of the relevant evolution equation.

\subsection{Color octet and negative signature amplitude} \label{secSignat}

The statement of gluon reggeization means that the scattering amplitude in the Regge limit, dominated 
by the gluon exchange (color octet) in the $t$-channel, and antisymmetric under $s \leftrightarrow u$ (negative signature),
takes the form
\be \label{regform}
A^{(-)}_{\bf 8}(s,t)_{\; r_A r_B, r_{1} r_{2}} = F^a_A(t)_{\; r_{1} r_A} \, 
\left[ \left(\frac{s}{-t}\right)^{\alpha(t)} - \left(\frac{-s}{-t}\right)^{\alpha(t)} \right] \,
F^a_B(t)_{\; r_{2} r_B},  
\ee
where $F^a_A(t)$ and $F^a_B(t)$ are the impact factors, which depend on the properties of the scattered particles. 
The Regge trajectory $\alpha(t)$ determines the dependence on the energy 
$\sqrt{s}$. It is a universal function of $t$ meaning that 
it does not depend on the particles involved in the scattering. 
\begin{figure} \center
\scalebox{1.0}[1.0]{\includegraphics*{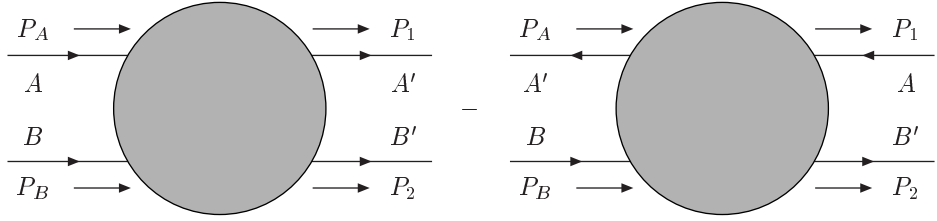}}
\caption{\label{signature} Projection of the amplitude onto the 
negative signature channel.} 
\end{figure}
In order to project onto the negative signature channel, besides the amplitude for $qq' \rightarrow qq'$ scattering, 
Eq. (\ref{qqqq}), we consider the amplitude for the process
\be \label{qqb}
{\bar q} (p_A, r_{1}, \lambda_A) + q'(p_B, r_B, \lambda_B) \rightarrow 
{\bar q} (p_A-q, r_A, \lambda_{1}) + q' (p_B + q, r_{2}, \lambda_{2}).
\ee
The amplitude for scattering (\ref{qqb}), $\tilde A(s,t)$, can be obtained from the amplitude 
for the process (\ref{qqqq}), $A(s,t)$, by the crossing symmetry $s \leftrightarrow u$. This means that 
$\tilde A (s,t) = A(u,t) \simeq A(-s,t)$. Hence, if we define
\be \label{aminus}
A^{(-)}(s,t) \equiv \frac{1}{2} \, \left(A(s,t) - {\tilde A}(s,t) \right),
\ee
as indicated in Fig. \ref{signature}, the amplitude $A^{(-)}$ will have a negative signature by construction.
Since $s \sim -u$ in the Regge limit, the logarithmic derivative of 
this amplitude is:
\be \label{amin}
\frac{\partial A^{(-)}}{\partial \ln s} (s,t) = \frac{1}{2} \, \left(
s \frac{\partial A}{\partial s}(s,t) - u \frac{\partial {\tilde A}}
{\partial u}(s,t) \right),
\ee

We now isolate the negative signature contribution to the amplitude for 
process (\ref{qqqq}) at NLL. 
The projection of the set of two gluon exchange diagrams, 
Fig. \ref{nllContribution}b, onto the negative signature takes the form:
\bea \label{a2}
A^{(-)}_2 & = & \int \frac{{\rm d}^{D} k}{(2 \pi)^D} \, \left[J_A^{(2) \; a \, b} 
(k^+ = 0, k^-, k_{\perp}) - {\tilde J}_A^{(2) \; a \, b} 
(k^+ = 0, k^-, k_{\perp})\right] \nonumber \\
& \times & S(k^+, k^-, k_{\perp}) \, \times \, J_B^{(2) \; a \, b} (k^- = 0, k^+, k_{\perp}),
\eea
where $S(k)$ has been defined in Eq. (\ref{sk}) 
and ${\tilde J}_A^{(2)}$ is the jet function corresponding to the 
anti-quark in the process (\ref{qqb}) moving in the plus direction.
In Appendix \ref{appa} we show that the jet functions $J_A^{(n)}$ and $\tilde J_A^{(n)}$ are related by
the symmetry:
\be \label{sym}
J_A^{(n)}(k_i^+ \rightarrow -k_i^+, k_i^- \rightarrow -k_i^-, k_{i \, \perp} \rightarrow k_{i \, \perp}) 
= (-1)^{n-1} \, (-1)^{\lambda_A + \lambda_{1}} \, 
\tilde J_A^{(n)}(k_i^+, k_i^-, k_{i \, \perp}), 
\ee
in the Regge limit. Applying this identity to the jet function with two external gluons ($n = 2$), the helicity 
conserving part of the jet function ($\lambda_A + \lambda_{1} = 2 \lambda_A = \pm 1$) obeys:
\be \label{sym1}
\tilde J_A^{(2)}(k^+=0,-k^-,k_{\perp}) = J_A^{(2)}(k^+=0,k^-,k_{\perp}). 
\ee
We apply this result to analyze Eq. (\ref{a2}). Using Eqs. (\ref{sk}) and 
(\ref{sym1}), we easily see that in the Glauber region 
$k^- \ll k^+ \sim k_{\perp}$ the integrand in Eq. (\ref{a2}) is over an antisymmetric function of $k^-$ and therefore the 
$k^-$ integral vanishes. This indicates that there is no contribution from the Glauber region when analyzing a soft gluon 
with momentum $k$ in Fig. \ref{nllAmplitude}b and therefore this gluon can be factored from the jet functions 
$J_A^{(2)}$ and $J_B^{(2)}$. As a consequence, the high-energy behavior of the amplitude for the process (\ref{qqqq}), 
projected onto the negative signature channel, is determined by the jet functions with one external gluon only, $J_A^{(1)}$ and 
$J_B^{(1)}$. It also means that $A^{(-)}_2$ is automatically in the color octet channel.

Combining the contributions from the one and the two gluon exchange diagrams 
shown in Fig. \ref{nllContribution}, the negative signature amplitude 
$A^{(-)}_{\bf 8}$ takes the form:
\bea \label{am}
A^{(-)}_{\bf 8} & = & - \frac{1}{t} \, \left(J_A^{(1) \, a}(p_A,q,\eta) - \tilde J_A^{(1) \, a}(p_A,q,\eta)\right) \, 
\left(1 + \frac{1}{t} \, \Pi_{+-}(q,\eta) + \frac{i \pi}{2} \, \alpha^{(1)}(t)\right) \nonumber \\ 
& \times & J_B^{(1) \, a}(p_A,q,\eta),
\eea
where $\Pi_{+-}(q,\eta)$ is the one-loop gluon self-energy and 
$\alpha^{(1)}(t)$ is the one-loop contribution to the gluon trajectory 
given in Eq. (\ref{trajec1}).   
Eq. (\ref{am}) indicates that in order to determine the high energy 
behavior of the scattering amplitude, we need to determine the 
high-energy behavior of $J_A^{(1)}$ and ${\tilde J}_A^{(1)}$. 
If we show that $J_A^{(1)}$ and ${\tilde J}_A^{(1)}$ satisfy:
\bea \label{nllreg}
p_A^+ \, \frac{\partial J_A^{(1)}}{\partial p_A^+} & = & \alpha(t) \, 
J_A^{(1)}, \nonumber \\
p_A^+ \, \frac{\partial {\tilde J}_A^{(1)}}{\partial p_A^+} & = & \alpha(t) \, 
{\tilde J}_A^{(1)},
\eea
at NLL, then the solution to the evolution equation (\ref{nllreg}) used in 
Eqs. (\ref{amin}) and (\ref{am}) implies the Regge ansatz, Eq. (\ref{regform}).
This would provide a proof of gluon reggeization at NLL to all orders 
in perturbation theory in QCD, Ref. \cite{gluonTraj}. 
We study the evolution equation (\ref{nllreg}) in the next section. 

\subsection{NLL evolution equation for $J_A^{(1)}$} \label{secNll}

\begin{figure} \center
\scalebox{0.7}[0.9]{\includegraphics*{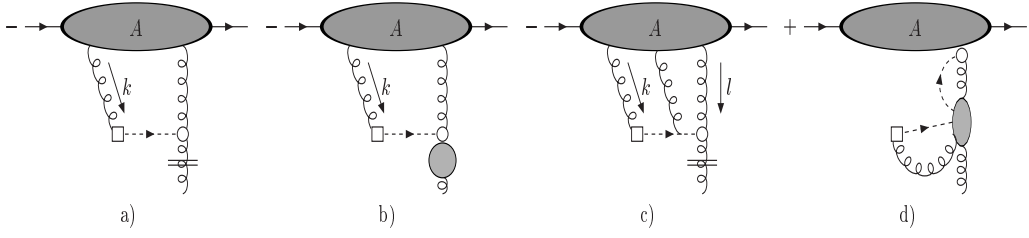}}
\caption{\label{nllEvolF} Diagrammatic representation of the evolution equation for jet $J_A^{(1)}$ at NLL.} 
\end{figure}

We now derive the evolution equation for jet function $J_A^{(1)\, a}$ at NLL.
We restrict ourselves to $J_A^{(1)}$ since ${\tilde J}_A^{(1)}$ 
can be analyzed in the same way.  
Applying the general evolution equation, Eq. (\ref{evolG}), for $n = 1$ and 
using the results in Fig. \ref{variationJ}, we arrive at the contributions to 
$(\partial / \partial \, \ln p_A^+ - 1) J_A^{(1)\, a}$ 
shown in Fig. \ref{nllEvolF}.
The diagram in Fig. \ref{nllEvolF}d, is already in a factorized form 
so it does not need to be analyzed further. The blob in the bottom part
of this diagram represents all possible two loop corrections.

In Fig. \ref{nllEvolF}b, the bottom dark blob denotes the one loop gluon-self energy. The gluon with momentum $k$ can be analyzed the same way
as in the case of LL, Sec. \ref{llAmplitude}, with a result in a factorized 
form. 

In order to prove reggeization of a gluon at NLL,
we need to show that the soft gluon with momentum $k$ in Fig. \ref{nllEvolF}a and the soft gluons with momenta $k$ and $l$ in Fig.
\ref{nllEvolF}c can be factored from the jet functions $J_A^{(2)}$ and 
$J_A^{(3)}$, respectively. Let us now analyze these particular diagrams.
The contribution to Fig. \ref{nllEvolF}a can be written
\be \label{fig1a}
{\rm Fig. \; \ref{nllEvolF}a} = - i g_s f_{a b c} \, \int \frac{{\rm d}^D k}{(2 \pi)^D} S_{\mu \, \nu}(k, q) 
\, J_{(A) \; b \, c}^{(2) \; \mu \, \nu}(p_A, q; k). 
\ee
\begin{figure} \center
\scalebox{0.9}[0.9]{\includegraphics*{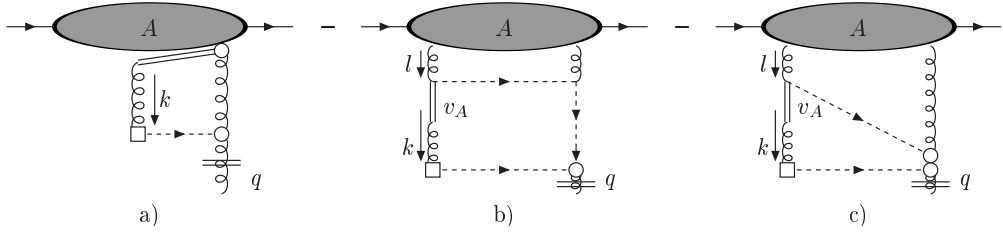}}
\caption{\label{nllKgluonF} Contributions to Fig. \ref{nllEvolF}a from the $K$ gluon at NLL.} 
\end{figure}
Next, we use the following identity for the integrand in Eq. (\ref{fig1a})
\bea \label{sjIdent}
S_{\mu \, \nu}(k) \, J_{(A) \; b \, c}^{(2) \; \mu \,\nu}(k) & = & 
S_{+ \, +}(k) \, J_{(A) \; b \, c}^{(2) \; + \, +}(k^+ = 0, k^-, k_{\perp}) \nonumber \\
& + & \left[ S_{+ \, +}(k) \, J_{(A) \; b \, c}^{(2) \; + \, +}(k) \, - \, 
S_{+ \, +}(k) \, J_{(A) \; b \, c}^{(2) \; + \, +}(k^+ = 0, k^-, k_{\perp}) \right] \nonumber \\
& + & \left[S_{ \mu \, \nu}(k) \, J_{(A) \; b \, c}^{(2) \; \mu \,\nu}(k) \, - \,  
S_{+ \, +}(k) \, J_{(A) \; b \, c}^{(2) \; + \, +}(k) \right].   
\eea
The first term in Eq. (\ref{sjIdent}), after being inserted into Eq. (\ref{fig1a}), can be analyzed using the $K$-$G$ decomposition, since
there is no contribution from the Glauber region due to the antisymmetry of $S_{+ \, +}(k^+, k^- = 0, k_{\perp})$ under the 
transformation $k^+ \rightarrow - k^+$. 
The contributions from the $K$ gluon, after applying the Ward identities, 
can be expressed in a form shown in Fig. \ref{nllKgluonF}. The diagram in Fig. \ref{nllKgluonF}a is the same as in the 
case of LL. At NLL, we also have to insert the gluon and ghost self-energies where appropriate. This diagram is in a 
factorized form and therefore consistent with a gluon reggeization at NLL. 

As for diagrams in Figs. \ref{nllKgluonF}b and \ref{nllKgluonF}c, 
the gluon with momentum $l$ can be factored from the jet function 
$J_A^{(2)}$ as shown in Appendix \ref{appb}.
Thus the contributions from the $K$ gluon in Figs. \ref{nllKgluonF}a
- \ref{nllKgluonF}c can be written in a factorized form and they are in 
an agreement with the gluon reggeization at NLL. 
Next, we analyze the contributions from the $G$ gluon with momentum $k$ corresponding to the first term in Eq. (\ref{sjIdent})
and to Fig. \ref{nllEvolF}a. They come from the diagrams shown in Figs. \ref{nllEvolaF}a, \ref{nllEvolaF}c and \ref{nllEvolaF}d 
since the $G$ gluon must be attached to a soft line. In Appendix \ref{appb}, 
we show that the soft gluons with momenta $l$ and $k-l$ can be factored from 
the jet function $J_A^{(3)}$ in Fig. \ref{nllEvolaF}a, the soft gluons with 
momenta $l$ can be factored from the jet functions $J_A^{(2)}$ in Figs. 
\ref{nllEvolaF}c and \ref{nllEvolaF}d and therefore the 
contribution from the $G$ gluon corresponding to Fig. \ref{nllEvolF}a 
is in a factorized form.   
   
Now, we examine the gluons with momenta $k$ and/or $q-k$ in the second and the third terms of Eq. (\ref{sjIdent}), 
after used in Eq. (\ref{fig1a}). They must attach to soft lines and therefore the typical contributions to these terms 
come from diagrams shown in Figs. \ref{nllEvolaF}a - \ref{nllEvolaF}d. 
\begin{figure} \center
{\includegraphics*{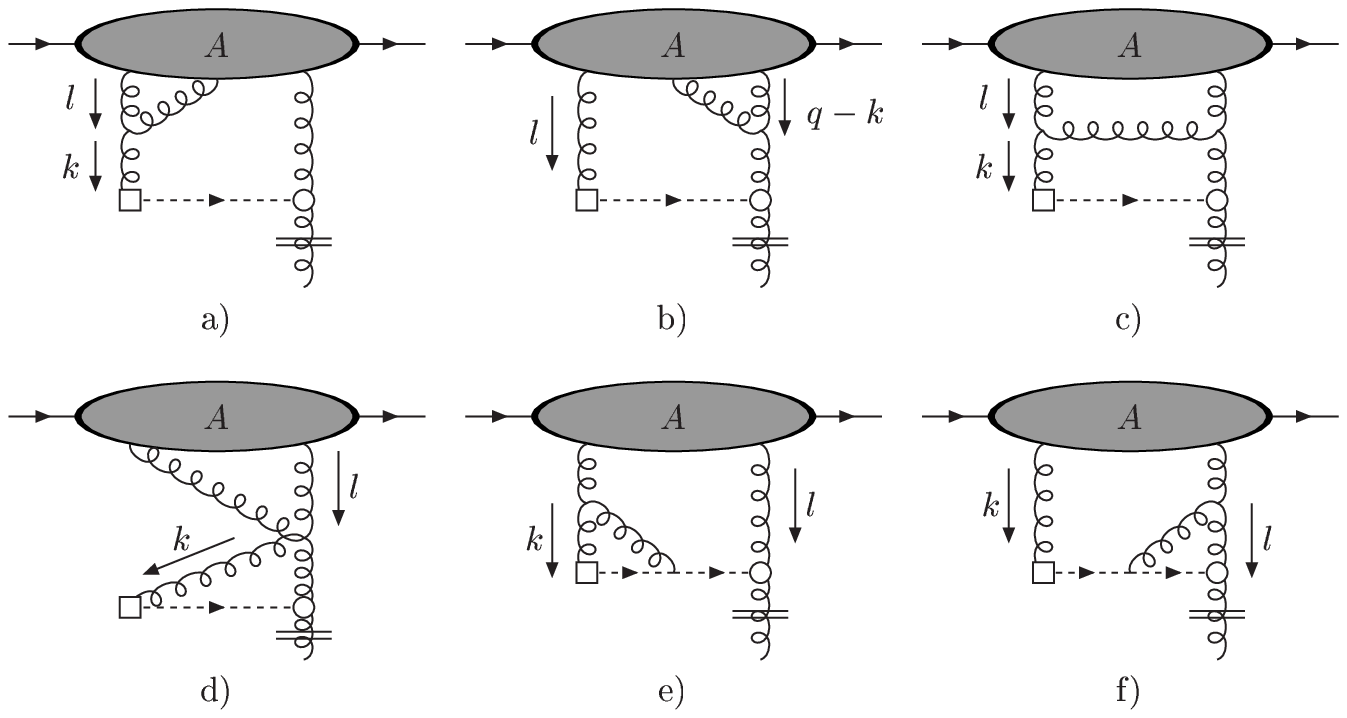}}
\caption{\label{nllEvolaF} Soft contributions to Figs. \ref{nllEvolF}a and \ref{nllEvolF}c.} 
\end{figure}
First we analyze the contributions to the second term in Eq. (\ref{sjIdent}).
As shown in Appendix \ref{appb}, the contribution to the diagram in 
Fig. \ref{nllEvolaF}c is in a factorized form. However, 
there are contributions to the diagrams in Fig. \ref{nllEvolaF}a and 
Fig. \ref{nllEvolaF}b, coming from the region when all soft gluons 
external to $J_A^{(3)}$ are Glauber, which cannot be written in a simple 
factorized form. These contributions, Eq. (\ref{p1Ares}) and Eq. (\ref{p2Ares}), are rather in 
a form of a convolution in transverse momentum space. 
    
As far as the last term in Eq. (\ref{sjIdent}) is concerned, its contributions come from the diagrams in Figs. 
\ref{nllEvolaF}a - \ref{nllEvolaF}d. As we demonstrate in Appendix \ref{appb}, the soft gluons external to $J_A^{(3)}$ in 
Figs. \ref{nllEvolaF}a and \ref{nllEvolaF}b, as well as the soft gluons 
external to $J_A^{(2)}$ in Figs. \ref{nllEvolaF}c and \ref{nllEvolaF}d 
can be factored from the the jet functions. 

Finally, we need to examine the diagram in Fig. \ref{nllEvolF}c which can be written as
\bea \label{fig1d}
{\rm Fig. \; \ref{nllEvolF}c} & \equiv & \int \frac{{\rm d}^D k}{(2 \pi)^D} \frac{{\rm d}^D l}{(2 \pi)^D} 
S_{b \, c \, d}(k, l, q) \nonumber \\ 
& \times & J_A^{(3) \; b \, c \, d}(p_A, q; k^+ = 0, k^-, k_{\perp}, l^+ = 0, l^-, l_{\perp}) \nonumber \\
& + & \int \frac{{\rm d}^D l}{(2 \pi)^D} S_{b \, c}(l, q) \, J_A^{(2) \; b \, c}(p_A, q; l^+ = 0, l^-, l_{\perp}) \nonumber \\
& + & \int \frac{{\rm d}^D k}{(2 \pi)^D} S_{b \, c}(k, q) \, J_A^{(2) \; b \, c}(p_A, q; k^+ = 0, k^-, k_{\perp}).
\eea
The contribution to the first term in Eq. (\ref{fig1d}) comes from a diagram shown 
in Fig. \ref{nllEvolF}c where the soft gluons with momenta $k$, $l$ and $q-k-l$ attach to jet lines inside the jet function
$J_A^{(3)}$. At NLL we can set $k^+ = l^+ = 0$ inside $J_A^{(3)}$ and also the soft gluons attach to the plus components of 
the jet's vertices. As shown in Appendix \ref{appb}, in the region when both gluons with momenta $k$ and $l$ are Glauber, 
we cannot factor out these gluons and the contribution from this double Glauber region is not in a factorized form.

The contribution to the last two terms in Eq. (\ref{fig1d}) comes from the graphs shown in Figs. \ref{nllEvolaF}e and
\ref{nllEvolaF}f, respectively. 
The gluon with momentum $q-k-l$ attaches to soft gluons with momenta $k$, Fig. \ref{nllEvolaF}e, and $l$, Fig. \ref{nllEvolaF}f, 
respectively. In Appendix \ref{appb} we demonstrate that the gluons external to $J_A^{(2)}$, can be factored from the jet 
functions $J_A^{(2)}$.

As a result of the previous reasoning, we arrive at a conclusion that the 
evolution equation for the jet function $J_A^{(1)}$ at NLL takes the form:
\bea \label{nlljet}
p_A^+ \, \frac{\partial J_A^{(1) \, a}}{\partial p_A^+}(p_A,q) & = & {\cal K}^{(1,1)}(t) \, J_A^{(1) \, a}(p_A,q) \, + \, 
\int \frac{{\rm d}^{D-2} k_{\perp}}{(2 \pi)^{D-2}} \frac{{\rm d}^{D-2} l_{\perp}}{(2 \pi)^{D-2}} \, 
{\cal K}^{(1,3)}_{a b c d}(q,k_{\perp},l_{\perp}) \nonumber \\ 
& \times & {\Gamma}_A^{(3) \; b \, c \, d}
(p_A, q; k_{\perp}, l_{\perp}).  
\eea
The kernel ${\cal K}^{(1,1)}$ is given by all the diagrams where it was possible to factor the soft gluons 
from the jet functions. This piece is in an agreement with gluon reggeization at NLL according to Eq. (\ref{nllreg}).
Besides this term, there is a contribution in Eq. (\ref{nlljet}) corresponding to the configurations when 
all soft gluons external to the jet function $J_A^{(3)}$ are Glauber. Combining the results of Eqs. (\ref{p1Ares}), 
(\ref{p2Ares}) and (\ref{j1res}) in Appendix \ref{appb}, 
we can find out the explicit form of the kernel ${\cal K}^{(1,3)}$. The presence of this term in Eq. (\ref{nlljet}) suggests
the breakdown of gluon reggeization at NLL level. 
We have to note the following, however. 
At two loop level, these non-factorizing terms vanish.
The non-factorizing piece in every diagram comes from the double 
pinched region, corresponding to the two delta function terms originating 
from the two denominators on the fermion line present in the tree-level 
jet function.
They, however, do not produce any dependence on transverse momenta and 
therefore this term is symmetric in external momenta. 
Hence if we interchange the two external momenta flowing out of $J_A^{(3)}$ 
and into the three point gluon-gluon-gluon or ghost-gluon-ghost vertex, 
Figs. \ref{nllEvolaF}a, \ref{nllEvolaF}b and \ref{nllEvolF}c, the integrand 
becomes an antisymmetric function under this exchange and the integral 
vanishes. 
However, when the jet function contains loop corrections, it will, in general, 
not be symmetric under the interchange of external momenta only. 
In order to get the symmetry, we need to interchange the colors as well.
Nevertheless, the above argument shows that the standard analysis and the 
calculations which have been performed at the two loop level, Refs. 
\cite{emission} - \cite{ducaFadin}, are not in a contradiction with our 
results.   
  
\section{Conclusions}

As an illustration of the general algorithm we have demonstrated it
in an action at NLL in both the amplitude and the evolution equations. 
First, in Sec. \ref{llAmplitude}, we have performed the resummation of the 
amplitude at LL. We have found the standard expression for the gluon Regge 
trajectory, Eq. (\ref{trajec1}). 
Then, in Sec. \ref{nllAmplitude}, we have identified the class of diagrams 
contributing to the amplitude at NLL level. They involved one and 
two gluon exchange diagrams, Fig. \ref{nllContribution}. 
The resummation of the two gluon exchange diagram in Sec. \ref{nllAmplitude}, 
needed to be performed at LL. 
It has lead to the celebrated BFKL equation, Eq. (\ref{evolGamma2Result}). 
The one gluon exchange contribution had to be resummed at NLL.
Together with this resummation, we have addressed the question of 
gluon reggeizaiton at NLL in Sec. \ref{gluonReg}. 
We have found that the majority of the terms in the evolution kernel 
are in an agreement with this conjecture to all orders in perturbation theory. 
However, we have also identified contributions violating the Regge ansatz 
starting at the three loop level and calculated the corresponding evolution 
kernels, Eqs. (\ref{p1Ares}), (\ref{p2Ares}), (\ref{j1res}). 
We have not been able to prove that these contributions decouple from 
the jets.
A further study of symmetries involving jets with three external gluons 
should shed more light on the presence or absence of the non-factorized 
terms, Ref. \cite{tibor2}.


\part{Resummation in Dijet Events}

\chapter{Event Shape / Energy Flow Correlations} \label{ch3}

\section{Introduction}

The agreement of
theoretical predictions  with experiment for jet cross sections is often
impressive.
This is especially so for inclusive jet cross sections at
high $p_T$, using fixed-order factorized perturbation theory
and parton distribution functions \cite{tevajet}.  A
good deal is also known about the substructure of jets,
through the theoretical and  experimental study
of multiplicity distributions and fragmentation functions \cite{DKTrev},
and of event shapes \cite{eshape,farhi,irsproof}.   Event shape
distributions \cite{thrustresum,broaden1,broaden2} in
particular offer a bridge between the perturbative, short-distance and
the nonperturbative, long-distance dynamics  of QCD \cite{ptnpdist}.

Energy flow \cite{flow} into angular regions between energetic
jets gives information that is in some ways complementary
to what we learn from event shapes.  In perturbation theory,
the distribution of particles in the final state reflects
interference between radiation from different jets \cite{DKTrev},
and there is ample evidence for perturbative
antenna patterns in interjet radiation at both $\rm e^+e^-$ \cite{elexp}
and hadron colliders \cite{EKSWjet,D0}.  Energy flow between
jets must also encode the mechanisms that neutralize color in the
hadronization process, and the transition of QCD from weak
to strong coupling.
Knowledge of the interplay
between energy and color flows \cite{KOS,BKS1} may help
identify the underlying
event in hadron collisions \cite{underlying},
to distinguish QCD bremsstrahlung
from signals of new physics.  Nevertheless, the systematic computation of
energy flow into interjet regions has turned out to be
subtle \cite{DS} for reasons that we will review below,
and requires a careful construction of the class of
jet events.  It is the purpose of this work to provide
such a construction, using event shapes as a tool.

In this paper, we introduce correlations
between event shapes and energy flow, ``shape/flow correlations",
    that are sensitive primarily to radiation from the highest-energy
jets.  So long as
the observed energy is not too small, in a manner to be quantified
below, we may control logarithms of the ratio of energy flow
to jet energy \cite{BKS1,BKSproc}.

The energy flow observables that we discuss
below are distributions
associated with radiation into a chosen interjet angular region,
$\O$.  Within $\O$ we identify a kinematic quantity $Q_\O\equiv
\varepsilon Q$, at c.m.\ energy $Q$, with
$ \varepsilon\ll 1$.  $Q_\O$ may be the sum of energies, transverse
energies or related
observables for the particles
emitted into $\O$.  Let us denote by $\bar\O$ the complement
of $\O$.  We are interested in the distribution of $Q_\O$
for events with a fixed number of jets in $\bar \O$.
This set of events may be represented  schematically as
\begin{equation}
A + B \rightarrow \mbox{  Jets }  + X_{\bar{\O}}
    + R_\O (Q_\O)\, .
\label{event}
\end{equation}
Here $X_{\bar\O}$ stands for radiation into the regions
between $\O$ and the jet axes, and $R_\O$ for
radiation into $\O$.

\begin{figure}[htb]
\begin{center}
\epsfig{file=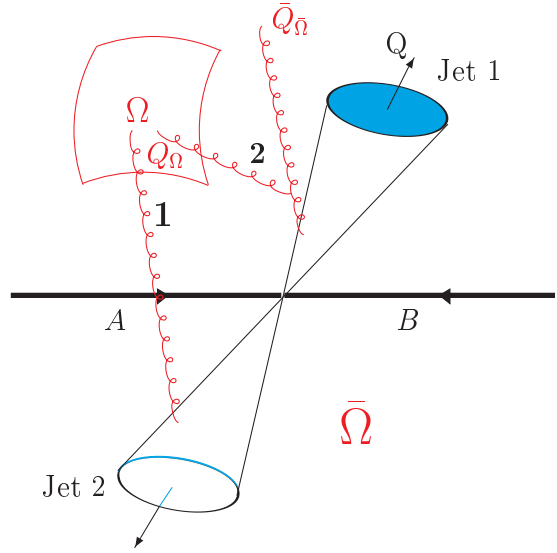,height=8cm,clip=0}
\caption{Sources of global and non-global logarithms in dijet events.
Configuration 1, a
primary emission, is the source of global logarithms.  Configuration 2
can give non-global logarithms.}
\label{eventfig}
\end{center}
\end{figure}

The subtlety associated with the computation of energy flow
concerns the origin of logarithms, and is illustrated by
Fig. \ref{eventfig}.
Gluon 1 in Fig.\ \ref{eventfig} is
an example of a primary gluon,
emitted directly from
the hard partons  near a jet axes.
Phase space integrals for primary emissions contribute single logarithms
per loop: $(1/Q_\O)\as^n \ln^{n-1} (Q/Q_\O) =
(1/\varepsilon Q)\as^n\ln^{n-1}(1/\varepsilon)$, $n\ge 1$, and
these logarithms exponentiate in a straightforward fashion \cite{BKS1}.
At fixed $Q_\O$
for Eq.\ (\ref{event}), however, there is another source of
potentially large logarithmic
corrections in $Q_\O$.  These are illustrated by gluon 2
in the figure, an example of
secondary radiation in $\O$, originating a parton emitted
by one  of the leading jets that define the event into intermediate region
    $\bar{\O}$.
As observed by Dasgupta and
Salam \cite{DS}, emissions into $\O$ from such secondary
partons   can also result in logarithmic corrections, of the form
$(1/Q_\O)\as^n \ln^{n-1}(\bar{Q}_{\bar{\O}}/Q_\O)$, $n\ge 2$,
where $\bar{Q}_{\bar{\O}}$ is the maximum energy
emitted into $\bar{\O}$.  These logarithms arise
from strong ordering in the energies of the primary
and secondary radiation
because real and virtual enhancements
associated with secondary emissions do not
cancel each other fully at fixed $Q_\O$.

If the cross section is
fully inclusive outside of $\O$, so that no restriction
is placed on the radiation into $\bar{\O}$,
$\bar{Q}_{\bar{\O}}$ can approach $Q$, and
the secondary logarithms can become as important as
the primary logarithms.   Such a cross section, in
which only radiation into a fixed portion of phase
space ($\O$) is specified, was termed ``non-global" by
Dasgupta and Salam, and the associated logarithms
are also called non-global \cite{DS,nonglobal,BanfiMarchSmye}.

In effect, a  non-global definition of energy
flow is not restrictive
enough to limit final states to a specific set of jets, and
non-global logarithms are produced by jets of intermediate energy,
emitted in directions between region $\O$ and
the leading jets.
Thus, interjet energy flow
does not always originate directly from the leading jets, in the absence of
a systematic criterion for suppressing intermediate radiation.
Correspondingly, non-global logarithms
reflect color flow at all scales, and do not
exponentiate in a simple manner.
Our aim in this paper is to formulate a set of observables
for interjet radiation in which non-global logarithms
are replaced by calculable corrections, and which
reflect the flow of color at short distances.
By restricting the sizes of event shapes,
we will limit radiation in
region $\bar{\O}$, while retaining the chosen jet structure.

An important observation  that
we will employ below is that non-global logarithms are not produced
by secondary emissions that are very close to a jet
direction, because a jet of parallel-moving
particles emits soft radiation coherently.  By
fixing the value of an event shape near the
   limit of narrow jets, we avoid
final states with large energies in $\bar{\O}$ away
from the jet axes.
At the same time, we will identify limits in which non-global logarithms
reemerge as leading corrections, and where the
methods introduced to study nongobal effects in Refs.\ 
\cite{DS,nonglobal,BanfiMarchSmye} provide
important insights.

To formalize these observations,
we study below correlated observables for $e^+e^-$
annihilation into two jets.
(In Eq.\ (\ref{event}) $A$ and $B$ denote positron and electron.) In
$e^+e^-$ annihilation dijet
events, the underlying color flow pattern is
simple, which enables us to concentrate  on the
energy flow within the event.  We will
introduce a class of event shapes, $\bar{f}(a)$
      suitable for measuring energy
flow into only part of phase space, with $a$ an
adjustable parameter.
To avoid large non-global
logarithmic corrections we weight events by
   $\exp[-\nu \bar{f}]$, with $\nu$
the Laplace transform conjugate variable.

For the restricted set
of events with narrow jets, energy flow is proportional
to the lowest-order cross section for gluon
radiation into the selected region.  The resummed cross section,
however, remains sensitive to color flow at short distances
through anomalous dimensions associated with coherent
interjet soft emission.  In a sense, our results show that
an appropriate selection of jet events automatically
suppresses nonglobal logarithms, and confirms
the observation of coherence in interjet radiation
\cite{DKTrev,EKSWjet}.

In the next section, we
introduce the event shapes that
we will correlate with energy flow, and describe their relation to
the thrust and jet broadening.
Section 3 contains the details of the
factorization procedure that
characterizes the cross
section in the two-jet limit.  This is followed in Sec.\ 4
by a derivation of the resummation of logarithms of the event shape
and energy flow, following the method introduced
by Collins and Soper \cite{ColSop}.
We then go on in Sec.\ 5 to exhibit
analytic results at leading logarithmic accuracy in $Q_\O/Q$
and next-to-leading logarithm in the event shape.
Section 6 contains
representative numerical results.
We conclude with a summary and a brief
outlook on further applications.

\section{Shape/Flow Correlations}

\subsection{Weights and energy flow in dijet events}

In the notation of Eq.\ (\ref{event}), we will study an event shape
distribution for the process
\be
e^+ + e^- \rightarrow J_1(p_{J_1}) + J_2(p_{J_2})  +
X_{\bar{\O}} \left(\bar{f}\right) + R_\O (Q_\O)\, ,
\label{crossdef}
\ee
at c.m.\ energy $Q\gg Q_\O\gg \Lambda_{\QCD} $.
Two jets with momenta $p_{J_c},\, c = 1,\,2$ emit
soft radiation (only) at wide angles.  Again,
$\O$ is a region between the
jets to be specified below, where the total energy or the transverse
energy $Q_\O$ of the soft radiation is measured,
and $\bar{\O}$ denotes the remaining
phase space (see Fig. \ref{event}).  Radiation into $\bar{\O}$
is constrained by event shape $\bar{f}$.  We
refer to cross sections at fixed values (or transforms) of $\bar{f}$ and
$Q_\O$
as shape/flow  correlations.

To impose the two-jet condition on the states of Eq.\ (\ref{crossdef}) we
choose weights that suppress states with substantial radiation into
$\bar{\O}$
away from the  jet axes.
We now introduce a class of event shapes $\bar{f}$, related
to the thrust, that
enforce the two-jet condition in a natural way.

These event shapes interpolate between and extend the familiar
thrust  \cite{farhi} and jet broadening \cite{broaden1,broaden2},
through an adjustable parameter $a$.
For each state $N$ that defines process (\ref{crossdef}),
we separate $\bar{\O}$ into two regions, $\bar{\O}_c$, $c=1,2$,
containing jet axes, $\hat{n}_c(N)$.  To be specific, we
let $\bar\O_1$ and $\bar\O_2$ be two hemispheres that
cover the entire space except for their intersections with
region $\O$.  Region $\bar\O_1$ is centered on $\hat n_1$,
and $\bar\O_2$ is the opposite hemisphere.
We will specify the
method that determines the jet axes $\hat n_1$ and
$\hat n_2$ momentarily. To identify a meaningful jet, of course, the
total energy
within $\bar\O_1$ should be a large fraction of the available energy,
of the order of $Q/2$
in dijet events.
In $\rm e^+e^-$ annihilation, if there is a well-collimated
jet in $\bar\O_1$ with nearly half the
total energy, there will automatically be one in $\bar{\O}_2$.

We are now ready to define
the contribution from particles in region $\bar\O_c$ to the $a$-dependent
event shape,
\be
\bar f_{\bar{\O}_c}(N,a) =
\frac{1}{\sqrt{s}}\
\sum_{\hat n_i\in \bar \O_c}\ k_{i,\,\perp}^a\, \o_i^{1-a}\,
\left(1-\hat n_i\cdot \hat n_c\right)^{1-a} \, ,
\label{barfdef}
\ee
where $a$ is any real number less than two, and
where $\sqrt{s}=Q$ is the c.m.\ energy.  The
sum is over those particles of state $N$  with direction $\hat n_i$
that flow into
$\bar\O_c$, and their transverse momenta $k_{i,\perp}$ are measured 
relative to $\hat{n}_c$.
The jet axis $\hat n_1$ for jet 1 is identified as that axis
that minimizes the specific thrust-related quantity $\bar 
f_{\bar{\O}_1}(N,a=0)$.
When $\bar{\O}_c$ in Eq.\ (\ref{barfdef}) is extended to all of phase space,
the case $a=0$ is then essentially $1-T$, with $T$ the thrust, while
$a=1$ is related to the
jet  broadening.

Any choice $a<2$ in (\ref{barfdef}) specifies an infrared safe event shape
variable, because the
contribution of any particle $i$ to the event shape behaves as
$\theta_i^{2-a}$ in the collinear limit, $\theta_i=\cos^{-1} (\hat n_i
\cdot \hat
n_c ) \rightarrow 0$.  Negative values of $a$ are clearly permissible, and
the limit $a\rightarrow -\infty$
corresponds to the total cross section.
At the other limit, the factorization and resummation
techniques that we discuss below will apply
only to $a<1$. For
$a> 1$, contributions to the event shape (\ref{barfdef}) from energetic
particles near the jet axis are generically larger than
contributions from soft, wide-angle radiation, or equal for
$a=1$.  When this is the case, the
analysis that we present below must be modified, at
least beyond the level of leading logarithm \cite{broaden2}.

In summary, once $\hat n_1$ is fixed, we have divided the phase space into
three regions:
\begin{itemize}
\item Region $\O$, in which we
measure, for example,
the energy flow,
\item Region $\bar \O_1$, the entire hemisphere centered on
$\hat n_1$, that is, around jet 1, except its intersection with $\O$,
\item Region $\bar \O_2$, the complementary hemisphere, except its
intersection with $\O$.
\end{itemize}
In these terms, we define
the complete event shape variable $\bar f(N,a)$ by
\ba
\bar f(N,a) &=& \bar f_{\bar{\O}_1}(N,a)+\bar f_{\bar{\O}_2}(N,a)\, , 
\label{2jetf}
\ea
with ${\bar f}_{\bar{\O}_c}$, $c=1,2$ given by (\ref{barfdef}) in terms of
the axes $\hat n_1$ of jet 1 and $\hat n_2$ of jet 2.
We will study the correlations of this set of event
shapes with the energy flow into $\O$, denoted as
\be
f(N) =  {1\over \sqrt s}\ \sum_{\hat n_i\in\O} \o_i\, .
\label{eflowdef}
\ee

The  differential cross section
for such dijet events at fixed values of $\bar f$ and $f$ is now
\ba
{d \bar{\sigma}(\varepsilon,\bar{\varepsilon},s, a)\over d \varepsilon
\,d\bar{\varepsilon}\, d\hat n_1}
&=&
{1\over 2s}\ \sum_N\;
|M(N)|^2\, (2\pi)^4\, \delta^4(p_I-p_N) \nonumber\\
&\ & \hspace{10mm} \times
\delta( \varepsilon-f(N))\, \delta(\bar{\varepsilon} -\bar f(N,a))\;
\delta^2  (\hat n_1 -\hat n(N))\, ,
\label{eventdef}
\ea
where we sum over all final states $N$ that contribute to the
weighted event, and where $M(N)$ denotes
the corresponding amplitude for ${\rm e^+e^-}\rightarrow N$.
The total momentum is $p_I$, with $p_I^2=s\equiv Q^2$.
As mentioned in the introduction, for much of our analysis,
we will work with the Laplace
transform of (\ref{eventdef}),
\ba
{d {\sigma}(\varepsilon,\nu,s, a)\over d \varepsilon
\, d\hat n_1}
&=&
\int_0^\infty d{\bar{\varepsilon}}\ {\rm e}^{-\nu{\bar{\varepsilon}}}\
{d \bar{\sigma}(\varepsilon,\bar\varepsilon,s, a)\over d \varepsilon
\,d\bar{\varepsilon}\, d\hat n_1}
\nonumber\\
&=&
{1\over 2s}\ \sum_N\;
|M(N)|^2\, {\rm e}^{-\nu\bar f(N,a)}\
(2\pi)^4\, \delta^4(p_I-p_N) \nonumber\\
&\ & \hspace{10mm} \times
\delta( \varepsilon-f(N))\;
\delta^2  (\hat n_1 -\hat n(N))\, .
\label{transeventdef}
\ea
Singularities of the form
$(1/\bar\varepsilon)\, \ln^n(1/\bar\varepsilon)$ in
the cross section (\ref{eventdef})
give rise to logarithms $\ln^{n+1}\nu$ in the transform
(\ref{transeventdef}).

      Since we are investigating energy
flow in two-jet cross sections, we fix the
constants $\varepsilon$ and $\bar{\varepsilon}$ to be
both much less than unity:
\be
0 < \varepsilon,\bar{\varepsilon} \ll 1.
\label{elasticlim}
\ee
We refer to this as  the elastic limit for the two jets.
In the elastic limit, the dependence of the directions of the
jet axes on soft radiation is weak.  We will return to
this dependence below.
Independent of soft radiation, we can
always choose our coordinate system such
that the
transverse momentum of jet 1 is
zero,
\be
p_{J_1,\, \perp} = 0\, ,
\ee
with $\vec p_{J_1}$ in the $x_3$ direction.  In the limit $\bar
\varepsilon, \varepsilon\rightarrow 0$, and in the overall c.m.,
$p_{J_1}$ and $p_{J_2}$ then approach light-like vectors in the plus and
minus directions:
\ba
p_{J_1}^\mu &\rightarrow&  \left(\sqrt{\frac{s}{2} },0^-,0_\perp \right)
\nonumber\\
p_{J_2}^\mu &\rightarrow&  \left(0^+,\sqrt{\frac{s}{2} },0_\perp \right)\, .
\label{lightlike}
\ea
As usual, it is convenient to work in light-cone coordinates,
$p^\mu  =  \left( p^+ , p^-, p_\perp \right)$, which we normalize as
$p^\pm  =  (1/\sqrt{2})(p^0 \pm p^3)$.
For small $\varepsilon$ and $\bar{\varepsilon}$, the cross section
(\ref{eventdef}) has
corrections in $\ln (1/\varepsilon)$ and
$\ln (1/\bar{\varepsilon})$, which we will organize in the following.

\subsection{Weight functions and jet shapes}

In Eq.\ (\ref{barfdef}), $a$ is a parameter that allows us to study
various event
shapes within the same formalism; it helps to control the
approach to the two-jet limit.   As noted above,
$a< 2$ for infrared safety, although the factorization
that we will discuss below applies beyond leading logarithm
only to $1>a>-\infty$.  A
similar weight function with a non-integer power has been discussed in
a related context for $2>a>1$ in
\cite{manoharwise}.
To see how the parameter $a$ affects the shape of the jets, let us
reexpress
the weight function for jet 1 as
\ba
\bar f_{\bar{\O}_1}(N,a) = \frac{1}{\sqrt{s}}\
\sum_{\hat n_i\in \bar \O_1} \o_i \sin^a \theta_i \left( 1 -
\cos \theta_i  \right)^{1-a}, \label{fbarexp}
\ea
where $\theta_i$ is the angle of the momentum of final state
particle $i$ with respect to jet axis $\hat n_1$.
As $a \rightarrow 2$ the weight vanishes only  very slowly for
$\theta_i\rightarrow 0$, and at fixed $\bar f_{\bar{\O}_1}$, the
jet becomes very narrow. On the other hand, as $a \rightarrow -
\infty$, the event
shape vanishes more and more rapidly in the forward direction, and the
cross
section at fixed $\bar f_{\bar{\O}_1}$ becomes more
and more inclusive in the radiation into $\bar{\O}_1$.

In this paper, as in Ref.\ \cite{BKS1},
we seek to control corrections in the single-logarithmic variable
$\alpha_s(Q) \ln (1/\varepsilon)$,
with $\varepsilon=Q_\O/Q$.  Such a resummation is most
relevant when
\be
\alpha_s(Q) \ln \left({1 \over \varepsilon}\right) \ge 1 \rightarrow
\varepsilon \le  \exp\left({- 1\over \alpha_s(Q) }\right)\, .
\label{einequal}
\ee
Let us compare these logarithms to non-global
effects in shape/flow correlations.
At $\nu=0$ and for $a\rightarrow -\infty$,
the cross section becomes inclusive outside $\O$.  As we show below,
the non-global logarithms discussed in Refs.\ \cite{BKS1, DS}
appear in shape/flow correlations as logarithms of the form
$\alpha_s(Q)\, \ln(1/(\varepsilon \nu))$, with $\nu$ the moment variable
conjugate to the event shape.  To treat these logarithms
as subleading for small $\varepsilon$ and (relatively)
large $\nu$, we require that
\ba
\alpha_s(Q)\, \ln \left({1 \over \varepsilon \nu}\right) < 1  \rightarrow
\varepsilon >
{1\over\nu}\;
\exp\left({- 1\over \alpha_s(Q) }\right)     \, .
\label{enuinequal}
\ea
For large $\nu$, there is a substantial range of $\varepsilon$
in which both (\ref{einequal}) and (\ref{enuinequal}) can
hold.  When $\nu$ is large, moments of the
correlation are dominated precisely by events with
strongly two-jet energy flows, which is the natural
set of events in which to study the influence of color
flow on interjet radiation.  (The peak of
the thrust cross section is at $(1-T)$ of order one-tenth
at LEP energies, corresponding to $\nu$ of order ten,
so the requirement of large $\nu$ is not overly restrictive.)
  In the next subsection, we
show how the logarithms of $(\varepsilon \nu )^{-1}$ emerge in a
low order example. This analysis
also assumes that $a$ is not large in absolute value. The event shape
at fixed angle decreases exponentially with $a$, and we
shall see that higher-order corrections can be proportional to $a$.
We always treat $\ln \nu$ as much larger than $|a|$.

\subsection{Low order example}
\label{sec:loe}

\begin{figure} \center
\includegraphics*{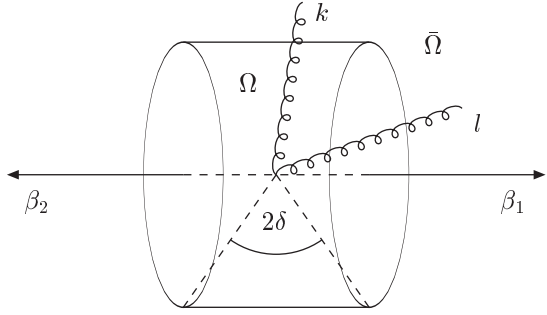}
\caption{\label{kinematics} A kinematic configuration that gives rise to
the
non-global logarithms. A soft gluon with momentum $k$ is radiated
into the region $\Omega$, and an
energetic gluon with momentum $l$ is radiated into $\bar\Omega$.
Four-vectors
$\beta_1$ and $\beta_2$, define the directions of jet 1 and jet 2,
respectively.}
\end{figure}

In this section, we check the general ideas developed above with
the concrete
example of a two-loop cross section for the process
(\ref{crossdef}). This is the lowest order in which a
non-global logarithm occurs, as observed in \cite{DS}. We
normalize this cross section to the Born cross section for
inclusive dijet production. A similar
analysis for the same geometry has been carried out in \cite{DS} and
\cite{nonglobalAS}.

The kinematic configuration we consider is shown in Fig. \ref{kinematics}.
Two fast partons, of velocities
    $\vec{\beta_1}$ and $\vec{\beta_2}$, are treated in eikonal
approximation.
In addition, gluons are emitted into the final state.
A soft gluon with momentum $k$ is radiated into region $\Omega$ and
an energetic gluon with momentum
$l$ is emitted into the region $\bar{\Omega}$.
We consider the cross section at fixed energy,
$\o_k\equiv \varepsilon\sqrt{s}$.
As indicated above, non-global logarithms arise from
strong ordering of the energies of the gluons,
which we choose as $ \ol \gg \ok $.
In this region, the gluon $l$ plays the role of a ``primary"
emission, while $k$ is a ``secondary" emission.

\begin{figure} \center
\includegraphics*{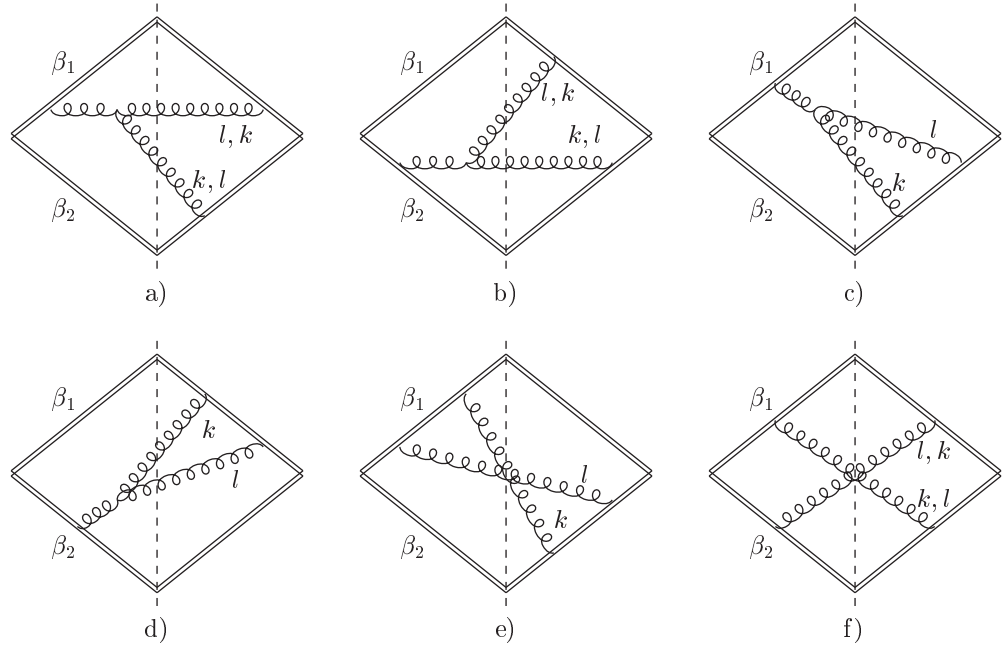}
\caption{\label{diagrams} The relevant two-loop cut diagrams
corresponding to the emission of
two real gluons in the final state contributing to the eikonal cross
section.
The dashed line represents the final state, with
contributions to the amplitude
to the left, and to the complex conjugate amplitude to the right.}
\end{figure}

For our calculation, we take the angular region $\Omega$
to be a ``slice" or ``ring'' in polar
angle of width $2 \delta$, or
equivalently, (pseudo) rapidity interval $(-\eta,\eta)$, with
\be 
\Delta \eta =2\eta= \ln\left(\frac{1+\sd}{1-\sd}\right)\, ,
\label{deltaeta1}
\ee
The lowest-order diagrams for
this process are those shown
in Fig. \ref{diagrams}, including distinguishable diagrams
in which the momenta $k$ and $l$ are interchanged.

The diagrams of Fig.\ \ref{diagrams}
give rise to color structures $C_F^2$ and $C_FC_A$,
but terms proportional to $C_F^2$  may be associated with a factorized
contribution to the cross section, in which the
gluon $k$ is emitted coherently by the combinations
of the gluon $l$ and the eikonals.
To generate the $C_FC_A$ part, on the other hand, gluon $k$
must ``resolve"
gluon $l$ from the eikonal lines,
giving a result that depends on
the angles between $\vec l$ and the
eikonal directions.

The computation of the diagrams is outlined
in Appendix \ref{eikapp}; here we quote the results.  We adopt the notation
$c_l\equiv \cos\theta_l,\,s_l \equiv \sin\theta_l$, with $\theta_l$
the angle of momentum $\vec l$ measured relative
to $\vec \beta_1$, and similarly for
$k$.   We take, as indicated above, a Laplace transform
with respect to the shape variable, and
identify the logarithm in the conjugate variable $\nu$.
We find that the logarithmic $C_FC_A$-dependence
of Fig.\ \ref{diagrams} may be written as a dimensionless
eikonal cross section in
terms of one energy and two polar angular integrals as
\bea
\label{ps0}
{d\sigma_{\rm eik}\over d\, \varepsilon}
    & = & C_F C_A \left(\frac{\alpha_s}{\pi}\right)^2 \,
\frac{1}{\varepsilon} \,
\int_{-\sd}^{\sd}
\mathrm{d} c_k \, \int_{\sd}^{1}
\mathrm{d} c_l \, \int_{\varepsilon \sqrt{s}}^{\sqrt{s}} \frac{\mathrm{d}
\ol}{\ol} \, e^{-\nu \, \ol \,
(1-c_l)^{1-a} \, s_l^a / Q}
\nonumber \\
& & \times\ \left[ \frac{1}{c_k + c_l} \, \frac{1}{1+c_k}
\left(\frac{1}{1+c_l} + \frac{1}{1-c_k}\right) - \frac{1}{s_k^2} \,
\frac{1}{1+c_l} \right]\, .
\eea
In this form, the absence of collinear singularities
in the $C_FC_A$ term at $\cos\theta_l=+ 1$ is manifest, independent of $\nu$.
Collinear singularities in the $l$ integral completely
factorize from the $k$ integral, and are proportional
to $C_F^2$.
The logarithmic dependence
on $\varepsilon$ for $\nu > 1$ is readily found to be
\ba
\label{psa}
{d\sigma_{\rm eik}\over d\, \varepsilon} = C_F C_A
\left(\frac{\alpha_s}{\pi}\right)^2 \, \frac{1}{\varepsilon} \,
\ln\left(\frac{1}{\varepsilon \nu}\right)\, C(\Delta \eta)\, ,
\ea
where  $C(\Delta \eta)$ is a finite function of the
angle $\delta$, given explicitly in
Appendix \ref{eikapp}.

We can contrast  this result to what happens when $\nu=0$,
that is, for an inclusive,  non-global cross section.  In this case,
recalling that $\varepsilon=Q_\O/Q$,
we find in place of Eq.\ (\ref{psa}) the non-global logarithm
\ba
\label{psb}
{d\sigma_{\rm eik}\over d\, \varepsilon} = C_F C_A
\left(\frac{\alpha_s}{\pi}\right)^2 \, \frac{1}{\varepsilon} \,
\ln\left(\frac{Q}{Q_\O}\right)\, C(\Delta \eta)\, .
\ea
As anticipated, the effect of the transform is to
replace the non-global logarithm in $Q/Q_\O$, by a logarithm
of $1/( \varepsilon \nu)$.  We are now ready to generalize this
result, starting from the factorization properties of
the cross section near the two-jet limit.

\section{Factorization of the Cross Section}

In this section we study the factorization of the
correlations (\ref{eventdef}).  The analysis
is based on a general approach
that begins with the all-orders treatment of singularities
in perturbative cross sections \cite{power,GStasi},
and derives factorization from the analyticity and
gauge properties of high energy Green functions and cross
sections \cite{pQCD}. The functions
that appear in factorized cross sections are expressible
in terms of QCD matrix elements \cite{pdfdef}, and
the matrix elements that we will encounter are familiar
from related analyses for heavy quark and jet production
\cite{hqjet}.  We refer in several places below
to standard arguments discussed in more detail
in \cite{GStasi,pQCD}.
The aim
of this section, and the reason why a
careful analysis is necessary, is to identify
the specific dimensionless
combinations of kinematic variables
on which the factorized matrix elements may depend.
We will use these dependences  in the following section,
when we discuss the resummation properties of
our correlations.

\subsection{Leading regions near the two-jet limit}

In order to resum logarithms of $\varepsilon$ and $\bar{\varepsilon}$
(or equivalently $\nu$, the Laplace conjugate of $\bar{\varepsilon}$) we
have first to
identify their origin in momentum space when
$\varepsilon,\bar{\varepsilon}\rightarrow 0$. Following the procedure
and terminology
of \cite{power}, we identify ``leading regions" in
the momentum integrals of cut diagrams, which can give rise
to logarithmic enhancements of the
cross section associated with lines approaching the
mass shell.  Within these regions, the lines of a cut diagram
fall into the following subdiagrams:

\begin{itemize}
\item A hard-scattering, or ``short-distance" subdiagram $H$, where all
components of line momenta are far off-shell, by order $Q$.
\item Jet subdiagrams, $J_1$ and $J_2$, where energies are
fixed and momenta are collinear
to the outgoing primary partons and the jet
directions that emerge from the hard scattering. (For
$\varepsilon=\bar{\varepsilon}=0$,
the sum of all energies in each jet is one-half the total energy.)
To characterize the momenta of the lines within the jets,
we introduce a scaling variable, $\lambda\ll 1$.  Within
jet 1, momenta $\ell$ scale as $(\ell^+ \sim Q,\ell^-\sim \lambda
Q,\ell_\perp \sim \lambda^{1/2}Q)$.
\item A soft subdiagram, $S$ connecting the jet functions $J_1$ and
$J_2$, in which the components of
   momenta $k$ are  small compared to $Q$ in all components,
scaling as $(k^\pm \sim \lambda Q,k_\perp \sim \lambda Q)$.
\end{itemize}

An arbitrary final state $N$ is the
union of substates associated with these subdiagrams:
\be
N=N_s \oplus N_{J_1} \oplus N_{J_2}\, .
\ee
As a result, the  event shape $\bar f$ can
also be written as a sum of contributions from the soft
and jet subdiagrams:
\ba
\bar f(N,a) &=& \bar f^N(N_s,a) + \bar f^N_{\bar{\O}_1}(N_{J_1},a) +
\bar f^N_{\bar{\O}_1}(N_{J_2},a)\, .
\label{fbarf}
\ea
The superscript $N$ reminds us that the contributions of
final-state particles associated with
the soft and jet functions depend implicitly on the
full final state, through the determination of the
jet axes, as discussed in Sec.\ 2.
In contrast, the energy flow weight, $f(N)$, depends only on
particles emitted at wide angles, and is hence insensitive
to collinear radiation:
\ba
f(N) = f(N_s)\, .
\ea

When we sum over all diagrams
that have a fixed final state, the contributions from these leading regions
may be factorized into a set of functions, each of which corresponds to
one of the generic hard, soft and jet subdiagrams.  The arguments for this
factorization at
leading power have been discussed extensively \cite{ColSop,pQCD,ColSte}.
The cross section becomes a convolution in
$\bar \varepsilon$, with the sums over states linked
by the delta function which fixes $\hat n_1$, and by momentum
conservation,
\ba
{d \bar{\sigma}(\varepsilon,\bar{\varepsilon},s,a)\over d \varepsilon
\,d\bar{\varepsilon}\, d\hat n_1}
&=& {d \sigma_0 \over d\hat{n}_1}\
H(s,\hat{n}_1)\
\sum_{N_s,N_{J_c}}\;
\int d\bar{\varepsilon}_s\, {\cal S}(N_s)\, \delta(\varepsilon-f(N_s))
\nonumber \\
&\ & \quad \times \, \delta(\bar{\varepsilon}_s-\bar{f}^N(N_s,a))
\nonumber\\
&\ & \quad \times \prod_{c=1}^2\,
   \int  d\bar{\varepsilon}_{J_c}\, {\cal J}_c (N_{J_c}) \,
\delta(\bar{\varepsilon}_{J_c}-\bar{f}^N_{\bar{\O}_c}(N_{J_c},a))\nonumber\\
&\ & \hspace{15mm} \times\
(2\pi)^4\, \delta^4(p_I-p(N_{J_2})-p(N_{J_1})-p(N_s))
\nonumber\\
&\ & \hspace{15mm} \times\
\delta^2(\hat n_1 -\hat n(N))\;
\delta(\bar{\varepsilon}-\bar{\varepsilon}_{J_1}-\bar{\varepsilon}_{J_2}-\bar{
\varepsilon}_s)
\nonumber\\
&=& {d \sigma_0 \over d\hat{n}_1}\; \delta(\varepsilon)\,
\delta(\bar{\varepsilon})+{\cal O}(\alpha_s)\, .
\label{sigmafact}
\ea
Here  $d\sigma_0/d\hat{n}_1$ is the Born
cross section for the production of a single
particle (quark or antiquark) in direction
$\hat{n}_1$, while the short-distance function
$H(s,\hat{n}_1)=1+{\cal O}(\alpha_s)$, which
    describes corrections to the hard scattering,
is an expansion in $\alpha_s$ with finite coefficients.
The functions ${\cal J}_c (N_{J_c}),\ {\cal S}(N_s)$
describe
the internal dynamics of the jets and wide-angle soft
radiation, respectively.  We will specify these functions below.
We have suppressed their dependence on a factorization scale.
Radiation at wide angles from the jets will be well-described
by our soft functions ${\cal S}(N_s)$, while
we will construct
the jet functions ${\cal J}_c (N_{J_c})$ to be independent of
$\varepsilon$, as  in
Eq.\ (\ref{sigmafact}).

So far, we have specified our sums over states in Eq.\ (\ref{sigmafact})
only when all lines in $N_s$ are
soft, and all lines in $N_{J_c}$ have momenta that are collinear, or
nearly collinear
to $p_{J_c}$.   As $\varepsilon$ and $\bar{\varepsilon}$ vanish, these are the
only final-state momenta that are kinematically possible.
Were we to restrict ourselves to these configurations
only, however, it would not be straightforward to make the individual
sums over $N_s$ and $N_{J_c}$ infrared safe.  Thus, it is necessary to
include soft partons in $N_s$ that are emitted near the jet directions,
and soft partons in the $N_{J_c}$ at wide angles.
We will show below how to define the functions ${\cal J}_c (N_{J_c}),\ {\cal
S}(N_s)$
so that they generate factoring, infrared safe functions that
avoid double counting.
We know on the basis of the arguments of Refs.\ \cite{ColSop,pQCD,ColSte}
that corrections to
the factorization of soft from jet functions are suppressed by
powers of the weight functions $\varepsilon$ and/or $\bar \varepsilon$.

\subsection{The factorization in convolution form}

Although formally factorized, the jet and soft functions
in Eq.\ (\ref{sigmafact}) are still  linked in a potentially complicated
way through their dependence on the jet
axes.  Our strategy is to simplify this complex dependence
to a simple convolution in contributions to $\bar\varepsilon$,
accurate to leading power in $\varepsilon$ and $\bar\varepsilon$.

First, we note that the cross section of Eq.\ (\ref{sigmafact})
is singular for vanishing $\varepsilon$ and $\bar{\varepsilon}$, but is a
smooth function of $s$ and $\hat{n}_1$.  We may therefore make any
approximation that changes $s$ and/or $\hat{n}_1$ by an amount
that vanishes as a power of $\varepsilon$ and $\bar{\varepsilon}$ in the
leading regions.

Correspondingly, the amplitudes for jet $c$ are singular in
$\bar{\varepsilon}_{J_c}$,
but depend smoothly on the jet energy and direction, while
the soft function is singular in both $\varepsilon$ and
$\bar{\varepsilon}_s$,
but depends smoothly on the jet directions.  As a result,
at fixed values of $\varepsilon$ and $\bar\varepsilon$ we
may approximate the jet directions and energies by
their values at $\varepsilon=\bar{\varepsilon}=0$ in the soft and jet
functions.

Finally, we may make any approximation that affects
the value of $\varepsilon$ and/or $\bar\varepsilon_{J_c}$ by
amounts that vanish faster than linearly for $\bar\varepsilon\rightarrow 0$.
It is at this stage that we will require that $a<1$.

With these observations in mind,
we enumerate the replacements
and approximations by which we
reduce Eq.\ (\ref{sigmafact}), while retaining leading-power accuracy.

\begin{enumerate}

\item  To simplify the definitions  of the jets in Eq.\ (\ref{sigmafact}),
we make the replacements $\bar{f}^N_{\bar{\O}_c}(N_{J_c},a)
\rightarrow \bar f_c(N_{J_c},a)$ with
\ba
\bar f_c(N_{J_c},a) \equiv
\frac{1}{\sqrt{s}} \sum_{{\rm all}\
\hat n_i
\in N_{J_c} }\
k_{i,\,\perp}^a\, \o_i^{1-a}\, \left(1-\hat n_i\cdot \hat n_c
\right)^{1-a}\, .
\label{fbar2jet1}
\ea
The jet weight function $\bar{f}_c(N_{J_c},a)$ now depends only on particles
associated with $N_{J_c}$.
   The contribution to $\bar{f}_c(N_{J_c},a)$
from particles within region $\bar{\O}_c$,
is exactly the same here as in the weight (\ref{barfdef}),
but we now include particles
in all other directions.
     In this way, the independent sums over final states of the
jet amplitudes will be naturally infrared
safe.  The value of $\bar f_c(N_{J_c},a)$ differs from
the value of $\bar{f}^N_{\bar{\O}_c}(N_{J_c},a)$, however, due to
radiation outside
$\bar\O_c$, as indicated by the new subscript.  This radiation is hence at
wide angles to the jet axis.  In the elastic limit (\ref{elasticlim}), it is
also constrained to be soft.  Double counting in contributions
to the total event shape, $\bar f(N,a)$, will be avoided by an appropriate
definition
of the soft function below.
The sums over states are still not yet fully independent,
however, because the jet directions $\hat n_c$ still depend
on the full final state $N$.

\item
Next, we turn our attention to the condition that fixes the jet
direction $\hat n_1$.
Up to corrections in the orientation of $\hat n_1$
that vanish as powers of $\varepsilon$ and $\bar{\varepsilon}$,
we may neglect the dependence of $\hat{n}_1$
on $N_s$ and $N_{J_2}$:
\be
\delta(\hat n_1-\hat n(N))  \rightarrow  \delta(\hat n_1 -\hat n(N_{J_1}))\, .
\label{nnJone}
\ee
In Appendix \ref{approxapp}, we show that
this replacement also leaves the value of
$\bar\varepsilon$ unchanged, up to corrections that vanish as
$\bar\varepsilon^{2-a}$.  Thus, for $a<1$, (\ref{nnJone}) is
acceptable to leading power.
For $a<1$, we can therefore
identify the direction of jet 1 with $\hat{n}_1$.
These approximations simplify Eq.\ (\ref{sigmafact})
by eliminating the implicit dependence of
the jet and soft weights on the full final state.  We may
now treat $\hat n_1$ as an independent vector.

\item  In the leading regions, particles
that make up each final-state jet are associated with states $N_{J_c}$,
while $N_s$  consists of soft particles only.
In the momentum conservation delta function, we
can neglect the four-momenta of lines in $N_s$,
whose energies all vanish as $\varepsilon,\bar{\varepsilon}\rightarrow 0$:
\be
\delta^4(p_I-p(N_{J_2})-p(N_{J_1})-p(N_s))
\rightarrow
\delta^4(p_I-p_{J_2}-p_{J_1}).
\ee

\item
Because the cross section is a smooth function of
the jet energies and directions, we may also
neglect the masses of the jets within the
momentum conservation delta function, as in
Eq.\ (\ref{lightlike}).  In this approximation,
we derive in the c.m.,
\ba
\delta^4(p_I-p_{J_2}-p_{J_1})
&\rightarrow&
\delta(\sqrt{s} - \o(N_{J_1})-\o(N_{J_2}))
\, \delta(|\vec p_{J_1}|-|\vec p_{J_2}|)
\, {1\over |\vec p_{J_1}|^2} \nonumber \\
& & \times \, \delta^2(\hat n_1 + \hat n_2)\nonumber\\
&\rightarrow& {2\over s}\,
\delta\left({\sqrt{s}\over 2} - \o(N_{J_1})\right)
\, \delta\left({\sqrt{s}\over 2} - \o(N_{J_2})\right) \nonumber \\
& & \times \, \delta^2(\hat n_1 + \hat n_2)\, .
\ea
Our jets are now back-to-back:
\be
\hat n_2 \rightarrow -\hat n_1\, .
\label{fbar2soft}
\ee

\end{enumerate}

Implementing these replacements and approximations for $a<1$,
we rewrite the cross section Eq.\
(\ref{sigmafact}) as
\ba
{d \bar{\sigma}(\varepsilon,\bar{\varepsilon},s,a)\over d \varepsilon\,
d\bar{\varepsilon}\, d\hat n_1}
&=&
{d \sigma_0 \over d\hat{n}_1}\
H(s,\hat{n}_1,\mu)\;
\int  d\bar{\varepsilon}_s\,
\bar{S}(\varepsilon,\bar{\varepsilon}_s,a,\mu) \,
\nonumber\\
&\ & \times
\prod_{c=1}^2\, \int  d\bar{\varepsilon}_{J_c}\,
\bar{J}_c(\bar{\varepsilon}_{J_c},a,\mu)\,
\delta(\bar{\varepsilon}- \bar{\varepsilon}_{J_1}-\bar{\varepsilon}_{J_2}-
\bar{\varepsilon}_s)\, ,
\label{factor}
\ea
with (as above) $H=1+{\cal O}(\alpha_s)$.  Referring to
the notation of Eqs.\ (\ref{sigmafact}) and (\ref{fbar2jet1}),
the functions $\bar{S}$ and $\bar{J}_c$ are:
\ba
\bar{S}(\varepsilon,\bar{\varepsilon}_s,a,\mu)
&=&
\sum_{N_s}\; {\cal S}(N_s,\mu)\, \delta(\varepsilon-f(N_s)) \,
\delta(\bar{\varepsilon}_s-\bar f(N_s,a))
\label{firstSdef}
\\
\bar{J}_c(\bar{\varepsilon}_{J_c},a,\mu)
&=&
\frac{2}{s}  (2\pi)^6\,   \sum_{N_{J_c}}
{\cal J}_c(N_{J_c},\mu) \, \delta(\bar{\varepsilon}_{J_c}-\bar 
f_c(N_{J_c},a))\,
\delta\left({\sqrt{s}\over 2} - \o(N_{J_c})\right) \nonumber \\ 
& & \times \, \delta^2(\hat n_1 \pm \hat n(N_{J_c})),
\nonumber\\
\label{firstJdef}
\ea
with the plus sign in the angular delta function
for jet 2, and the minus for jet 1. The weight functions for the jets are
given by Eq.\
(\ref{fbar2jet1}) and induce dependence on
the parameter $a$.   We have introduced the factorization scale
$\mu$, which we set equal to the
renormalization scale.

We note that we must construct the soft functions $\bar{S}(N_s,\mu)$
to cancel the contributions of final-state particles from
each of the $\bar{J}_c(N_{J_c},\mu)$ to the weight $\varepsilon$,
as well as the contributions of the jet functions to $\bar\varepsilon$
from soft radiation outside their respective regions $\bar\O_c$.
Similarly, the jet amplitudes
must be constructed to include collinear enhancements only in their
respective jet directions.  Explicit constructions that satisfy these
requirements will be
specified in the following subsections.

To disentangle the convolution in (\ref{factor}), we take Laplace
moments with respect to $\bar{\varepsilon}$:
\ba
     {d \sigma(\varepsilon,\nu,s,a)\over d \varepsilon \,d\hat n_1}
& = &  \int_0^\infty d\bar{\varepsilon}\, e^{- \nu \,\bar{\varepsilon}}\,
{d \bar{\sigma}(\varepsilon,\bar{\varepsilon},a)\over d \varepsilon\,
d\bar{\varepsilon}\,
d\hat n_1}
\label{trafo} \nonumber
\\
& = & {d \sigma_0 \over d\hat{n}_1}\
H(s,\hat{n}_1,\mu)\;  S(\varepsilon,\nu,a,\mu) \,\prod_{c=1}^2\,
J_c(\nu,a,\mu).
\label{trafosig}
\ea
Here and below  unbarred quantities are the transforms in
$\bar{\varepsilon}$,
and barred quantities denote untransformed functions.
\be
S(\varepsilon,\nu,a,\mu) =   \int_0^\infty d\bar{\varepsilon}_s \,e^{- \nu
\,\bar{\varepsilon}_s}
\bar{S}(\varepsilon,\bar{\varepsilon}_s,a,\mu),
\label{trafodef}
\ee
and similarly for the jet functions.

\begin{figure}[htb]
\begin{center}
\epsfig{file=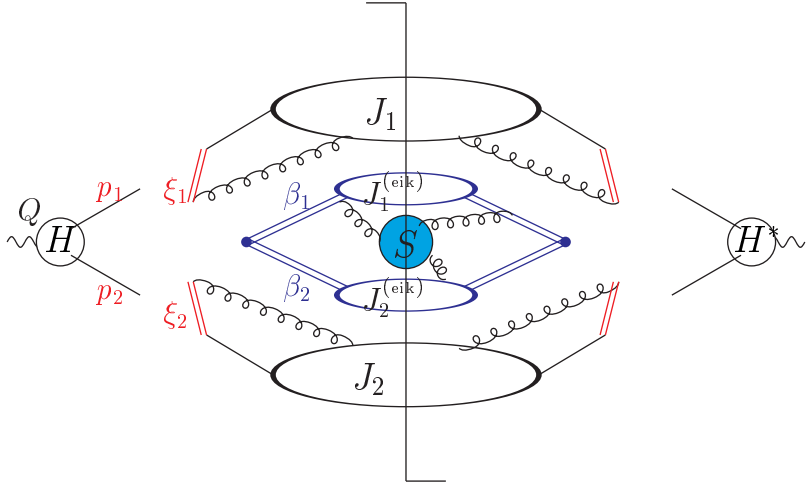,height=7cm,clip=0}
\caption{Factorized cross section (\ref{factor})
after the application of Ward identities. The vertical line
denotes the final state cut.} \label{factorized}
\end{center}
\end{figure}

In the following subsections, we give explicit constructions for the functions
participating in the factorization formula (\ref{factor}),
which satisfy the requirement of infrared safety,
and avoid double counting.
An  illustration of the cross section factorized into these functions is
shown in Fig.\ \ref{factorized}.
    As discussed above,
non-global logarithms will emerge when $ \varepsilon \nu$ becomes small
enough.

\subsection{The short-distance function}
\label{sec:sdf}

The power counting described in \cite{power} shows
that in  Feynman  gauge
the subdiagrams of Fig.\ \ref{factorized} that contribute to
$H$ in Eq.\ (\ref{factor}) at leading
power in $\varepsilon$ and
$\bar \varepsilon$ are connected to
each of the two jet subdiagrams
   by a single on-shell
quark  line,  along with a possible set of on-shell, collinear gluon lines
that carry scalar polarizations.
The hard subdiagram is
not connected directly to the soft subdiagram in any leading region.

The couplings of the scalar-polarized gluons that connect the jets with
short-distance subdiagrams
may be simplified with the help of Ward
identities (see, e.\,g.\ \cite{pQCD}).  At each order of
perturbation theory, the coupling of scalar-polarized gluons
from either jet to the short-distance function is equivalent
to their coupling to a path-ordered exponential of
the gauge field, oriented in any direction that is not
collinear to the jet.  Corrections are infrared safe, and
can be absorbed into the short-distance function.
Let $h(p_{J_c},\hat{n}_1,{\mathcal{A}})$ represent
the set of all short-distance contributions to diagrams
that couple any number of scalar-polarized gluons to the jets,
in the amplitude for the production of any final state.  The argument
   ${\mathcal{A}}$ stands for the fields that create the
scalar-polarized gluons linking the short-distance function
to the jets.
On a diagram-by-diagram basis, $h$ depends on
the momentum of each of the scalar-polarized gluons.
After the sum over all diagrams, however,
   we can make the replacement:
\be
h(p_{J_c},\hat{n}_1,{\mathcal{A}}^{({\rm q,\bar{q}})})
\rightarrow
\Phi^{({\rm \bar{q}})}_{\xi_2} (0,-\infty;0)\, h_2(p_{J_c},\hat{n}_1,\xi_c)\,
\Phi^{({\rm q})}_{\xi_1} (0,-\infty;0) \, ,
\ee
where $h_2$ is a short-distance function that depends only on the
total momenta $p_{J_1}$ and $p_{J_2}$.  It also depends on vectors $\xi_c$
that characterize  the path-ordered
exponentials $\Phi(0,-\infty;0)$:
\be
\Phi^{(\rm f)}_{\xi_c} (0,-\infty;0)  =  P e^{-i g \int_{-\infty}^{0} d
\lambda\; \xi_c \cdot {\mathcal{A}}^{(\rm f)} (\lambda \xi_c )}\, ,
\label{patho}
\ee
where the superscript $(\rm f)$ indicates that the vector potential
takes values in
representation $\rm f$, in our case the representation of a quark or
antiquark.
These operators will be associated with
gauge-invariant definitions of the jet functions below.
To avoid spurious collinear singularities,
we choose the vectors $\xi_c$, $c=1,2$, off the light cone.
   In the full cross section (\ref{trafosig}) the
$\xi_c$-dependence cancels, of course.

The dimensionless short-distance function $H=\left|h_2\right|^2$ in
Eq.\ (\ref{factor})
depends on $\sqrt{s}$ and $p_{J_c}\cdot \xi_c$, but not
on any variable that vanishes with $\varepsilon$ and $\bar{\varepsilon}$:
\be
H(p_{J_c},\xi_c,\hat{n}_1,\mu) =  H \left(
\frac{\sqrt{s}}{\mu},\frac{p_{J_c} \cdot
\xic}{\mu},\hat{n}_1,\as(\mu) \right)\, ,
\label{harddef}
\ee
where
\be
\xic \equiv \xi_c / \sqrt{|\xi_c^2|}\, .
\ee
Here we have observed that each diagram is independent of the overall
scale of the eikonal vector $\xi^\mu_c$.

\subsection{The jet functions}\label{jets}

The jet functions and the soft functions in Eq.\ (\ref{factor})
can be defined in terms of specific matrix elements, which
absorb the relevant contributions to leading regions in
the cross section, and which are infrared safe.
Their perturbative expansions
   specify the functions ${\cal S}$ and ${\cal J}_c$ of
Eq.\ (\ref{firstJdef}).  We begin with
our definition of the jet functions.

The jet functions, which absorb enhancements collinear to the two
outgoing particles produced in the primary hard scattering, can be
defined in terms of matrix elements
in a manner reminiscent of parton distribution or decay functions 
\cite{pdfdef}.
To be specific, we consider the
quark jet function:
\ba
\bar{J'}_c^\mu (\bar{\varepsilon}_{J_c},a,\mu)
&=&
   {2\over s}\, \frac{(2\pi)^6}{\Ncol} \; \sum\limits_{N_{J_c}}
{\rm Tr} \left[\gamma^\mu
\left<0 \left|
\Phi^{(\rm q)}_{\xi_c}{}^\dagger(0,-\infty;0) q(0) \right| N_{J_c} \right>
\right. \nonumber \\ 
& &\, \hspace{15mm} \times \, \left. \left< N_{J_c}\left| \bar{q}(0) 
\Phi^{(\rm q)}_{\xi_c}(0,-\infty;0) \right| 0 \right>\right] \nonumber \\
& &\, \hspace{15mm} \times \, \delta(\bar{\varepsilon}_{J_c}-\bar
f_c(N_{J_c},a)) \,
\delta\left({\sqrt{s}\over 2} - \o(N_{J_c})\right) \nonumber \\
& &\, \hspace{15mm} \times \, \delta^2(\hat n_c - \hat n(N_{J_c}))
\, ,
\label{jetdef}
\ea
where $\Ncol$ is the number of colors, and
where $\hat n_c$ denotes the direction of the momentum of
   jet $c$, Eq.\ (\ref{firstJdef}),
with $\hat{n}_2 = - \hat n_1$.
    $q$ is the quark field, $\Phi_{\xi_c}^{(\rm q)}(0,-\infty;0)$
a path-ordered exponential in the notation of (\ref{patho}),
and the trace is taken over color and Dirac indices.
We have chosen the normalization so that the
jet functions $\bar{J}'{}^\mu$ in (\ref{jetdef}) are dimensionless
and begin at lowest order with
\ba
\bar{J'}_c^\mu{}^{(0)} (\bar{\varepsilon}_{J_c},a,\mu) = \beta_{c}^\mu\,
\delta({\bar \varepsilon}_{J_c})\, ,
\label{norm}
\ea
with $\beta_c^\mu$
the lightlike velocities corresponding to the jet momenta in Eq.\ 
(\ref{lightlike}):
\be
\beta_1^\mu=\delta_{\mu +}\ , \quad \beta_2^\mu = \delta_{\mu -}\, .
\label{betadef}
\ee
The scalar jet functions of Eq.\
(\ref{firstJdef}) are now obtained by projecting out
the component of $J'_c{}^\mu$ in the jet direction:
\be
\bar{J}_c (\bar{\varepsilon}_{J_c},a,\mu) =  \bar{\beta}_c \cdot
\bar{J'}_c
(\bar{\varepsilon}_{J_c},a,\mu) = \delta(\bar{\varepsilon}_{J_c}) +{\cal
O}(\alpha_s)\, ,
\label{Jnorm}
\ee
where
$\bar{\beta}_1=\beta_2$, $\bar{\beta_2}=\beta_1$ are the lightlike vectors
in the directions opposite to $\beta_1$ and $\beta_2$, respectively.
By construction, the
$\bar{J}_c$ are linear in $\bar{\beta}_c$.

To resum the jet functions in the variables $\bar{\varepsilon}_{J_c}$,
it is convenient to reexpress the weight functions
(\ref{fbar2jet1}) in combinations of light-cone momentum
components that are invariant under boosts in the $x_3$ direction,
\ba
\bar{f}_1\left(N_{J_1},a\right) & = & \frac{1}{s^{1-a/2}}
\sum_{\hat n_i \in N_{J_1} }\
k_{i,\,\perp}^a\, \left(2 p_{J_1}^+k_i^-\right)^{1-a},
   \label{fbarLC1}
\nonumber \\
\bar{f}_2\left(N_{J_2},a \right) & = & \frac{1} {s^{1-a/2}}
\sum_{\hat n_i \in N_{J_2} }\
k_{i,\,\perp}^a\, \left(2 p_{J_2}^-k_i^+\right)^{1-a}.
\label{fbarLC2}
\ea
Here we have used the relation $\sqrt{s}/2 = \o_{J_c}$, valid for
both jets in the c.m.  At the same time, we make the identification,
\be
{1\over s} \delta\left({\sqrt{s}\over 2} - \o(N_{J_c})\right)\,
\delta^2(\hat n_c - \hat n(N_{J_c}))
=
{1\over 4}\, \delta^3\left(\vec p_{J_c}-\vec p(N_{J_c})\right)\, ,
\ee
which again holds in the c.m.\ frame.  The spatial components
of each $p_{J_c}$ are thus fixed.  Given that we are at small
$\bar{\varepsilon}_{J_c}$,
the jet functions may be thought of as  functions of
the light-like jet momenta $p_{J_c}^\mu$ of Eq.\ (\ref{lightlike})
and of $\bar{\varepsilon}_{J_c}$.  Because the vector jet function is
constructed
to be dimensionless, $\bar{J}'{}_c^\mu$ in Eq.\ (\ref{jetdef})
is proportional to $\beta_c$
rather than $p_{J_c}$.  Otherwise, it is free of explicit
$\beta_c$-dependence.

The jet functions can now be written in terms of boost-invariant
arguments,
homogeneous of degree zero in $\xi_c$:
\ba
\bar J_c\left(\bar{\varepsilon}_{J_c},a,\mu\right) &=&
\bar{\beta}_c{\,}_\mu \ \Bigg [\
   \beta_c^\mu
\, \bar{J}_c^{(1)} \left(\frac{p_{J_c} \cdot \xic}{\mu},
\bar{\varepsilon}_{J_c} \, \frac{\sqrt{s}}{\mu} \, \left(
\frac{\sqrt{s}}{2 p_{J_c} \cdot \xic} \right)^{1-a}, a,\as(\mu)
\right)
\nonumber
\\
&\ & \hspace{-5mm} +\  \, \frac{2\, \xi_c^\mu\ \beta_c\cdot \xi_c
}{\left|\xi_c\right|^2} \,
\bar{J}_c^{(2)} \left(\frac{p_{J_c} \cdot \xic}{\mu},
\bar{\varepsilon}_{J_c} \, \frac{\sqrt{s}}{\mu} \, \left(
\frac{\sqrt{s}}{2 p_{J_c} \cdot \xic} \right)^{1-a}, a, \as(\mu)
\right) \Bigg ]\, , \nonumber \\
\label{primedef2}
\ea
where ${\bar J}^{(1)}$ and ${\bar J}^{(2)}$ are independent functions, and
where we have suppressed possible dependence on
${\hat \xi}_{c, \, \perp}$.
For jet $c$, the weight $\bar{\varepsilon}_{J_c}$ is fixed by
$\delta(\bar{\varepsilon}_{J_c}-\bar{f}_c(N_{J_c},a))$,
where on the right-hand side of  the expression for the weight (\ref{fbarLC1}),
the sum over each particle's momentum involves the overall factor
$(2 p_{J_c}^\pm/\sqrt{s})^{1-a}$.
After integration over final states at fixed $\bar{\varepsilon}_{J_c}$,
the jet can thus depend on the vector $p_{J_c}^\mu$.
At the same time, it is easy to see from the definition
of the weight that $p_{J_c}^\mu$ can only appear
in the combination $(1/\bar{\varepsilon}_{J_c} \sqrt{s})^{1/(1-a)}\,
(2 p_{J_c}^\mu/\sqrt{s})$.
This vector can combine with $\xi_c$ to form an invariant, and
all $\xi_c$-dependence comes about in this way.

Expression (\ref{primedef2}) can be further simplified by noting that
\be
2\, \bar{\beta}_c \cdot  \xi_c\ {\beta}_c \cdot  \xi_c    =
\xi_c^2 + \xi_{c,\,\perp}^2\,   \, .
\ee
Choosing $\xi_{c,\,\perp} = 0$, we find a single combination,
\ba
\bar J_c\left(\bar{\varepsilon}_{J_c},a,\mu\right)
=
\bar J_c\left( \frac{p_{J_c} \cdot \xic}{\mu},
   \bar{\varepsilon}_{J_c} \, \frac{\sqrt{s}}{\mu} \, \left( \zeta_c
\right)^{1-a}, a, \as(\mu)
\right)\, ,
\ea
where, in the notation of Eq.\ (\ref{primedef2}), $\bar J_c=\bar
J_c^{(1)}+\bar J_c^{(2)}$, and we have defined
\ba
\zeta_c\equiv {\sqrt{s} \over 2 p_{J_c}\cdot \xic} \, .
\label{zetadef}
\ea
In these terms, the Laplace  moments of the jet function inherit
dependence on the
moment variable $\nu$ through
\ba
J_c \left(\nu,a,\mu \right)
&=& \int_0^\infty d\bar{\varepsilon}_{J_c} \; {\rm e}^{-\nu 
\bar{\varepsilon}_{J_c}}\, \bar
J_c\left(\bar{\varepsilon}_{J_c},a,\mu\right)
\nonumber\\
& \equiv &
J_c\left(\frac{p_{J_c} \cdot \xic}{\mu}, \frac{\sqrt{s}}{\mu \nu} \,
\left(\zeta_c \right)^{1-a},
a,\as(\mu) \right),
\label{primedef}
\ea
where the unbarred and barred quantities denote transformed and
untransformed functions, respectively. We have constructed
the jet functions to be independent of $\varepsilon$, since the radiation
into $\O$
is at wide angles from the jet axes and can therefore be completely
factored from the collinear radiation. This radiation at wide angles
is contained in the soft function, which will be defined below in
a manner that avoids double counting in the cross section.

\subsection{The soft function}

Given  the definitions for the jet functions in the
previous subsection, and the factorization (\ref{factor}),
we may in principle calculate  the soft function $S$
order by order in perturbation theory.
We can derive a more explicit definition of the soft function,
however, by relating it to an eikonal
analog of Eq.\  (\ref{factor}).

As reviewed  in  Refs.\ \cite{BKS1,pQCD},
soft radiation at wide angles from the jets decouples
from the collinear lines within the jet.
As a result, to
compute amplitudes for wide-angle radiation, the jets
may be replaced by nonabelian phases, or Wilson lines.
We therefore construct a dimensionless
quantity, $\sigma^{(\mbox{\tiny eik})}$,
in which gluons are radiated by path-ordered exponentials
$\Phi$, which mimic the color flow of outgoing quarks,
\be
\Phi^{({\mathrm f})}_{\beta_c} (\infty,0;x)  =  P e^{-i g \int_{0}^{\infty} d
\lambda \beta_c \cdot {\mathcal{A}}^{({\mathrm f})} (\lambda \beta_c + x )},
\ee
with $\beta_c$ a light-like velocity in either of the jet directions.
For the two-jet cross section at measured
$\varepsilon$ and $\bar \varepsilon_{\mbox{\tiny eik}}$, we define
\ba
\bar{\sigma}^{(\mbox{\tiny
eik})}\left(\varepsilon,\bar{\varepsilon}_{\mbox{\tiny eik}}, a ,\mu
\right)\!\!\!
&\!\! \equiv  \!\!\!& \!\!\!{1\over \Ncol}\
\sum_{N_{\mbox{\tiny eik}}} \left< 0
\left| \Phi^{(\bar {\rm q})}_{\beta_2}{}^\dagger(\infty,0;0)
\Phi^{(\rm q)}_{\beta_1}{}^\dagger(\infty,0;0)
\right| {N_{\mbox{\tiny eik}}} \right>
\nonumber \\
& \ & \hspace{5mm} \times\;
\left< N_{\mbox{\tiny eik}}
\left| \Phi^{(\rm q)}_{\beta_1}(\infty,0;0)
\Phi_{\beta_2}^{(\bar{\rm q})}(\infty,0;0)
\right| 0 \right>  \nonumber \\
& \ & \hspace{5mm} \times\; 
\delta\left(\varepsilon - f(N_{\mbox{\tiny eik}})\right)
\delta\left(\bar{\varepsilon}_{\mbox{\tiny eik}} -
\bar{f}(N_{\mbox{\tiny eik}},a) \right)
\nonumber\\
&=& \delta(\varepsilon)\,  \delta(\bar{\varepsilon}_{\mbox{\tiny
eik}}) +{\cal O}(\alpha_s)\,
.
\label{eikdef}
\ea
  The sum is over all final states $N_{\mbox{\tiny eik}}$ in
the eikonal cross section. The renormalization
scale in this cross section, which will also serve as a factorization
scale, is denoted $\mu$.  Here the event shape function
$\bar{\varepsilon}_{\mbox{\tiny eik}}$
is defined by $\bar{f}(N_{\rm eik},a)$ as in Eqs.\  (\ref{barfdef}) and
(\ref{2jetf}),
distinguishing between the hemispheres around the jets.
As usual, $\Ncol$  is the number of colors,
and a trace over color is understood.

The eikonal cross section (\ref{eikdef}) models the soft
radiation away from the jets, including the radiation into $\O$,
accurately.
It also contains enhancements
for configurations collinear to the jets, which, however,
are  already taken into account by the partonic jet
functions in (\ref{factor}).  Indeed, (\ref{eikdef}) does not reproduce
the
partonic cross section accurately for collinear radiation.
It is also easy to verify at lowest
order that even at fixed $\bar{\varepsilon}_{\mbox{\tiny eik}}$
the eikonal cross section (\ref{eikdef}) is
ultraviolet divergent in dimensional regularization,
unless we also impose a cutoff on
the energy of real gluon emission collinear to $\beta_1$
or $\beta_2$.

The construction of the soft function $S$
from $\bar{\sigma}^{(\mbox{\tiny eik})}$ is nevertheless possible
   because the eikonal cross
section (\ref{eikdef}) factorizes in the same manner
as the cross section itself, into eikonal jet functions
and a soft function.  The essential point \cite{KOS} is that
the soft function in the factorized eikonal cross section
is the same as in the original cross section (\ref{factor}).
The eikonal jets organize all collinear enhancements
in (\ref{eikdef}), including the spurious ultraviolet
divergences.  These eikonal jet functions are defined
analogously to their partonic counterparts, Eq.\ (\ref{jetdef}),
but now with ordered exponentials replacing the quark fields,
\ba
\bar{J}_c^{(\mbox{\tiny eik})}\left(\bar{\varepsilon}_c,a,\mu \right)
& \equiv  & {1\over \Ncol}\,
\sum_{N_c^{(\mbox{\tiny eik})}}
\left<0 \left| \Phi^{({\mathrm f}_c)}_{\xi_c}{}^\dagger(0,-\infty;0)
\Phi_{\beta_c}^{({\mathrm f}_c)}{}^\dagger(\infty,0;0) \right|
N_c^{(\mbox{\tiny eik})} \right>
\nonumber \\
& \ & \hspace{5mm} \times \, \left< N_c^{(\mbox{\tiny eik})} \left|
\Phi^{({\mathrm f}_c)}_{\beta_c}(\infty,0;0)
\Phi_{\xi_c}^{({\mathrm f}_c)}(0,-\infty;0) \right|  0 \right> 
\nonumber \\ 
& \ & \hspace{5mm} \times \, \delta\left(\bar{\varepsilon}_c -  
\bar{f}_c(N_c^{(\mbox{\tiny eik})},a) \right)
\nonumber\\
&=& \delta(\bar{\varepsilon}_c) +{\cal O}(\alpha_s)\, ,
\label{eikjetdef}
\ea
where ${\mathrm f}_c$ is a quark or antiquark, and where the
trace over color is understood.
The weight functions are given as above, by Eq.\ (\ref{fbar2jet1}),
with the sum over particles in all directions.

In terms of the eikonal jets, the eikonal cross section (\ref{eikdef})
factorizes as
\ba
\bar{\sigma}^{(\mbox{\tiny
eik})}\left(\varepsilon,\bar{\varepsilon}_{\mbox{\tiny eik}},a,\mu \right)
& \equiv  &
\int  d \bar{\varepsilon}_s \,
\bar{S}\left(\varepsilon,\bar{\varepsilon}_s,a,\mu \right)
\prod\limits_{c = 1}^2 \int d \bar{\varepsilon}_c \,
\bar{J}_c^{(\mbox{\tiny eik})}\left(\bar{\varepsilon}_c,a,\mu \right)
\nonumber \\
& \ & \hspace{5mm} \times \, \delta \left(\bar{\varepsilon}_{\mbox{\tiny eik}} 
- \bar{\varepsilon}_s-\bar{\varepsilon}_1-\bar{\varepsilon}_2 \right),
\label{eikfact}
\ea
where we pick the factorization scale equal to the renormalization scale
$\mu$.  As for the full cross section, the convolution
in (\ref{eikfact}) is simplified by a Laplace
transformation (\ref{primedef}) with respect to
$\bar{\varepsilon}_{\mbox{\tiny eik}}$,
which allows us to solve for the soft function as
\be
S \left(\varepsilon,\nu,a,\mu\right) =
\frac{\sigma^{(\mbox{\tiny eik})}\left(\varepsilon,\nu,a,\mu \right) }
{\prod\limits_{c = 1}^2 J_c^{(\mbox{\tiny eik})}\left(\nu,a,\mu\right) }
=\delta(\varepsilon)+{\cal O}(\alpha_s)\, .
\label{s0}
\ee
In this ratio, collinear logarithms
in $\nu$ and the unphysical ultraviolet divergences and their
associated cutoff dependence cancel between the eikonal
cross section and the eikonal jets, leaving a soft
function that is entirely free of collinear enhancements.
The soft function retains $\nu$-dependence through soft
emission, which is also restricted by the weight function
$\varepsilon$.  In addition, because soft radiation
within the eikonal jets can be factored from its collinear
radiation, just as in the partonic jets, all
logarithms in $\nu$ associated with wide-angle radiation
are identical between the partonic and eikonal jets,
and factor from logarithmic corrections associated with
collinear radiation in both cases.
As a result, the inverse eikonal jet
functions cancel contributions from the wide-angle soft radiation of
the partonic jets in the
transformed cross section (\ref{trafosig}).

Given the definition
of the energy flow weight function $f$, Eq.\ (\ref{eflowdef}), the 
soft function is not
boost invariant.  In addition, because it is  free of
collinear logs, it can have at most a single
logarithm per loop.  Its dependence on $\varepsilon$
is therefore only through ratios of the dimensional
quantities $\varepsilon\sqrt{s}$  with the renormalization
(factorization) scale.

As in the case of the partonic jets, Eq.\ (\ref{primedef}),  we need to
identify
the variable through which $\nu$ appears in the soft
function.
We note that dependence on the velocity vectors $\beta_c$
and the factorization vectors $\xi_c$ must be scale invariant
in each, since they arise only from eikonal lines and vertices.
The eikonal jet functions cannot depend explicitly on the scale-less, lightlike
eikonal velocities $\beta_c$, and  $\sigma^{\rm (eik)}$
is independent of the $\xi_c$.  Dependence on the factorization
vectors $\xi_c$ enters only
   through the weight functions, (\ref{fbarLC1}) for the eikonal
jets, in a manner analogous to the case of the partonic jets. This results in
a dependence on $(\zeta_c)^{1-a}$, as above, with $\zeta_c$ defined in
Eq. (\ref{zetadef}).  In summary, we may characterize the arguments 
of the soft function in
transform space as
\be
S\left(\varepsilon,\nu,a,\mu \right) =
S\left(\frac{\varepsilon \sqrt{s}}{\mu},\varepsilon\nu,
\frac{\sqrt{s}}{\mu \nu} \, \left( \zeta_c \right)^{1-a},
a, \as(\mu)
\right)\, .
\label{Sargs}
\ee

\section{Resummation}

We may summarize the results of the previous
section by rewriting the transform of the factorized cross section
(\ref{trafosig})
in terms of the hard, jet and soft functions identified above,
which depend on the kinematic variables and the moment $\nu$
according to Eqs.\ (\ref{harddef}), (\ref{primedef}) and (\ref{Sargs})
respectively,
\ba
\frac{d \sigma \left(\varepsilon,\nu,s ,a\right)}{d\varepsilon\,d
\hat{n}_1 }
&=&
   {d \sigma_0 \over d\hat{n}_1}\ H \left(
\frac{\sqrt{s}}{\mu},\frac{p_{J_c} \cdot \xic}{\mu},\hat{n}_1,\as(\mu)
\right) \nonumber \\
& \ & \hspace{15mm} \times\ \prod_{c=1}^2\;
J_c\left(\frac{p_{J_c} \cdot \xic}{\mu}, \frac{\sqrt{s}}{\mu \nu} \,
(\zeta_c)^{1-a},
a,\as(\mu) \right)\,
\nonumber \\
& \ &\hspace{15mm} \times\
    S\left(\frac{\varepsilon \sqrt{s}}{\mu}, \varepsilon \nu,
\frac{\sqrt{s}}{\mu \nu}\left(\zeta_c \right)^{1-a},
a, \as(\mu)
\right)\, .
    \label{factorcom}
\ea
The natural scale for the strong coupling
in the short-distance function $H$ is $\sqrt{s}/2$.
Setting $\mu = \sqrt{s}/2$, however, introduces large logarithms of
$\varepsilon$ in the soft function and large logarithms of $\nu$ in both the
soft and jet functions.
The purpose of this section is to control these logarithms by
the identification and solution of  renormalization
group and evolution equations.

The information necessary to perform the resummations is already present in
  the factorization (\ref{factorcom}).
The cross section itself is independent of the factorization scale
\ba
\mu \frac{d}{d \mu} \frac{d \sigma \left(\varepsilon,\nu,s, a \right)}{d
\varepsilon d\hat{n}_1 }
& = & 0\, ,
\label{muev} \
\ea
   and of the choice of
the eikonal directions, $\xic$, used in the factorization,
\ba
\frac{\partial}{\partial \ln  \left(p_{J_c} \cdot \hat{\xi}_c\right) }
\frac{d \sigma \left(\varepsilon,\nu,s, a \right)}{d \varepsilon d\hat{n}_1 }
& = & 0\, .
\label{xiev}
\ea
The arguments of this section
closely follow the analysis of Ref.\ \cite{CLS}.
We will see that the dependence of
jet and soft functions on the parameter $a$
that characterizes
the event shapes (3) is reflected
in the resummed correlations, so that
the relationship between correlations with
different values of $a$ is both calculable
and nontrivial.

\subsection{Energy flow}

As a first step, we use the renormalization group equation (\ref{muev})
to organize dependence on the energy flow variable $\varepsilon$.
Applying Eq.\ (\ref{muev}) to the factorized correlation (\ref{factorcom}), we
derive the following consistency conditions, which are themselves
renormalization
group equations:
\ba
\mu \frac{d}{d \mu}\;
\ln\, S\left(\frac{\varepsilon \sqrt{s}}{\mu}, \varepsilon \nu,
\frac{\sqrt{s}}{\mu \nu} (\zeta_c)^{1-a},
a, \as(\mu) \right)
& = & -
\gamma_s\left(\as(\mu)\right),
\label{softmu}
\\
\mu \frac{d}{d \mu}\;
\ln\, J_c\left(\frac{ p_{J_c} \cdot \xic }{\mu },
   \frac{\sqrt{s}}{\mu \nu} \, (\zeta_c)^{1-a} ,
a,\as(\mu) \right) & = & - \gamma_{J_c}\left(\as(\mu)\right),
\label{jetmu}
\\
\mu \frac{d}{d \mu}\; \ln\,  H\left( \frac{\sqrt{s}}{\mu},\frac{p_{J_c} \cdot
\xic}{\mu},\hat{n}_1,\as(\mu) \right)
&=& \gamma_s\left(\alpha_s(\mu)\right)
+ \sum_{c=1}^2\gamma_{J_c}\left(\alpha_s(\mu)\right)\, . \nonumber \\
\label{Hmu}
\ea
The anomalous dimensions $\gamma_d$, $d=s,\, J_c$ can
depend only on variables held in common between  at least two
of the functions.  Because each function is infrared safe,
while ultraviolet divergences are present only in virtual
diagrams, the anomalous dimensions cannot depend on
the parameters $\nu$, $\varepsilon$ or $a$.  This leaves
as arguments of the $\gamma_d$ only
the  coupling $\as(\mu)$, which we exhibit, and $\zeta_c$, which
we suppress for now.

Solving Eqs. (\ref{softmu})
and (\ref{jetmu}) we find
\ba
S\left(\frac{\varepsilon \sqrt{s}}{\mu}, \varepsilon \nu,
\frac{\sqrt{s}}{\mu \nu} \left(\zeta_c \right)^{1-a},
a, \as(\mu) \right)
& = &
S\left(\frac{\varepsilon \sqrt{s}}{\mu_0}, \varepsilon \nu,
\frac{\sqrt{s}}{\mu_0 \nu} \left( \zeta_c \right)^{1-a},
a, \as(\mu_0) \right) \nonumber \\ 
& \times & e^{-\int\limits_{\mu_0}^\mu \frac{d
\lambda}{\lambda} \gamma_s\left(\as(\lambda)\right)}, \\
\label{softevol}
J_c \left(\frac{ p_{J_c} \cdot \xic }{\mu},
   \frac{\sqrt{s}}{\mu \nu} \, \left(\zeta_c \right)^{1-a} , a, \as(\mu)
\right)
& = &
J_c \left( \frac{p_{J_c} \cdot \xic }{\tilde{\mu}_0 },
\frac{\sqrt{s}}{\tilde{\mu}_0 \nu} \, \left(\zeta_c \right)^{1-a} ,
a,\as(\tilde{\mu}_0) \right) \nonumber \\
& \times & \,e^{-\int\limits_{\tilde{\mu}_0}^\mu \frac{d \lambda}{\lambda}
\gamma_{J_c}\left(\as(\lambda)\right)}\, , 
\label{jetevol}
\ea
for the soft and jet functions.  As suggested above, we will eventually pick
$\mu\sim \sqrt{s}$
to avoid large logs in $H$.
Using these expressions in Eq. (\ref{factorcom}) we can avoid
logarithms of $\varepsilon$ or $\nu$ in the soft function, by evolving from
$\mu_0 = \varepsilon \sqrt{s}$ to the factorization scale $\mu \sim \sqrt{s}$.
No choice of $\tilde{\mu}_0$, however, controls all logarithms of $\nu$ in
the jet functions.  Leaving $\tilde \mu_0$ free, we find for the
cross section (\ref{factorcom}) the
intermediate result
\ba
    \label{resume}
\frac{d \sigma \left(\varepsilon, \nu,s ,a\right)}{d\varepsilon \, d
\hat{n}_1 }
&=&
{d \sigma_0 \over d\hat{n}_1}\ H \left( \frac{\sqrt{s}}{\mu},\frac{p_{J_c}
\cdot \xic}{\mu},\hat{n}_1,\as(\mu) \right)\,
\nonumber\\
&\ & \hspace{-15mm} \times\; S\left(1, \varepsilon \nu, (\zeta_c)^{1-a}, a
,\as(\varepsilon \sqrt{s})
\right)\,
\exp\left\{ -\int\limits_{\varepsilon \sqrt{s}}^{\mu} \frac{d
\lambda}{\lambda} \, \gamma_s\left(\as(\lambda)\right)\right\}
\\
&\ & \hspace{-15mm} \times\;
J_c \left( \frac{p_{J_c} \cdot \xic }{\tilde{\mu}_0 },
\frac{\sqrt{s}}{\tilde{\mu}_0 \nu} \, \left(\zeta_c \right)^{1-a} ,
a,\as(\tilde{\mu}_0) \right)
    \,\exp\left\{-\int\limits_{\tilde{\mu}_0}^\mu \frac{d \lambda}{\lambda}
\gamma_{J_c}\left(\as(\lambda)\right)\right\} \, . \nonumber
\ea
We have avoided introducing logarithms of $\varepsilon$ into the jet functions,
which originally only depend on $\nu$, by evolving the soft and the
jet functions independently.
The choice of $\mu_0=\varepsilon\sqrt{s}$ or
$\sqrt{s}/\nu$ for the soft function is to some extent a matter of
convenience,
since the two choices differ by logarithms of $\varepsilon\nu$.
In general, if we choose $\mu_0=\sqrt{s}/\nu$, logarithms
of $\varepsilon\nu$ will appear multiplied by coefficients that reflect
the
size of region $\O$.  An example is Eq.\ (\ref{ps0}) above.
When $\O$ has a small angular size, $\mu_0=\sqrt{s}/\nu$ is generally the
more natural choice, since then logarithms in $\varepsilon\nu$ will
enter with small weights.  In contrast, when $\O$ grows to cover most
angular directions, as in
the study of rapidity gaps \cite{Oderda}, it is more
natural to choose $\mu_0 = \varepsilon\sqrt{s}$.

\subsection{Event shape transform}

The remaining unorganized ``large" logarithms in (\ref{resume}),
are in the jet functions,
which we will resum by using the consistency equation (\ref{xiev}).
The requirement that the cross section be independent of $p_{J_c}\cdot
\hat{\xi}_c$
implies that the jet, soft and hard functions obey equations
analogous to (\ref{softmu})--(\ref{Hmu}), again in terms of the variables
that they hold in common \cite{CLS}.  The same results may
be derived following the method of Collins and Soper
   \cite{ColSop}, by defining the jets in an axial gauge,
and then studying their variations under boosts.

For our purposes, only the equation satisfied by the
jet functions \cite{ColSop,CLS} is necessary,
\ba
    \frac{\partial }{\partial \ln \left(p_{J_c} \cdot \xic\right)}
\ln\ J_c \left( \frac{p_{J_c} \cdot \xic}{\mu},
\frac{\sqrt{s}}{\mu \nu} \, (\zeta_c)^{1-a} ,a,\as(\mu)
\right)
& \ & \nonumber
\\
&\ & \hspace{-70mm} =
    K_c\left(\frac{\sqrt{s}}{\mu\,
\nu}(\zeta_c)^{1-a},
a,\as(\mu)
\right)   +  G_c\left(\frac{p_{J_c} \cdot \xic}{\mu},\as(\mu)\right)   \, .
    \label{KGend}
\ea
The functions $K_c$ and $G_c$
compensate
the $\xi_c$-dependence of the soft and hard functions, respectively,
which determines the kinematic variables upon which they may depend.
In particular, notice the combination of $\nu$- and $\xi_c$-dependence
required by the arguments of the jet function, Eq.\ (\ref{primedef}).

Since the definition of our jet functions (\ref{jetdef}) is gauge
invariant,
we can derive the kernels $K_c$ and $G_c$ by an explicit  computation of
   ${\partial\, J_c}/{\partial \ln
\left(p_{J_c} \cdot \hat{\xi}_c\right)}$ in any gauge.
The multiplicative renormalizability of the jet function, Eq. (\ref{jetmu}),
with an anomalous dimension that is independent of $p_{J_c}\cdot \xic$
ensures that the right-hand side of Eq.\ (\ref{KGend}) is
a renormalization-group invariant.  Thus, $K_c+G_c$ are renormalized
additively, and satisfy \cite{ColSop}
\ba
\mu \frac{d}{d \mu}\
K_c\left(\frac{\sqrt{s}}{\mu\, \nu}\left(\zeta_c\right)^{1-a},
a,\as(\mu) \right) & = & - \gamma_{K_c}
\left(\as(\mu)\right),
\nonumber\\
\mu \frac{d}{d \mu}G_c\left(\frac{p_{J_c} \cdot \xic }{\mu},\as(\mu)\right)
   & = &  \gamma_{K_c}
\left(\as(\mu)\right) \, .
\label{Gevol}
\ea
Since $G_c$ and hence
$\gamma_{K_c}$, may be computed from
virtual diagrams, they do not depend on $a$, and $\gamma_{K_c}$ is the
universal Sudakov anomalous dimension \cite{ColSop,sudgam}.

With the help of these evolution equations, the terms $K_c$ and $G_c$
in Eq. (\ref{KGend}) can be reexpressed as
\cite{cssdy}
\ba
    K_c\left(\frac{\sqrt{s}}{\mu\, \nu}\left(\zeta_c\right)^{1-a},a,\as(\mu)
\right)   +  G_c\left(\frac{p_{J_c} \cdot \xic}{\mu},\as(\mu)\right)
&\ &  \nonumber \\
&\ & \hspace{-90mm} =
K_c\left(\frac{1}{c_1},a,\as\left(c_1 \,
\frac{\sqrt{s}}{\nu}\left(\zeta_c\right)^{1-a}\right)
\right)
+  G_c\left(\frac{1}{c_2},\as\left(c_2 \, p_{J_c} \cdot \xic \right) \right)
\nonumber \\
& \ & \hspace{-90mm} - \, \int\limits_{c_1 {\sqrt{s}}\, 
\left(\zeta_c\right)^{1-a}/{\nu} }^{ c_2\,p_{J_c} \cdot \xic }
\frac{d  \lambda'}{\lambda'} \gamma_{K_c}\left(\as\left(\lambda'\right)
\right) \nonumber \\
&\ & \hspace{-90mm} =
- B'_c\left(c_1,c_2,a,  \as\left(c_2 \, p_{J_c} \cdot \xic \right) \right)
-
2 \int\limits_{c_1 {\sqrt{s}}\, \left(\zeta_c\right)^{1-a}/{\nu} }^{ c_2 \,
p_{J_c}
\cdot \xic }
\frac{d  \lambda'}{\lambda'} A'_c\left(c_1, a, \as\left(\lambda'\right)
\right)\, , \nonumber \\
\label{ABabbr}
\ea
where in the second equality we have shifted the argument of
the running coupling in $K_c$, and have introduced the notation
\ba
B'_c\left(c_1,c_2,a, \as\left(\mu \right) \right)
& \equiv & -
K_c\left(\frac{1}{c_1},a, \as\left(\mu \right)  \right) -
G_c\left(\frac{1}{c_2} ,\as\left(\mu \right)\right),
\nonumber \\
2 A'_c\left( c_1, a, \as\left(\mu \right) \right) & \equiv &  \gamma_{K_c}
\left(\as(\mu) \right) + \beta(g(\mu)) \frac{\partial}{\partial
g(\mu)} K_c\left(\frac{1}{c_1},a, \as(\mu)\right). \nonumber \\
\label{ABdef}
\ea
The primes on the functions $A'_c$ and $B'_c$ are to distinguish
these anomalous dimensions from their somewhat more familiar versions
given below.

The solution to Eq. (\ref{KGend}) with $\mu = \tilde{\mu}_0$ is
\ba
J_c \left( \frac{p_{J_c} \cdot \xic }{\tilde{\mu}_0},
   \frac{\sqrt{s}}{\tilde{\mu}_0 \nu} \, \left(\zeta_c
   \right)^{1-a} ,a,\as(\tilde{\mu}_0)
\right)
&=&
J_c \left( \frac{\sqrt{s}}{2 \,\zeta_0 \,\tilde{\mu}_0},
   \frac{\sqrt{s}}{\tilde{\mu}_0 \nu} \, \left(\zeta_0
   \right)^{1-a} ,a,\as(\tilde{\mu}_0)
\right)   \nonumber \\
&\ & \hspace{-80mm}
    \times \, \exp \left\{ -\int\limits_{\sqrt{s}/(2\zeta_0) }^{p_{J_c} \cdot
\xic}
    \frac{d \lambda}{\lambda} \left[B'_c\left(c_1,c_2,a,
\as\left(c_2 \lambda \right) \right)  +  2 \int\limits_{c_1 \frac{s^{1-a/2}
}{ \nu
(2\,\lambda)^{1-a} } }^{c_2\, \lambda}\frac{d \lambda'}{\lambda'} A'_c\left(
c_1,
a,\as\left(\lambda'\right) \right) \right] \right\}\, , \nonumber \\
   \label{orgsol}
\ea
where we evolve from $\sqrt{s}/(2\,\zeta_0)$ to $p_{J_c} \cdot \xic =
\sqrt{s}/(2 \,\zeta_c)$ (see Eq.\ (\ref{zetadef})) with
\be
\zeta_0 = \left(\frac{\nu}{2}\right)^{1/(2-a)}. \label{zeta0}
\ee
After combining Eqs.\ (\ref{jetevol}) and (\ref{orgsol}),
the choice $\tilde{\mu}_0 = \sqrt{s}/(2\zeta_0) = \frac{\sqrt{s}}{\nu}
(\zeta_0)^{1-a}$
   allows us to control
all large logarithms in
the jet functions simultaneously:
\ba
J_c \left( \frac{p_{J_c} \cdot \xic}{\mu},
\frac{\sqrt{s}}{\mu \nu} \, (\zeta_c)^{1-a} ,a,\as(\mu)\right)
&=&
J_c \left(
1, 1,a,\as\left(\frac{\sqrt{s}}{2 \, \zeta_0} \right) \right)
\nonumber \\
&\ & \hspace{-70mm} 
\times \, \exp \left\{-\int\limits_{\sqrt{s}/(2 \zeta_0)}^\mu
\frac{d\lambda}{\lambda} \gamma_{J_c} \left(\as(\lambda)\right) \right\}
\, \nonumber \\
&\ & \hspace{-70mm}
    \times \, \exp \left\{ -\int\limits_{\sqrt{s}/(2\, \zeta_0) }^{p_{J_c}
\cdot \xic}
    \frac{d \lambda}{\lambda} \left[B'_c\left(c_1,c_2,a, \as\left(c_2 \lambda
\right) \right) + 2 \int\limits_{c_1 \frac{s^{1-a/2} }{ \nu
(2\,\lambda)^{1-a} } }^{c_2\, \lambda}\frac{d \lambda'}{\lambda'} A'_c\left(
c_1,
a,\as\left(\lambda'\right) \right) \right] \right\}\, . \nonumber \\
   \label{jetxiend}
\ea
As observed above, we treat $a$ as a fixed parameter, with $|a|$ 
small compared to
$\ln\, (1/\varepsilon)$ and $\ln\nu$.

\subsection{The resummed correlation}

Using Eq. (\ref{jetxiend}) in (\ref{resume}), and setting $\mu = \sqrt{s}/2$,
we find a fully resummed form for the correlation,
\ba
\frac{d \sigma \left(\varepsilon, \nu,s,a \right)}{d \varepsilon\, d
\hat{n}_1 }
&=&
{d \sigma_0 \over d\hat{n}_1}\ H \left(\frac{2 \, p_{J_c}
\cdot \xic}{\sqrt{s}},\hat{n}_1,\as\left(\frac{\sqrt{s}}{2}\right) \right)\,
   \nonumber \\
& \ & \hspace*{-2cm}
\times \, S\left(1, \varepsilon \nu, (\zeta_c)^{1-a}, a,\as(\varepsilon
\sqrt{s} ) \right)
\, \exp \left\{  - \int\limits_{\varepsilon \sqrt{s}}^{\sqrt{s}/2} \frac{d
\lambda}{\lambda} \gamma_s\left(\as(\lambda)\right) \right\} \nonumber \\
& \ & \hspace*{-2cm}
\times \, \prod_{c=1}^2\, J_c \left(1,1,a,\as\left(\frac{\sqrt{s}}{2 \,
\zeta_0}\right) \right)
    \exp \left\{ - \int\limits_{\sqrt{s}/(2 \, \zeta_0)}^{\sqrt{s}/2}
   \frac{d \lambda}{\lambda} \gamma_{J_c} \left(\as(\lambda)\right) \right\}
\nonumber \\
& \ & \hspace*{-4cm}
\times \, \exp \left\{ -\int\limits_{\sqrt{s}/(2\, \zeta_0) }^{p_{J_c} \cdot
\xic}
    \frac{d \lambda}{\lambda} \left[B'_c\left(c_1,c_2,a,
\as\left(c_2 \lambda \right) \right)  +  2 \int\limits_{c_1 \frac{s^{1-a/2}
}{ \nu
(2\,\lambda)^{1-a} } }^{c_2\, \lambda}\frac{d \lambda'}{\lambda'} A'_c\left(
c_1,
a,\as\left(\lambda'\right) \right) \right] \right\}\, . \nonumber \\
& & \label{evolend}
\ea

Alternatively, we can combine all jet-related exponents in Eq.
(\ref{evolend}) in the correlation.
As we will verify below in Section
\ref{gauge}, the cross section is independent of the choice of $\xi_c$.
As a result, we can choose
\be
p_{J_c} \cdot \xic = \frac{\sqrt{s}}{2}\, .
\label{xiid}
\ee
This choice allows us to combine $\gamma_{J_c}$ and
$B'_c$ in Eq. (\ref{evolend}),
\ba
\frac{d \sigma \left(\varepsilon, \nu,s,a \right)}{d \varepsilon\, d
\hat{n}_1 }
&=&
{d \sigma_0 \over d\hat{n}_1}\ H \left(1,
\hat{n}_1,\as\left(\frac{\sqrt{s}}{2}\right) \right)\,
   \nonumber \\
& \ & \hspace*{-3cm}
\times \, S\left(1, \varepsilon \nu, 1, a,\as(\varepsilon \sqrt{s} ) \right)
\, \exp \left\{  - \int\limits_{\varepsilon \sqrt{s}}^{\sqrt{s}/2} \frac{d
\lambda}{\lambda} \gamma_s\left(\as(\lambda)\right) \right\} \,
\prod_{c=1}^2\, J_c \left(1,1,a,\as\left(\frac{\sqrt{s}}{2 \, \zeta_0}\right)
\right) \nonumber \\
& \ & \hspace*{-3cm}
\times \, \exp \left\{ -\int\limits_{\sqrt{s}/(2\, \zeta_0) }^{\sqrt{s}/2}
    \frac{d \lambda}{\lambda} \left[ \gamma_{J_c} \left(\as(\lambda)\right) +
B'_c\left(c_1,c_2,a,
\as\left(c_2 \lambda \right) \right) \right. \right. \nonumber \\ 
& \ & \hspace{-2cm} \left. \left. + 2 \int\limits_{c_1 \frac{s^{1-a/2}}{ \nu
(2\,\lambda)^{1-a} } }^{c_2\, \lambda}\frac{d \lambda'}{\lambda'} A'_c\left(
c_1,
a,\as\left(\lambda'\right) \right) \right] \right\}\, , \nonumber \\
& & \label{evolendnoxi}
\ea
with $\zeta_0$ given by Eq. (\ref{zeta0}).

In Eqs. (\ref{evolend}) and (\ref{evolendnoxi}), the energy flow
$\varepsilon$ appears at the level
of one logarithm per loop, in $S$, in the first exponent.
Leading logarithms of $\varepsilon$ are
therefore resummed by knowledge of $\gamma_s^{(1)}$,
the one-loop soft anomalous dimension, where we employ the
standard notation,
\ba
\gamma_s(\as) = \sum_{n=0}^\infty \gamma_s^{(n)}\ \left({\as\over
\pi}\right)^n
\ea
for any expansion in $\as$.
At the same time, $\nu$ appears in up to two logarithms per loop,
characteristic of conventional Sudakov resummation.  To control
$\nu$-dependence at the same level as $\varepsilon$-dependence, it is
natural to work to next-to-leading logarithm in $\nu$,
by which we mean the level $\as^k\, \ln^k\nu$ in the exponent.  As usual, this
requires one loop in
$B'_c$ and $\gamma_{J_c}$, and two loops in the
Sudakov anomalous dimension $A'_c$, Eq.\ (\ref{ABdef}).
These functions are straightforward to calculate from their definitions
given in the previous sections. Only the soft
function $S$ in Eqs.\ (\ref{evolend}) and (\ref{evolendnoxi})
contains information on the geometry of $\O$. The exponents are
partially process-dependent, but geometry-independent.  In
Section \ref{sec:res}, we will derive
explicit expressions for these quantities, suitable for
resummation to leading logarithm in $\varepsilon$ and next-to-leading
logarithm in $\nu$.

\subsection{The inclusive event shape}

It is also of interest
to consider the cross section for $e^+e^-$-annihilation into two jets
without fixing the energy of radiation into $\O$, but with the final state
radiation into all of phase space weighted according to Eq.\
(\ref{2jetf}), schematically
\be
e^+ + e^- \rightarrow J_1(p_{J_1}, \bar{f}_{\bar{\O}_1}) + J_2(p_{J_2},
\bar{f}_{\bar{\O}_2})\, ,
\ee
where $\bar{\O}_1$ and  $\bar{\O}_2$ cover the entire sphere.
This cross section can be factorized and resummed in a completely
analogous manner. The
final state is a convolution in the contributions
of the jet and soft functions to $\bar{\varepsilon}$ as in Eq.
(\ref{factor}),
but with no separate restriction on energy flow into $\O$.
All particles contribute to the event shape.
We obtain an expression very analogous to
Eq.\ (\ref{evolend}) for this inclusive event shape in transform space,
which can be written in terms of the same jet functions
as before, and a new function $S^{\rm incl}$ for soft radiation as:
\ba
\frac{d \sigma^{\rm incl} \left(\nu,s,a \right)}{ d \hat{n}_1 }
&=&
{d \sigma_0 \over d\hat{n}_1}\ H \left(\frac{2 p_{J_c}
\cdot \xic}{\sqrt{s} },\hat{n}_1,\as(\sqrt{s}/2) \right)\,
\nonumber\\
&\  &  \hspace*{-2cm} \times\  S^{\rm incl}\left((\zeta_c)^{1-a},
a,\as\left(\frac{\sqrt{s}}{\nu}
\right) \right) \,
\exp \left\{  - \int\limits_{\sqrt{s} / \nu }^{\sqrt{s}/2} \frac{d
\lambda}{\lambda} \gamma_s\left(\as(\lambda)\right) \right\} \nonumber \\
&\  &  \hspace*{-2cm} \times\ \prod_{c=1}^2\, J_c
\left(1,1,a,\as\left(\frac{\sqrt{s}}{2 \, \zeta_0} \right) \right)
    \exp \left\{ - \int\limits_{\sqrt{s} / (2 \, \zeta_0)}^{\sqrt{s}/2}
\frac{d \lambda}{\lambda}  \gamma_{J_c}\left(\as(\lambda)\right) \right\}
\nonumber \\
& \ & \hspace*{-3.5cm}
\times \, \exp \left\{ -\int\limits_{\sqrt{s}/(2\, \zeta_0) }^{p_{J_c} \cdot
\xic}
    \frac{d \lambda}{\lambda} \left[B'_c\left(c_1,c_2,a,
\as\left(c_2 \lambda \right) \right)  +  2 \int\limits_{c_1 \frac{s^{1-a/2}
}{ \nu
(2\,\lambda)^{1-a} } }^{c_2\, \lambda}\frac{d \lambda'}{\lambda'} A'_c\left(
c_1,
a,\as\left(\lambda'\right) \right) \right] \right\}\, . \nonumber \\
    \label{globalend}
\ea
Here the soft function $S^{\rm incl}=1 + {\cal O}(\alpha_s)$.
The double-logarithmic dependence of the shape transform is
identical to our resummed correlation, Eq.\ (\ref{evolend}). We will
show below, in Sec. \ref{inclusiveNLL}, that Eq. (\ref{globalend}) coincides
at NLL with the known result  for the thrust \cite{thrustresum}
when we  choose $a = 0$.

\section{Results at NLL}
\label{sec:res}

\subsection{Lowest order functions and  anomalous dimensions}

In this section, we describe the low-order calculations and results that
provide explicit expressions for the resummed shape/flow correlations and
inclusive event shape distributions at next-to-leading
logarithm in $\nu$ and leading logarithm in $\varepsilon$
(we refer to this level collectively as NLL below).   We go on to 
verify that for
the case $a=0$ we rederive the known result for the resummed
thrust at  NLL, and we exhibit the expressions for the
correlation that we will evaluate in Sec.\ \ref{numerics}.

\subsubsection{The soft function}

The one-loop soft anomalous dimension
is readily calculated in Feynman gauge from
the combination of virtual  diagrams in $\sigma^{\rm (eik)}$, Eq. 
(\ref{eikdef}), and
$J^{\rm  (eik)}$, Eq. (\ref{eikjetdef}),
in  Eq.\ (\ref{s0}).  The calculation and the result are equivalent
to those of Ref.\ \cite{KOS}, where the soft function was
formulated in axial gauge,
\be
\gamma_s^{(1)}  = - 2 \, C_F
\left[ \sum_{c=1}^2 \ln \left(\beta_c \cdot \xic \right) - \ln \left(
\frac{\beta_1 \cdot \beta_2}{2} \right) - 1 \right]\, .
\label{softad}
\ee
The first, $\xi_c$-dependent logarithmic term is associated with the eikonal
jets, while the second is a finite remainder from the
combination of $\sigma^{\rm (eik)}$ and $J^{\rm  (eik)}$ in (\ref{s0}).
Whenever $\xi_{c,\,\perp}=0$, the logarithmic terms cancel
identically,  leaving only the final  term, which comes from
the $\xic$ eikonal self-energy diagrams in the eikonal
jet functions.

The soft function is normalized to $S^{(0)}(\varepsilon) =
\delta(\varepsilon)$
as can be seen from (\ref{s0}).
For non-zero $\varepsilon$,
$d\sigma /d \varepsilon$ is given at lowest order  by
\be
S^{(1)} \left( \varepsilon \neq 0, \Omega\right) =
C_F \frac{1}{\varepsilon}
\int\limits_\O  d \mbox{PS}_2\,
\frac{1}{2 \pi} \frac{\beta_1 \cdot \beta_2}{\beta_1
\cdot \hat{k} \,\beta_2 \cdot \hat{k} }\, ,
\label{oneLoopSoft}
\ee
where $\mbox{PS}_2$ denotes the two-dimensional angular phase space
to be integrated over region $\O$, and $\hat k \equiv k / \omega_k$.
We emphasize again that the soft function contains the only
geometry-dependence of the
cross section. Also, $S^{(1)}$ for $\varepsilon \neq 0$ is independent of
$\nu$
and $a$.

As an example, consider a cone with opening angle $2 \delta$,
centered at angle $\alpha$ from jet 1.  In this case,
the lowest-order soft function is given by
\be
S^{(1)} \left( \varepsilon \neq 0, \alpha, \delta \right) =
C_F \frac{1}{\varepsilon}
\ln \left(\frac{1-\cos^2 \alpha }{\cos^2 \alpha - \cos^2 \delta}\right).
\label{softcone}
\ee
Similarly, we may choose $\O$ as a ring
extending angle $\delta_1$ to the right and $\delta_2$ to
the left of the plane perpendicular to the jet directions
in the center-of-mass.  In this case, we obtain
\be
S^{(1)} \left( \varepsilon \neq 0, \delta_1, \delta_2 \right) =
    C_F \frac{1}{\varepsilon}
\ln \left(
\frac{(1+ \sin \delta_1)}{(1-\sin \delta_1)}\frac{(1+ \sin
\delta_2)}{(1-\sin \delta_2)} \right)
= C_F \frac{2}{\varepsilon}\, \Delta\eta\, ,
\label{deltaeta2}
\ee
with $\Delta\eta$ the rapidity spanned by the ring.
For a ring centered around the center-of-mass ($\delta_1 =
\delta_2 = \delta$) the angular integral reduces to the form that we
encountered in the example  of Sec.\ \ref{sec:loe},
and that we will use in our
numerical examples of Sec.\ \ref{numerics}, with $\Delta\eta$
given by Eq.\ (\ref{deltaeta1}).

\subsubsection{The jet functions}

Recall from Eq. (\ref{Jnorm}) that the lowest-order jet function is given
by $J_c^{(0)} = 1.$

The anomalous dimensions of the jet functions are found to be
\be
\gamma_{J_c}^{(1)} = - \frac{3}{2} \, C_F
\, ,
\label{jetAD}
\ee
the same for each of the jets.
The jet anomalous dimensions
are process-independent, but of course flavor-dependent. The same
anomalous dimensions for final-state quark jets appear in three- and
higher-jet cross sections.

\subsubsection{The $K$-$G$-decomposition}

The anomalous dimension for the $K$-$G$-decomposition
is, as noted above, the Sudakov anomalous dimension,
\ba
\gamma_{K_c}^{(1)} 
& = &   2 \, C_F
, \\
\gamma_{K_c}^{(2)} 
& = &   K \, C_F
, \ea
also independent of the jet-direction. The well-known coefficient $K$
(not to be confused with the functions $K_c$) is given by \cite{KT}
\be
K = \left( \frac{67}{18} - \frac{\pi^2}{6} \right) C_A - \frac{10}{9} T_F
N_f,
\ee
with the normalization $T_F = 1/2$ and $N_f$ the number of quark
flavors.

$K_c$ and $G_c$, the functions that describe the evolution
of the jet functions in Eq.\ (\ref{KGend}), are given at one loop by
\ba
K_c^{(1)} \left(\frac{s^{1-a/2}}{\mu \nu} \left(2 p_{J_c} \cdot
\xic\right)^{a-1}\!\!\!\!\!\!,a\right) & = & - C_F
\, \ln\left(e^{2 \gamma_E-(1-a)}  \frac{\mu^2
\nu^2}{s^{2-a}} \left(2 p_{J_c} \cdot \xic \right)^{2(1-a)} \right) , 
\nonumber \\
G_c^{(1)} \left(\frac{p_{J_c} \cdot \xic}{\mu} \right) & = & - C_F
\, \ln\left( e^{-1} \frac{\left(2 \, p_{J_c} \cdot \xic \right)^2}{\mu^2}
\right)\, .
\ea
Evolving them to the values of $\mu$ with which they
appear in the functions $A_c'$ and $B'_c$, Eq.\ (\ref{ABdef}),
they become
\ba
K_c^{(1)} \left(\frac{1}{c_1},a\right) & = & - C_F
\, \ln \left( e^{2 \gamma_E-(1-a)} c_1^2 \right) , \\
G_c^{(1)} \left(\frac{1}{c_2}\right) & = & - C_F
\ln \left( e^{-1} \frac{4}{c_2^2}
\right) .
\ea
Recall that  $G_c$ is computed from virtual diagrams
only, and  thus does not
depend on the weight function.  It therefore agrees with
the result found in \cite{ColSop}.
The soft-gluon contribution, $K_c$, which involves
real gluon diagrams, does depend on the cross section
being resummed.

With the definitions  (\ref{ABdef}) of $A'_c$ and
$B'_c$ we obtain
\ba
A_c^{\prime \,(1)}  & = & C_F  \label{A1prime}
, \\
A_c^{\prime\, (2)} \left(c_1,a\right) & = & \frac{1}{2} C_F
\left[ K + \frac{\beta_0}{2}
\ln \left( e^{2 \gamma_E -1 +a } c_1^2 \right) \right], \\
B_c^{\prime\, (1)} \left(c_1,c_2,a\right) & = & 2 C_F
\ln \left(
e^{ \gamma_E -1 +a/2 } \frac{2 \,c_1}{c_2} \right).
\ea
Here $\beta_0$ is the one-loop coefficient of the QCD beta-function,
$\beta_0 = \frac{1}{3} \left( 11 \Ncol - 4 T_F N_f \right)$ ($\beta(g)
= - g \frac{\as}{4 \pi} \beta_0 + \mathcal{O}(g^3)$).

\subsubsection{The hard scattering, and the Born cross section}

At NLL only the lowest-order hard scattering function contributes, which
is normalized to
\be
H^{(0)}(\alpha_s(\sqrt{s}/2)) = 1\,.
\ee
At this order the hard function is independent of the eikonal vectors
$\xi_c$, although it acquires $\xi_c$-dependence at higher order
through the factorization described in Sec.\ \ref{sec:sdf}.
For completeness,  we also
give the electromagnetic Born cross section $\frac{d\sigma_0}{d \hat n_1}$,
at fixed polar and azimuthal angle:
\be
\frac{d \sigma_0}{d \hat n_1} =
\Ncol \left( \sum_{\rm f} Q_{\rm f}^2 \right) \frac{\alpha_{\rm em}^2}{4 s}
\left( 1 + \cos^2 \theta \right),
\label{bornCross}
\ee
where $\theta$ is the c.m.\ polar angle of $\hat{n}_1$,
$e \, Q_{\rm f}$ is the charge of quark flavor $\rm f$, and 
$\alpha_{\rm em} = e^2/(4 \pi)$
is the fine
structure constant.

\subsection{Checking the $\xi_c$-dependence} \label{gauge}

It is instructive to verify how dependence on
the eikonal vectors $\xi_c$ cancels in the exponents of the
resummed cross section (\ref{evolend}) at the
accuracy at which we work, single logarithms of
$\varepsilon$, and single and double logarithms of $\nu$.
In these exponents,
$\xi_c$-dependence
enters only through the
combinations
$(\beta_c \cdot \xic)$ and
$(p_{J_c} \cdot \xic)$.

Let us introduce the following notation for the exponents in Eq.
(\ref{evolend}), to
which we will return below:
\ba
E_1 & \equiv &  - \int\limits_{\varepsilon \sqrt{s}}^{\sqrt{s}/2} \frac{d
\lambda}{\lambda} \gamma_s\left(\as(\lambda)\right) - \sum_{c=1}^2
\int\limits_{\sqrt{s}/(2 \, \zeta_0)}^{\sqrt{s}/2}
   \frac{d \lambda}{\lambda} \gamma_{J_c} \left(\as(\lambda)\right),
   \label{E1} \\
E_2 &  \equiv &  - \sum_{c=1}^2 \int\limits_{\sqrt{s}/(2\, \zeta_0) 
}^{p_{J_c} \cdot \xic}
    \frac{d \lambda}{\lambda} \left[B'_c\left(c_1,c_2,a,
\as\left(c_2 \lambda \right) \right)  +  2 \int\limits_{c_1 \frac{s^{1-a/2}
}{ \nu
(2\,\lambda)^{1-a} } }^{c_2\, \lambda}\frac{d \lambda'}{\lambda'} A'_c\left(
c_1,
a,\as\left(\lambda'\right) \right) \right]. \nonumber \\ 
\label{E2}
\ea
At NLL, explicit $\xi_c$ dependence is
found only in $\gamma_s$, Eq. (\ref{softad}), for $E_1$,
and in the upper limit of the $\lambda$ integral of $E_2$.
We then find that
\ba
{\partial \over  \partial \ln\beta_c\cdot\xic}\left(E_1+E_2\right)
=
2C_F\, \, \int\limits_{\varepsilon \sqrt{s}}^{\sqrt{s}/2}
\frac{d
\lambda}{\lambda}\; \frac{\as(\lambda)}{\pi}
-2C_F \int_{c_1{s^{1-a/2}\over
\nu(2p_{J_c}\cdot \xic)^{1-a}}}^{c_2\,p_{J_c}\cdot\xic}
{d\lambda'\over\lambda'}\;
\frac{\as(\lambda')}{\pi} +{\rm NNLL}
\, . \nonumber \\
\label{gaugevar}
\ea
Here the second term stems entirely from $A^{\prime\,(1)}$, Eq.
(\ref{A1prime});
other contributions of $E_2$ are subleading.
The $\xi_c$-dependence in the exponents begins only at the level that
we do not resum, at $\as\ln(1/\varepsilon \nu)$,
which is compensated by corrections in $S(\varepsilon\nu,\as)$.
The remaining contributions are of NNLL order,
that is, proportional to $\as^k(\sqrt{s})
   \ln^{k-1} \left (\nu \, \beta_c \cdot \xic \right)$,
as may be verified by expanding the running couplings.
Thus, as required by the factorization procedure,
the relevant $\xi_c$-dependence cancels between
the resummed soft and jet functions, which give rise
to the first and second integrals, respectively, in Eq.\ (\ref{gaugevar}).

\subsection{The inclusive event shape at NLL}\label{inclusiveNLL}

We can simplify the differential event shape, Eq.\ (\ref{globalend}),
by absorbing the soft anomalous dimension $\gamma_s$ into
the remaining terms.  We will find a form that can be compared
directly to the classic  NLL  resummation for the thrust
($a=0$).   This is done by rewriting the integral over
the soft anomalous dimension as
\ba
\int_{\sqrt{s}/\nu}^{\sqrt{s}/2} {d\lambda\over \lambda}\;
\gamma_s\left(\as(\lambda)\right)
&=&
\int_{\sqrt{s}/\left[2(\nu/2)^{1/(2-a)}\right]}^{\sqrt{s}/2}
{d\lambda\over \lambda}\;
\gamma_s\left(\as(\lambda)\right) \nonumber \\
&\ & \hspace{-40mm} +
\int_{\sqrt{s}/\nu}^{\sqrt{s}/\left[2(\nu/2)^{1/(2-a)}\right]}
{d\lambda\over \lambda}\;
\gamma_s\left(\as(\lambda)\right) =
\int_{\sqrt{s}/\left[2(\nu/2)^{1/(2-a)}\right]}^{\sqrt{s}/2}
{d\lambda\over \lambda}\;
\gamma_s\left(\as(\lambda)\right) \nonumber \\ 
&\ & \hspace{-40mm} + (1-a)
\int_{\sqrt{s}/\left[2(\nu/2)^{1/(2-a)}\right]}^{\sqrt{s}/2}
{d\lambda\over \lambda}\;
\gamma_s\left(\as\left( \frac{s^{1-a/2}}{\nu (2 \lambda)^{1-a}}
\right)\right) \nonumber \\
&\ & \hspace{-40mm} =
(2-a)\int_{\sqrt{s}/\left[2(\nu/2)^{1/(2-a)}\right]}^{\sqrt{s}/2}
{d\lambda\over
\lambda}\;
\gamma_s\left(\as(\lambda)\right) \nonumber\\
&\ & \hspace{-40mm}
- (1-a)\int_{\sqrt{s}/\left[2(\nu/2)^{1/(2-a)}\right]}^{\sqrt{s}/2}
{d\lambda\over \lambda}\; \int_{s^{1-a/2}/\left[\nu (2 \lambda)^{1-a}
\right]}^\lambda
\frac{d \lambda'}{\lambda'}
\beta(g(\lambda'))\, {\partial \over \partial g}\,
\gamma_s\left(\as(\lambda')\right)\, . \nonumber \\
\label{split}
\ea
In the first equality we split the $\lambda$ integral so that the limits of
the first term match those of the $B'_c$ integral of Eq.\ (\ref{globalend}).
In the second equality we have changed variables in the
second term according to
\be
\lambda \rightarrow \left({s^{1-{a\over 2}}\over
2^{1-a}\nu\lambda}\right)^{1\over 1-a}\, ,
\ee
so that the limits of the second integral also match.
In the third equality of Eq. (\ref{split}),
   we have reexpressed the running coupling at the old
scale
$\lambda$
   in terms of the new scale.
This is a generalization of the procedure of Ref.\ \cite{CT91},
applied originally to the threshold-resummed
Drell-Yan cross section \cite{DYold}.

Using Eq.\ (\ref{split}), and
identifying $p_{J_c} \cdot \xic$ with $\sqrt{s}/2$
(Eq. (\ref{xiid})) in the inclusive event shape
distribution, Eq. (\ref{globalend}),
we can rewrite this distribution at NLL as
\ba
\frac{d \sigma^{\rm incl} \left(\nu,s,a \right)}{ d \hat{n}_1 }
&=&
{d \sigma_0 \over d\hat{n}_1}\
\nonumber \\
&   & \hspace*{-27mm} \times\ \prod_{c=1}^2\,
    \exp \left\{ - \int\limits_{\sqrt{s} /
\left[2(\nu/2)^{1/(2-a)}\right]}^{\sqrt{s}/2}
\frac{d \lambda}{\lambda}  \left[ B_c
\left(c_1,c_2,a,\as\left(\lambda\right)\right) 
\right. \right. \nonumber \\
& & \hspace*{-27mm} + \left. \left. 
2 \int\limits_{c_1 \frac{s^{1-a/2} }{ \nu
(2\,\lambda)^{1-a} } }^{c_2\, \lambda}\frac{d \lambda'}{\lambda'} 
A_c\left(c_1,a,\as\left(\lambda'\right) \right) \right] \right\}\, , 
\nonumber \\
\label{globalendcat}
\ea
where we have rearranged the contribution of $\gamma_s$ as:
\ba
A_c \left( c_1, a,\as\left(\mu \right) \right) & \equiv &
A'_c \left( c_1, a,\as\left(\mu \right) \right)
- \frac{1}{4} (1-a)\,  \beta(g(\mu))\, {\partial \over \partial g}\,
\gamma_s\left(\as(\mu)\right) ,  \nonumber \\
   B_c \left(c_1,c_2,a,\as\left(\mu\right)\right) & \equiv &
\gamma_{J_c} \left(\as(\mu) \right)
   + \left( 1 - \frac{a}{2} \right) \gamma_s \left(\as(\mu) \right) +
   B'_c \left(c_1,c_2,a,\as\left(\mu\right)\right). \nonumber \\
   \ea
Next, we replace
   the lower limit of the $\lambda'$-integral
by an explicit $\theta$-function. Then we exchange orders of integration,
and change variables in the term containing $A$
from the dimensionful variable $\lambda$ to the dimensionless combination
\be
u = {2\lambda\lambda'\over s}\, .
\ee
We find
\ba
\frac{d \sigma^{\rm incl} \left(\nu,s,a \right)}{ d \hat{n}_1 }
&=&
{d \sigma_0 \over d\hat{n}_1}\  \prod_{c=1}^2\,
    \exp \left\{ - \int\limits_{\sqrt{s} /
\left[2(\nu/2)^{1/(2-a)}\right]}^{\sqrt{s}/2}
\frac{d \lambda}{\lambda}
   B_c \left(c_1,c_2,a,\as\left(\lambda\right)\right) \right\}
\nonumber\\
&\ & \hspace*{-3.5cm} \times\
\prod_{c=1}^2\,
    \exp \left\{ - 2 \int_0^{\sqrt{s}} \frac{d \lambda'}{\lambda'} \;
\int_{\lambda'{}^2/s}^{\lambda'/\sqrt{s}}\, {d u \over u}\;
\theta\left( c_1^{-1}\nu\, {\lambda'{}^a u^{1-a}\over s^{a/2}}-1
\right)\;
A_c\left( c_1,
a,\as\left(\lambda'\right) \right) \right\} \, . \nonumber \\
    \label{globalendcat2}
\ea
Here, the $\theta$-function vanishes for small $\lambda'$, and the 
remaining effects of
replacing the lower boundary of the $\lambda'$ integral by 0 are
next-to-next-to-leading logarithmic.

A further change of variables allows us to write the NLL resummed event shapes
in a form familiar from the NLL resummed thrust.
In the first line of Eq. (\ref{globalendcat2}), we
replace $\lambda^2 \rightarrow u s/4$. In the second line we relabel
$\lambda' \rightarrow \sqrt{q^2}$,
and exchange orders of integration.
Finally, choosing
\ba
c_1 & = & e^{-\gamma_E}, \nonumber \\
c_2 & = & 2,
\label{cipick}
\ea
we find at NLL
\ba
\frac{d \sigma^{\rm incl} \left(\nu,s,a \right)}{ d \hat{n}_1 }
&=&
{d \sigma_0 \over d\hat{n}_1}\  \nonumber \\ 
& \times & \prod_{c=1}^2\,
    \exp \left\{ \int\limits_0^1 \frac{d u}{u} \left[ \,
    \int\limits_{u^2 s}^{us} \frac{d q^2}{q^2} A_c\left(\as(q^2)\right)
    \left( e^{- u^{1-a} \nu \left(q^2/s\right)^{a/2} }-1 \right)
\right. \right.\nonumber \\
    & & \qquad \qquad
    \qquad
    + \frac{1}{2} B_c\left(\as(u s/4)\right) \left( e^{-u
\left(\nu/2\right)^{2/(2-a)} e^{-\gamma_E}} -1 \right)
    \bigg] \Bigg\}, \nonumber \\
\label{thrustcomp}
\ea
and reproduce the well-known coefficients
\ba
A_c^{(1)}  & = & C_F
, \\
A_c^{(2)}  & = & \frac{1}{2} C_F  K, \\
B_c^{(1)} & = & - \frac{3}{2} \, C_F,
\ea
independent of $a$.  In Eq.\ (\ref{thrustcomp}), we have made use of the
relation
\be
e^{-x/ y} - 1 \approx - \theta \left(x - y\, e^{-\gamma_E} \right),
\ee
which is valid at NLL in the logarithmic integrals.
With these choices, when $a = 0$ we reproduce the
NLL resummed thrust cross section \cite{thrustresum}.

The choices of
the $c_i$ in Eq.\ (\ref{cipick}) cancel all purely soft NLL
components ($\gamma_s$ and $K_c$). The
remaining double logarithms stem from simultaneously soft and collinear
radiation, and single logarithms arise from collinear configurations only.
At NLL, the cross section is determined by the
anomalous dimension $A_c$, which is the coefficient
of the singular $1/[1-x]_+$ term in the
nonsinglet evolution kernel \cite{oneover1x}, and the
quark anomalous dimension.
All radiation in dijet events
thus appears to be emitted coherently by the two jets \cite{thrustresum}.
This, however, is not necessarily true
beyond next-to-leading logarithmic accuracy for dijets, and
is certainly not the case for multijet events \cite{KOS}.  Similar 
considerations apply to
the resummed correlation, Eq.\ (\ref{evolend}).

\subsection{Closed expressions}

Given the explicit results above, the integrals in  the
exponents of the resummed correlation, Eq.\ (\ref{evolend}), may
be easily performed in closed form.
We give the analytic results for the exponents
of Eq. (\ref{evolend}), as defined in
Eqs. (\ref{E1}) and  (\ref{E2}). As in Eq. (\ref{xiid}),
we identify $p_{J_c} \cdot \xic$ with
$\sqrt{s}/2$.
\ba
e^{E_1(a)} & = & \left(
\frac{ \as(\sqrt{s}/ 2)}{\as(\varepsilon \sqrt{s})}\right)^{\frac{4
C_F}{\beta_0}}
\left( \frac{\as\left(\frac{\sqrt{s}}{2 \,\zeta_0}\right)}{ \as(\sqrt{s}/2)}
   \right)^{\frac{6
C_F}{\beta_0} } , \label{E1result} \\
e^{E_2(a)} & = &   \left( \frac{ \as(c_2 \, \sqrt{s}/2)}
{\as\left(\frac{c_2 \,\sqrt{s}}{2 \, \zeta_0} \right)} \right)^{\frac{4
C_F}{\beta_0} \kappa_1(a)}
    \left( \frac{  \as\left(\frac{c_1 \, \sqrt{s}}{2 \,\zeta_0 }\right)} {
\as\left(\frac{c_1 \, \sqrt{s}}{\nu}\right)}
\right)^{\frac{1}{a-1} \frac{4 C_F}{\beta_0} \kappa_2(a)} \nonumber \\
& \times & \left( \frac{ \as(c_2 \, \sqrt{s}/2)}
{\as\left(\frac{c_1 \, \sqrt{s}}{2 \,\zeta_0 }\right)} 
\right)^{\frac{1}{2-a} \, \frac{8C_F}{\beta_0} \, \ln (\nu / 2)}, 
\nonumber \\
& &  \label{E2result}
\ea
with
\ba
\kappa_1(a) \!\!& = &\!\! \ln \left(\frac{4}{c_2^2 e}\right) +
\frac{4\pi}{\beta_0} \left[\as\left(\frac{c_2 \, \sqrt{s}}{2 \,\zeta_0}\right)
\right]^{-1}
- \frac{2 K}{\beta_0} \nonumber \\ 
& - & \frac{\beta_1}{2 \beta_0^2} \ln \left( \left(\frac{\beta_0}{4 \pi
e}\right)^2
\as\left(\frac{c_2\, \sqrt{s}}{2}\right) \as\left(\frac{c_2\, \sqrt{s}}{2 \,
\zeta_0}\right)\right), \nonumber \\
& & \label{C1} \\
\kappa_2(a) \!\!& = & \!\!(1 - a - 2 \gamma_E) + \frac{4\pi}{\beta_0}
\left[\as\left(\frac{\sqrt{s}}{\nu}\right)\right]^{-1}
- \frac{2 K}{\beta_0} \nonumber \\
& - & \frac{\beta_1}{2 \beta_0^2} \ln \left( \left(\frac{\beta_0}{4 \pi
e}\right)^2
\as\left(\frac{c_1 \, \sqrt{s}}{\nu}\right) \as\left(\frac{c_1 \, \sqrt{s}}{2
\, \zeta_0}\right)\right). \nonumber \\
& & \label{C2}
\ea
   We have used the two-loop running coupling, when appropriate,
to derive Eqs.\ (\ref{E1result}) - (\ref{C2}).
The results are expressed in terms of the one-loop running coupling
\be
\as(\mu) = \frac{2 \pi}{\beta_0} \frac{1}{\ln \left(
\frac{\mu}{\Lambda_{\mbox{\tiny QCD} } }  \right)}\, ,
\ee
and the first two coefficients in the expansion of the QCD beta-function,
$\beta_0$ and
\be
\beta_1 = \frac{34}{3} \, C_A^2 - \left(\frac{20}{3} C_A  + 4 C_F\right) \,
T_F \, N_f\, .
\ee
Combining the expressions for the exponents, Eqs.\ (\ref{E1result}) and
(\ref{E2result}), for the Born cross section,
Eq.\ (\ref{bornCross}), and for the soft function, Eq.\ (\ref{oneLoopSoft}), in
Eq.\ (\ref{evolend}),
the complete differential cross section, at LL in $\varepsilon$ and at NLL
in $\nu$, is given by
\ba
\frac{d \sigma \left(\varepsilon, \nu,s ,a\right)}{d\varepsilon\, d
\hat{n}_1 }
&=&
\Ncol \left( \sum_{\rm f} Q_{\rm f}^2 \right) \frac{\pi \alpha_{\mbox{\tiny
em}}^2}{2 s} \left(1+ \cos^2 \theta \right) \nonumber \\ 
& \times & C_F \frac{\as(\varepsilon
\sqrt{s})}{\pi} \frac{1}{\varepsilon} \int\limits_\O  d \mbox{PS}_2\,
\frac{1}{2 \pi} \frac{\beta_1 \cdot \beta_2}{\beta_1
\cdot \hat{k} \, \beta_2 \cdot \hat{k} } \nonumber \\
& \times &
\left(
\frac{ \as\left(\frac{\sqrt{s}}{2}\right)}{\as(\varepsilon
\sqrt{s})}\right)^{\frac{4 C_F}{\beta_0}}
\left( \frac{\as\left(\frac{\sqrt{s}}{2 \,\zeta_0}\right)}{
\as\left(\frac{\sqrt{s}}{2}\right)}
   \right)^{\frac{6
C_F}{\beta_0} }
   \nonumber \\
&  \times & \left( \frac{ \as\left(c_2 \, \frac{\sqrt{s}}{2}\right)}
{\as\left(\frac{c_2 \,\sqrt{s}}{2 \, \zeta_0} \right)} \right)^{\frac{4
C_F}{\beta_0} \kappa_1(a)}
    \left( \frac{  \as\left(\frac{c_1 \, \sqrt{s}}{2 \,\zeta_0 }\right)} {
\as\left(\frac{c_1 \, \sqrt{s}}{\nu}\right)}
\right)^{\frac{1}{a-1} \frac{4 C_F}{\beta_0} \kappa_2(a)} \nonumber \\
& \times & \left( \frac{ \as\left(c_2 \, \frac{\sqrt{s}}{2}\right)}
{\as\left(\frac{c_1 \, \sqrt{s}}{2 \, \zeta_0 }\right)} \right)^{
\frac{1}{2-a} \, \frac{8
C_F}{\beta_0} \, \ln \left(\frac{\nu}{2}\right)}\hspace*{-2.0cm} .
\label{resulto1}
\ea
These are the expressions that we will evaluate in the
next section.  We note that this is not the only possible closed form for the
resummed correlation at this level of accuracy.  When a full
next-to-leading order calculation for this set of event shapes
is given, the matching procedure of
\cite{thrustresum} may be more convenient.

\section{Numerical Results}
\label{numerics}

Here we show some representative examples of numerical results for the
correlation, Eq.\ (\ref{resulto1}). We pick the constants $c_i$ as in
Eq.\ (\ref{cipick}),
unless stated otherwise. The effect of different choices
is nonleading, and is numerically small, as we will see below.
In the following we choose the
region $\O$ to be a ring between the jets, centered in their
center-of-mass, with a width of $\Delta \eta = 2$, or equivalently, opening
angle $\delta \approx 50$ degrees (see Eq.\ (\ref{deltaeta1})).
The analogous cross section for
   a cone centered at 90 degrees from the jets
(Eq.\ (\ref{softcone})) has a similar behavior.
In the following, the center-of-mass energy
$Q=\sqrt{s}$ is  chosen to be $100$ GeV.

\begin{figure}[htb]
\vspace*{7mm}
\begin{center}
a) \hspace*{7.5cm} b) \hspace*{4cm} \vspace*{-6mm} \\
\epsfig{file=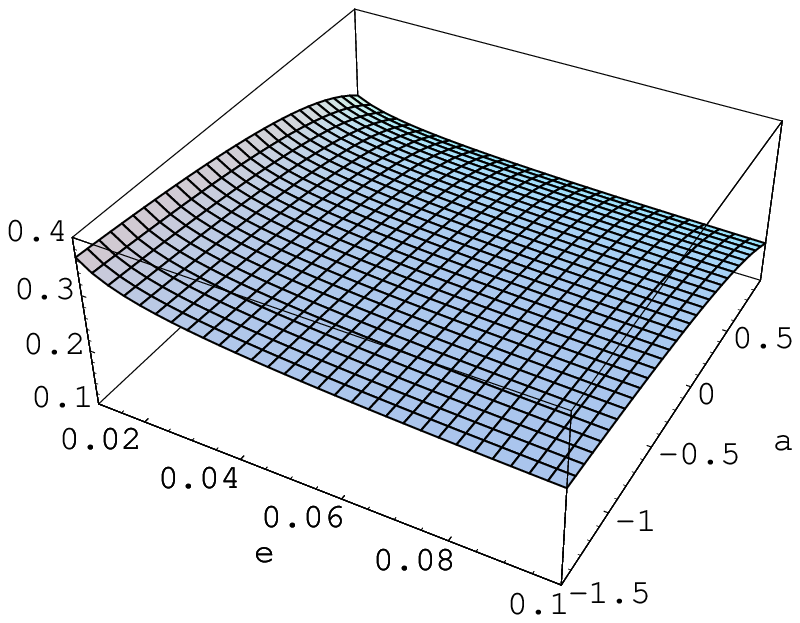,width=6.8cm,clip=0}  \mbox{
}\hspace*{5mm}  \epsfig{file=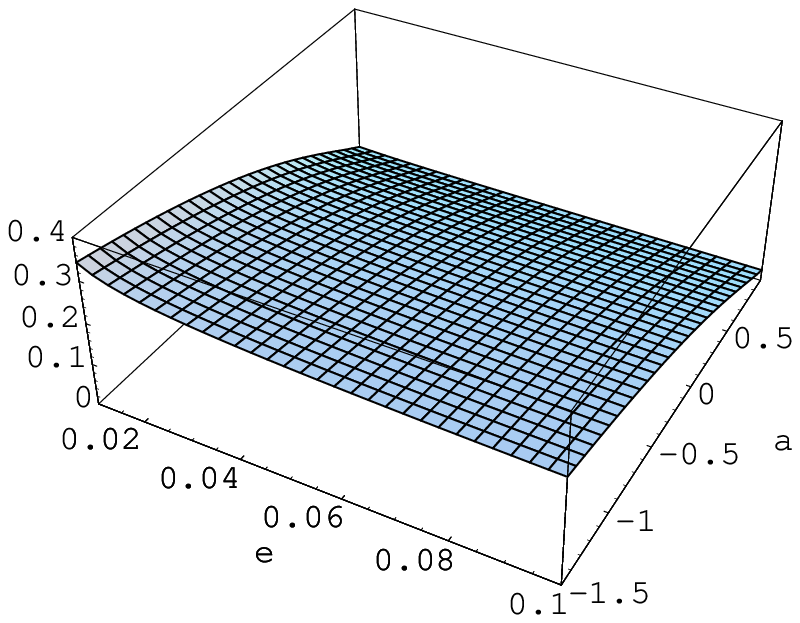,width=6.8cm,clip=0}
\caption{Differential cross section $\frac{\varepsilon d
\sigma/(d\varepsilon d\hat n_1)}{d \sigma_0/d \hat n_1}$,
normalized by the Born cross section, at $Q = 100$ GeV,
as a function of
$\varepsilon$ and $a$ at fixed $\nu$: a) $\nu = 10$, b) $\nu =
50$. $\O$ is a ring (slice) centered around the jets, with a
width of $\Delta \eta = 2$.} \label{num1}
\end{center}
\end{figure}

Fig. \ref{num1} shows the dependence of the differential cross
section (\ref{evolend}), multiplied by $\varepsilon$
and normalized by the Born cross section,
$\frac{\varepsilon d \sigma/(d\varepsilon d\hat n_1)}{d
\sigma_0/d \hat n_1}$, on the measured energy $\varepsilon$ and
on the parameter $a$, at fixed $\nu$.
In Fig. \ref{num1} a), we plot $\frac{\varepsilon
d \sigma/(d\varepsilon d\hat n_1)}{d \sigma_0/d \hat n_1}$ for
$\nu = 10$, in Fig. \ref{num1} b) for $\nu  = 50$.
As $\nu$ increases, the radiation into the
complementary region $\bar \O$ is more restricted,
as illustrated by the comparison of Figs.
\ref{num1} a)  and b). Similarly, as $a$ approaches 1, the
cross section falls, because the jets are restricted to
be very narrow.  On the other hand,
as $a$ assumes more and more negative values at
fixed $\varepsilon$, the correlations (\ref{evolend}) approach a
constant value.  For $a$ large and negative, however, non-global
dependence on
$\ln\varepsilon$ and $|a|$ will emerge from higher order corrections
in the soft function, which we do not include in Eq.\ (\ref{resulto1}).

\begin{figure}[htb]
\begin{center}
\epsfig{file=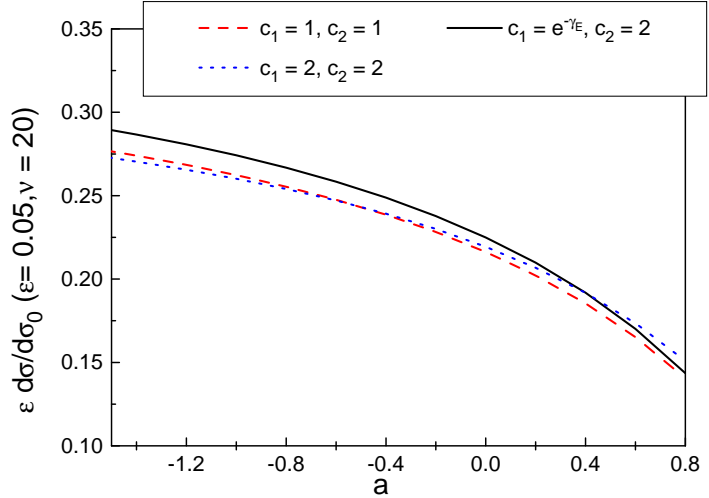,height=10cm,angle=270,clip=0}
\caption{Differential cross section $\frac{\varepsilon d
\sigma/(d\varepsilon d\hat n_1)}{d \sigma_0/d \hat n_1}$,
normalized by the Born cross section, at $Q = 100$ GeV,
as a function of
$a$ at fixed $\nu = 20$ and $\varepsilon = 0.05$.
   $\O$ is chosen as in Fig. \ref{num1}. Solid line: $c_1 = e^{-\gamma_E},\,
c_2 = 2$,
as in Eq.  (\ref{cipick}), dashed line: $c_1 = c_2 = 1$, dotted line:
$c_1 = c_2 = 2$. } \label{numci}
\end{center}
\end{figure}

In Fig.\ \ref{numci} we investigate the sensitivity of the resummed
correlation,
Eq.\ (\ref{resulto1}), to our choice of the constants $c_i$. The 
effect of these constants is
of next-to-next-to-leading logarithmic order in the event shape. We plot the
differential cross section $\varepsilon
\frac{\varepsilon d \sigma/(d\varepsilon d\hat n_1)}{d
\sigma_0/d \hat n_1}$, at $Q = 100$ GeV, for fixed $\varepsilon = 0.05$ and
$\nu = 20$, as a function of $a$. The effects of changes in the $c_i$ are
of the order of a few percent for moderate values of $a$.

\begin{figure}[htb]
\vspace*{6mm}
\begin{center}
a) \hspace*{7.5cm} b) \hspace*{4cm} \vspace*{-6mm} \\
\epsfig{file=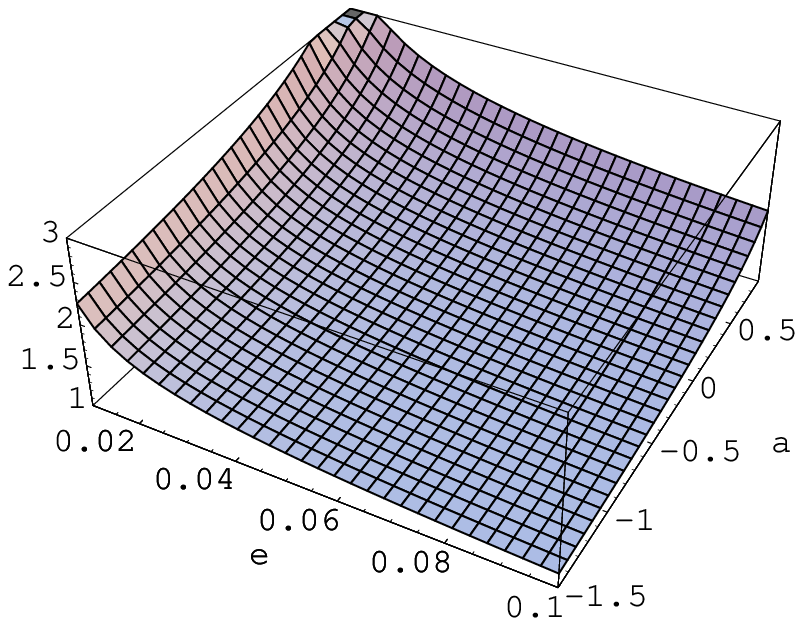,width=6.7cm,clip=0}  \mbox{
}\hspace*{5mm}  \epsfig{file=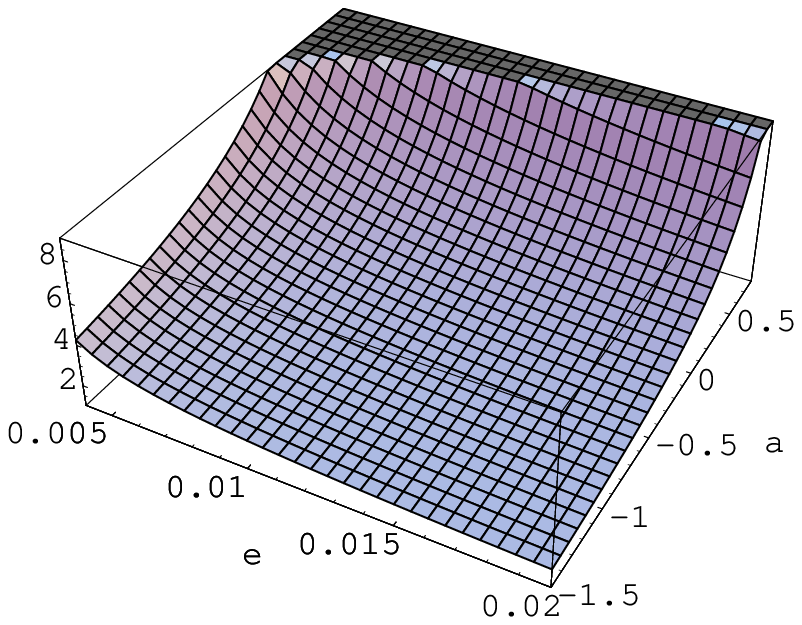,width=6.7cm,clip=0}
\caption{Ratios of differential cross sections  for quark to
gluon jets $\frac{C_A}{C_F} \left(\frac{\varepsilon d
\sigma^q/(d\varepsilon d\hat n_1)}{d \sigma_0^q/d \hat n_1}\right)
\left(\frac{\varepsilon d
\sigma^g/(d\varepsilon d\hat n_1)}{d \sigma_0^g/d \hat n_1}\right)^{-1}$ at
$Q = 100$ GeV
as a
function of $\varepsilon$ and $a$ at fixed $\nu$: a) $\nu = 10$,
b) $\nu = 50$. $\O$ as in Fig. \ref{num1}, $c_1$ and $c_2$ as in Eq.
(\ref{cipick}).}
\label{num2}
\end{center}
\end{figure}

Finally, we illustrate the sensitivity of these results to the flavor
of
the primary partons. For this purpose we study the corresponding 
ratio of the shape/flow
correlation to the cross section for gluon jets produced
by a hypothetical color singlet source. Fig.\ \ref{num2} displays the ratio
of the differential cross section
$d\sigma^q(\varepsilon,a)/(d\varepsilon d \hat n_1)$, Eq.\ (\ref{resulto1}),
normalized
by the lowest-order cross section, to the
analogous quantity with gluons as
primary partons in the outgoing jets, again at
$Q = 100$ GeV.
This ratio is multiplied by $C_A/C_F$ in the figure to compensate for
the difference in the normalizations of the lowest-order soft functions.
Gluon jets have wider angular extent, and hence are
suppressed relative to quark jets with increasing  $\nu$ or $a$,
as can be seen by
comparing Figs. \ref{num2} a) and b). Fig. \ref{num2} a) shows the ratio
at $\nu = 10$, and Fig. \ref{num2} b) at $\nu = 50$.
These results suggest sensitivity to
the more complex color and flavor flow
characteristic of hadronic scattering \cite{KOS,BKS1}.

\section{Summary and Outlook}

We have
introduced a general class of inclusive event shapes in $e^+e^-$ dijet events
which
reduce to the thrust and the jet broadening distributions as special
cases.
We have derived analytic expressions in transform
space, and have shown the equivalence of our formalism at NLL
with the well-known result for the thrust \cite{thrustresum}.
Separate studies of this class of event shapes
in the untransformed space,
at higher orders, and for nonperturbative effects
\cite{ptnpdist} are certainly of interest.
We reserve these studies
for future work.

We have introduced a set of correlations of interjet energy
flow for the general class of  event shapes, and have shown
that for these quantities it is possible to control
the influence of secondary radiation and nonglobal logarithms.
These correlations are sensitive mainly to radiation
emitted directly from the primary hard scattering, through transforms in
the weight functions that suppress
secondary, or non-global, radiation.
  We have presented analytic  and
numerical studies of these shape/flow correlations at leading
logarithmic order in the
flow variable and at next-to-leading-logarithmic order in the
event shape.
   The application of our formalism
to multijet events and to scattering with initial state hadrons is
certainly possible, and may shed light on the relationship
between color and energy flow in hard scattering processes with non-trivial
color exchange.


\appendix
\chapter{} \label{cha}

\section{Power counting with contracted vertices} \label{contracted vertices}

In this appendix we will include the possibility of contracted 
vertices in the reduced diagram
in Fig. \ref{reduced}a. These are associated with internal lines 
(collapsed to a point) which are off-shell by
$\sqrt{s}$. Our analysis closely follows \cite{sterman78} and \cite{sen81}.

If we go back to the argument that led us to Eq. (\ref{os}) for the 
superficial degree of IR
divergence for the soft part, we see that the same reasoning as in 
the case of elementary vertices applies to the case of
contracted vertices since the result (\ref{os}) has been obtained by 
means of dimensional counting.

The analysis of contracted vertices connecting jet lines only is, 
however,  more subtle.
We have to demonstrate that the suppression factors corresponding to 
the contracted vertices are at least as great as
the ones for the elementary vertices. The expression (\ref{oa3}) 
tells us that we can restrict ourselves to the two and
three point vertices.  For these cases, we analyze the full two and 
three-point subdiagrams, by
studying the tensor structures that are found after integration over 
their internal loop momenta.

Before we discuss all the possible structures, we state some results 
which will be essential for the upcoming analysis.
The first one is the simple Dirac matrix identity
\be \label{slash}
\not{\!a} \, \not{\!b} \, \not{\!a} = 2 (a \cdot b) \, \not{\!a} - 
a^2 \, \not{\!b}.
\ee
The other two follow from Eqs. (\ref{propagator}) and 
(\ref{propagComp}) for the gluon propagator in Coulomb gauge,
and hold for any jet momenta scaling as
$l_A \, \sim \, l_A^{\prime} \, \sim \, 
\sqrt{s}(1^+,\lambda^-,{\lambda}^{1/2})$ collinear to the
momentum $p_A$ defined in Eq. (\ref{momentaDefinition})
\bea \label{contractPropagator}
l_A^{\prime \; \alpha} \, N_{\alpha \beta}(l_A, \eta) & = & {\cal 
O}({\lambda}^{1/2} \, \sqrt{s}), \nonumber \\
{\bar l}_A^{\prime \; \alpha} \, N_{\alpha \beta}(l_A, \eta) & = & 
{\cal O}({\lambda}^{1/2} \, \sqrt{s}),
\eea
for all components of $\beta$.
We now proceed to discuss the particular cases.

{\it Ghost self-energy:} The most general covariant structure is, 
using $p \cdot \bar p = {\bar p}^2$,
\footnote {In the rest of this subsection we are concerned the 
momentum factors only, and we omit dependence on the color structure.}
\be \label{ghostSelfEnergy}
\Pi(p, \bar p) = p \cdot {\bar p} \, f(p^2 / {\mu}^2, \, {\bar p}^2 / 
{\mu}^2, \, {\alpha}_s(\mu)),
\ee
where $\mu$ is a scale introduced by a UV/IR regularization of 
Feynman diagrams and $p$ is the momentum of an internal jet line.
Strictly speaking, the covariants should be formed
from the vectors $p$ and $\eta$, but since $p$ has nonzero light-cone 
components, we can use
Eq. (\ref{barVector}), to express $\eta$
in terms of $\bar p$.
The maximum degree of divergence for the ghost self-energy occurs 
when the internal lines become either parallel to the external 
momentum $p$ or soft. The most general pinch singular
surface consists of a subdiagram of collinear lines moving in a 
direction of the external ghost.
This subdiagram can interact with itself through the exchange of soft quanta.
Power counting arguments similar to the ones given in Sec. 
\ref{elementary vertices} show, however, that there is no IR 
divergence for these pinch singular points. This shows that the 
dimensionless function $f$ in Eq. (\ref{ghostSelfEnergy}) is IR 
finite. Hence the combination [tree level ghost propagator] - [ghost 
self-energy] - [tree level ghost propagator], $[1/(p \cdot {\bar p})] 
\, \Pi (p, \bar p) \,
[1/(p \cdot {\bar p})]$, is suppressed at least as much as a single 
tree level ghost
propagator, $1/(p \cdot {\bar p})$. Therefore the contracted two 
point ghost vertex within a jet subdiagram
contributes at least the same suppression as a single tree level 
ghost propagator.

{\it Gluon self-energy:} With external momentum $p$, its most general 
tensor decomposition has the form
\be \label{gluonSelfEnergy}
{\Pi}_{\mu\nu}(p, \bar p) = g_{\mu\nu} \, p^2 \, f_1 + p_{\mu}p_{\nu} 
\, f_2 + {\bar p}_{\mu}{\bar p}_{\nu} \, f_3 +
(p_{\mu}{\bar p}_{\nu} + {\bar p}_{\mu}p_{\nu}) \, f_4 \, .
\ee
As verified by explicit one-loop calculations in Refs. 
\cite{leibbrandt96} and \cite{leibbrandt98}
the gluon self-energy in Coulomb gauge is not transverse.
In Eq. (\ref{gluonSelfEnergy}), the $f_i = f_i \, ( p^2 / {\mu}^2, \, 
{\bar p}^2 / {\mu}^2, \, {\alpha}_s(\mu))$
are dimensionless functions.
Contracting ${\Pi}_{\mu\nu}$ with tree level gluon propagators, and 
using Eq. (\ref{gluonProperties}),
the last two terms in Eq. (\ref{gluonSelfEnergy}) drop out and the 
first and the second terms give at least one factor of
$p^2$ in the numerator, which cancels one  of the $(1/p^2)$ 
denominator factors.
Since the maximum degree of IR divergence for the gluon self-energy 
occurs when all the internal lines
become either collinear to the external momentum $p$ or soft, we can 
use the results of the power counting
of Sec. \ref{elementary vertices} to demonstrate that the 
dimensionless functions $f_i$ are at
worst logarithmically divergent.  Therefore the combination: gluon 
jet line - 2 point gluon contracted vertex - gluon jet
line, behaves the same way as a gluon jet line for the purpose of the 
jet power counting.

{\it Fermion self-energy:} In the massless fermion limit, the most 
general matrix structure of the
fermion self-energy is
\be \label{sigma}
\Sigma(p,{\bar p}) = \not{\!p} \, g_1 + \not{\!{\bar p}} \, g_2,
\ee
with dimensionless functions $g_{i}=g_i(p^2/{\mu}^2, \, {\bar 
p}^2/{\mu}^2, \, {\alpha}_s(\mu)), \; i=1,2$.
When we sandwich the fermion self-energy between the tree level 
fermion denominators, the first term in Eq. (\ref{sigma}) behaves
the same way as the tree level fermion  propagator, modulo 
logarithmic enhancements due to the function $g_1$.
The second term, however, is absent from the fermion self-energy as 
was shown in Ref. \cite{sen81} using the method of
induction and Ward identities. The idea was to study a variation of 
the fermion self-energy by making an infinitesimal
Lorentz boost on the external momentum. This implies a relationship 
between the $(r+1)$ and the $r$-loop self
energy. Assuming that the term proportional to $\not{\!\!{\bar p}}$ 
is absent from the $r$-loop expansion Sen shows that
it is also absent from the $(r+1)$-loop expansion. So the first term 
in Eq. (\ref{sigma}) is the only possible structure
of the fermion self-energy when its external momentum is jet like and 
approaches mass shell.

Now let us investigate the 3 point functions. \\
{\it Fermion-gluon-fermion vertex function:} ${\Gamma}_{\mu}$, can 
depend on vectors that scale as $l_A, \, l_A^{\prime}$
in Eq. (\ref{contractPropagator}), provided all momenta external to 
the contracted vertex are collinear to momentum $p_A$
given in Eq. (\ref{momentaDefinition}). It has one Lorentz index, 
$\mu$, and neglecting the fermion masses, it contains an odd number 
of gamma matrices. This implies that the most general tensor and 
gamma matrix expansion of ${\Gamma}_{\mu}$ involves
\begin{enumerate}
\item ${\gamma}_{\mu}$,
\item ${\gamma}_{\mu}\not{\!l_A}\not {\!{\bar l}_A} \, / \, (l_A 
\cdot {\bar l}_A)$ and all permutations of ${\gamma}_{\mu}, \, 
\not{\!l_A}, \, \not {\!{\bar l}_A}$,
\item  $\not{\!l_A} \, l_A^{\mu} \, / \, l_A^2$, \, $\not{\!{\bar 
l}_A} \, l_A^{\mu} \, / \, ({\bar l}_A \cdot l_A)$, \,
$\not{\!l_A} \, {\bar l}_A^{\mu} \, / \, (l_A \cdot {\bar l}_A)$, \, 
$\not{\!{\bar l}_A} \, {\bar l}_A^{\mu} \, / \, {{\bar l}_A}^2$.
\end{enumerate}
The differences between the listed set of structures and other 
possible combinations are ${\cal O}({\lambda}^{1/2} \, \sqrt{s})$, as
  can be shown using Eqs. (\ref{slash})-(\ref{contractPropagator}).
The listed gamma matrix structures are multiplied by dimensionless 
functions, which can depend on the combinations
$l_A^2, \; {\bar l}_A^2, \; l_A^{\prime \; 2}, \; {\bar l}_A^{\prime 
\; 2}$, besides the renormalization scale
and the running coupling. Using the arguments similar to the ones 
leading to Eq. (\ref{oa4}),
we easily verify that the above mentioned dimensionless functions 
are at most logarithmically divergent.
Next we analyze the possible Dirac structures.
\begin{enumerate}
\item The first case has the same structure as the elementary vertex, 
and therefore causes the same suppression as the
elementary vertex.
\item The fermion-gluon-fermion composite 3-point vertex is 
sandwiched between the factors $\not{\!l_A^{\prime}}$ and 
$\not{\!l_A}$, originating from the numerators of the fermion 
propagators external to the composite vertex. Therefore the terms 
from case 2 where $\not{\!l_A}$ is on the first or third position in 
the string of the gamma matrices provide a suppression 
$\sqrt{l_A^2}$. On the other hand in the case, when $\not{\!l_A}$ is 
in the middle of this string of three gamma matrices, we encounter 
the combination
$\not{\!l_A^{\prime}} \, {\gamma}_{\mu} \not{\!l_A}$ after taking 
into account the numerators of the external fermions.
Using Eq. (\ref{slash}), we can immediately recognize that this 
combination provides a suppression ${\lambda}^{1/2}$.
\item Based on the preceding arguments it is obvious that also the 
structures included in item 3 supply at least the same suppression
factor as the elementary vertex.
\end{enumerate}
Therefore, we conclude that the composite 3-point 
fermion-gluon-fermion vertex behaves as the elementary vertex for the 
purposes
of the jet power counting.

{\it Three gluon vertex:} $V_{\mu\nu\rho}$, with external momenta 
collinear to momentum $p_A$.
This vertex can depend on momenta
$l_A, \, {\bar l}_A$ defined above and the metric tensor $g_{\alpha\beta}$.
Taking into account the dimension of the 3 gluon Green function, its 
only possible tensor structure involves combinations
of the form $[g_{\mu \nu} \, l_A^{\rho} + \mathrm{perm.} + {\cal 
O}({\lambda}^{1/2} \, \sqrt{s})]$ and
$[l_A^{\mu} \, l_A^{\nu} \, l_A^{\rho} / l_A^2 + {\cal 
O}({\lambda}^{1/2} \, \sqrt{s})]$,
with all possible replacements of $l_A \rightarrow {\bar l}_A$. These 
tensor structures are multiplied by dimensionless functions.
The former is the same as in the case of an elementary vertex and it 
therefore supplies the same suppression factor
as the elementary vertex. The latter also provides the same 
suppression as the elementary vertex, since the two momenta,
say $l_A^{\mu}, \, l_A^{\nu}$, after being contracted with the 
propagators of the external gluons, give suppression factors,
as in Eq. (\ref{contractPropagator}), which cancel the $1/l_A^2$ 
enhancement. The leftover momentum $l_A^{\rho}$
provides the same suppression factor as the elementary vertex.
Using the collinear power counting of Sec. \ref{elementary vertices}, 
one can immediately see that the
IR divergence of the dimensionless functions multiplying these tensor 
structures is not worse than logarithmic.
Hence, there is a suppression factor ${\lambda}^{1/2}$ associated 
with every contracted 3 gluon vertex.

{\it Ghost - gluon - antighost three point vertex:} When all lines 
external to the contracted vertex
are of the order $l_A$, the most general tensor structure for this 
contracted vertex is
\be \label{ghostVertex}
l_A^{\mu} \, h_1 + {\bar l}_A^{\mu} \, h_2 + {\cal O}({\lambda}^{1/2} 
\, \sqrt{s}),
\ee
with dimensionless functions $h_i = h_i \, ( l_A^2 / {\mu}^2, \, 
{\bar l}_A^2 / {\mu}^2, \, {\alpha}_s(\mu))$, $i = 1, 2$,
which are at most logarithmically IR divergent. Using Eq. 
(\ref{contractPropagator}), we see that when the momenta in Eq.
(\ref{ghostVertex}) are contracted with the tree level gluon 
propagator, we get a suppression of the order of the transverse
jet momentum, and that this contracted vertex gives the same 
suppression as the elementary three point vertex, at least.

\section{Varying the Gauge-Fixing Vector} \label{variation}

In this appendix we study the effect of an infinitesimal boost, 
performed on the gauge fixing vector $\eta$, on an expectation of a
time ordered product of fields, denoted by $O$, taken between physical states.
The gauge-fixing and the ghost terms in the QCD lagrangian are
\bea \label{lagrangian}
{\cal L}_{\mathrm{g.f.}} (x) & = & - \frac{1}{2 \xi} \, g_a^2 (x), \nonumber \\
{\cal L}_{\mathrm{ghost}} (x) & = & - b_a(x) \, \delta_{\mathrm{BRS}} 
\, g_a(x) \, / \delta \Lambda,
\eea
respectively. In Eq. (\ref{lagrangian}), $\delta \Lambda$ is a 
Grassmann parameter defining the BRS transformation, $b_a(x)$ is
an antighost field and
\be
g_a(x) \equiv - {\bar \partial} \cdot A_a (x) \equiv - (\partial - 
(\eta \cdot \partial) \, \eta) \cdot A_a(x).
\ee
Let us consider an infinitesimal boost with velocity $\delta \beta$ 
on a gauge fixing vector $\eta $ performed in the plus-minus plane
\be
\eta \rightarrow \eta' \equiv \eta + {\tilde \eta} \, \delta \beta,
\ee
where the vectors $\eta$ and ${\tilde \eta}$ are defined in Eqs. 
(\ref{eta}) and (\ref{etaTilde}), respectively.
Since only the gauge fixing and the ghost terms in the QCD lagrangian 
depend on $\eta$, we can write to accuracy
${\cal O}(\delta \beta ^2)$
\bea
\delta <{O}> \, & \equiv & \, <{O}(\eta')> - <{O}(\eta)> \; = \; 
<{\tilde \eta}^{\alpha} \frac{\partial \, {O}}
{\partial {\eta}^{\alpha}} \, \delta \beta> \nonumber \\
& = & - \frac{i}{\xi} \, \int {\mathrm d}^4 x <{O}
(\eta) \, g_a (x) \, \delta g_a (x)> \nonumber \\ 
& - & i \int {\mathrm d}^4 x < {O} 
(\eta) \, b_a(x) \, \delta \, (\delta_{\mathrm{BRS}} \, g_a (x) / \delta 
\Lambda)>. \nonumber \\
& &
\eea
Using the BRS invariance of the QCD lagrangian and the BRS 
transformation rule for an antighost field
\be
\delta_{\mathrm{BRS}} b_a(x) / \delta \Lambda \, = \, \frac{1}{\xi} \, g_a(x),
\ee
we arrive at
\be \label{var}
\delta <{O}> = -i \int {\mathrm d}^4 x <(\delta_{\mathrm{BRS}} {O} / 
\delta \Lambda) \, b_a(x) \, \delta \, g_a(x)>.
\ee
Taking a variation of $g_a(x)$ in Eq. (\ref{var}), we obtain
\be \label{varEquation}
\delta <{O}> = -i \int {\mathrm d}^4 x <(\delta_{\mathrm{BRS}} \, {O} 
/ \delta \Lambda) \, b_a(x) \,
(({\tilde \eta} \cdot \partial) \, \eta + (\eta \cdot \partial) \, 
{\tilde \eta}) \cdot A_a(x)>.
\ee
Substituting for $O$ a product of $n$ gluon fields, we can use Eq. 
(\ref{varEquation}), together with the rule
for the BRS transformation of a gluon field
\be \label{aBRS}
\delta_{\mathrm{BRS}} \, A_{\mu}^a(x) / \delta \Lambda = 
\partial_{\mu} c^a(x) + g_s f^{abc}A_{\mu}^b(x)c^c(x),
\ee
with $c^a(x)$ representing the ghost field, to derive the gauge 
variation for a connected
Green function.   However, our jet functions are one-particle irreducible
in external soft lines and we therefore cannot apply Eq. 
(\ref{varEquation}) directly, and must find an analog for
this subset of diagrams.
The modification of Eq.\ (\ref{varEquation}) due to the restriction 
to 1PI diagrams is, however, not difficult to identify.

Let us consider the graphical analog of the derivation
of Eq.\ (\ref{varEquation}) just outlined.  The variation in
$\eta$ may be implemented as a change in the gluon propagator and, in
Coulomb gauge, the ghost-gluon
interaction, which is also $\eta$-dependent.  This is the viewpoint
that was taken in axial gauge
in Ref.\ \cite{coso}.
At lowest order in the variation,
the modified gluon propagator produces scalar-polarized gluon lines,
which decouple through
repeated applications
of tree-level Ward identities to the sum over all diagrams.  The
relevant tree-level identities
are given in \cite{thooft}.
We need not describe these identities in
detail here.  We need only note that they are to be applied to any
diagram in which
a scalar polarized gluon appears at an internal vertex.  Every such application
produces a sum of diagrams, each of which
fall into one of two sets: 1)  diagrams in which an internal gluon
line is transformed
to a yet another ghost line ending in a scalar polarization, and 2)
diagrams in
which one gluon line is contracted to a point.  The new vertex
formed in the former case is the ghost term, and in the
latter case it is the ghost-gluon vertex of the BRS variation
(\ref{aBRS}).
Eq. (\ref{varEquation}) must result from the cancellation of all diagrams,
set 2), in which an internal gluon line is contracted.  Contracted external
lines provide the ghost-gluon terms,
and the ghost lines of set 1) eventually provide the ghost terms of 
the BRS variations (\ref{aBRS})
of external fields in Eq. (\ref{varEquation}).

The simplicity of the tree level Ward identities
puts strong limitations on the sets of diagrams that can combine
to form different diagrammatic contributions to Eq. (\ref{varEquation}).
For diagrams of set 1), the topology of the original diagram is
unchanged, and a 1PI diagram remains
1PI.  For diagrams of set 2), generally 1PI diagrams remain 1PI, except in the
special case of a diagram that is two-particle reducible, with these
two lines separated by a single propagator.
In this case, the contraction of the
internal line that separates the other two will bring those two lines together
at a single vertex, producing a diagram precisely of the topology
shown in Fig.\
\ref{variationJ}.   On the one hand, by Eq.\ (\ref{varEquation}) all
such diagrams must cancel
in the full perturbative sum. On the
other hand, the same topology results from a diagram that is
one-particle {\it reducible}
with respect to a single line, which is then contracted as a result of the
tree-level Ward identity.
The latter diagram, however, is not included in
the set of 1PI diagrams with which we work.  The application of the
Ward identity
to 1PI diagrams only, therefore, results in terms that would
cancel this special set of one-particle reducible diagrams,
in which the only line that spoils irreducibility is
contracted to a point.  These are the diagrams
shown in Fig.\ \ref{variationJ}, in which the ghost-gluon vertex of
Eq.\ (\ref{aBRS})
is inserted between one-particle irreducible  subdiagrams in all
possible ways.  The ghost line ending at this composite vertex is
continuously connected to the variation of a gluon propagator,
according to Eq.\ (\ref{varEquation}).
The full composite vertex of the Ward identity in Eq.\ (\ref{varEquation})
appears only at true external lines of the 1PI jet.  This vertex is given by
the momentum factor in Eq. (\ref{invPropag}) and is represented by 
the double line crossing a gluon line
in Fig. \ref{feynmanRulesF} below.
Diagrams that are reducible in one or more internal lines can be 
treated in a similar manner.
The ``left-over" terms in the Ward identities for each set of
diagrams of definite reducibility properties
(1PI, 2PI, etc.), must cancel in the full sum, reproducing the
identity for Green functions, Eq.\ (\ref{varEquation}).

\section{Tulip-Garden Formalism} \label{tulipGarden}

\begin{figure}
\scalebox{0.9}[0.9]{\includegraphics*{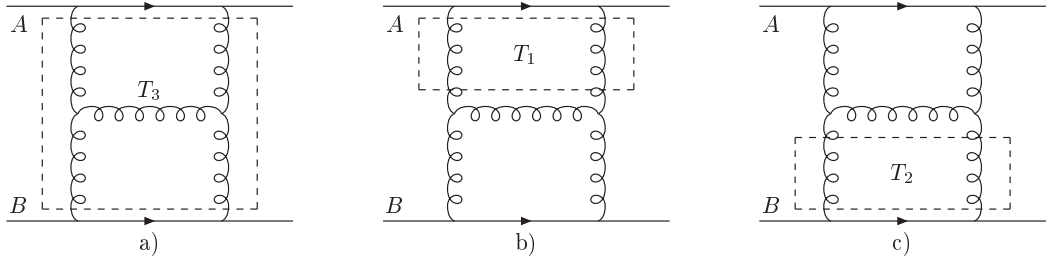}}
\caption{ \label{tulip} Two-loop diagram illustrating the idea of
the tulip-garden formalism. $T_1, T_2, T_3$ are the possible tulips.}
\end{figure}

In this appendix we illustrate how a given Feynman diagram 
contributing to the process (\ref{qqqq}) in the
leading power can be systematically written in the form 
(\ref{fact1}). For concreteness let us consider a two
loop diagram where the quarks interact via the exchange of a one rung 
gluon ladder as in Fig. \ref{tulip}.
The important contributions of this diagram come from the regions 
when all of the exchanged gluons are soft,
Fig. \ref{tulip}a or when the gluons attached to the $A$ quark line 
are soft, while the rest of the gluons carries
momenta parallel to the $-$ direction (they belong to jet $B$), Fig. 
\ref{tulip}b, or when the two gluon lines
attached to the $B$ quark line are soft and the other gluons are 
collinear to the $+$
direction (they belong to jet $A$), Fig. \ref{tulip}c.
The possible central soft exchange parts are called tulips. In our 
case the possible tulips are denoted as
$T_1, T_2, T_3$ in Fig. \ref{tulip}. The garden is defined as a 
nested set of tulips
\{$T_1, \ldots, T_n$\} such that $T_{i} \subset T_{i+1}$ for $i=1, 
\ldots, n-1$.
In Fig. \ref{tulip}, $\{T_1\}$, $\{T_2\}$, $\{T_3\}$, $\{T_1, T_3\}$, 
$\{T_2, T_3\}$ are the possible gardens.

For a given tulip we make the soft approximation, consisting of 
attaching a soft gluon to
jet $A$ via the $-$ component of its polarization only and to jet $B$ 
via the the $+$ component of its polarization.
The result of this soft approximation for a given Feynman diagram $F$ 
corresponding to a tulip $T$ is denoted $S(T)F$.
It has obviously the form of Eq. (\ref{fact1}).
Following the prescription given in Refs. \cite{coso} and 
\cite{sen83} we write the contribution to a given diagram $F$ in the 
form
\be \label{tg}
F = \sum_{G} (-1)^{n+1} S(T_1) \ldots \, S(T_n) \, F + F_R,
\ee
where the sum over inequivalent gardens, as defined bellow, $G$ in 
Eq. (\ref{tg}) is understood.
The meaning of this expression is the following. For a given garden 
consisting of a set of
tulips $\{T_1, \ldots, T_n\}$, we start with the largest tulip $T_n$ 
and make the soft approximation
for the gluon lines coming out of it. Then for $T_{n-1}$ we proceed 
the same way as for $T_n$.
If some of the lines coming out of $T_{n-1}$ are identical to the 
ones coming out of $T_n$ we leave them untouched.
For instance, if we consider a garden $\{T_2, T_3\}$ from Fig. \ref{tulip},
we first perform the soft approximation on tulip $T_3$ and then 
proceed to tulip $T_2$.
However the lines coming out of $T_2$ and $T_3$ which attach to the 
$B$ quark line are identical
so when performing $S(T_2) S(T_3) F$ we leave these gluon lines out 
of the game and make soft
approximations only on the gluon lines attaching to the ladder's 
rung. Two gardens are equivalent
if the soft approximation is identical for both of them. $F_R$ is 
defined by Eq. (\ref{tg}).
The contribution to $F_R$ comes from the integration region where 
$|\vec{k}| \gtrsim \sqrt{s}$
for all gluons coming out of the central soft part. As a result, the 
contribution to $F_R$ is
suppressed by positive powers of $\sqrt{-t}/\sqrt{s}$.
Therefore we can ignore the contribution from $F_R$ within the 
accuracy at which we are working.

We can now rewrite Eq. (\ref{tg}),  as
\be
F = \sum_{T} \left( \sum_{G, T_n = T} (-1)^{n+1} \, S(T_1) \ldots \, 
S(T_{n-1}) \right)
     S(T) \, F + F_R.
\ee
This expression is indeed in the form of Eq. (\ref{fact1}) since the 
term $S(T) F$
is of that form and the subtractions $\sum (-1)^{n+1} S(T_1) \ldots 
\, S(T_{n-1})$ modify only the
soft function $S$ in Eq. (\ref{fact1}), but  do not alter the form of 
the equation.
We can therefore conclude that the contribution to a given Feynman 
diagram in leading power can be
expressed in the first factorized form given by Eq. (\ref{fact1}).

\section{Feynman Rules} \label{feynmanRules}

\begin{figure} \center
\includegraphics*{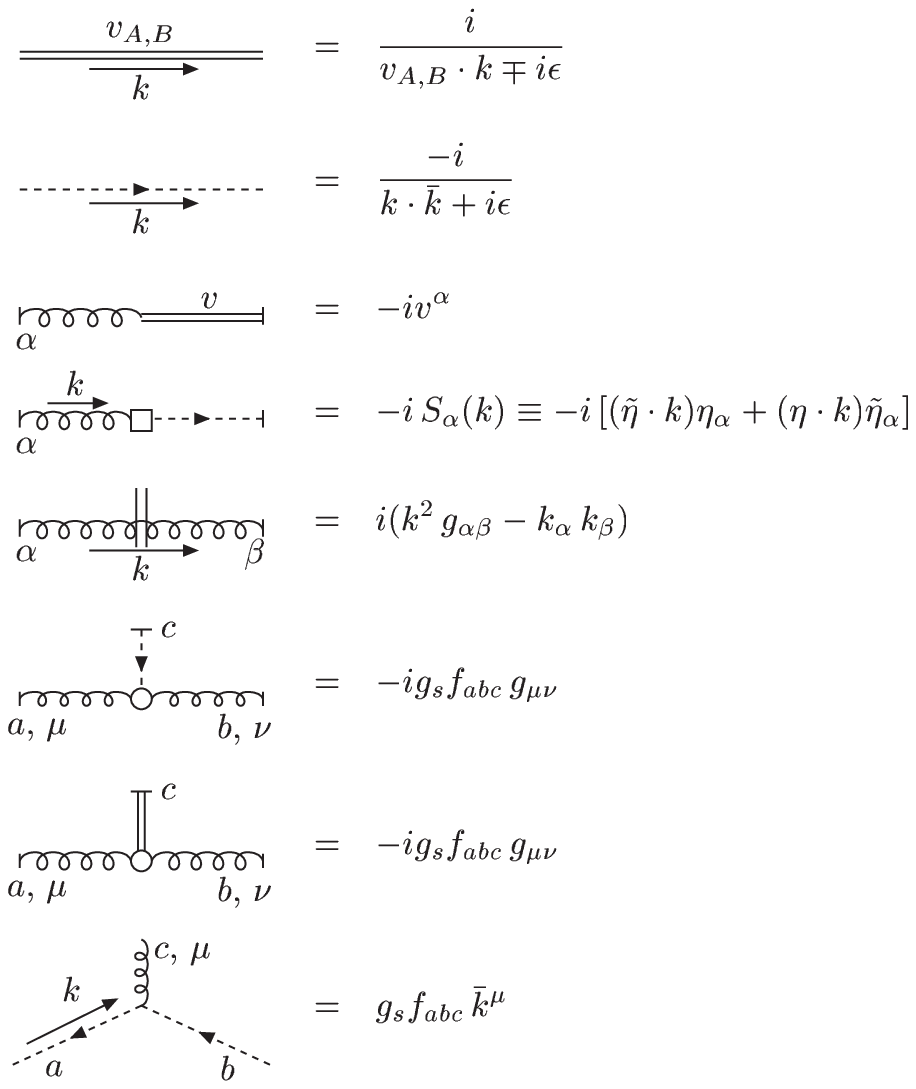}
\caption{ \label{feynmanRulesF} Feynman rules for the eikonal lines, 
ghost lines and special vertices.}
\end{figure}

In Fig. \ref{feynmanRulesF}, we list the Feynman rules for the lines 
and the vertices encountered in the text.
The double lines are eikonal lines, while the dashed lines represent 
ghosts. The four vectors $\eta, \, \tilde{\eta}$
are defined in Eqs. (\ref{eta}) and (\ref{etaTilde}), respectively.
The conventions for the gluon-ghost and gluon-eikonal vertices (third and 
second from the bottom of Fig. \ref{feynmanRulesF}) are the following.
We start with a color index of a gluon external to the diagram
defining the evolution kernel, see for instance Fig. \ref{trajll}a, 
then proceed to the gluon internal to the diagram and finally to the
ghost/eikonal line  in order to assign the color indices of 
$f_{abc}$. For the three point antighost - gluon - ghost
vertex at the bottom of Fig. \ref{feynmanRulesF}, 
we start with an antighost (arrow flowing out of the vertex) 
then proceed to the ghost and finally we reach the gluon line.

\section{Origin of Glauber Region} \label{glauberRegion}

\begin{figure} \center
\includegraphics*{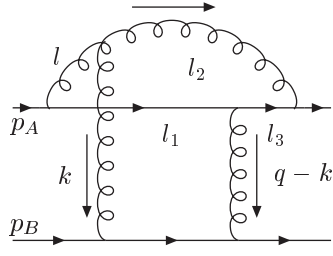}
\caption{\label{glauberExample} Two loop diagram demonstrating the 
origin of the Glauber (Coulomb) region.}
\end{figure}

In this appendix we exhibit the origin of the Glauber (Coulomb) region
using the two-loop diagram shown in Fig.
\ref{glauberExample}. Consider a situation when the upper gluon loop 
is a part of $J_A$. Momentum $k$ of the exchanged
gluon flows through jet lines  with momenta $l_2 = l - k$ and $l_3 = 
p_A - l - q + k$.
The components of $k$ can be pinched by
double poles coming from the denominators of the gluon propagators 
$k^2 + i\epsilon$ and $(q-k)^2 + i\epsilon$. In
addition to these pinches, the component $k^-$  can be pinched by the 
singularities of the jet lines $l_2$ and $l_3$,
at values
\bea \label{glauberPoles}
k^- & = & l^- - \frac{l_{2 \, \perp}^2 - i\epsilon}{2 l_2^+}, \nonumber \\
k^- & = & l^- + \frac{l_{3\perp}^2 - i\epsilon}{2 l_3^+}.
\eea
The two poles given by Eq. (\ref{glauberPoles}) are located in 
opposite half planes since in the region considered
$l_2^+, l_3^+ > 0$. This indicates that
we must consider the possibility that the
different components of the soft momentum $k$ can scale differently.
For instance, we can have  $k^+ \sim k_{\perp} \sim \sigma \sqrt{s}$ 
and $k^- \sim \lambda \sqrt{s}$ where $\lambda \ll \sigma \ll 1$.
Indeed, the power counting performed in Sec. \ref{elementary 
vertices} shows that the singularities originating
from these
regions can produce a logarithmic enhancement.  We also note that it 
is only minus components
that are pinched in this way by the lines in $J_A$, and plus 
components by the lines in $J_B$.

\chapter{} \label{chb}

\section{Symmetry property of the jet functions} \label{appa}
Here we prove Eq. (\ref{sym}). In order to do so, we first shift the 
momentum of ${\tilde J_A}^{(n)}$ in the transverse direction: 
$p_A \rightarrow p_A + q$. 
This does not effect the logarithmic behavior of the jet function, 
but it makes easier the comparison of diagrams from 
$J_A^{(n)}$ and ${\tilde J_A}^{(n)}$.
       
\begin{figure} \center
{\includegraphics*{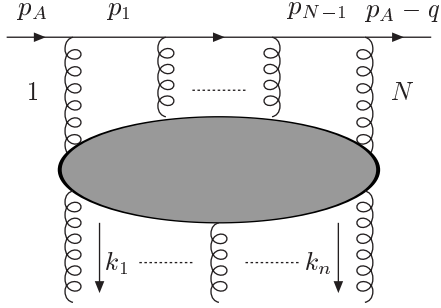}}
\caption{\label{jet} General class of diagrams contributing to $J_A^{(n)}$.}
\end{figure}

\begin{figure} \center
{\includegraphics*{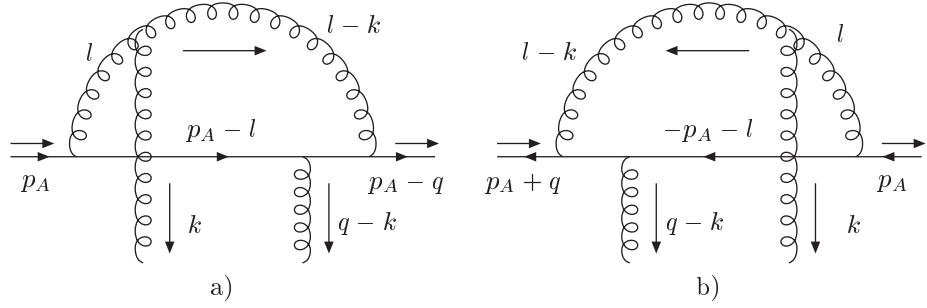}}
\caption{\label{jetexmp} Identification of a particular one loop 
diagram between the jet functions $J_A^{(2)}$ and ${\tilde J_A}^{(2)}$.}
\end{figure}

The general diagram contributing to the jet function
$J_A^{(n)}$ has the form shown in Fig. \ref{jet}. There are $N$ 
gluons connecting the quark line to the rest of the diagram, which contains 
$L$ loops. 
For every diagram contributing to $J_A^{(n)}$, we identify the corresponding 
diagram from ${\tilde J}_A^{(n)}$. The guide is that they should be a mirror 
image of each other, but they should have the same color structure. 
For instance, in the case of a one loop contribution to $J_A^{(2)}$ shown 
in Fig. \ref{jetexmp}a the related diagram from ${\tilde J}_A^{(2)}$ 
is the one in Fig. \ref{jetexmp}b. 

Let us now make a transformation in ${\tilde J_A}^{(n)}(p_A+q,\eta)$:
\be
k_i^{\pm} \rightarrow -k_i^{\pm}, \; k_{i \, \perp} \rightarrow 
k_{i \, \perp} \;\; {\rm for} \; {\rm all} \;\; i = 1,\ldots,n \, . 
\ee
In the denominators of the propagators this transformation can be compensated 
by a similar transformation of loop momenta 
$l_j^{\pm} \rightarrow -l_j^{\pm}$ and $l_{j \, \perp} \rightarrow 
l_{j \, \perp}$ for $j = 1,\ldots, (L+N-1)$. The three-point vertices 
in the $L$-loop subdiagram generate a factor $(-1)^{v_3}$ under this 
set of transformations. 
Using the relationship between the number of loops, vertices and 
external lines on the $L$-loop subdiagram:
\be
v_3 = N + n + 2(L - v_4 -1),
\ee 
we see that
\be \label{v3}
(-1)^{v_3} = (-1)^{N+n}. 
\ee
We also have to identify the relationship between the strings of 
gamma matrices on the fermion line
\be
J_A^{(n, 0)} \equiv \bar{u}(p_A - q,\lambda_{1}) \; \gamma_{\mu_N} \left(\; 
\prod_{i = N - 1}^{1} \not{\!p}_i \, \gamma_{\mu_i} \;\right) \, u(p_A,\lambda_A),
\ee
in $J_A^{(n)}$, and 
\be
{\tilde J_A}^{(n, 0)} \equiv \bar{v}(p_A + q,\lambda_{A}) \; 
\gamma_{\mu_N} \left(\; 
\prod_{i = N - 1}^{1} (- \not{\!p}_i) \, \gamma_{\mu_i} \;\right) \, 
v(p_A,\lambda_{1}),
\ee
in ${\tilde J}_A^{(n)}$.
Using the charge conjugation matrix $C$ and its properties, Ref. 
\cite{georgeqft}:
\bea
\gamma_{\mu}^{T} & = & - \; C \gamma_{\mu} C^{-1}, \nonumber \\
C^T & = & - \; C, \nonumber \\ 
\bar{v}_{\alpha} (p,\lambda) & = & (-1)^{\lambda + 1/2} \; C_{\alpha \beta} 
u_{\beta}(p,\lambda), \nonumber \\
\bar{u}_{\alpha} (p,\lambda) & = & (-1)^{\lambda + 1/2} \; C_{\alpha \beta} 
v_{\beta}(p,\lambda), 
\eea
we obtain
\be
{\tilde J_A}^{(n, 0)} = 
(-1)^{N} \, (-1)^{2N - 1} (-1)^{\lambda_A + \lambda_{1}} \, 
\bar{u}(p_A,\lambda_{1}) \; \left(\; \prod_{i = 1}^{N-1} 
\gamma_{\mu_i} \, \not{\!p}_i \;\right) \gamma_{\mu_N} \; u(p_A + q,\lambda_A).
\ee
In the Regge limit, we can neglect the momentum $q$ in the numerators. 
Since all the momenta $p_i$ are collinear to $p_A$, ${\tilde J_A}^{(n, 0)}$ is 
symmetric under the exchange of its external Lorentz indices. 
Therefore, up to the power-suppressed corrections, we have:
\be \label{symj}
{\tilde J_A}^{(n, 0)} = (-1)^{N-1} \, (-1)^{\lambda_A + \lambda_{1}} \, 
J_A^{(n, 0)}   
\ee
Combining Eqs. (\ref{v3}) and (\ref{symj}), we immediately recover 
Eq. (\ref{sym}).

\section{Evolution kernel} \label{appb}

In this appendix we provide the calculational details of the results given 
in Sec. \ref{secNll}.

The contribution to Fig. \ref{nllKgluonF}b is:

\bea \label{eqFig2b}
{\rm Fig. \; \ref{nllKgluonF}b} & \equiv & i g_s^3 \, t \, f_{acb} \, f_{fbd} \, f_{cfe} 
\, \int \frac{{\rm d}^D k}{(2 \pi)^D} \frac{{\rm d}^D l}{(2 \pi)^D} \, J_A^{(2) \, d \, e}(p_A, q; l^+ = 0, l^-, l_{\perp})
\nonumber \\
& \times & \, \left( \frac{N^{- \, \mu}(l)}{l^2} (\bar l - \bar k)_{\mu} \frac{N^{- \, \nu}(q - l)}{(q-l)^2} 
(\bar q - \bar k)_{\nu} \frac{k^+}{(\bar l - \bar k)^2} \right) (k^+ = 0) \nonumber \\ 
& \times & \, \frac{N^{- \, \rho}(k)}{k^2} \frac{S_{\rho}(k)}{k \cdot \bar k} \frac{1}{k^- - i\epsilon} 
\frac{1}{(\bar q - \bar k)^2} = 0. 
\eea
Note that the overall sign in Eq. (\ref{eqFig2b}) reflects the minus sign explicit in Fig. \ref{nllKgluonF}b.

We analyze the diagram in Fig. \ref{nllKgluonF}c in a similar way. 
It takes the form:
\bea \label{eqFig2c}
{\rm Fig. \; \ref{nllKgluonF}c} & \equiv & - i g_s^3 \, t \, f_{acb} \, f_{fbd} \, f_{cfe} 
\, \int \frac{{\rm d}^D k}{(2 \pi)^D} \frac{{\rm d}^D l}{(2 \pi)^D} \; J_A^{(2) \, d \, e}(p_A, q; l^+ = 0, l^-, l_{\perp})
\nonumber \\
& \times & \, \left( \frac{N^{- \, \mu}(l)}{l^2} (\bar l - \bar k)_{\mu} \frac{N^{- \, +}(q - l)}{(q-l)^2} 
\frac{1}{(\bar l - \bar k)^2} \right) (k^+ = 0) \nonumber \\ 
& \times & \, \frac{N^{- \, \rho}(k)}{k^2} \frac{S_{\rho}(k)}{k \cdot \bar k} \frac{1}{k^- - i\epsilon}.
\eea
Note, again, that the overall sign in Eq. (\ref{eqFig2c}) reflects the minus sign explicit in Fig. \ref{nllKgluonF}c.
The imaginary part of the eikonal propagator in Eq. (\ref{eqFig2c}) gives vanishing contribution since the resulting integrand
is an odd function under $k^+ \rightarrow - k^+$. Hence only the principal value part contributes. 
In the Glauber region, $l^- \ll l^+ \sim l_{\perp}$, the integral vanishes due to the antisymmetry of the integrand 
under the transformation $l^+ \rightarrow - l^+$, $k^{\pm} \rightarrow - k^{\pm}$. 
In the region $l^- \sim l^+ \sim l_{\perp}$ we can factor
the gluon with momentum $l$ from the jet $J_A^{(2)}$.
 
The contribution to Fig. \ref{nllEvolaF}a from a $G$ gluon with momentum $k$ 
corresponding to the first term in Eq. (\ref{sjIdent}) and to Fig. \ref{nllEvolF}a is: 
\bea \label{g1}
{\rm (Fig. \; \ref{nllEvolaF}a)_G} & \equiv & g_s^2 \, t \, f_{ade} \, f_{ebc} \, \int \frac{{\rm d}^D k}{(2 \pi)^D} 
\frac{{\rm d}^D l}{(2 \pi)^D} \, S_G(k,l,q) \nonumber \\ 
& \times & J_A^{(3) \; b \, c \, d}(p_A, q; k^+ = 0, k^-, k_{\perp}, l^+ = 0, l^-, l_{\perp}),
\eea
with the soft function $S_G$ in Eq. (\ref{g1}) defined as
\bea \label{sg1}
S_G(k,l,q) & \equiv & \left(\frac{N^{- \, \mu}(l)}{l^2} \, V_{\mu \rho \nu}(l, -k, k-l) \, 
\frac{N^{\nu \, -}(k-l)}{(k-l)^2} \left(g^{\rho \, +} - \frac{k^{\rho}}{k^-} \right) \right) \, (k^+ = 0) \nonumber \\
& \times & \frac{N^{- \, \alpha}(k)}{k^2} \, \frac{S_{\alpha}(k)}{k \cdot \bar k}   
\, \frac{N^{- \, +}(q-k)}{(q-k)^2}.
\eea
Since, by construction $k^- \sim k^+ \sim k_{\perp}$ for an external $G$ gluon 
with momentum $k$, we can factor the gluon with 
momentum $q-k$ from $J_A^{(3)}$ in Fig. \ref{nllEvolaF}a. 
Then the resulting integrand in Eq. (\ref{g1}); $S_G(k,l,q)/k^-$, with $S_G$ defined in Eq. (\ref{sg1}); is an 
antisymmetric function under the transformation $k^{\pm} \rightarrow - k^{\pm}$, $l^+ \rightarrow - l^+$ 
in the Glauber region $l^- \ll l^+ \sim l_{\perp}$. This indicates that the gluon with momentum $l$ can be factored from
$J_A^{(3)}$. Therefore the contribution to Fig. \ref{nllEvolaF}a from a $G$ gluon agrees with a gluon reggeization at NLL.    
 
The contribution to Fig. \ref{nllEvolaF}c is: 
\bea \label{g2}
{(\rm Fig. \; \ref{nllEvolaF}c)_G} & \equiv & i g_s^3 \, t \, f_{acb} \, f_{fbd} \, f_{cfe} 
\times \, \int \frac{{\rm d}^D k}{(2 \pi)^D} \frac{{\rm d}^D l}{(2 \pi)^D} \; 
J_A^{(2) \, d \, e}(p_A, q; l^+ = 0, l^-, l_{\perp}) \nonumber \\
& \times & \, \left[ \frac{N^{- \, \mu}(l)}{l^2} V_{\mu \rho \sigma}(l,-k, k-l) \left(g^{\rho \, +} - 
\frac{k^{\rho}}{k^-}\right) \frac{N^{\sigma \, \lambda}(l-k)}{(l-k)^2} \right. \nonumber \\
& \times & \, \left. V_{\nu \lambda -}(q-l,l-k,k-q) \, \frac{N^{\nu \, -}(q - l)}{(q-l)^2} \right] (k^+ = 0) \nonumber \\ 
& \times & \, \frac{N^{- \, \alpha}(k)}{k^2} \frac{S_{\alpha}(k)}{k \cdot \bar k} \frac{N^{- \, +}(q - k)}{(q-k)^2}. 
\eea
In the Glauber region the $l^+$ integral vanishes due to the antisymmetry of the integrand. In the region 
$l^- \sim l^+ \sim l_{\perp}$ the gluon with momentum $l$ factors out and the contribution takes a factorized form.

The diagram in Fig. \ref{nllEvolaF}d for a $G$ gluon with momentum $k$ is: 
\bea \label{g3}
{(\rm Fig. \; \ref{nllEvolaF}d)_G} & \equiv & - i g_s^3 \, t \, f_{aed} \,  
\times \, \int \frac{{\rm d}^D k}{(2 \pi)^D} \frac{{\rm d}^D l}{(2 \pi)^D} \; 
J_A^{(2) \, b \, c}(p_A, q; l^+ = 0, l^-, l_{\perp}) \nonumber \\
& \times & \, \frac{N^{- \, \mu}(q-l)}{(q-l)^2} \, W_{\mu \rho - \nu}^{bedc} \, 
\left(g^{\rho \, +} - \frac{k^{\rho}}{k^-}\right) (k^+ = 0) \, \frac{N^{\nu \, -}(l)}{l^2} \nonumber \\
& \times & \, \frac{N^{- \, \alpha}(k)}{k^2} \, \frac{S_{\alpha}(k)}{k \cdot \bar k} 
\, \frac{N^{- \, +}(q - k)}{(q-k)^2}, \nonumber \\ 
\eea
where 
\bea \label{w}
W_{\mu \rho - \nu}^{bedc} & \equiv & f_{pbe} \, f_{pdc} \left(g_{\mu -}g_{\rho \nu} - g_{\mu \nu}g_{\rho -}\right) +
f_{pbd} \, f_{pce} \left(g_{\mu \nu}g_{\rho -} - g_{\mu \rho}g_{\nu -}\right) 
\nonumber \\ 
& + & f_{pbc} \, f_{ped} \left(g_{\mu \rho}g_{\nu -} - 
g_{\mu -}g_{\rho \nu}\right),
\eea
is the Lorentz and the color part of the four point gluon vertex.
In the Glauber region the integral over $l^+$ vanishes due to the antisymmetry of the integrand in Eq. (\ref{g3}). 
In the region $l^- \sim l^+ \sim l_{\perp}$ we can factor the gluon with momentum $l$ from the jet function $J_A^{(2)}$.
Therefore the contribution from Fig. \ref{nllEvolaF}d is in accordance with 
gluon reggeization.

The contribution to Fig. \ref{nllEvolaF}a from the second term in Eq. (\ref{sjIdent}) used in Eq. (\ref{fig1a}) is: 
\bea \label{p1}
{\rm (Fig. \; \ref{nllEvolaF}a)_{\rm soft}} & \equiv & g_s^2 \, t \, f_{ade} \, f_{ebc} \, \int \frac{{\rm d}^D k}{(2 \pi)^D} 
\frac{{\rm d}^D l}{(2 \pi)^D} \, S_s^{(a)}(k,l,q) \nonumber \\ 
& \times & J_A^{(3) \; b \, c \, d}(p_A, q; k^+ = 0, k^-, k_{\perp}, 
l^+ = 0, l^-, l_{\perp}),
\eea
with the soft function $S_s^{(a)}$ defined by
\bea \label{p1s} 
S_s^{(a)}(k,l,q) & \equiv & 
\left[ \left(\frac{N^{- \, \mu}(l)}{l^2} \, V_{\mu \rho \nu}(l, -k, k-l) \, \frac{N^{\nu \, -}(k-l)}{(k-l)^2} \right)(k^+) \, - 
\right. \nonumber \\
& & \left. \left(\frac{N^{- \, \mu}(l)}{l^2} \, V_{\mu \rho \nu}(l, -k, k-l) \, \frac{N^{\nu \, -}(k-l)}{(k-l)^2} \right)(k^+ = 0) 
\right] \nonumber \\
& \times & \frac{N^{- \, \alpha}(k)}{k^2} \, \frac{S_{\alpha}(k)}{k \cdot \bar k} \, \frac{N^{- \, +}(q-k)}{(q-k)^2}.
\eea
Eqs. (\ref{p1}) and (\ref{p1s}) immediately follow form Eqs. (\ref{g1}) and (\ref{sg1}).  
We apply the following identity
\bea \label{sg1Ident}
S(l^-,k^-) & = & S(l^- = 0, k^- = 0) \, \theta(M-|l^-|) \, \theta(M-|k^-|) \nonumber \\
& + & [S(l^-, k^- = 0) - S(l^- = 0, k^- = 0) \, \theta(M-|l^-|) ] \, \theta(M-|k^-|) \nonumber \\
& + & [S(l^-=0, k^-) - S(l^- = 0, k^- = 0) \, \theta(M-|k^-|)]  \, \theta(M-|l^-|) \nonumber \\
& + & [ \{S(l^-, k^-) - S(l^-, k^- = 0) \, \theta(M-|k^-|) - \} \nonumber \\
& & \{S(l^- = 0, k^-) - S(l^- = 0, k^- = 0) \, \theta(M-|k^-|)\} \, \theta(M-|l^-|)], \nonumber \\
\eea
to the soft function $S_s^{(a)}$, Eq. (\ref{p1s}), to analyze the contribution to Eq. (\ref{p1}). 
After using the first term of Eq. (\ref{sg1Ident}) for the soft function $S_s^{(a)}$ in Eq. (\ref{p1}), we arrive at
\bea \label{p1A}
{\rm (Fig. \; \ref{nllEvolaF}a)_{\rm soft}^{(1)}} & \equiv & g_s^2 \, t \, f_{ade} \, f_{ebc} \, 
\int \frac{{\rm d}^{D-2} k_{\perp}}{(2 \pi)^D} \frac{{\rm d}^{D-2} l_{\perp}}{(2 \pi)^D} \nonumber \\ 
& \times & \left( \int {\rm d}k^+ \, {\rm d}l^+ \, S_s^{(a)}(k^+, l^+, k^-=0, l^-=0, k_{\perp}, l_{\perp},q) \right) \nonumber \\
& \times & {\Gamma}_A^{(3) \; b \, c \, d}(p_A, q; 
k_{\perp}, l_{\perp}).
\eea
Performing the $k^+$ and $l^+$ integrals, we obtain
\bea \label{p1Ares}
{\rm (Fig. \; \ref{nllEvolaF}a)_{\rm soft}^{(1)}} & \equiv & - \frac{{\alpha}_s}{8 \pi} \, t \, f_{ade} \, f_{ebc} \, 
\int \frac{{\rm d}^{D-2} k_{\perp}}{(2 \pi)^{D-2}} \frac{{\rm d}^{D-2} l_{\perp}}{(2 \pi)^{D-2}} \, 
\frac{(\kp - \qklp)}{\kp^2 \, \qklp^2} \nonumber \\ 
 & & \hspace{-3.5cm} \times \; \frac{(\kp - \qlp + \qklp) \, (\kp + \lp+ 2\qlp+ \qklp)}{\qlp \, (\lp + \qlp)^2 \, (\kp+\lp+\qklp)} \nonumber \\ 
& & \hspace{-3.5cm} \times \; \frac{1}{(\kp+\qlp+\qklp)} \; {\Gamma}_A^{(3) \; b \, c \, d}(p_A, q; k_{\perp}, l_{\perp}).
\eea 
When analyzing the second (third) term in Eq. (\ref{sg1Ident}), after used in Eq. (\ref{p1}), we can factor the gluon with momentum
$l$ ($k$) from the jet function $J_A^{(3)}$. The resulting $l^{\pm}$, $k^+$ ($k^{\pm}$, $l^+$) integral is over an antisymmetric
function under the transformation $l^{\pm} \rightarrow - l^{\pm}$, $k^+ \rightarrow - k^+$ 
($k^{\pm} \rightarrow - k^{\pm}$, $l^+ \rightarrow - l^+$) and therefore it vanishes. In the case of the last term in 
Eq. (\ref{sg1Ident}), we can factor both gluons with momenta $k$ and $l$ from the jet $J_A^{(3)}$. So this part is 
in agreement with gluon reggeization.

The contribution to Fig. \ref{nllEvolaF}b from the second term in Eq. (\ref{sjIdent}) used in Eq. (\ref{fig1a}) is: 
\bea \label{p2}
{\rm (Fig. \; \ref{nllEvolaF}b)_{\rm soft}} & \equiv & - g_s^2 \, t \, f_{aeb} \, f_{ced} \, \int \frac{{\rm d}^D k}{(2 \pi)^D} 
\frac{{\rm d}^D l}{(2 \pi)^D} \, S_s^{(b)}(k,l,q) \nonumber \\
& \times & J_A^{(3) \; b \, c \, d}(p_A, q; k^+ = 0, k^-, k_{\perp}, l^+ = 0, l^-, l_{\perp}), 
\nonumber \\
\eea
with the soft function $S_s^{(b)}(k,l,q)$ defined by
\bea \label{p2s}
S_s^{(b)}(k,l,q) & \equiv & \left[ \left(\frac{N^{- \, \mu}(k-l)}{(k-l)^2} \, V_{\mu - \nu}(k-l, l-q, q-k) \, 
\frac{N^{\nu \, -}(q-k)}{(q-k)^2} \right)(k^+) \, - \right. \nonumber \\
& & \left. \left(\frac{N^{- \, \mu}(k-l)}{(k-l)^2} \, V_{\mu - \nu}(k-l, l-q, q-k) \, \frac{N^{\nu \, -}(q-k)}{(q-k)^2} \right)(k^+ = 0) 
\right] \nonumber \\
& \times & \frac{N^{- \, \alpha}(l)}{l^2} \, \frac{S_{\alpha}(l)}{l \cdot \bar l} \, \frac{N^{- \, +}(q-l)}{(q-l)^2}.
\eea
We apply the identity (\ref{sg1Ident}) to the soft function $S_s^{(b)}$, Eq. (\ref{p2s}), to analyze the contribution to Eq. (\ref{p2}). 
After using the first term of Eq. (\ref{sg1Ident}) in Eq. (\ref{p2}), we arrive at a contribution
\bea \label{p2A}
{\rm (Fig. \; \ref{nllEvolaF}b)_{\rm soft}^{(1)}} & \equiv & - g_s^2 \, t \, f_{aeb} \, f_{ced} \, 
\int \frac{{\rm d}^{D-2} k_{\perp}}{(2 \pi)^D} \frac{{\rm d}^{D-2} l_{\perp}}{(2 \pi)^D} \nonumber \\
& \times & \left( \int {\rm d}k^+ \, {\rm d}l^+ \, S_s^{(b)}(k^+, l^+, k^-=0, l^-=0, k_{\perp}, l_{\perp},q) \right) \nonumber \\
& \times & {\Gamma}_A^{(3) \; b \, c \, d}(p_A, q; k_{\perp}, l_{\perp}).
\eea
Performing the $k^+$ and $l^+$ integrals, we obtain
\bea \label{p2Ares}
{\rm (Fig. \; \ref{nllEvolaF}b)_{\rm soft}^{(1)}} & \equiv & \frac{{\alpha}_s}{8 \pi} \, t \, f_{abe} \, f_{edc} \, 
\int \frac{{\rm d}^{D-2} k_{\perp}}{(2 \pi)^D} \frac{{\rm d}^{D-2} l_{\perp}}{(2 \pi)^D} \nonumber \\ 
& & \hspace{-3cm} \times \; \frac{(2\kp + \lp +\qkp + \qklp)}{\kp \, \lp^2 \, \qklp^2 \, (\lp+\qkp+\qklp)} \nonumber \\ 
& & \hspace{-3cm} \times \; \frac{\left(\lp^3 - \lp \left(\qkp^2 + \qklp^2\right) - 2 \qkp^2 \qklp\right)}{(\kp + \lp + \qklp)^2 \, 
(\kp+\qkp)^2} \nonumber \\ 
& & \hspace{-3cm} \times \; {\Gamma}_A^{(3) \; b \, c \, d}(p_A, q; k_{\perp}, l_{\perp}). \eea 
When analyzing the second (third) term in Eq. (\ref{sg1Ident}), when 
applied to Eq. (\ref{p2}), we can factor the gluon with momentum
$l$ ($k$) from the jet function $J_A^{(3)}$. The resulting $l^{\pm}$, $k^+$ ($k^{\pm}$, $l^+$) integral is over a function antisymmetric under the 
transformation $l^{\pm} \rightarrow - l^{\pm}$, $k^+ \rightarrow - k^+$ 
($k^{\pm} \rightarrow - k^{\pm}$, $l^+ \rightarrow - l^+$) and therefore it vanishes. In the case of the last term in 
Eq. (\ref{sg1Ident}), we can factor both gluons with momenta $k$ and $l$ from the jet $J_A^{(3)}$. So this part is 
in agreement with the gluon reggeization.

The contribution to Fig. \ref{nllEvolaF}c from the second term in Eq. (\ref{sjIdent}) used in Eq. (\ref{fig1a}) is: 
\bea \label{p3}
{(\rm Fig. \; \ref{nllEvolaF}c)_{\rm soft}} & \equiv & i g_s^3 \, t \, f_{acb} \, f_{fbd} \, f_{cfe} 
\times \, \int \frac{{\rm d}^D k}{(2 \pi)^D} \frac{{\rm d}^D l}{(2 \pi)^D} \; 
J_A^{(2) \, d \, e}(p_A, q; l^+ = 0, l^-, l_{\perp}) \nonumber \\
& & \hspace*{-4cm} \times \, \left[ \left( \frac{N^{- \, \mu}(l)}{l^2} V_{\mu - \sigma}(l,-k, k-l) \, 
\frac{N^{\sigma \, \lambda}(l-k)}{(l-k)^2} \, V_{\nu \lambda -}(q-l,l-k,k-q) \, \frac{N^{\nu \, -}(q - l)}{(q-l)^2} \right) (k^+) 
\right. \nonumber \\
& & \hspace*{-4cm} - \, \left. \left( \frac{N^{- \, \mu}(l)}{l^2} V_{\mu - \sigma}(l,-k, k-l) 
\frac{N^{\sigma \, \lambda}(l-k)}{(l-k)^2} V_{\nu \lambda -}(q-l,l-k,k-q) \, \frac{N^{\nu \, -}(q - l)}{(q-l)^2} \right) 
(k^+ = 0) \right] \nonumber \\ 
& & \hspace*{-4cm} \times \, \frac{N^{- \, \alpha}(k)}{k^2} \frac{S_{\alpha}(k)}{k \cdot \bar k} \frac{N^{- \, +}(q - k)}{(q-k)^2}. \nonumber \\
\eea
This follows immediately from Eq. (\ref{g2}).
In the Glauber region, $l^- \ll l^+ \sim l_{\perp}$, the integrand is an antisymmetric function of $k^{\pm}$ and $l^+$ and therefore 
the $k^{\pm}$, $l^+$ integral vanishes. In the region $l^- \sim l^+ \sim l_{\perp}$, we can factor the gluon with momentum $l$
from the jet function $J_A^{(2)}$. Hence this contribution is in agreement with gluon reggeization. 

The contribution to Fig. \ref{nllEvolaF}a from the third term in Eq. (\ref{sjIdent}) used in Eq. (\ref{fig1a}) is: 
\bea \label{s1}
{\rm (Fig. \; \ref{nllEvolaF}a)_{\rm soft'}} & \equiv & g_s^2 \, t \, f_{ade} \, f_{ebc} \, \int \frac{{\rm d}^D k}{(2 \pi)^D} 
\frac{{\rm d}^D l}{(2 \pi)^D} \, S_{s'}^{(a)} \nonumber \\ 
& \times & J_A^{(3) \; b \, c \, d}(p_A, q; k^+ = 0, k^-, k_{\perp}, l^+ = 0, l^-, l_{\perp}), \nonumber \\
\eea
with the soft function $S_{s'}^{(a)}$ defined by 
\bea \label{s1s} 
S_{s'}^{(a)} & \equiv &  \frac{N^{- \, \mu}(l)}{l^2} \, \left( V_{\mu \rho \nu}(l, -k, k-l) - g_{+ \rho} 
\, V_{\mu - \nu}(l, -k, k-l) \right) \, \frac{N^{\nu \, -}(k-l)}{(k-l)^2} \nonumber \\
& \times & \frac{N^{\rho \, \alpha}(k)}{k^2} \, \frac{S_{\alpha}(k)}{k \cdot \bar k} \, \frac{N^{- \, +}(q-k)}{(q-k)^2}.
\eea
This follows immediately from Eqs. (\ref{g1}) and (\ref{sg1}).
We apply the identity (\ref{sg1Ident}) to the soft function $S^{(a)}_{s'}$, Eq. (\ref{s1s}), to analyze the contribution 
to Eq. (\ref{s1}). 
After using the first term of Eq. (\ref{sg1Ident}) in Eq. (\ref{s1}), 
we arrive at
\bea \label{s1A}
{\rm (Fig. \; \ref{nllEvolaF}a)_{\rm soft'}^{(1)}} & \equiv & g_s^2 \, t \, f_{ade} \, f_{ebc} \, 
\int \frac{{\rm d}^{D-2} k_{\perp}}{(2 \pi)^D} \frac{{\rm d}^{D-2} l_{\perp}}{(2 \pi)^D} \nonumber \\ 
& \times & \left( \int {\rm d}k^+ \, {\rm d}l^+ \, S_{s'}^{(a)}(k^+, l^+, k^-=0, l^-=0, k_{\perp}, l_{\perp},q) \right) \nonumber \\
& \times & {\Gamma}_A^{(3) \; b \, c \, d}(p_A, q; k_{\perp}, l_{\perp}).
\eea
The integrand $S_{s'}^{(a)}$ in Eq. (\ref{s1A}) is an antisymmetric function of $l^+$ and $k^+$ and therefore the integral
over $k^+$ and $l^+$ in Eq. (\ref{s1A}) vanishes. In order to show this antisymmetry, we have used that the component
\be 
N^{+ \, \alpha}(k^- = 0, k^+, k_{\perp}) \, S_{\alpha}(k^- = 0, k^+, k_{\perp}) = 0,
\ee
in the Glauber region, $k^- \sim 0$. We have also used the Ward identity
\be 
l^{\mu} k^{\rho} V_{\mu \rho \nu}(l, -k, k-l) (k-l)^{\nu} = 0,
\ee
when analyzing the contribution from the transverse polarization of the gluon with momentum $k$ in Fig. \ref{nllEvolaF}a. 
When studying the second (third) term in Eq. (\ref{sg1Ident}), after applied 
to Eq. (\ref{s1}), we can factor the gluon with momentum
$l$ ($k$) from the jet function $J_A^{(3)}$. The resulting $l^{\pm}$, $k^+$ ($k^{\pm}$, $l^+$) integral is over an antisymmetric
function under the transformation $l^{\pm} \rightarrow - l^{\pm}$, $k^+ \rightarrow - k^+$ 
($k^{\pm} \rightarrow - k^{\pm}$, $l^+ \rightarrow - l^+$) and therefore it vanishes. In the case of the last term in 
Eq. (\ref{sg1Ident}), we can factor both gluons with momenta $k$ and $l$ from the jet $J_A^{(3)}$.  
Therefore the contribution from ${\rm (Fig. \; \ref{nllEvolaF}a)_{{\rm soft}'}}$, Eq. (\ref{s1}), is in an agreement with 
gluon reggeization at NLL.

The contribution to Fig. \ref{nllEvolaF}b from the third term in Eq. (\ref{sjIdent}) used in Eq. (\ref{fig1a}) is: 
\bea \label{s2}
{\rm (Fig. \; \ref{nllEvolaF}b)_{\rm soft'}} & \equiv & - g_s^2 \, t \, f_{aeb} \, f_{ced} \, \int \frac{{\rm d}^D k}{(2 \pi)^D} 
\frac{{\rm d}^D l}{(2 \pi)^D} \, S_{s'}^{(b)} \nonumber \\
& \times & J_A^{(3) \; b \, c \, d}(p_A, q; k^+ = 0, k^-, k_{\perp}, l^+ = 0, l^-, l_{\perp}), 
\eea
with the soft function $S_{s'}^{(b)}$ defined by
\bea \label{s2s} 
S_{s'}^{(b)} & \equiv & \frac{N^{- \, \mu}(k-l)}{(k-l)^2} \, \left( V_{\mu \rho \nu}(k-l, l-q, q-k) - g_{+ \rho} \, 
V_{\mu - \nu}(k-l, l-q, q-k) \right) \nonumber \\ 
& \times & \frac{N^{\nu \, -}(q-k)}{(q-k)^2} \, \frac{N^{- \, \alpha}(l)}{l^2} \, \frac{S_{\alpha}(l)}{l \cdot \bar l} \, \frac{N^{\rho \, +}(q-l)}{(q-l)^2}.
\eea
This follows from Eqs. (\ref{p2}) and (\ref{p2s}).
We apply the identity (\ref{sg1Ident}) to the soft function, Eq. (\ref{s2s}), to analyze the contribution to Eq. (\ref{s2}). 
After using the first term of Eq. (\ref{sg1Ident}) in Eq. (\ref{s2}), we arrive at
\bea \label{s2A}
{\rm (Fig. \; \ref{nllEvolaF}b)_{\rm soft'}^{(1)}} & \equiv & - g_s^2 \, t \, f_{aeb} \, f_{ced} \, 
\int \frac{{\rm d}^{D-2} k_{\perp}}{(2 \pi)^D} \frac{{\rm d}^{D-2} l_{\perp}}{(2 \pi)^D} \nonumber \\ 
& \times & \left( \int {\rm d}k^+ \, {\rm d}l^+ \, S_{s'}^{(b)}(k^+, l^+, k^-=0, l^-=0, k_{\perp}, l_{\perp},q) \right) \nonumber \\
& \times & {\Gamma}_A^{(3) \; b \, c \, d}(p_A, q; k_{\perp}, l_{\perp}).
\eea
The integrand $S_{s'}^{(b)}$ in Eq. (\ref{s2A}) is an antisymmetric function of $l^+$ and $k^+$ and therefore the integral
over $k^+$ and $l^+$ in Eq. (\ref{s2A}) vanishes. In order to show this antisymmetry, we have used that the component 
$N^{+ +}(q-l) = 0$ in the Glauber region, $l^- \sim 0$. We have also used the Ward identity
\be
(k-l)^{\mu} (l-q)^{\rho} V_{\mu \rho \nu}(k-l, l-q, q-k) (q-k)^{\nu} = 0,
\ee
when analyzing the contribution from the transverse polarization of the gluon with momentum $q-l$ in Fig. \ref{nllEvolaF}b. 
When analyzing the second (third) term in Eq. (\ref{sg1Ident}), applied to Eq. (\ref{s2}), we can factor the gluon with momentum
$l$ ($k$) out the jet function $J_A^{(3)}$. The resulting $l^{\pm}$, $k^+$ ($k^{\pm}$, $l^+$) integral is over an antisymmetric
function under the transformation $l^{\pm} \rightarrow - l^{\pm}$, $k^+ \rightarrow - k^+$ 
($k^{\pm} \rightarrow - k^{\pm}$, $l^+ \rightarrow - l^+$) and therefore it vanishes. In the case of the last term in 
Eq. (\ref{sg1Ident}), we can factor both gluons with momenta $k$ and $l$ from the jet $J_A^{(3)}$.  
Therefore the contribution from ${\rm (Fig. \; \ref{nllEvolaF}b)_{\rm soft'}}$, Eq. (\ref{s2}), 
is in an agreement with gluon reggeization at NLL.

The contribution to Fig. \ref{nllEvolaF}c from the third term in Eq. (\ref{sjIdent}) used in Eq. (\ref{fig1a}) is: 
\bea \label{s3}
{(\rm Fig. \; \ref{nllEvolaF}c)_{\rm soft'}} & \equiv & i g_s^3 \, t \, f_{acb} \, f_{fbd} \, f_{cfe} 
\times \, \int \frac{{\rm d}^D k}{(2 \pi)^D} \frac{{\rm d}^D l}{(2 \pi)^D} 
\nonumber \\ 
& & \hspace{-2.cm} \times \; J_A^{(2) \, d \, e}(p_A, q; l^+ = 0, l^-, l_{\perp}) \, \frac{N^{- \, \mu}(l)}{l^2} \nonumber \\
& & \hspace{-2.cm} \times \; \left( V_{\mu \beta \sigma}(l,-k, k-l) -  
g_{- \beta} \, V_{\mu - \sigma}(l,-k, k-l) \right) 
\frac{N^{\sigma \, \lambda}(l-k)}{(l-k)^2} \nonumber \\
& & \hspace{-2.cm} \times \; \left( V_{\nu \lambda \gamma}(q-l,l-k,k-q) - g_{- \gamma} \, V_{\nu \lambda -}(q-l,l-k,k-q) \right) 
\, \nonumber \\ 
& & \hspace{-2.cm} \times \; \frac{N^{\nu \, -}(q - l)}{(q-l)^2} \, 
\frac{N^{\beta \, \alpha}(k)}{k^2} \frac{S_{\alpha}(k)}{k \cdot \bar k} 
\frac{N^{\gamma \, +}(q - k)}{(q-k)^2}. 
\eea
In the Glauber region, $l^- \ll l^+ \sim l_{\perp}$, the integrand is an antisymmetric function of $k^{\pm}$ and $l^+$ and therefore 
the $k^{\pm}$, $l^+$ integral vanishes. In the region $l^- \sim l^+ \sim l_{\perp}$, we can factor the gluon with momentum $l$
out the jet function $J_A^{(2)}$. Hence this contribution is in agreement with gluon reggeization. 

The diagram in Fig. \ref{nllEvolaF}d from the third term in Eq. (\ref{sjIdent}) used in Eq. (\ref{fig1a}) is: 
\bea \label{s4}
{(\rm Fig. \; \ref{nllEvolaF}d)_{soft'}} & \equiv & - i g_s^3 \, t \, f_{aed} \,  
\times \, \int \frac{{\rm d}^D k}{(2 \pi)^D} \frac{{\rm d}^D l}{(2 \pi)^D} \; 
J_A^{(2) \, b \, c}(p_A, q; l^+ = 0, l^-, l_{\perp}) \nonumber \\
& \times & \, \frac{N^{- \, \mu}(q-l)}{(q-l)^2} \, \left( W_{\mu \rho \sigma \nu}^{bedc} - g_{- \rho} \, g_{- \sigma} \,
W_{\mu - - \nu}^{bedc} \right) \, \frac{N^{\nu \, -}(l)}{l^2} \,
\frac{N^{\rho \, \alpha}(k)}{k^2} \nonumber \\
& \times & \frac{S_{\alpha}(k)}{k \cdot \bar k} \, \frac{N^{\sigma \, +}(q - k)}{(q-k)^2},
\eea
In the Glauber region, $l^- \ll l^+ \sim l_{\perp}$, the integrand is an antisymmetric function of $k^{\pm}$ and $l^+$ and therefore 
the $k^{\pm}$, $l^+$ integral vanishes. In the region $l^- \sim l^+ \sim l_{\perp}$, we can factor the gluon with momentum $l$
out the jet function $J_A^{(2)}$. Hence this contribution is in agreement with gluon reggeization. 

The contribution from the first term in Eq. (\ref{fig1d}) is:
\bea \label{eqFig1cColl}
{\rm (Fig. \; \ref{nllEvolF}c)_{jet}} & \equiv & g_s^2 \, t \, f_{ade} \, f_{ebc} \, \int \frac{{\rm d}^D k}{(2 \pi)^D} 
\frac{{\rm d}^D l}{(2 \pi)^D} \, S_{\rm jet}(k,l,q) \nonumber \\
& \times & J_A^{(3) \; b \, c \, d}(p_A, q; k^+ = 0, k^-, k_{\perp}, 
l^+ = 0, l^-, l_{\perp}), \nonumber \\
\eea
with a soft function $S_{\rm jet}$ defined  
\bea \label{softFun}
S_{\rm jet}(k, l, q) & \equiv & \frac{N^{- \, \mu}(k)}{k^2} \, \frac{S_{\mu}(k)}{k \cdot \bar k} 
\, \frac{N^{- \, \alpha}(q-l-k)}{(q-l-k)^2} \, \frac{(\bar q - \bar l)_{\alpha}}{(l-q) \cdot (\bar l - \bar q)} 
\, \frac{N^{- \, +}(l)}{l^2}. \nonumber \\
\eea
Next we use an identity (\ref{sg1Ident}) for $S_{\rm jet}(k^-, l^-)$, defined in Eq. (\ref{softFun}).
After performing the $k^+$ and $l^+$ integrals in the first term of Eq. (\ref{sg1Ident}), when used in Eq. (\ref{eqFig1cColl}),
we obtain the following contribution:
\bea \label{j1res}
{\rm (Fig. \; \ref{nllEvolF}c)_{\rm jet}^{(1)}} & \equiv & - \frac{{\alpha}_s}{8 \pi} \, t \, f_{ade} \, f_{ebc} \, 
\int \frac{{\rm d}^{D-2} k_{\perp}}{(2 \pi)^{D-2}} \frac{{\rm d}^{D-2} l_{\perp}}{(2 \pi)^{D-2}} \; 
{\Gamma}_A^{(3) \; b \, c \, d}(p_A, q; k_{\perp}, l_{\perp}) \nonumber \\  
& & \hspace{-3cm} \times \; \left[\frac{(\kp + \lp + \qlp + 2 \qklp)}{\kp \, \qklp \, (\lp+\qlp) \, (\kp + \qkp +\qklp)^2} \right. \nonumber \\
& & \left. \hspace{-3cm} \times \; \frac{(\kp^2 - \qlp^2 + \qklp^2)}{(\kp + \qlp +\qklp)^2 \, \qklp^2} \right. \nonumber \\ 
& & \left. \hspace{-3cm} \; + \left(\frac{\lp+2\qklp}{(\lp+\qklp) \, (\kp+\lp+\qklp)^2} \right. \right. \nonumber \\
& & \left. \left. \hspace*{-3cm} - \frac{\qlp+2\qklp}{(\qlp+\qklp) \, (\kp+\qlp+\qklp)^2}\right) \right. \nonumber \\
& & \left. \hspace*{-3cm} \times \; \frac{(\kp^2 - \qlp^2 - \qklp^2)}{\kp \, (\lp^2 - \qlp^2) \, \qklp^3} \; \right]. 
\eea
In the second (third) term of Eq. (\ref{sg1Ident}), we can factor the gluon with momentum $l$ ($k$) 
from the jet function $J_A^{(3)}$. The resulting $l^{\pm}$, $k^+$ ($k^{\pm}$, $l^+$) integrals are over antisymmetric functions
and therefore they vanish. In the last term of Eq. (\ref{sg1Ident}), we can factor both gluons with momenta $k$ and $l$ from the jet
function $J_A^{(3)}$. Hence this contribution is in a factorized form, in agreement with the gluon reggeization.

The contribution to the second term in Eq. (\ref{fig1d}) comes from a diagram in Fig. \ref{nllEvolaF}e:
\bea \label{eqFig3e}
{\rm Fig. \; \ref{nllEvolaF}e} & = & i g_s^3 \, t \, f_{aeb} \, f_{bcf} \, f_{fdc}
\times \, \int \frac{{\rm d}^D k}{(2 \pi)^D} \frac{{\rm d}^D l}{(2 \pi)^D} 
\; J_A^{(2) \, d \, e}(p_A, q; l^+ = 0, l^-, l_{\perp}) \nonumber \\
& \times & \, \frac{N^{- \, \mu}(q-l)}{(q-l)^2} \, V_{\mu \nu \rho}(q-l,-k, k+l-q) \, \frac{N^{\nu \, \alpha}(k)}{k^2} 
\, \frac{S_{\alpha}(k)}{k \cdot \bar k} \nonumber \\
& \times & \, \frac{N^{\rho \, \sigma}(q-l-k)}{(q-l-k)^2} \, \frac{(\bar l - \bar q)_{\sigma}}{(l-q) \cdot (\bar l - \bar q)} 
\, \frac{N^{- \, +}(l)}{l^2}. 
\eea
In the Glauber region $l^- \ll l^+ \sim l_{\perp}$, the integrand is an antisymmetric function under the transformation
$k^{\pm} \rightarrow - k^{\pm}$ and $l^+ \rightarrow - l^+$, so the integral over 
$k^{\pm}$ and $l^+$ vanishes. In the region $l^- \sim l^+ \sim l_{\perp}$ we can factor the gluon with momentum $l$ from the jet
function $J_A^{(2)}$ using the $K$-$G$ decomposition. Therefore the contribution from Fig. \ref{nllEvolaF}e takes a factorized 
form. 

The same reasoning applies to the diagram in Fig. \ref{nllEvolaF}f, 
which is due to the third term in Eq. (\ref{fig1d}):
\bea \label{eqFig3f}
{\rm Fig. \; \ref{nllEvolaF}f} & = & i g_s^3 \, t \, f_{abc} \, f_{cdf} \, f_{fbe}
\times \, \int \frac{{\rm d}^D k}{(2 \pi)^D} \frac{{\rm d}^D l}{(2 \pi)^D} 
\; J_A^{(2) \, d \, e}(p_A, q; k^+ = 0, k^-, k_{\perp}) \nonumber \\
& \times & \, \frac{N^{- \, \alpha}(k)}{k^2} \, \frac{S_{\alpha}(k)}{k \cdot \bar k} \,   
\frac{N^{- \, \mu}(q-k)}{(q-k)^2} \, V_{\mu \nu \rho}(q-k,k+l-q,-l) \, \frac{N^{\rho \, +}(l)}{l^2} \nonumber \\
& \times & \, \frac{N^{\nu \, \sigma}(q-l-k)}{(q-l-k)^2} \, \frac{(\bar l - \bar q)_{\sigma}}{(l-q) \cdot (\bar l - \bar q)}.
\eea
It also factorizes in agreement with the gluon reggeization hypothesis.


\chapter{} \label{chc}

\section{Eikonal Example}
\label{eikapp}

In this appendix, we give details of the
calculation of the logarithmic behavior in the
diagrams of Fig.\ \ref{diagrams}.
We choose the reference frame such
that the momenta of the final
state particles are given by:
\ba \label{momenta}
{\beta}_1 & = & (1,0,0, 1), \nonumber \\
{\beta}_2 & = & (1,0,0,-1), \nonumber \\
l & = & \ol (1, s_l, 0, c_l), \nonumber \\
k & = & \ok (1, s_k \cfi, s_k \sfi, c_k).
\ea
Here we define $s_{l,k} \equiv \sin
{\theta}_{l,k}$ and $c_{l,k} \equiv
\cos {\theta}_{l,k}$. $\theta_l$ is the angle between the vectors
$\vec{l}$ and $\vec{\beta_1}$,  $\theta_k$ is the
angle between the vectors $\vec{k}$ and $\vec{\beta_1}$
and $\phi$ is the azimuthal angle
of the gluon with momentum $k$ relative to the plane defined by
$\beta_1$, $\beta_2$ and $l$. The
available phase space in polar angle for the radiated gluons is
${\theta}_k \in (\pi/2 - \delta, \,
\pi/2 + \delta)$ and ${\theta}_l \in (0,\, \pi/2 - \delta) \cup
(\pi/2 + \delta, \, \pi)$.

Using the diagrammatic rules for eikonal lines and
vertices, as listed for example in \cite{pQCD}, we can write down the
expressions corresponding to each diagram separately. For example,
diagram \ref{diagrams} a) gives
\ba \label{a}
a) \, + \, (k \leftrightarrow l)\!\! &\!\! =\!\! &\!\! \left[ f_{abc} 
\mathrm{Tr}(t_a
t_b t_c) \right] \left( -i g_s^4 \,
\beta_1^{\alpha} \beta_2^{\beta} \beta_1^{\gamma} \right) \,
V_{\alpha \beta \gamma}(k+l, -k, -l) \nonumber \\
& \times & \frac{1}{\beta_1
\cdot (k+l)} \, \frac{1} {2 k \cdot l} \, \frac{1}{\beta_1 \cdot l}
\, \frac{1}{\beta_2 \cdot k} \nonumber \\
& + &  \, (k \leftrightarrow l).
\eea
$V_{\alpha \beta \gamma}(k+l, -k, -l) =
[ (2k+l)_{\gamma} g_{\alpha \beta} + (l-k)_{\alpha} g_{\beta \gamma}
- (2l + k)_{\beta} g_{\alpha
\gamma}]$ is the
momentum-dependent part of the three gluon vertex.
Using the color identity $f_{abc} \mathrm{Tr}(t_a t_b t_c) = i C_F \Ncol
C_A /2$, and the approximation $\beta_j \cdot l \gg \beta_j \cdot
k$ for $j=1,2$, which is valid due
to the strong ordering of the final state gluon energies, we arrive at
\be
a) + (k \leftrightarrow l) = \frac{1}{4} \, C_F \Ncol C_A \, g_s^4 \,
\frac{\beta_1 \cdot \beta_2}{k
\cdot l}
\left( \frac{1}{\beta_1 \cdot k \, \beta_2 \cdot l} +
\frac{2}{\beta_1 \cdot l \, \beta_2 \cdot k}
\right).
\ee
We proceed in a similar manner for the rest of the diagrams. The
results are:
\ba \label{b-e}
b) + (k \leftrightarrow l) & = & \frac{1}{4} \, C_F \Ncol C_A \, g_s^4 \,
\frac{\beta_1
\cdot \beta_2}{k \cdot l}\left(
\frac{2}{\beta_1 \cdot k \, \beta_2 \cdot l} + \frac{1}{\beta_1 \cdot
l \, \beta_2
\cdot k}\right), \nonumber \\
c) & = & \frac{1}{4} \, C_F \Ncol C_A \, g_s^4 \, \frac{\beta_1 \cdot \beta_2}
{k \cdot l}\frac{1}{\beta_1 \cdot l} \frac{1}{\beta_2 \cdot k}, \nonumber
\\
d) & = & \frac{1}{4} \, C_F \Ncol C_A \, g_s^4 \, \frac{\beta_1 \cdot \beta_2}
{k \cdot l}\frac{1}{\beta_1 \cdot k} \frac{1}{\beta_2 \cdot l}, \nonumber
\\
e) & = & C_F \Ncol (C_F - C_A/2) \, g_s^4 \, \frac{(\beta_1 \cdot \beta_2)^2}
{\beta_1 \cdot l \, \beta_2 \cdot l} \frac{1}{\beta_1 \cdot k \,
\beta_2 \cdot k}, \nonumber \\
f) + (k \leftrightarrow l) & = & C_F \Ncol (C_F - C_A/2) \, g_s^4 \,
\frac{(\beta_1 \cdot \beta_2)^2}{\beta_1 \cdot l \, \beta_2 \cdot l}
\frac{2}{\beta_1 \cdot k \, \beta_2 \cdot k}.
\ea
The color factors in the last two equations of
(\ref{b-e})
are obtained from the identity $\mathrm{Tr}(t_a t_b t_a t_b) = C_F \Ncol
(C_F - C_A/2)$. Combining the terms proportional to
the color factor $C_F \Ncol C_A$, and including the complex conjugate
diagrams, we find for the squared amplitude
\ba \label{mm}
|M|^2 & = & 2 \, g_s^4 \, C_F \Ncol C_A \, \beta_1 \cdot \beta_2 
\nonumber \\ 
& \times & \left(\frac{1}{k \cdot \l \, \beta _1
\cdot k \, \beta_2 \cdot l} +
\frac{1}{k \cdot \l \, \beta _1 \cdot l \, \beta_2 \cdot k} -
\frac{\beta_1 \cdot \beta_2}{\beta_1
\cdot l \, \beta_2 \cdot l \, \beta_1 \cdot k \, \beta_2 \cdot k} \right).
\nonumber \\
\ea
Having determined the amplitude, we need to integrate $|M|^2$ over
the phase space corresponding
to the geometry given in Fig. \ref{kinematics}. Specifically, we have
to evaluate:
\be \label{ps}
I \equiv \frac{1}{\Ncol} \int {\mathrm d}{\bar \varepsilon} \, e^{-\nu \,
{\bar \varepsilon}}
\, \int_{\Omega}
\frac{{\mathrm d}^3 k}{(2\pi)^3 \, 2 \omega_k} \,
\int_{\bar{\Omega}} \frac{{\mathrm d}^3 l}{(2\pi)^3 \, 2 \omega_l} \,
\delta(\varepsilon - \omega_k
/ \sqrt{s}) \, \delta({\bar \varepsilon} - {\bar f}(l,a))
\, |M|^2,
\ee
where the weight function ${\bar f}(l,a)$ is given, as in Eqs.\
(\ref{2jetf})
and (\ref{fbarexp}), by
\be \label{fb}
{\bar f}(l,a) = \left\{ \begin{array}{l@{\quad:\quad}l}
\frac{\omega_l}{\sqrt{s}} \, (1-c_l)^{1-a} \, s_l^a & \theta_l \in
(0, \, \pi/2 - \delta) \\
\frac{\omega_l}{\sqrt{s}} \, (1+c_l)^{1-a} \, s_l^a & \theta_l \in
(\pi/2 + \delta, \, \pi),
\end{array} \right.
\ee
with $a < 1$.

\begin{figure} \center
\epsfig{file=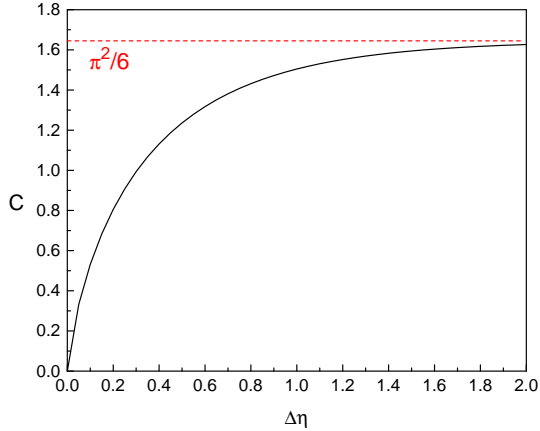,height=8cm,angle=270,clip=0}
\caption{\label{crossSec_vs_rapidity} $C(\Delta \eta)$, as defined in 
(\ref{Cdef}),
as a function of rapidity
width
$\de$ of the region $\Omega$. The dashed line is its limiting value,
$C (\de \rightarrow \infty) = \pi^2/6$.}
\end{figure}

Using the equalities: $\beta_1 \cdot \beta_2 = 2$, $\beta_1 \cdot l =
\ol(1-c_l)$,
$\beta_2 \cdot l = \ol(1 + c_l)$, $\beta_1 \cdot k = \ok(1 - c_k)$,
$\beta_2 \cdot k = \ok(1 + c_k)$ and $k \cdot l =
\ok \ol (1 - c_k c_l - s_k s_l \cfi)$ in Eq. (\ref{mm}), performing
the integration
   over $\phi$, and changing the integration variable
$c_l \rightarrow - c_l$ in the angular region $\theta_l \in (\pi/2 +
\delta, \, \pi)$, we easily arrive at the following
three-dimensional integral:
\bea \label{ps1}
I & = & C_F C_A \left(\frac{\alpha_s}{\pi}\right)^2 \,
\frac{1}{\varepsilon} \,
\int_{-\sd}^{\sd}
\mathrm{d} c_k \, \int_{\sd}^{1}
\mathrm{d} c_l \, \int_{\varepsilon \sqrt{s}}^{\sqrt{s}} \frac{\mathrm{d}
\ol}{\ol} \, e^{-\nu \, \ol \,
(1-c_l)^{1-a} \, s_l^a / \sqrt{s}}
\nonumber \\
& & \left[ \frac{1}{c_k + c_l} \, \frac{1}{1+c_k}
\left(\frac{1}{1+c_l} + \frac{1}{1-c_k}\right) - \frac{1}{s_k^2} \,
\frac{1}{1+c_l} \right].
\eea
We are interested in the $(1/\varepsilon)\ln(1/\varepsilon)$ behavior of $I$.
This is
obtained after performing the $\ol$ integral with
the replacement $e^{-\nu \ol (1-c_l)^{1-a} \, s_l^a  / \sqrt{s}} \rightarrow
\theta(1 - \nu \ol (1-c_l)^{1-a} \, s_l^a  /
\sqrt{s})$. Remainders do
not contain terms proportional to $\ln \varepsilon$.
In this approximation, the $c_l$ integration can be carried out, and
we obtain the integral representation for the
term containing $(1/\varepsilon)\ln(1/\varepsilon)$:
\ba \label{ps2}
I & = & 2 \, C_F C_A \left(\frac{\alpha_s}{\pi}\right)^2 \,
\frac{1}{\varepsilon} \,
\ln\left(\frac{1}{\varepsilon \nu}\right) \nonumber \\
& \times & \left[ \int_{0}^{\sd} \,
\frac{\mathrm{d} c_k}{s_k^2} \, \ln \left(\frac{s_k^2}{s_k^2 -
\cos^2\delta} \right) -
\ln \left(\frac{2}{1 + \sd} \right) \,
\ln \left(\frac{1+\sd}{1-\sd} \right) \right]. \nonumber \\
& &
\ea
The  potential non-global logarithm of $\varepsilon$ is replaced by 
$\ln(\varepsilon \nu)$.
The angular integral over $c_k$ can be expressed in terms of
dilogarithmic functions. The final expression for the term proportional to
$\ln(\varepsilon \nu)/\varepsilon$ takes the form:
\ba \label{ps3}
I & = & C_F C_A \left(\frac{\alpha_s}{\pi}\right)^2 \,
\frac{1}{\varepsilon} \,
\ln\left(\frac{1}{\varepsilon \nu}\right) \,
    \left[\frac{\pi^2}{6} +
\ln \left(\frac{\cot\delta \, (1 + \sd)}{4} \right) \, \ln
\left(\frac{1+\sd}{1-\sd} \right) \right. \nonumber \\
& + & \left. \mathrm{Li}_2\left(\frac{1-\sd}{2}\right) - 
\mathrm {Li}_2\left(\frac{1+\sd}{2}\right) -
\mathrm{Li}_2\left(-\frac{2 \sd}{1-\sd}\right) -
\mathrm{Li}_2\left(\frac{1-\sd}{1+\sd}\right) \right]. \nonumber \\
\ea
Equivalently, we can express our results in terms of the rapidity
width of the region $\Omega$, Eq. (\ref{deltaeta1}),
and we obtain
\ba \label{psRapidity}
I & = & C_F C_A \left(\frac{\alpha_s}{\pi}\right)^2 \,
\frac{1}{\varepsilon} \,
\ln\left(\frac{1}{\varepsilon \nu}\right) \,
\left[\frac{\pi^2}{6} + \de \left(\frac{\de}{2} - 
\ln \left(2 \sinh (\de)\right)\right) \right. \nonumber \\ 
& + & \left. \mathrm{Li}_2 \left ( \frac{e^{-\de/2}}{2 \, \cosh (\de/2)} 
\right) - \mathrm {Li}_2 \left( \frac{e^{\de/2}}{2 \, \cosh
(\de/2)} \right) \right. \nonumber \\ 
& - & \left. \mathrm{Li}_2 \left( -2 \sinh (\de/2) \, e^{\de/2} \right) 
- \mathrm{Li}_2 (e^{-\de}) \right].
\ea
The coefficient
\be
C (\Delta \eta) \equiv - \left( \frac{\pi}{ \alpha_s} \right)^2 \, 
\frac{\varepsilon \,
I}{C_F C_A \ln (\varepsilon
\nu)} \label{Cdef}
\ee
as a
function of $\de$ is shown in
Fig. \ref{crossSec_vs_rapidity}.
Naturally, $C$ is a monotonically increasing function of $\Delta\eta$.
For $\de \rightarrow 0$,
\be
C \sim {\mathcal O}(\de \, \ln \de)\, ,
\ee
and the cross section vanishes, as expected. On the other hand, as the size of
region $\Omega$ increases, $C$ rapidly saturates and reaches its limiting
value \cite{DS}
\be
\lim_{\de \rightarrow \infty} \, C = \frac{\pi^2}{6}\, .
\ee

\section{Recoil} \label{approxapp}

In this appendix, we return to the justification
of the technical step represented by Eq.\ (\ref{nnJone}).
According to this approximation, we may
compute the jet functions by identifying
axes that depend only upon particles in the
final states
$N_{J_c}$ associated with those functions, rather
than the full final state $N$.
Intuitively, this is a reasonable estimate, given
that the jet axis should be determined by
a set of energetic, nearly collinear particles.
When we make this replacement, however,  the contributions
to the event shape from energetic particles near the jet axis may
change.  This change is neglected in going from
the original factorization, Eq.\ (\ref{sigmafact}), to the
factorization in convolution form, Eq.\ (\ref{factor}),
which is the starting point for the resummation
techniques that we employ in this paper.
The weight functions $\bar f^N(N_i,a)$
in Eq.\ (\ref{sigmafact}) are defined
relative to the unit vector $\hat n_1$ corresponding to
$a=0$, the thrust-like event shape.
The factorization of Eq.\ (\ref{sigmafact})
applies to any $a<2$, but as indicated by the superscript,
individual contributions to $\bar f^N(N_i,a)$ on the
right-hand side continue to depend on the full
final state $N$, through the identification of the jet axis.

To derive the factorization
of Eq.\ (\ref{factor}) in a simple convolution
form, we must be able to
treat the thrust axis, $\hat n_1$, as a fixed vector
for each of the states $N_s$, $N_{J_c}$.  This is possible
if we can neglect the effects of recoil from soft,
wide-angle radiation on the direction of the axis.
Specifically, we must be able to make the replacement
\be
\bar f_{\bar{\O}_c}^N(N_{J_c},a) \rightarrow \bar f_c(N_{J_c},a)\, , 
\label{replace}
\ee
where $\bar f_c(N_{J_c},a)$ is the event shape variable
for jet $c$, in which the axis $\hat n_c$ is
specified by state $N_{J_c}$ {\it only}.  Of course, this
replacement changes the value of the weight, $\bar\varepsilon$,
$\bar f_{\bar{\O}_c}^N(N_{J_c},a) \ne \bar f_c(N_{J_c},a)$.
As we now show, the error induced by this
replacement is suppressed by a power
of $\bar \varepsilon$ so long as $a<1$.  In general,
the error is nonnegligible for $a\ge 1$.
The importance of recoil for jet broadening, at
$a=1$, was pointed out in \cite{broaden2}.  We now
discuss how the neglect of such radiation
affects the jet axis
(always determined from $a=0$)
and hence the value of the event shape for arbitrary $a<2$.

The jet axis is
found by minimizing $\bar f(a=0)$
in each state.
The largest influence on the axis ${\hat n}_c$ for jet $c$
is, of course, the set of fast, collinear particles
within the state $N_{J_c}$ associated with the jet function
in Eq.\ (\ref{sigmafact}).
Soft, wide-angle radiation, however,
does affect the precise direction
of the axis.  This is what we mean by `recoil'.

Let us denote by $\o_s$ the energy of the soft wide-angle radiation that is
neglected in the factorization
(\ref{factor}).  Neglecting this soft radiation in
the determination of the jet axis
will result in an axis $\hat n_1(N_{J_c})$, which differs from the
axis $\hat n_1 (N)$
determined from the complete final state $(N)$ by an angle $\Delta_s\phi$:
\be
     \angle\!\!\!) \left(\hat n_1(N), \hat n_1(N_{J_c}) \right) \equiv
\Delta_s \phi
\sim  {\o_s \over Q}\, .
\label{deltaphi}
\ee
At the same time, the soft, wide-angle radiation also contributes
to the total event shape
$\bar f(N,a) \sim (1/Q)k_\perp^a (k^-)^{1-a}$ at the
level of
\be
\bar\varepsilon_s \sim {\o_s\over Q} \, ,
\ee
because for such wide-angle radiation, we may take $k_s^-\sim 
k_{s,\perp}\sim\omega_s$.
In summary, the neglect of wide-angle soft radiation rotates the jet axis
by an angle that is of the order of the contribution
of the same soft radiation to the event shape.

In the factorization (\ref{factor}), the contribution of
each final-state particle is taken into account,
just as in Eq.\ (\ref{sigmafact}).  The question
we must answer is how the rotation of the jet axis affects
these contributions, and hence the value of the event
shape.

For a wide-angle particle, the rotation of the jet
axis by an angle of order $\Delta_s\phi$
in Eq.\ (\ref{deltaphi}) leads to a
negligible change in its contributions to the event shape, because
its angle to the axis is a number of order unity, and the
jet axis is rotated only
by an angle of order $\bar\varepsilon_s$.
Contributions from soft radiation are therefore stable
under the approximation (\ref{nnJone}).
The only source of large corrections is then associated
with energetic jet radiation,
because these particles are nearly collinear to the jet axis.

It is easy to see from the form
of the shape function in terms of angles, Eq.\ (\ref{fbarexp}),
that for any value of parameter $a$, a particle
of energy $\omega_i$ at a small angle $\theta_i$ to
the jet axis $\hat n_1 (N)$ contributes to the
event shape at the level
\be
\bar\varepsilon_i \sim  {\omega_i \over Q}\theta_i{}^{2-a}\, .
\ee
The rotation of the jet
axis by the angle $\Delta_s\phi$ due to neglect of soft radiation
may be as large as, or larger than,
$\theta_i$. Assuming the latter, we find a shift in the
$\bar\varepsilon_i$ of order
\be
\delta \bar\varepsilon_i \equiv \bar\varepsilon_i \left( \hat n_1(N) \right)
- \bar\varepsilon_i \left( \hat n_1(N_{J_c}) \right)
\sim {\omega_i\over Q}\,
\left(\Delta_s\phi\right)^{2-a} \sim
{\omega_i\over Q}\, \left({\o_s\over Q}\right)^{2-a} \sim
{\omega_i\over Q}\, \bar\varepsilon_s{}^{2-a}\, .
\ee
The change in $\bar\varepsilon_i$
   is thus suppressed by at least a factor $\bar\varepsilon_s{}^{1-a}$
compared to $\bar\varepsilon_s$, which is the
contribution of the wide-angle soft radiation
to the event shape.  The contributions of nearly-collinear,
energetic radiation to the event shape thus change
significantly under the replacement (\ref{nnJone}),
but so long as $a<1$, these contributions are
power-suppressed in the value
of the event shape, both before and after the approximation
that leads to a rotation of the axis.
For this reason, when $a<1$ (and only when $a<1$), the value
of the event shape is stable whether or not
we include soft radiation in the determination
of the jet axes, up to corrections that are suppressed
by a power of the event shape.  In this case, the
transition from Eq.\ (\ref{sigmafact}) to Eq.\ (\ref{factor})
is justified.

\end{document}